\documentclass[12pt,twoside,openright]{report} 
\usepackage[T1]{fontenc}
\usepackage[latin1]{inputenc}
\usepackage[francais,english,german]{babel}

\usepackage{epsf}
\usepackage{makeidx}
\usepackage{amssymb,amsfonts,amsmath,amscd,latexsym,epsfig}
\input{epsf}
  
\usepackage{fancyheadings}
\usepackage{these}
\makeindex
  
{\count255=\time\divide\count255 by 60
\xdef\hourmin{\number\count255}
\multiply\count255 by-60\advance\count255 by\time
\xdef\hourmin{\hourmin:\ifnum\count255<10 0\fi\the\count255}}
\def\ps@draft{\let\@mkboth\@gobbletwo
    \def\@oddhead{}
    \def\@oddfoot{\hbox to 7 cm{\tiny \versionno
       \hfil}\hskip -7cm\hfil\rm\thepage \hfil {\tiny\draftdate}}
    \def\@evenhead{}\let\@evenfoot\@oddfoot}
\def\draftdate{\number\month/\number\day/\number\year\ \ \ \hourmin }

\long\def\query#1{\hskip 0pt{\vadjust{\everypar={}\small\vtop to 0pt{\hbox{}%
     \vskip -13pt\rlap{\hbox to 49.0pc{\hfil{\vtop{\hsize=8pc\tolerance=6000%
     \hfuzz=.5pc\rightskip=0pt plus 3em\noindent#1}}}}\vss}}}}%
\newcommand\version[1] {\typeout{}\typeout{#1}\typeout{}
                   \vskip3mm \centerline{\fbox{{\tt DRAFT -- #1 -- }
                   {\small\draftdate}}} \vskip3mm}

\def\start{\documentclass[12pt]{article}
                   \usepackage{amssymb,amsfonts,amsmath,amscd,latexsym
                 }\setlength{\textwidth}{17cm} \setlength{\textheight}{25.2cm}
 \hoffset -20mm \topmargin= -13mm \begin{document}\version\versionno}

\def \etc {{\noindent \bf etc }}

\def \rem#1 {}

\def \de#1 {{\it #1 }}

\def \vs { \noindent | \linebreak | \linebreak | \linebreak | \linebreak | }

\def \Fac#1  {#1 }
\def \esp {\ \ \ \  }

\def \bs {\bigskip}

\def \Fig#1#2#3 {
\begin{figure}
\begin{center}
\scalebox{.6}{\includegraphics{#1.eps}}
\label{#1}
\end{center}
\caption{#3}
\end{figure}
}

\def \en#1 {\ensuremath{#1}\ {}}

\def \H  {\en{H_3} }
\def \SL {\en{SL(2,\R)} }
\def \SU {\en{SU(2)} }
\def \AAA {\en{AdS_3} }
\def \AA {\en{AdS_2} }
\def \SLC {\en{SL(2,\C)} }
\def \SLU {\en{SL(2,\R)/U(1)} }
\def \SUU {\en{SU(2)/U(1)} }
\def \sl {\en{sl(2,\R)} }
\def \su {\en{su(2)} }
\def \au {\en{\widehat{u}(1)} }
\def \asu {\en{\widehat{su}(2)} }
\def \asl {\en{\widehat{sl}(2,\R)} }
\def \aslc {\en{\widehat{sl}(2,\C)} }
\def \lieg {\en{\bf g} }
\def \alieg {\en{\bf \hat{g}} } 
\def \lieh {\en{\bf h} }
\def \alieh {\en{\bf \hat{h}} } 

\def \im  {\en{ {\rm Im}(1-\oAd) }}
\def \ke  {\en{ {\rm Ker}(1-\oAd) }}
\def \Id  {\en{ {\rm Id} }}

\def \MM {{{\cal M}-{\cal M'}}}

\def \vs {\\ }

\def \et {{\acute{e}tal\acute{e}}}

\def \LL { L_0+\bar{L}_0-\frac{c}{24} }

\def \Fun {{\rm Fon }}

\def \cyl {{\rm cyl}}
\def \ouv {{\rm ouv}}
\def \cig {{\rm cigare } }
\def \tr  {{\rm trompette }  }
\def \cl  {{\rm cloche } }
\def \sgn {{\rm signe }}
\def \cst {{\rm constante}}

\def \B  {Br}

\def \L {{\cal L}_{\rm BI}}

\def \r {r}                
\def \t {t}                   

\def \half {\frac{1}{2}}
\def \halfpi {\frac{\pi}{2}}
\def \halfi {\frac{i}{2}}
\def \halfib {\frac{i}{2b^2}}
\def \ip {i\frac{\pi}{2}}
\def \rr {\frac{r}{\pi b^2}}

\def \xm {X^\mu}

\def \j {-\half+iP}

\def \ap {\alpha'}

\def \mn {_{\mu\nu}}

\def \p {\partial}

\def \a {\alpha}
\def \b {\beta}
\def \e {\epsilon}
\def \o {\omega}
\def \om {\omega}
\def \g {\gamma}
\def \bg {\bar{\gamma}}
\def \vf {\varphi}

\def \F {F}

\def \rar {\rightarrow}

\def \db {\dot{\beta}}

\def \hg {\hat{g}}
\def \hA {\hat{A}}
\def \hB {\hat{B}}
\def \hR {\hat{R}}

\def \bz {\bar{z}}
\def \bp {\bar{\partial}}
\def \bJ {\bar{J}}
\def \bA {\bar{A}}

\def \Ad { {\rm Ad}_g }
\def \oAd { \omega \circ {\rm Ad}_g }

\def \Tr {{\rm Tr } }

\def \id {\rm Id}

\def \- {^{-1}}

\def \Z {\mathbb{Z}}
\def \N {\mathbb{N}}
\def \R {\mathbb{R}}
\def \C {\mathbb{C}}

\def \llangle {\left\langle}
\def \rrangle {\right\rangle}
\def \llang   {\langle\langle}
\def \rrang   {\rangle\rangle}

\def \tq {\tilde{q}}

\def \KK {\p_\alpha K \p^\alpha K }
\def \Pdd {P^{\alpha\beta}\p_\alpha\vf\p_\beta\vf}

\def \bea {\begin{eqnarray}}
\def \eea {\end{eqnarray}}
\def \ber {\begin{eqnarray*}}
\def \eer {\end{eqnarray*}}
\def \nn  {\nonumber}
\def \be  {\begin{equation}}
\def \ee  {\end{equation}}

\begin{document}
\pagenumbering{roman}

\addtolength{\oddsidemargin}{.5cm}

\enlargethispage{3cm}
\thispagestyle{empty}

\vspace*{1.2cm}
\centerline{\scshape 
\large Centre de Physique Th{\'e}orique -- {\'E}cole Polytechnique}
\vskip 1.1cm
\centerline{\Large \bfseries TH{\`E}SE DE DOCTORAT DE L'{\'E}COLE POLYTECHNIQUE}
\vskip .8cm
\centerline{\large Sp{\'e}cialit{\'e} : \bfseries\scshape PHYSIQUE TH{\'E}ORIQUE}
\vskip .5cm
\vskip 1.2cm
\centerline{pr{\'e}sent{\'e}e par}
\vskip .6cm
\centerline{\Large \bf Sylvain RIBAULT}
\vskip 1cm
\centerline{pour obtenir le grade de}
\vskip .6cm
\centerline{\Large \bf Docteur de l'{\'E}cole Polytechnique}
\vskip 1.5cm
\centerline{Sujet :}
\vskip 1cm
\centerline{\LARGE \bfseries \itshape Cordes et D-branes }
\vskip .3cm
\centerline{\LARGE \bfseries \itshape dans les espaces-temps courbes }
\vskip 3cm
\noindent
Soutenue le 20 juin 2003 devant le jury compos{\'e} de~:
\vskip 1cm
\begin{tabular}{cll}
MM. & Costas Bachas, & \\
    & Marios Petropoulos,& directeur de th{\`e}se, \\
    & Augusto Sagnotti, & rapporteur, \\
    & Volker Schomerus, & \\
    & Paul Windey, & rapporteur ,\\
\&  & Jean-Bernard Zuber. & pr{\'e}sident du jury.\\
\end{tabular}
\topmargin=2cm

\clearemptydoublepage

\addtolength{\oddsidemargin}{-.5cm}


\renewcommand{\baselinestretch}{1.3} \normalsize
\thispagestyle{empty}
\enlargethispage{5mm}
\centerline{\large \itshape Remerciements}
\vskip 1cm

Je suis reconnaissant {\`a} I. Antoniadis, M. Petropoulos et {\`a} tous
les membres du Centre de Physique Th{\'e}orique de l'{\'E}cole
Polytechnique 
de
m'avoir accueilli. Je remercie mon directeur de th{\`e}se,
M. Petropoulos, de m'avoir suivi et encourag{\'e} tout au long de la
th{\`e}se, et notamment de m'avoir donn{\'e} la possibilit{\'e} de modifier
l'orientation des recherches dans une direction moins
ph{\'e}nom{\'e}nologique qu'initialement pr{\'e}vu. 

Je voudrais remercier les personnes avec qui j'ai collabor{\'e}
au cours de mes recherches, que ces collaborations aient donn{\'e} lieu
{\`a} publication ou non~: C. Bachas, P. Bordalo, A. Fotopoulos,
M. Petropoulos, V. Schomerus, C. Schweigert.
En particulier je suis
reconnaissant {\`a} C. Bachas pour ses encouragements. 

Je voudrais aussi mentionner tous ceux qui, outre les rapporteurs
A. Sagnotti et P. Windey,
ont bien voulu me donner
leur avis sur tout ou partie de ce m{\'e}moire~: P. Bordalo, B. Dum{\'e}zil,
A. Fotopoulos, S. Fredenhagen, P. Grange, E. Kohlprath,
H. Partouche, M. Petropoulos
et P. Vanhove. 

Enfin, je suis reconnaissant aux membres du jury
de leur int{\'e}r{\^e}t pour cette th{\`e}se.

\clearpage

\thispagestyle{empty}
\enlargethispage{5mm}
\centerline{\large \itshape Introduction }
\vskip 1cm

Le but de cette th{\`e}se est d'{\'e}tudier les
D-branes dans les espaces courbes. Cette {\'e}tude doit contribuer {\`a} une
meilleure 
compr{\'e}hension de la th{\'e}orie des cordes, dont les principes
physiques fondamentaux restent {\`a} d{\'e}couvrir.

\paragraph{Statut de la th{\'e}orie des cordes. }

La th{\'e}orie des cordes est une tentative d'unification des interactions
fondamentales
de la physique, dont l'id{\'e}e de base est de quantifier des objets
{\'e}tendus plut{\^o}t que des particules ponctuelles. Si ces objets {\'e}tendus
ont une taille typique bien inf{\'e}rieure aux {\'e}chelles observables,
la th{\'e}orie des cordes n'en conduit pas moins
{\`a} la pr{\'e}diction (ou plut{\^o}t la {\it r{\'e}trodiction}) de
l'existence d'une force gravitationnelle ob{\'e}issant {\`a} l'{\'e}chelle
macroscopique {\`a} la th{\'e}orie de la relativit{\'e} g{\'e}n{\'e}rale, et {\`a}
des th{\'e}ories de jauge dont la structure est similaire {\`a} celles qui
r{\'e}gissent les interactions {\'e}lectrofaible et forte. 

Cependant, la
th{\'e}orie des cordes ne pr{\'e}dit pas de mani{\`e}re univoque 
le Mod{\`e}le Standard. En effet,
sa formulation d{\'e}pend du choix d'un {\'e}tat du vide,
d{\'e}termin{\'e} en partie par la donn{\'e}e d'un
espace-temps {\`a} dix
dimensions, et de nombreux choix sont coh{\'e}rents dans le cadre de
la th{\'e}orie. M{\^e}me si l'on se restreint aux choix compatibles avec
la g{\'e}om{\'e}trie observ{\'e}e de l'espace-temps, avec 3+1 grandes
dimensions presques plates, il en reste un nombre immense, qui
correspondent {\`a} une grande diversit{\'e} de th{\'e}ories de jauges. 
L'{\'e}tude de tous ces vides en vue d'en s{\'e}lectionner un qui
correspondrait {\`a} l'univers observ{\'e}, serait une t{\^a}che
extr{\^e}mement fastidieuse pour autant qu'elle soit r{\'e}alisable. 

Le manque de pr{\'e}dictivit{\'e} de la th{\'e}orie est bien illustr{\'e} par
les recherches portant sur la r{\'e}alisation de l'espace de de Sitter en
th{\'e}orie des cordes. Alors qu'il semblait exister des obstacles
fondamentaux {\`a} cette r{\'e}alisation, une construction {\`a} base de
D-branes d'un vide de de Sitter a {\'e}t{\'e} propos{\'e}e en janvier
dernier par Kachru, Kallosh, Linde et Trivedi. La
th{\'e}orie des cordes semble donc pouvoir s'accommoder de la
positivit{\'e} de la constante cosmologique, comme de n'importe quelle
autre observation physique. 

La n{\'e}cessit{\'e} de formuler la th{\'e}orie dans un espace-temps donn{\'e}
 est {\'e}galement
insatisfaisante au plan th{\'e}orique, pour une th{\'e}orie qui 
cherche {\`a} quantifier la th{\'e}orie de la relativit{\'e} g{\'e}n{\'e}rale. En effet,
si l'invariance par reparam{\'e}trisation correspondant au principe
d'{\'e}quivalence {\'e}tait explicitement respect{\'e}e, les degr{\'e}s de libert{\'e}
 fondamentaux d'une telle th{\'e}orie de la gravitation quantique ne
 devraient pas {\^e}tre d{\'e}finis dans une g{\'e}om{\'e}trie donn{\'e}e, mais
 engendrer eux-m{\^e}mes la g{\'e}om{\'e}trie d'espace-temps
comme un concept {\it effectif},
 valable seulement {\`a} une {\'e}chelle (de longueur) suffisamment grande. 

Cependant, malgr{\'e} la multiplicit{\'e} des probl{\`e}mes ouverts, la
th{\'e}orie des cordes a r{\'e}ussi {\`a} calculer avec pr{\'e}cision
l'entropie de certains trous noirs extr{\'e}maux en les r{\'e}alisant au moyen de
D-branes. Ces D-branes sont des objets non perturbatifs qui ont jou{\'e}
un r{\^o}le essentiel dans les d{\'e}veloppements r{\'e}cents de la
th{\'e}orie, en particulier dans la formulation de conjectures de
dualit{\'e} qui nous ont donn{\'e} acc{\`e}s {\`a} certains r{\'e}gimes non
perturbatifs de la th{\'e}orie des cordes. De plus, les S-branes ou branes
de genre espace, qui g{\'e}n{\'e}ralisent les D-branes, sont {\`a} la base
de progr{\`e}s r{\'e}cents dans le domaine de la th{\'e}orie des cordes
d{\'e}pendant du temps.
Les D-branes pourraient donc
contribuer {\`a} la formulation ultime de la th{\'e}orie, au m{\^e}me titre
que les cordes qui ont servi {\`a} la d{\'e}finir en premier lieu et qui
lui ont donn{\'e} son nom.
Cependant, leur d{\'e}finition d{\'e}pend encore de la notion de
g{\'e}om{\'e}trie, comme l'indique l'{\'e}tymologie~: \flqq brane\frqq\
provient de \flqq membrane\frqq\footnote{ Le D, lui, est l'initiale de
  Dirichlet~; on d{\'e}finit en effet les D-branes {\`a} partir de
  conditions au bord de Dirichlet pour des cordes ouvertes. }, ces
membranes sont des sous-vari{\'e}t{\'e}s de l'espace-temps o{\`u} les cordes
ouvertes peuvent aboutir. Toutefois, en tant qu'objets fondamentaux de la
gravitation quantique, les D-branes devraient {\^e}tre d{\'e}finies
ind{\'e}pendamment du choix d'un vide. Un premier pas dans cette
direction consiste {\`a} 
{\'e}tudier leur comportement dans
divers espaces-temps courbes.

\paragraph{Espaces courbes et D-branes.}
L'{\'e}tude des cordes et des D-branes n'est techniquement accessible
que par un d{\'e}veloppement 
autour d'{\'e}tats du vide poss{\'e}dant certaines sym{\'e}tries. 
La coh{\'e}rence m{\^e}me de la th{\'e}orie exige qu'elle soit
supersym{\'e}trique, et ses vides sont d'autant plus accessibles au calcul
qu'ils
respectent plus de supersym{\'e}tries. 
Nous nous int{\'e}resserons dans cette th{\`e}se principalement
{\`a} une autre classe de sym{\'e}tries, d{\'e}finies par un groupe
continu $G$ d'isom{\'e}tries de l'espace-temps et du vide en g{\'e}n{\'e}ral 
(il ne s'agit pas ici de sym{\'e}tries de jauge). 

La th{\'e}orie des cordes dans un groupe compact $G$ peut {\^e}tre
r{\'e}solue au moyen d'une th{\'e}orie conforme rationnelle. Cette
th{\'e}orie d{\'e}crit la dynamique de la surface d'univers
(bidimensionnelle) d'une corde~; la rationalit{\'e}
correspond au fait que le nombre d'{\'e}tats physiques est fini, {\`a}
cause de la compacit{\'e} du groupe $G$. Certaines D-branes dans ce groupe,
dites sym{\'e}triques, peuvent  {\^e}tre construites au moyen de
r{\'e}sultats dus {\`a} Cardy et valables pour les th{\'e}ories conformes
rationnelles avec bord. Ces constructions s'{\'e}tendent {\`a} des
D-branes pr{\'e}sentant moins de sym{\'e}tries, et {\`a} des mod{\`e}les quotients
$G/H$.

Dans le cas des groupes compacts, la g{\'e}om{\'e}trie des D-branes et leur dynamique
sont bien comprises. Elles peuvent {\^e}tre
{\'e}tudi{\'e}es au moyen de th{\'e}ories effectives telles que la th{\'e}orie de
Born-Infeld, et on peut d{\'e}terminer en d{\'e}tail la fa{\c c}on dont un
objet intrins{\`e}quement \flqq dispers{\'e}\frqq\ ou \flqq flou\frqq\
acquiert une localisation g{\'e}om{\'e}trique pr{\'e}cise dans la limite
semi-classique. 
On comprend la dynamique des D-branes avec suffisamment de d{\'e}tail pour pouvoir
esp{\'e}rer r{\'e}aliser les diff{\'e}rentes charges pr{\'e}dites par la
K-th{\'e}orie, dont des arguments g{\'e}n{\'e}raux assurent qu'elle
classifie les charges des D-branes.

La situation est beaucoup plus compliqu{\'e}e dans le cas des groupes
non compacts, {\`a} cause de l'existence d'un continuum d'{\'e}tats
asymptotiques. La th{\'e}orie des cordes elle-m{\^e}me n'est bien comprise
que dans le cas le plus simple de \SL. Ce cas est par ailleurs
extr{\^e}mement int{\'e}ressant, car le recouvrement
universel de \SL\ est \AAA, qui appara{\^\i}t dans des vides
int{\'e}ressants de la th{\'e}orie des cordes. \AAA\ fournit ainsi un
exemple de la conjecture $AdS/CFT$, qui
r{\'e}alise dans le cadre de la th{\'e}orie des cordes
le principe g{\'e}n{\'e}ral de la dualit{\'e} holographique entre la
physique dans un espace Anti-de Sitter et une th{\'e}orie conforme sur son
bord.

Pour {\'e}tudier les D-branes dans \AAA, on peut d'abord g{\'e}n{\'e}raliser les
id{\'e}es g{\'e}om{\'e}triques issues du cas compact. Par exemple, on peut
s'int{\'e}resser {\`a} des D-branes pr{\'e}servant le m{\^e}me genre de
sym{\'e}tries, ce qui permet de d{\'e}terminer leur g{\'e}om{\'e}trie dans la
limite semi-classique, pour autant que ces D-branes existent.
Pour aller plus loin,
il faut recourir {\`a} des th{\'e}ories conformes non rationnelles. 
Dans le cadre de ces th{\'e}ories, on peut formuler des contraintes 
qui d{\'e}terminent (de mani{\`e}re unique) les propri{\'e}t{\'e}s des D-branes
\AA\ dans \AAA, {\`a} d{\'e}faut de prouver leur existence.
Cependant, le caract{\`e}re minkowskien de \SL\ provoque
des complications suppl{\'e}mentaires, li{\'e}es {\`a} l'existence d'{\'e}tats
d'{\'e}nergie n{\'e}gative.

On peut {\'e}viter ce probl{\`e}me en
{\'e}tudiant les D-branes dans le quotient euclidien \SLU. La proc{\'e}dure
de d{\'e}finition d'un tel quotient est en effet la m{\^e}me que dans le
cas compact, et on s'attend {\`a} l'existence d'une correspondance entre
les D-branes dans \SLU\ et les D-branes dans \SL. 
Ce quotient \SLU\ pr{\'e}sente par ailleurs d'autres attraits~: 
il appara{\^\i}t tout comme \AAA\ dans des vides int{\'e}ressants de la
th{\'e}orie des cordes (des espaces cr{\'e}{\'e}s par des NS5-branes), qui
r{\'e}alisent leur propre version de la correspondance holographique (qui
les relie non pas {\`a} des th{\'e}ories conformes, mais {\`a} la \flqq Petite
Th{\'e}orie des Cordes\frqq).

Dans cette th{\`e}se, j'exposerai mes contributions aux recherches
r{\'e}centes sur ces sujets, dont certaines ont fait l'objet de
publications reproduites en appendice. Ces contributions mettent en
oeuvre des
techniques g{\'e}om{\'e}triques ainsi que des constructions de th{\'e}orie
conforme avec bord; elles comparent les r{\'e}sultats de ces deux approches 
pour fournir une
image coh{\'e}rente de certaines D-branes dans les groupes compacts, le groupe
\SL, le quotient \SLU, et d'autres espaces courbes.

\paragraph{Plan de cette th{\`e}se. }

Nous commencerons par rappeler \flqq quelques id{\'e}es de th{\'e}orie des
cordes\frqq (chapitre 1), et introduirons quelques propri{\'e}t{\'e}s
essentielles des D-branes en partant d'une analogie avec les trous
noirs charg{\'e}s de la th{\'e}orie d'Einstein-Maxwell. Nous expliquerons aussi la
d{\'e}finition des D-branes {\`a} partir de conditions au bord de la
th{\'e}orie des cordes ouvertes, apr{\`e}s avoir d{\'e}duit ces conditions
au bord d'un principe variationnel. Cependant, un expos{\'e} plus
d{\'e}taill{\'e} des techniques de la th{\'e}orie conforme avec bord sera
report{\'e} au chapitre 3. En passant, nous mentionnerons quelques
r{\'e}sultats marquants de la th{\'e}orie des cordes ferm{\'e}es que nous
utiliserons par la suite, notamment les dualit{\'e}s S et T. 
Enfin, nous pr{\'e}senterons l'action de Born-Infeld, qui d{\'e}crit la
dynamique des D-branes dans le r{\'e}gime g{\'e}om{\'e}trique.

Cette action de Born-Infeld nous servira {\`a} {\'e}tudier les D-branes
dans les groupes compacts et dans \SL(chapitre 2). Nous pr{\'e}senterons
{\`a} cette occasion les r{\'e}sultats publi{\'e}s dans \cite{brs, pr}. Apr{\`e}s
avoir rappel{\'e} quelques propri{\'e}t{\'e}s des cordes ferm{\'e}es
dans les groupes, sans omettre les complications propres {\`a} \SL, nous
analyserons en d{\'e}tail la g{\'e}om{\'e}trie, la stabilit{\'e}, et les
positions (quantifi{\'e}es) des D-branes sym{\'e}triques dans les groupes
compacts.
Nous mentionnerons les difficult{\'e}s que pose la g{\'e}n{\'e}ralisation de
ces r{\'e}sultats aux groupes non-compacts, et traiterons explicitement
l'exemple de \SL. Nous rencontrerons alors une D-brane
dont la g{\'e}om{\'e}trie est \AA, dont nous expliquerons pourquoi elle
est particuli{\`e}rement int{\'e}ressante. Nos m{\'e}thodes semi-classiques
ne produisant que des r{\'e}sultats partiels, il deviendra n{\'e}cessaire
de se tourner vers la th{\'e}orie conforme au bord.

C'est donc du point de vue de la th{\'e}orie conforme que nous
reconsid{\`e}rerons les D-branes dans les groupes (chapitre 3). Nous
exposerons des r{\'e}sultats d{\'e}j{\`a} pr{\'e}sents dans la litt{\'e}rature, 
ainsi que des
contributions personnelles
\cite{deuxbranes} {\`a} l'{\'e}tude des D-branes \AA, et une
synth{\`e}se finale inspir{\'e}e de \cite{corfu}. Apr{\`e}s avoir expos{\'e}
des r{\'e}sultats bien connus sur les D-branes sym{\'e}triques dans
les groupes compacts, dont on verra explicitement qu'ils {\'e}taient
correctement reproduits par l'analyse semi-classique du chapitre
pr{\'e}c{\'e}dent, nous entrerons dans le domaine de la th{\'e}orie conforme
non rationnelle. Nous r{\'e}sumerons les quelques r{\'e}sultats solides
d{\'e}j{\`a} connus au sujet des D-branes \AA\ dans la version euclidienne
d'\AAA\ (en omettant 
de nombreuses contributions moins convaincantes), et exploiterons ces
m{\'e}thodes pour d{\'e}terminer le spectre des cordes ouvertes
s'{\'e}tirant entre deux D-branes \AA\ distinctes. Ce calcul sera
confirm{\'e} par l'{\'e}valuation de l'amplitude du cylindre
correspondante. Il nous fournira l'occasion d'{\'e}valuer la port{\'e}e
des hypoth{\`e}ses d'analyticit{\'e} qui sont {\`a} la base des r{\'e}sultats
connus
sur les
th{\'e}ories conformes non rationnelles. 

Le r{\'e}sultat du calcul du spectre des cordes ouvertes {\'e}tir{\'e}es
entre certaines D-branes \AA\
sugg{\`e}re une interpr{\'e}tation naturelle en termes de \flqq D-branes dans
\SLU\frqq\ (chapitre 4). Les r{\'e}sultats que nous pr{\'e}senterons dans
ce chapitre n'ont
pas encore {\'e}t{\'e} publi{\'e}s, {\`a} l'exception bien s{\^u}r des faits
standard sur les mod{\`e}les de Wess-Zumino-Witten jaug{\'e}s et leurs
D-branes, que nous rappellerons. Pour {\'e}tudier les D1-branes et les
D2-branes dans \SLU, dont l'existence est r{\'e}v{\'e}l{\'e}e par l'analyse
semi-classique, nous supposerons qu'elles \flqq descendent\frqq\ des
D-branes \AA\ dans l'\AAA\ Euclidien trait{\'e}es dans le chapitre
pr{\'e}c{\'e}dent. Nous pr{\'e}ciserons bien s{\^u}r en quoi consiste cette
\flqq descente\frqq, avant d'en d{\'e}duire les {\'e}tats de bord et le
spectre des D-branes en question. Nous verrons notamment comment les
complications potentielles dues aux {\'e}tats discrets sont
{\'e}vit{\'e}es. 

Une application naturelle de ces constructions concerne les D3-branes
dans certains espaces de NS5-branes (chapitre 5). 
Nous pr{\'e}senterons des r{\'e}sultats sur ces D3-branes en partie publi{\'e}s dans
\cite{d3ns5}.
En chemin, nous ferons une digression
technique pour reformuler g{\'e}om{\'e}triquement 
la dynamique semi-classique des D-branes. Cette reformulation {\'e}tait
{\`a} l'origine motiv{\'e}e par le probl{\`e}me de l'{\'e}valuation des corrections
d{\'e}rivatives {\`a} l'action de Born-Infeld, et nous servira {\`a} {\'e}tudier les
D3-branes dans le contexte g{\'e}n{\'e}ral d'espaces cr{\'e}{\'e}s par des
NS5-branes. La dynamique de Born-Infeld {\'e}quivaut dans ce cas {\`a} une
condition de pr{\'e}servation de la supersym{\'e}trie, ce qui nous
conduira {\`a}
une relation g{\'e}n{\'e}rale entre ces
D3-branes et certaines D1-branes, que nous attribuerons {\`a} l'effet
Myers. Enfin, nous indiquerons lesquelles de ces D3-branes sont
reli{\'e}es {\`a} nos D-branes dans \SLU. 

Le chapitre 6 sera consacr{\'e} {\`a} la conclusion et 
pr{\'e}sentera des perspectives ouvertes par les r{\'e}sultats de cette
th{\`e}se. 

\bs
\bs

{\it \noindent La pr{\'e}sente introduction est suivie d'un Lexique, qui
  rappelle la traduction anglaise de certains termes techniques peu
  courants dans la litt{\'e}rature francophone. }

\clearpage

\thispagestyle{empty}
\enlargethispage{5mm}
\centerline{\large \itshape Lexique }
\vskip 1cm

\begin{center}
\begin{tabular}{ll}
{\it une th{\'e}orie coh{\'e}rente } & a consistent theory
\\
{\it une condition de collage } & a gluing condition
\\
{\it le bord } & the boundary
\\
{\it la masse } & the bulk
\\
{\it la sym{\'e}trie de croisement } & crossing symmetry
\\
{\it une coupure } & a branch cut
\\
{\it un {\'e}tat tordu } & a twisted state 
\\
{\it la Petite Th{\'e}orie des Cordes } & Little String Theory
\\
{\it un espace de NS5-branes } & an NS5-brane background
\end{tabular}
\end{center}

\clearpage




\clearemptydoublepage


\renewcommand{\baselinestretch}{1.3} \normalsize
\selectlanguage{francais}
\tableofcontents 
\cleardoublepage
\renewcommand{\thepage}{\arabic{page}}

\chapter{ Quelques id{\'e}es de th{\'e}orie des cordes }

Nous allons pr{\'e}senter dans ce chapitre les notions essentielles 
de th{\'e}orie des
cordes dont nous aurons besoin par la suite. 
Nous ne donnerons pas de r{\'e}f{\'e}rences, l'essentiel de ces notions
{\'e}tant bien connu et trait{\'e} dans le livre de Polchinski \cite{livre}.
Un autre texte d'introduction, plus sp\'ecifiquement orient\'e vers
les cordes ouvertes qui nous int\'eresseront particuli\`erement, est
l'article d'Angelantonj et Sagnotti \cite{openstrings}.

Notre pr{\'e}sentation sera
beaucoup plus fonctionnelle qu'historique~: ainsi, alors que la
th{\'e}orie des cordes est n{\'e}e d'une tentative de d{\'e}crire les
interactions fortes, nous insisterons plus sur les aspects
gravitationnels dont la decouverte est pourtant ult{\'e}rieure.

L'attraction gravitationnelle se manifeste g{\'e}om{\'e}triquement par la
courbure de l'espace-temps. Les espace-temps courbes peuvent {\^e}tre
remarquables 
pour leurs propri{\'e}t{\'e}s alg{\'e}briques, comme
les sym{\'e}tries de groupes, ou pour des
propri{\'e}t{\'e}s analytiques, comme des singularit{\'e}s et
des horizons. Ces singularit{\'e}s et ces horizons sont d{\'e}j{\`a}
pr{\'e}dits par la th{\'e}orie de la relativit{\'e} g{\'e}n{\'e}rale, que la
th{\'e}orie des cordes englobe, tout comme elle entend englober le
Mod{\`e}le Standard, qui d{\'e}crit les autres interactions
fondamentales. Nous allons donc bri{\`e}vement introduire ces ph{\'e}nom{\`e}nes
gravitationnels dans le cadre de la relativit{\'e} g{\'e}n{\'e}rale, avant
de passer {\`a} la th{\'e}orie des cordes.

\section{ Th{\'e}orie des cordes et supergravit{\'e} }

\subsection{ Des g{\'e}n{\'e}ralisations supersym{\'e}triques de la
  Relativit{\'e} G{\'e}n{\'e}rale }

La th{\'e}orie de la Relativit{\'e} G{\'e}n{\'e}rale d{\'e}crit la force de
gravitation au moyen de la \de{m{\'e}trique de l'espace-temps}, un tenseur
sym{\'e}trique {\`a} deux indices $g\mn$. Les fluctuations de la
m{\'e}trique d{\'e}crivent les ondes gravitationnelles, qui transmettent
l'attraction gravitationnelle, et la structure de tenseur permet
d'{\'e}crire des {\'e}quations invariantes par reparam{\'e}trisation, ce qui
r{\'e}alise le principe d'{\'e}quivalence. Ainsi, la dynamique de la
m{\'e}trique est dict{\'e}e par la seule action invariante par
reparam{\'e}trisation que l'on puisse construire avec au maximum deux
d{\'e}riv{\'e}es de la m{\'e}trique, 
\footnote{ Dans cette {\'e}quation, l'espace-temps {\`a} quatre dimensions
  est muni de coordonn{\'e}es $x^\mu$, la m{\'e}trique a pour signature
  $(-1,1,1,1)$ donc son d{\'e}terminant est n{\'e}gatif, et $G$ est la
  constante de la gravitation. }
\bea
S^{\rm RG}(g\mn)= \frac{1}{16\pi G} \int d^4x^\mu 
\sqrt{-\det g\mn } \left( R(g\mn)-2\Lambda \right). 
\label{actRG}
\eea
Nous n'avons pas omis la \de{ constante cosmologique} $\Lambda$, que nous
nous empresserons cependant d'oublier car nous ne traiterons pas de
cosmologie. L'expression $R(g\mn)$ d{\'e}signe la \de{ courbure
  scalaire} de la m{\'e}trique $g\mn$, c'est-{\`a}-dire la courbure
associ{\'e}e {\`a} la \de{ connection m{\'e}trique}
v\bea
\Gamma ^\rho\mn (g) =\half g^{\rho\sigma} 
\left( \p_\mu g_{\nu\sigma} + \p_\nu g_{\mu\sigma} - \p_\sigma
  g_{\mu\nu}
\right).
\label{conmet}
\eea
On peut tenir compte de la pr{\'e}sence de mati{\`e}re dans l'espace-temps
en ajoutant des termes {\`a} l'action $S^{\rm RG}$. Par exemple,
le champ {\'e}lectromagn{\'e}tique $F\mn = \nabla_\mu A_\nu - \nabla_\nu
A_\mu $ est d{\'e}crit par
l'action d'Einstein-Maxwell,
\bea
S^{\rm EM} = \frac{1}{16\pi G} \int d^4x^\mu 
\sqrt{-\det g\mn } R(g\mn) -\frac{1}{4} \int d^4x^\mu \sqrt{-\det g\mn
  } F\mn F^{\mu\nu}.
\label{actEM}
\eea
Notons que les coefficients des deux termes de cette action sont
ind{\'e}pendants. 

Une bonne fa{\c c}on d'{\'e}tudier les propri{\'e}t{\'e}s physiques d'un
espace-temps d{\'e}fini par sa m{\'e}trique $g\mn$ est de consid{\'e}rer les
trajectoires de particules-test, c'est-{\`a}-dire de points mat{\'e}riels
dont on n{\'e}glige qu'ils courbent l'espace-temps sur leur passage. Une
telle particule de masse $m$ et de charge $q$, se mouvant sur une
ligne de l'espace-temps param{\'e}tr{\'e}e par $x^\mu(\tau)$, est d{\'e}crite par
l'action 
\footnote{ Les tenseurs et formes induits sur la ligne d'univers de la
  particule-test sont d{\'e}sign{\'e}s par $\hg$ et $\hA$.}
\bea
S^{\rm test}(x^\mu(\tau))= m \int d\tau \sqrt{-\det \hg } + q \int
d\tau \hA .
\label{acttest}
\eea
Le fait qu'il s'agisse d'une particule-test signifie que l'on utilise
seulement $S^{\rm EM}$ pour d{\'e}terminer la configuration des champs
$g\mn$ et $A_\mu$, et non $S^{\rm EM} + S^{\rm test}$. Autrement dit,
$m$ et $q$ sont suffisamment petits. 

\bs

La th{\'e}orie de la relativit{\'e} g{\'e}n{\'e}rale est cependant notoirement
difficile {\`a} quantifier. Techniquement, elle est
non-renormalisable. Comme sa structure est dict{\'e}e de mani{\`e}re
essentiellement unique par des principes de sym{\'e}trie, on ne peut la
modifier de fa{\c c}on fondamentale qu'en changeant ces m{\^e}mes
principes. Sous des hypoth{\`e}ses physiques raisonnables, 
la seule extension possible de 
la sym{\'e}trie locale de Lorentz est la supersym{\'e}trie. Les extensions
supersym{\'e}triques connues de la th{\'e}orie de la relativit{\'e}
g{\'e}n{\'e}rale comportent d'ailleurs des champs de jauge et sont donc
des extensions de la th{\'e}orie d'Einstein-Maxwell. Par exemple,
l'action du secteur bosonique 
des th{\'e}ories de supergravit{\'e} de type II {\`a} dix
dimensions est
\footnote{ Nous omettons ici les termes de Chern-Simons. Nous ignorons 
aussi des subtilit{\'e}s li{\'e}es {\`a} l'auto-dualit{\'e} de
  la 5-forme $dC^{(3+1)}$ dans le cas de la th{\'e}orie de type IIB. }
\bea
S^{\rm IIA / IIB} &=& \frac{1}{G_{(10)}}\int d^{10}x^\mu \sqrt{- \det g}\
e ^{-2\Phi} \left(R(g)+4 (\p\Phi)^2 -\frac{1}{12} (dB)^2 \right)
\nn  \\
&-& \frac{1}{G_{(10)}}\int d^{10}x^\mu \sqrt{- \det g}\sum_{p\ {\rm
    pair/impair }} \frac{1}{2(p+2)!}(dC^{(p+1)})^2.
\label{acttypeII}
\eea
Dans ces th{\'e}ories de supergravit{\'e}, la supersym{\'e}trie se traduit
d'abord par la pr{\'e}sence, outre la m{\'e}trique, d'autres champs
bosoniques~: le dilaton $\Phi$, la deux-forme de NS-NS $B\mn$, et les
champs de Ramond-Ramond $C^{(p+1)}$ (o{\`u} $p+1$ est le degr{\'e}). Nous
avons omis tout le secteur fermionique~: $S^{\rm IIA / IIB}$ est
l'action des configurations dont les champs fermioniques sont nuls.
Une autre cons{\'e}quence notable de la supersym{\'e}trie est
l'{\'e}galit{\'e} des coefficients des deux termes de l'action. Alors que
les couplages gravitationnels et de jauge {\'e}taient ind{\'e}pendants
dans l'action d'Einstein-Maxwell, les secteurs analogues de la
supergravit{\'e} ont le m{\^e}me couplage, que nous avons not{\'e}
$G_{(10)}$. Plus pr{\'e}cis{\'e}ment, nous faisons ici une analogie entre
les champs de Ramond-Ramond $C^{(p+1)}$ et le potentiel
{\'e}lectromagn{\'e}tique $A_\mu$, et nous consid{\'e}rons le dilaton et la
forme $B\mn$ comme faisant partie du secteur gravitationnel, au
m{\^e}me titre que la m{\'e}trique. Cette
interpr{\'e}tation sera justifi{\'e}e par la structure du spectre des
cordes ferm{\'e}es, les transformations de T-dualit{\'e} et autres effets
de cordes. En outre, une
interpr{\'e}tation g{\'e}om{\'e}trique du champ $B\mn$ est sugg{\'e}r{\'e}e par
l'{\'e}galit{\'e} 
\bea
R(g)-\frac{1}{12}(dB)^2=R(\Gamma)\ \ , \ \ \Gamma ^\rho\mn = \Gamma
(g)^\rho\mn -\half (dB)^\rho\mn.
\label{contor}
\eea
Autrement dit, la contribution du champ $B\mn$ dans l'action de la
supergravit{\'e} peut {\^e}tre prise en compte au moyen d'un terme de
torsion ajout{\'e} {\`a} la connexion m{\'e}trique. 

\bs

L'alg{\`e}bre de supersym{\'e}trie des th{\'e}ories de supergravit{\'e} de
type II est engendr{\'e}e par 32 g{\'e}n{\'e}rateurs (au m{\^e}me titre que
les translations sont engendr{\'e}es par 10 g{\'e}n{\'e}rateurs). Une
solution donn{\'e}e des {\'e}quations du mouvement de ces th{\'e}ories
pr{\'e}serve ou non une partie de cette alg{\`e}bre. La pr{\'e}servation de
certaines supersym{\'e}tries se traduit par des conditions sur les
champs bosoniques $G\mn,B\mn...$ (en effet, ces champs interviennent
dans les transformations de supersym{\'e}trie des champs fermioniques,
que nous avons suppos{\'e}s nuls). Par exemple, on peut rechercher des
solutions de la supergravit{\'e} de type II telles que tous les champs
soient nuls sauf la m{\'e}trique, qui se factorise en une m{\'e}trique
plate {\`a} 3+1 dimensions et une m{\'e}trique compacte Euclidienne {\`a} 6
dimensions. Si, en plus de l'invariance de Lorentz de la m{\'e}trique
plate, nous voulons que la configuration pr{\'e}serve une alg{\`e}bre de
supersym{\'e}trie ${\cal N}=2$ {\`a} 3+1 dimensions (et soit sans
singularit{\'e}s), 
la m{\'e}trique
compacte en question doit {\^e}tre celle d'un espace de
Calabi-Yau. Alors 8 des 32 supersym{\'e}tries des th{\'e}ories de type II
sont pr{\'e}serv{\'e}es. Ainsi, les sym{\'e}tries physiques de la th{\'e}orie
se traduisent par des structures g{\'e}om{\'e}triques. Mais les configurations
faisant appel {\`a} d'autres champs que la seule m{\'e}trique sont souvent
plus
difficiles {\`a} interpr{\'e}ter g{\'e}om{\'e}triquement.

\bs

Cependant, les th{\'e}ories de supergravit{\'e} se trouvent {\^e}tre
non-renormalisables elles aussi. 
Mais leur non-renormalisabilit{\'e} n'a
de signification que si on les consid{\`e}re comme des th{\'e}ories
fondamentales suppos{\'e}es d{\'e}crire la physique au niveau
microscopique. Nous allons maintenant traiter de th{\'e}ories des
cordes, dont les variables microscopiques sont des objets {\`a} une
dimension et dont les champs effectifs {\`a} basse {\'e}nergie sont
pr{\'e}cis{\'e}ment ceux des th{\'e}ories de supergravit{\'e}.

\subsection{ Les cordes fondamentales }

Nous allons consid{\'e}rer des th{\'e}ories des cordes, dont les champs
microscopiques sont des cordes {\`a} une dimension plong{\'e}es dans un
l'espace-temps. Lors de leur {\'e}volution au cours du temps, une corde
d{\'e}crit une \de{surface d'univers} bidimensionnelle. Dans la
th{\'e}orie d{\'e}crivant la dynamique de cette surface d'univers, les
champs d'espace-temps $g\mn, B\mn,$ etc jouent le r{\^o}le de couplages,
de m{\^e}me que la m{\'e}trique $g\mn$ et le potentiel
{\'e}lectromagn{\'e}tique $A_\mu$ dans la th{\'e}orie de la particule-test
(\ref{acttest}). 
Si nous choisissons une configuration o{\`u} tous ces
champs d'espace-temps sont nuls, sauf $g\mn$, $B\mn$ et $\Phi$, l'action de
notre corde-test est l'action de Polyakov
\footnote{La surface d'univers est param{\'e}tr{\'e}e par
  $X^\mu(\sigma^i)$ et par sa m{\'e}trique $\gamma_{ij}$, dont on note
  $R^{(2)}$ la courbure scalaire.}
\bea
S^{\rm P}(X^\mu(\sigma^i),\gamma_{ij}) = 
\frac{1}{4\pi\ap}\int d^2\sigma^i \sqrt{-\det \gamma
  }\left( \gamma^{ij} g\mn \p_i X^\mu \p_j X^\nu \right. &+ \epsilon ^{ij} B\mn
  \p_i X^\mu \p_j X^\nu  \nn
\\
&+\left. \ap \Phi R^{(2)}(\gamma) \right).
\label{actpoly}
\eea
Nous avons introduit un param{\`e}tre $\ap$, qui a la dimension d'une
surface~; $\frac{1}{2\pi\ap}$
s'interpr{\`e}te comme la \de{tension} de la corde-test, qui
g{\'e}n{\'e}ralise la masse d'un objet ponctuel. Pour
obtenir une th{\'e}orie supersym{\'e}trique, il faudrait tenir compte des
fermions qui sont les partenaires supersym{\'e}triques des champs
$X^\mu(\sigma_i)$, et ajouter les termes correspondants {\`a}
l'action. Avec ou sans supersym{\'e}trie, ce genre d'action d{\'e}crit un
\de{mod{\`e}le sigma}. 
Une propri{\'e}t{\'e} essentielle de ce mod{\`e}le est
l'invariance conforme sur la surface d'univers, qui g{\'e}n{\'e}ralise
l'invariance de l'action de la particule-test par reparam{\'e}trisation
de la ligne d'univers. Les transformations conformes
sont en g{\'e}n{\'e}ral d{\'e}finies comme les reparam{\'e}trisations
locales qui laissent la m{\'e}trique (ici $\gamma_{ij}$) invariante
modulo un changement d'{\'e}chelle (ici $\gamma_{ij}\rightarrow
\lambda(\sigma_i) \gamma_{ij}$). 
{\`A} deux dimensions,
la sym{\'e}trie conforme est d{\'e}crite par une alg{\`e}bre de dimension
infinie, l'alg{\`e}bre de Virasoro ${\cal V}$ engendr{\'e}e par $L_m,\
m\in \Z$:
\bea
[L_m,L_n]=(m-n)L_{m+n}+\frac{c}{12}m(m^2-1)\delta_{m+n}. 
\label{algvir}
\eea
Nous avons introduit ici la \de{charge centrale} $c$, un nombre qui
vaut z{\'e}ro au niveau classique, mais dont la valeur, apr{\`e}s quantification de
la th{\'e}orie, 
d{\'e}pend des champs d'espace-temps et en particulier de la dimension
de l'espace-temps. De m{\^e}me, s'il est facile de constater que le
prolongement supersym{\'e}trique de l'action de
Polyakov est invariant conforme au niveau classique pour tous champs
$g\mn,B\mn,...$, l'absence d'anomalie quantique {\`a} l'invariance conforme
pose des conditions sur les champs d'espace-temps. Ces conditions sont
l'annulation des fonctions $\beta$ associ{\'e}es aux couplages du mod{\`e}le sigma,
c'est-{\`a}-dire aux champs d'espace-temps. Ces fonctions $\beta$ 
se calculent perturbativement en $\ap$, et {\`a} l'ordre le plus bas,
leur annulation {\'e}quivaut aux {\'e}quations du mouvement d'une
th{\'e}orie de supergravit{\'e} (avec la condition que la dimension de
l'espace-temps soit dix). 

\bs

{\`A} une {\'e}chelle d'{\'e}nergie tr{\`e}s petite par rapport {\`a}
$\frac{1}{\sqrt{\ap}}$, la th{\'e}orie
des cordes sur une surface d'univers donn{\'e}e 
est donc d{\'e}crite par les champs et la dynamique de la
supergravit{\'e}. Des corrections apparaissent {\`a} l'{\'e}chelle
$\frac{1}{\sqrt{\ap}}$, qui s'identifie donc avec l'{\'e}chelle de la
gravit{\'e} quantique $M_{Pl}$ (la masse de Planck). Ces corrections sont
des termes additionnels dans l'action de la supergravit{\'e}, comportant
plus de d{\'e}riv{\'e}es (et, pour des raisons dimensionnelles, autant de
facteurs $\sqrt{\ap}$ que de d{\'e}riv{\'e}es suppl{\'e}mentaires), et de
nouveaux champs.

Les champs qui apparaissent {\`a} cette {\'e}chelle s'obtiennent {\`a}
partir du spectre du mod{\`e}le sigma associ{\'e} {\`a} l'espace plat. Ce
spectre contient tout d'abord des {\'e}tats de masse nulle qui
s'identifient aux champs de la supergravit{\'e}. En effet, ces champs
sont les couplages du mod{\`e}le sigma, et leur variation
infinit{\'e}simale correspond {\`a} une petite d{\'e}formation de ce
mod{\`e}le. Ainsi, une variation infinit{\'e}simale de la m{\'e}trique, un
\de{graviton}, est un champ de spin deux et de masse nulle. On
l'observe effectivement dans le spectre. Ensuite, le spectre contient
des champs massifs dont les plus l{\'e}gers ont pour masse
$\frac{1}{\sqrt{\ap}}$ (il s'agit de la seule {\'e}chelle de masse disponible
dans la d{\'e}finition du mod{\`e}le sigma). La masse n'est pas born{\'e}e,
il y a une infinit{\'e} de champs massifs dans le spectre. Tous ces
champs sont trop massifs pour {\^e}tre observables dans les
acc{\'e}l{\'e}rateurs de particules, mais contribuent {\`a}
la coh{\'e}rence de la th{\'e}orie des cordes (par opposition {\`a} la
non-renormalisabilit{\'e} de la supergravit{\'e}). Si l'on cherche {\`a}
relier la th{\'e}orie des cordes au mod{\`e}le standard, on doit donc
identifier les champs de ce dernier avec
des champs de masse nulle, suppos{\'e}s acqu{\'e}rir une masse
(petite par rapport {\`a} $M_{Pl}$), par exemple par la brisure de la
supersym{\'e}trie. 

L'int{\'e}grale de chemin de la th{\'e}orie des cordes porte en principe sur toutes
les configurations possibles de la surface d'univers, et en
particulier sur des configurations de topologies
variables. La topologie d'une surface {\`a} deux
dimensions est trahie par la quantit{\'e} 
\bea
\int d^2\sigma ^i \sqrt{-\det \gamma }R^{(2)},
\label{topows}
\eea
qui appara{\^\i}t dans l'action eq. (\ref{actpoly}) si l'on suppose le
dilaton $\Phi$ constant. Chaque boucle de la surface d'univers apporte
donc un facteur $e ^\Phi$ {\`a} l'int{\'e}grale de chemin. Cette
quantit{\'e} $g_s=e ^\Phi$ s'appelle la \de{constante de couplage des
  cordes}\footnote{ Si $e ^\Phi$ n'est pas constant, on l'{\'e}crit avec
  un facteur constant $g_s$. } 
et r{\'e}git les interactions de l'hypoth{\'e}tique th{\'e}orie
d'espace-temps dont les champs seraient les {\'e}tats du spectre du
mod{\`e}le sigma, et dont la limite de basse {\'e}nergie serait une
th{\'e}orie de supergravit{\'e}. 

\subsection{ Dualit{\'e}s perturbatives et non-perturbatives
\label{dualites}
}

Consid{\'e}rons le mod{\`e}le sigma associ{\'e} {\`a} l'espace-temps
plat. Si la surface d'univers est Euclidienne et a la 
topologie d'une deux-sph{\`e}re, 
la sym{\'e}trie de reparam{\'e}trisation et l'invariance conforme  
permettent de choisir une m{\'e}trique plate,
\bea
\gamma_{ij}d\sigma ^i d\sigma ^j = dz\, d\bz.
\label{jaugews}
\eea
L'{\'e}quation du mouvement du champ $X^\mu(z,\bz)$ est alors $\p\bp X^\mu=0$. Les
solutions sont
\bea
X^\mu(z,\bz)=X^\mu_G(z)+X^\mu_D(\bz).
\label{solws}
\eea
Les solutions se \flqq factorisent\frqq\ donc en deux secteurs
ind{\'e}pendants (Gauche
et Droit). Cette propri{\'e}t{\'e} contribue {\`a} rendre la th{\'e}orie des
cordes en espace plat simple et accessible au calcul. Nous verrons
qu'elle tient {\'e}galement dans d'autres espaces-temps, comme les
vari{\'e}t{\'e}s de groupes.

En th{\'e}orie conforme, on consid{\`e}re dans un premier temps 
les variables $z$ et $\bz$ comme
ind{\'e}pendantes, ce qui permet d'{\'e}tudier s{\'e}par{\'e}ment les secteurs
G et D comme des th{\'e}ories bidimensionnelles, 
et on impose ensuite des contraintes reliant les deux secteurs pour
d{\'e}terminer les quantit{\'e}s physiques, dont le spectre. Ainsi,
chaque secteur jouit de sa propre alg{\`e}bre de Virasoro
(\ref{algvir}). 

\bs

Consid{\'e}rons maintenant, dans une direction $\mu=0$ donn{\'e}e, la
transformation de \de{T-dualit{\'e}}:
\bea
X^0(z,\bz)\rar X^0_G(z)-X^0_D(\bz).
\label{tdualws}
\eea
Ce simple changement de variable n'affecte pas le spectre du mod{\`e}le
sigma. Il en modifie cependant l'interpr{\'e}tation en termes
d'espace-temps. Par exemple, consid{\'e}rons les {\'e}tats les plus l{\'e}gers de
la th{\'e}orie des cordes dans un cercle de rayon $R$ d{\'e}crit par une
coordonn{\'e}e $X$ p{\'e}rodique, $X=X+2\pi R$. Ces {\'e}tats correspondent aux
configurations suivantes du champ $X(\sigma,\tau)$ du mod{\`e}le sigma~:
\bea
X=\frac{2\pi \ap}{R}n\tau\ ,\ n\in \Z,
\label{impulsion}
\eea
est un {\'e}tat d'impulsion quantifi{\'e}e $n/R$. Il ne d{\'e}pend pas de
$\sigma $ et a donc un sens m{\^e}me dans le cas d'une  particule
ponctuelle. En revanche, l'existence d'\de{{\'e}tats d'enroulement},
\bea
X=2\pi Rw\sigma\ , \ w\in \Z,
\label{enroulement}
\eea
est un effet purement cordiste, d{\^u} au fait que la corde peut {\`a} un
instant donn{\'e} s'enrouler autour d'un cercle donn{\'e}. On peut
facilement constater que la transformation de T-dualit{\'e} dans la
direction de $X$ {\'e}change les {\'e}tats d'implulsion avec les {\'e}tats
d'enroulement, tout en pr{\'e}servant la structure du spectre (et
notamment la masse des {\'e}tats). Cet {\'e}change revient {\`a} inverser le
rayon $R$ de la coordonn{\'e}e $X$,
\bea
R\rar \frac{2\pi\ap}{R}.
\label{invrayon}
\eea
De fa{\c c}on {\'e}quivalente, on constate que la T-dualit{\'e} dans une
direction $X^0$
change le signe de la composante $\delta g_{00}$ du graviton,
c'est-{\`a}-dire de la fluctuation de la m{\'e}trique, qui appara{\^\i}t
dans le spectre des cordes ferm{\'e}es. 
On peut g{\'e}n{\'e}raliser ces observations, et interpr{\'e}ter la
T-dualit{\'e} comme une transformation des champs d'espace-temps.
{\'E}tant donn{\'e}e une direction compacte $0$
correspondant {\`a} une sym{\'e}trie des champs d'espace-temps
\footnote{ C'est-{\`a}-dire $\frac{\p}{\p X^0}$ est un vecteur de
  Killing de la m{\'e}trique, $\Phi$ ne d{\'e}pend pas de $X^0$, etc. 
},
la transformation des champs de Neveu-Schwarz $g\mn,B\mn,\Phi$ 
peut s'{\'e}crire~:
\bea
\begin{array}{| r | c | l |}
\hline
g'_{00}=g_{00}\- \  & g'_{\mu 0}=g_{00}\- B_{\mu 0 }\ & g'_{\mu\nu}=g\mn
- g_{00}\- \left( g_{\mu 0}g_{\nu 0}-B_{\mu 0 }B_{\nu 0}\right)
\\
\hline
e ^{2\Phi'}= g_{00}\- e ^{2\Phi}\ & B'_{\mu 0}= g_{00}\- g_{\mu 0}\ & 
B'_{\mu\nu}=B\mn -g_{00}\- \left( B_{\mu 0}g_{\nu 0}- B_{\nu 0}g_{\mu
    0}\right)
\\
\hline
\end{array}
\label{ttransfo}
\eea
Cette transformation est non perturbative
du point de vue du mod{\`e}le sigma, car elle inverse le couplage
$g_{00}$, mais perturbative du point de vue de la th{\'e}orie des
cordes~: si $g_s=e ^\Phi$ est petit alors $g_s'= e ^{\Phi '}$
l'est aussi. La transformation (\ref{ttransfo}) est une sym{\'e}trie des
th{\'e}ories de supergravit{\'e}, en ce sens qu'elle {\'e}change les th{\'e}ories
de type IIA et IIB\footnote{Cela ne peut naturellement pas se voir dans les
transformations des champs du secteur de Neveu-Schwarz, qui est commun
aux deux th{\'e}ories.
}.
Mais, comme elle provient d'une
sym{\'e}trie du mod{\`e}le sigma, elle doit s'{\'e}tendre {\`a} toute la
th{\'e}orie des cordes. 
En particulier, cela veut dire que les
corrections d{\'e}rivatives {\`a} la supergravit{\'e} doivent v{\'e}rifier une
sym{\'e}trie de T-dualit{\'e}, qui {\`a} l'ordre le plus bas en $\ap$ se
r{\'e}duit {\`a} la transformation (\ref{ttransfo}). 

\bs

Dans un espace-temps donn{\'e}, si l'on consid{\`e}re l'ensemble des
T-dualit{\'e}s correspondant aux sym{\'e}tries, que l'on ajoute certaines
reparam{\'e}trisations et certaines transformations de jauge du champ
$B\mn$, et que l'on prend en compte les conditions de quantification
(par exemple la quantification de l'impulsion dans une direction
compacte), on engendre un goupe (discret) de T-dualit{\'e}. Par exemple,
le groupe de T-dualit{\'e} du tore {\`a} $d$ dimensions $T^d$ est
$O(d,d,\Z)$. 

\bs

La th{\'e}orie des cordes de type IIB jouit de plus d'une sym{\'e}trie de 
dualit{\'e} non perturbative conjectur{\'e}e, qui
inverse le couplage des cordes $g_s=e ^\Phi$. Comme la T-dualit{\'e},
cette S-dualit{\'e} est une sym{\'e}trie de la supergravit{\'e}. Rappelons
que la supergravit{\'e} de type IIB comporte des champs de Ramond-Ramond
$C^{(0)},C^{(2)},C^{(4)}$. Nous nous
contenterons de citer une transformation de S-dualit{\'e} dans le cas
o{\`u} $C^{(0)}=C^{(4)}=0$, 
\bea 
\begin{array}{| c | c |}
\hline
e ^{\Phi '}= e ^{-\Phi} &  B'\mn=C^{(2)}\mn
\\
\hline
g'\mn= e ^{-\Phi}g\mn &
C^{(2)}\mn{}'=-B^{(2)}\mn
\\
\hline
\end{array}
\label{stransfo}
\eea 
Si l'on admet que cette dualit{\'e} non perturbative est une sym{\'e}trie
de la th{\'e}orie des cordes, comme l'y incitent de nombreuses
v{\'e}rifications dans des cas particuliers, alors elle nous informe sur
le 
r{\'e}gime de 
fort couplage, inaccessible {\`a} la th{\'e}orie perturbative. En
particulier, les {\'e}tats perturbatifs correspondent par S-dualit{\'e}
{\`a} des objets non perturbatifs. Nous allons maintenant introduire
certains de ces objets.

\section{ Branes et trous noirs }

\subsection{ Solitons, singularit{\'e}s et objets {\'e}tendus }

Nous avons introduit les th{\'e}ories de supergravit{\'e} comme des
g{\'e}n{\'e}ralisations de la th{\'e}orie de la relativit{\'e}
g{\'e}n{\'e}rale. Dans cette derni{\`e}re, on conna{\^\i}t des
objets solitoniques, d{\'e}crits par des configurations singuli{\`e}res de
la m{\'e}trique : les trous noirs. Un trou noir peut se former lors de
l'effondrement gravitationnel d'un objet suffisamment dense. A
fortiori, un objet infiniment dense comme une particule ponctuelle
massive doit cr{\'e}er un trou noir. Ainsi, si l'on prend en compte
l'action de la particule-test $S^{\rm test}$ dans la dynamique de la
m{\'e}trique, les termes {\`a} ajouter aux {\'e}quations du mouvement
d'Einstein-Maxwell se comportent comme des sources, et les solutions
de ces {\'e}quations sont la m{\'e}trique et le champ
{\'e}lectromagn{\'e}tique d'un trou noir {\'e}ventuellement charg{\'e}. Par
exemple, la m{\'e}trique et le champ {\'e}lectromagn{\'e}tique du trou noir
de Reissner-Nordstr{\"o}m {\`a} 3+1 dimentions sont~:
\footnote{Les positions des horizons sont $r_\pm = G(M\pm
  \sqrt{M^2-Q^2})$, o{\`u} $G$ est la constante de la gravitation, $M$
  et $Q$ la masse et la charge du trou noir.
}
\bea
ds^2=-\frac{(r-r_+)(r-r_-)}{r^2}dt^2 + \frac{r^2}{(r-r_+)(r-r_-)}dr^2
+ r^2 d\Omega_2^2, \ F=\frac{Q}{r^2}dt\wedge dr.
\label{trounoir}
\eea
Ce trou noir correspond {\`a} un objet ponctuel, en ce sens qu'{\`a} tout
instant $t$ donn{\'e} la m{\'e}trique de l'espace {\`a} trois dimentions $t=\cst $
a une sym{\'e}trie sph{\'e}rique.

Il est aussi possible d'ajouter des sources aux {\'e}quations de la
supergravit{\'e}, et de trouver des solutions solitoniques
g{\'e}n{\'e}ralisant les trous noirs. Cependant, nous n'identifierons que
plus tard les objets qui, {\`a} l'instar des particules-test, engendrent
ces sources. Nous suivrons donc une logique inverse de celle que nous
avons d{\'e}crite pour la relativit{\'e} g{\'e}n{\'e}rale, en commen{\c c}ant par
{\'e}crire des solutions de la supergravit{\'e} qui ressembleront
formellement {\`a} (\ref{trounoir}).

\bs

Nous allons consid{\'e}rer des solutions de la supergravit{\'e} {\`a} 10
dimensions correspondant {\`a} des objets {\'e}tendus {\`a} $p$
dimensions. Autrement dit, les solutions pr{\'e}serveront la sym{\'e}trie
de Lorentz {\`a} $p+1$ dimensions et la sym{\'e}trie sph{\'e}rique {\`a} $9-p$
dimensions. De m{\^e}me que le trou noir ponctuel est {\'e}lectriquement
charg{\'e}, nos objets seront charg{\'e}s pour la forme $C^{(p+1)}$ (ainsi,
ils pourront exister en th{\'e}orie de type IIA ou IIB selon la parit{\'e} de
$p$). Les solutions de supergravit{\'e}, appel{\'e}es $p$-branes noires, sont donc~:
\bea
ds^2=H_p(r)^{-\half}(-dt^2+dx_1^2+\cdots+dx_p^2)+H_p(r)^\half (dr^2+r^2
d\Omega ^2_{8-p}), 
\label{metbrane}
\\
C^{(p+1)}=g_s\- (1-H_p(r)\- ) dt\wedge dx^1\wedge \cdots \wedge dx^p.
\label{champbrane}
\eea
Ces solutions d{\'e}pendent d'une fonction harmonique {\`a} sym{\'e}trie
sph{\'e}rique $(9-p)$-dimensionnelle, $H_p(r)$, qui est reli{\'e}e au dilaton
par $e ^\Phi = g_s H_p(r)^\frac{3-p}{4}$. Explicitement, cette fonction
vaut
\bea
H_p(r)=1+ \frac{Ng_s \ap ^{\frac{7-p}{2}}}{r^{7-p}}.
\label{Hbrane}
\eea
(Nous avons omis des facteurs num{\'e}riques dans le deuxi{\`e}me terme). 
L'entier $N$ repr{\'e}sente la charge de l'objet
par rapport au champ $C^{(p+1)}$~; on peut prouver la quantification de
cette charge par un argument g{\'e}n{\'e}ralisant la quantification de
Dirac. On peut calculer la masse de l'objet pour la th{\'e}orie de la
supergravit{\'e} en g{\'e}n{\'e}ralisant la notion de masse ADM~: on obtient
une masse inversement proportionnelle {\`a} $g_s$. Cela d{\'e}note la
nature non-perturbative de l'objet qui cr{\'e}e cette $p$-brane
noire. Par exemple, dans le cas $p=1$, il ne peut donc pas s'agir de
la corde fondamentale d{\'e}crite par le mod{\'e}le sigma. 
Nous identifierons plus tard cet objet comme une
\de{D$p$-brane}.

\bs

On conna{\^\i}t aussi des solutions plus compliqu{\'e}es, c'est-{\`a}-dire moins
sym{\'e}triques, qui d{\'e}crivent
des {\'e}tats li{\'e}s d'objets de dimensions diff{\'e}rentes. 
La superposition de D1 et D5-branes a ainsi la forme~:  
\bea
ds ^2 = H_1(r)^{-\half}H_5(r)^{-\half}(-dt^2+dx_1^2)+H_1(r)^\half
H_5(r)^{-\half} (dx_2^2+\cdots +dx_5^2) \nn 
\\
 + H_1(r)^\half H_5(r)^\half
(dr^2+r^2 d\Omega _3^2), 
\label{sold1d5}
\eea
avec la fonction $H_5(r)$ d{\'e}finie plus haut, une fonction $H_1(r)$ en
$\frac{1}{r^2}$ comme $H_5(r)$, 
les champs de Ramond-Ramond $C^{(2)}$ et $C^{(6)}$
comme pr{\'e}c{\'e}demment, et le dilaton $e ^\Phi = g_s H_1(r)^\half
H_5(r)^{-\half}$. Notons que les D1 et D5-branes sont reli{\'e}es par
une g{\'e}n{\'e}ralisation de la dualit{\'e} {\'e}lectrique-magn{\'e}tique,
puisque $dC^{(6)}$ et $dC^{(2)}$ sont reli{\'e}es par la dualit{\'e} de
Hodge {\`a} 10 dimensions. Par S-dualit{\'e}, ces deux
champs de Ramond-Ramond correspondent au champ $B$ et {\`a} son dual de
Hodge $d^{-1}\ast dB$. 
La solution de supergravit{\'e} S-duale {\`a} la solution (\ref{sold1d5})
est donc charg{\'e}e {\'e}lectriquement et magn{\'e}tiquement pour le champ
$B$, et s'{\'e}crit
\footnote{ $d\Omega_3^2$ est la m{\'e}trique de la 3-sph{\`e}re $S^3$ de rayon un, et
  $\epsilon_3$ sa forme volume. } 
\bea
g_s\, ds^2 &=& H_1(r)\- (-dt^2+dx_1^2) + (dx_2^2+\cdots +dx_5^2) 
 + H_5(r) (dr^2+r^2 d\Omega_3^2),
\label{metf1ns5}
\\
H=dB&=& N_5\ap \epsilon_3 + N_1\ap H_1(r)\- H_5(r) dt\wedge dx_1\wedge
dr,
\label{Bf1ns5}
\\
 e ^\Phi&=& g_s\- H_1(r)^{-\half}H_5(r)^\half.
\label{dilf1ns5}
\eea 
Cette solution nous int{\'e}ressera notamment pour sa \de{limite proche de
l'horizon}. Si en effet on l'examine pour $r$ petit, on obtient une
nouvelle solution de la supergravit{\'e} qui a la m{\^e}me forme mais o{\`u}
les fonctions $H_1(r)$ et $H_5(r)$ ont perdu leurs termes constants,
\bea
H_1(r)~:\ \ 1+\frac{N_1 g_s \ap}{r^2} \ \rar \ \frac{N_1 g_s \ap }{r^2}.
\label{nhg}
\eea
Cette solution est du type $AdS_3\times S^3\times \R^4$, o{\`u} $AdS_3$
est l'espace Anti-de Sitter. Nous nous int{\'e}resserons plus tard {\`a}
la th{\'e}orie des cordes dans l'espace $AdS_3$, qui est localement la
vari{\'e}t{\'e} du groupe \SL. 

La limite proche de l'horizon de cette solution n'est cependant
$AdS_3\times S^3\times \R^4$ que dans le cas o{\`u} $N_1\neq 0$. Dans le
cas contraire, on obtient {\`a} la place un espace $\R^{1,5}\times
\SU\times \R_\Phi$, o{\`u} $\R_\Phi$ est l'espace du \de{dilaton
  lin{\'e}aire}, d{\'e}fini par une m{\'e}trique plate unidimensionnelle et
un dilaton $\Phi(x)\propto x$. Nous retrouverons plus loin cette
limite proche de l'horizon d'une configuration de NS5-branes.

\subsection{ Quelques propri{\'e}t{\'e}s des $p$-branes noires }

Dans le but d'identifier les objets non-perturbatifs de la th{\'e}orie
des cordes qui peuvent cr{\'e}er les branes noires, 
nous allons en {\'e}tudier quelques propri{\'e}t{\'e}s.

Nous commencerons par une analogie avec le trou noir de
Reissner-Nordstr{\"o}m. Les termes de source capables d'engendrer cette
solution de la th{\'e}orie d'Einstein-Maxwell proviennent de l'action
(\ref{acttest}) 
d'une particule de masse $M$ et de charge $Q$.
 L'{\'e}tude des termes de source correspondant aux
$p$-branes noires dans les {\'e}quations de supergravit{\'e} peut de
m{\^e}me 
nous renseigner sur les objets en question.

Remarquons d'abord qu'une $p$-brane noire est sp{\'e}cifi{\'e}e par un
seul param{\`e}tre, que nous avons appel{\'e} $N$. Celui-ci d{\'e}termine
{\`a} la fois sa masse et sa charge pour les champs de
Ramond-Ramond, et doit {\^e}tre entier. Que la charge soit quantifi{\'e}e
se produit d{\'e}j{\`a} pour le trou noir de Reissner-Nordstr{\"o}m, mais
que la masse le soit est propre {\`a} la th{\'e}orie des cordes. En fait,
il s'agit l{\`a} d'une cons{\'e}quence de la supersym{\'e}trie, les
solutions que nous avons d{\'e}crites pr{\'e}servant une partie
des supersym{\'e}tries des th{\'e}ories de type II. Nous avons d{\'e}j{\`a}
rencontr{\'e} un ph{\'e}nom{\`e}ne analogue dans l'action (\ref{acttypeII})
des th{\'e}ories de supergravit{\'e} elles-m{\^e}mes, o{\`u} la force de
l'attraction gravitationnelle {\'e}tait {\'e}gale {\`a} celle des
interactions de jauge. Ce ph{\'e}nom{\`e}ne est une manifestation du
caract{\`e}re unifi{\'e} de la th{\'e}orie des cordes, qui ne d{\'e}pend
d'aucun param{\`e}tre physique libre ($\ap $ est une simple {\'e}chelle de
surface). 

\bs

L'action d'un l'objet-test ($N=1$) 
{\`a} $p+1$ dimensions, dont la masse {\'e}gale
la charge, est une int{\'e}grale sur la singularit{\'e} $r=0$ de la
$p$-brane noire, 
\bea
S^{\rm test}_{p+1}= T_p \int d^{p+1}x e ^{-\Phi}
\sqrt{- \det \hg } + T_p \int
C^{(p+1)}, 
\label{branetest}
\eea
qui est {\`a} comparer avec l'action (\ref{acttest}). Nous avons
introduit la quantit{\'e}
\bea
T_p = g_s\- \ (2\pi)^{-p} \ \ap^{-\frac{p+1}{2}}, 
\label{tension}
\eea
qui est la masse de l'objet, ou plut{\^o}t sa tension si $p\geq
1$. 
Le facteur $g_s\- $ traduit son
caract{\`e}re non perturbatif (solitonique).

Nous allons maintenant identifier cet objet comme une D-brane, que
nous d{\'e}finirons {\`a} partir des conditions au bord des cordes
ouvertes. Nous trouverons notamment une D-brane dont l'action effective
co{\"\i}ncide avec l'action $S^{\rm test}_{p+1}$. 

\section{ D-branes et bouts de cordes ouvertes }

\subsection{ Le bord de la surface d'univers }

Imaginons que la surface d'univers $\Sigma$ soit munie d'un bord $\p
\Sigma$. Nous le localiserons en $\sigma=0,\pi$ dans les coordonn{\'e}es
$(\sigma,\tau)$. 
Les propri{\'e}t{\'e}s physiques de ce bord peuvent {\^e}tre
sp{\'e}cifi{\'e}es au moyen de conditions au bord. Nous allons supposer
que ces conditions au bord d{\'e}coulent d'un principe variationnel
d{\'e}duit de l'action de Polyakov (\ref{actpoly}). 
Cependant, la pr{\'e}sence d'un bord
nous invite {\`a} ajouter {\`a} cette action un terme de bord. De m{\^e}me
que $\Sigma$ est coupl{\'e}e {\`a} la deux-forme $B\mn dx^\mu\wedge dx^\nu$, 
nous allons
coupler $\p \Sigma$ {\`a} une forme $A_\mu dx^\mu$. L'action avec terme de
bord est ainsi
\bea
S^\ouv = S^{\rm P} + \int _{\p \Sigma} A_\mu (X^\nu)  \p_\tau X^\mu
d\tau.
\label{actouv}
\eea
Appliquons le principe de moindre action. Les {\'e}quations du mouvement
dans la masse ne sont pas modifi{\'e}es par la pr{\'e}sence du terme de
bord, et la contribution de $\p \Sigma$ aux fluctuations de l'action
est, si l'on suppose la m{\'e}trique $\gamma$ de la surface d'univers plate,
\bea
(\delta S^\ouv)_{\rm bord} = \frac{1}{2\pi\ap} \int d\tau\ \delta
X^\mu 
\left(g\mn \p_\sigma X^\nu + (B\mn+2\pi\ap F\mn ) \p_\tau X^\nu\right),
\label{fluctactouv}
\eea
o{\`u} $F=dA$. Nous allons d{\'e}terminer dans quels cas $(\delta S^\ouv)_{\rm bord}$
s'annule pour des conditions au bord du type
\bea
(\p_\tau+\p_\sigma)X^\mu=C^\mu_\nu (\p_\tau-\p_\sigma)X^\nu,
\label{bordouv}
\eea
qui correspondent {\`a} \flqq coller\frqq\ les secteurs Gauche et
Droit au moyen d'une application lin{\'e}aire $C_\mu ^\nu$ (pas n{\'e}cessairement
constante). 
Remarquons que, dans (\ref{fluctactouv}), on
peut remplacer $\delta X^\mu$ par $\p_\tau X^\mu$, si les conditions
au bord sont suffisamment r{\'e}guli{\`e}res
\footnote{ En effet, ces conditions prennent la forme de l'annulation 
d'un produit de
  deux facteurs (\ref{fluctactouv}). 
Si le facteur $\delta X^\mu$ est nul et l'autre pas
  en un point, il doit en {\^e}tre de m{\^e}me au voisinage de ce
  point, donc $\p_\tau \delta X^\mu=0$, donc $X^\mu$ est localement
  constant sur le bord. 
}. En utilisant de plus les conditions au bord (\ref{bordouv}), on trouve 
\bea
(\p_\tau+\p_\sigma)X^\mu \left( C^\nu_\mu g_{\nu\rho} C^\rho_\lambda
  -g_{\mu\lambda}\right) (\p_\tau +\p_\sigma)X^\lambda =0.
\label{ccondition}
\eea
Cette condition doit {\^e}tre v{\'e}rifi{\'e}e pour tous les vecteurs
$(\p_\tau +\p_\sigma)X^\lambda $, donc
la matrice $C$ doit donc pr{\'e}server la m{\'e}trique $g\mn$, et
l'ensemble des solutions est $O(d,\R)$ o{\`u} $d$ est la dimension de
l'espace. Bien s{\^u}r, les conditions au bord doivent v{\'e}rifier
des {\'e}quations suppl{\'e}mentaires pour {\^e}tre coh{\'e}rentes
quantiquement, qui ne seront pas satisfaites par toutes les
fonctions {\`a} valeurs dans $O(d,\R)$.

La matrice $C$ peut avoir des valeurs propres $-1$. Les vecteurs
propres correspondants d{\'e}finissent des directions de type Dirichlet,
v{\'e}rifiant des conditions au bord $X^\mu=\cst$. Dans les autres
directions, on a $g\mn \p_\sigma X^\nu + (B\mn+2\pi\ap F) \p_\tau
X^\nu =0$, donc
\bea
C=\frac{ 1- g\- (B+2\pi\ap F) }{1+ g\- (B+2\pi\ap F) },
\label{cmatrice}
\eea
qui est bien une matrice orthogonale au sens d{\'e}fini plus
haut. Cependant, cette relation, et l'existence m{\^e}me du champ $F$,
n'ont de sens que dans les directions orthogonales aux directions de
type Dirichlet. Autrement dit, le champ $F$ n'existe que sur la
sous-vari{\'e}t{\'e} o{\`u} les conditions au bord de type Dirichlet forcent
les cordes ouvertes {\`a} aboutir. Cette sous-vari{\'e}t{\'e} s'appelle
naturellement une \de{D-brane}, ou \de{D$p$-brane} si l'on veut
indiquer sa dimension, $p+1$.

\bs

Donc, la th{\'e}orie des cordes ouvertes est d{\'e}finie par la
donn{\'e}e d'une D-brane, d{\'e}termin{\'e}e par un plongement $X^\mu(x ^i)$
d'une surface de dimension $p+1$ dans l'espace-temps,
et d'une deux-forme $F_{ij}(x^k)=d(A_i(x^k))$ sur cette
D-brane, de m{\^e}me que la th{\'e}orie des cordes ferm{\'e}es est d{\'e}finie
par la donn{\'e}e d'une m{\'e}trique, d'un champ $B$, etc. Nous
reviendrons sur les {\'e}quations que doivent v{\'e}rifier les
D-branes. Une chose est s{\^u}re~: elles doivent respecter l'invariance
de jauge $B\rar B+\Lambda,\ F\rar F-2\pi \ap \Lambda$ o{\`u}
$d\Lambda=0$. En effet, la dynamique des cordes ne d{\'e}pend que des
combinaisons invariantes de jauge $H=dB$ et $\omega=\hat{B}+2\pi\ap
F$, o{\`u} $\hat{B}$ est la forme induite sur la D-brane. Cette
invariance de jauge formalise la remarque que la matrice $C$,
{\'e}q. (\ref{cmatrice}), n'est d{\'e}finie que dans la direction de la
D-brane, et ne d{\'e}pend de $B$ et $F$ qu'{\`a} travers $B+2\pi\ap F$. 

Pour l'instant, 
les D-branes peuvent {\^e}tre 
sous-vari{\'e}t{\'e}s de l'espace-temps de dimensions quelconques. Mais
l'absence de contrainte sur leur dimension est due au fait que nous
avons consid{\'e}r{\'e} des cordes bosoniques. Si l'on traite le m{\^e}me
probl{\`e}me pour les th{\'e}ories des cordes supersym{\'e}triques de type
II, on obtient des D-branes de dimensions impaires (IIA) ou paires
(IIB). Ces dimensions concordent avec le degr{\'e} des formes de
Ramond-Ramond $C^{(p+1)}$ existant dans ces th{\'e}ories. En fait, les
D-branes sont des sources pour ces champs, auxquels elles se
couplent naturellement {\it via} des termes 
\bea
\int _{{\rm D}p}C^{(p+1)}.
\label{branecouple}
\eea
Cette observation concorde avec le comportement des D-branes
vis-{\`a}-vis de la T-dualit{\'e}. Par d{\'e}finition (\ref{tdualws}),
cette derni{\`e}re transforme une condition au bord de type Dirichlet ($\p_\tau
X^\mu =0$) en une
condition de type Neumann ($\p_\sigma X^\mu=0$) et
vice-versa. Ainsi, elle {\'e}change la coordonn{\'e}e de plongement
$X^\mu(x^k)$ d'une D-brane avec le champ $A ^\mu(x^k)$. Donc, la
T-dualit{\'e} augmente ou diminue d'un la dimension de la D-brane selon
qu'elle s'effectue dans une direction orthogonale ou parall{\`e}le {\`a}
celle-ci. Elle transforme donc les D-branes de la th{\'e}orie de type
IIA en D-branes de la th{\'e}orie de type IIB et vice-versa.

\bs

Mentionnons pour finir comment les D-branes de la th{\'e}orie de type IIB
se transforment par la S-dualit{\'e}, qui est une sym{\'e}trie de cette
th{\'e}orie. Cela peut se d{\'e}duire facilement des lois de
transformation des champs de fond auxquels se couplent les
D-branes. Ainsi, la D1-brane est une source pour $C^{(2)}\mn$, que la
S-dualit{\'e} (\ref{stransfo}) transforme en $B\mn$. Nous connaissons
d{\'e}j{\`a} un objet qui se couple {\`a} $B\mn$, c'est la corde
fondamentale elle-m{\^e}me (voir l'action de Polyakov (\ref{actpoly})),
parfois appel{\'e}e F1. Ainsi, les objets D1 et F1 sont {\'e}chang{\'e}s par
S-dualit{\'e}. La D3-brane est invariante, et la D5-brane se transforme
en une \de{NS5-brane} qui se couple magn{\'e}tiquement au champ $B\mn$. 

Par leur couplage aux champs de Ramond-Ramond et leurs transformations
par les dualit{\'e}s, les D-branes sont de bons candidats pour {\^e}tre
les sources des configurations de la supergravit{\'e} appel{\'e}es
$p$-branes noires dont nous avons discut{\'e}. Une confirmation
proviendra de l'{\'e}tude des actions effectives de ces D-branes.

\subsection{ L'action effective des D-branes }

Rappelons que les {\'e}quations du mouvement des th{\'e}ories de
supergravit{\'e} {\'e}quivalent {\`a} l'annulation des fonctions $\beta$ des
couplages du mod{\`e}le sigma. 
Depuis notre discussion de ce fait,
nous avons ajout{\'e} au mod{\`e}le sigma
des couplages au bord d{\'e}crits pas une D-brane, et l'invariance
conforme du mod{\`e}le avec bord requiert l'annulation des fonctions
$\beta$ associ{\'e}es {\`a} ces nouveaux couplages. {\`A} l'ordre le plus
bas en $\ap$, cette annulation s'interpr{\`e}te comme les {\'e}quations du
mouvement de l'action effective de la D-brane. Nous d{\'e}crivons
toujours une D$p$-brane au moyen de son plongement dans l'espace-temps,
$X^\mu(x^i)$, et du champ $F_{ij}$. Alors son action effective est une
fonctionnelle sur la vari{\'e}t{\'e} {\`a} $p+1$ dimensions param{\'e}tr{\'e}e par
$x^i$, qui
comprend deux types de termes~: une action de Born-Infeld, qui vaut 
\bea
S^{\rm BI}(X^\mu(x^i),F_{ij})= T_p \int dx^i\ e ^{-\Phi}
\sqrt{-\det(\hg +\hB + 2\pi \ap
  F)},
\label{actBI}
\eea
et des termes de Wess-Zumino,
\bea
S^{\rm WZ}(X^\mu(x^i),F_{ij})= T_p \int dx^i \widehat{ [e ^B \sum_n
  C^{(n)}] } e ^{2\pi \ap F},
\label{actWZ}
\eea
o{\`u} l'int{\'e}grale porte sur le terme de degr{\'e} $p+1$ du tenseur {\`a}
int{\'e}grer. Ainsi, seules les formes de Ramond-Ramond de degr{\'e} au
plus $p+1$ contribuent {\`a} l'action effective.

Ces expressions m{\'e}ritent quelques commentaires. D'abord, l'action de
Born-Infeld n'est pas valable seulement {\`a} l'ordre le plus bas en
$\ap$, puisque le terme $2\pi \ap F$ engendre des termes {\`a} tous les
ordres. En fait, le domaine de validit{\'e} de l'action effective
$S^{\rm BI}+S^{\rm WZ}$ est d{\'e}fini par la petitesse des d{\'e}riv{\'e}es
secondes des plongements $X^\mu$ et des d{\'e}riv{\'e}es premi{\`e}res 
du champ $F$. Les corrections {\`a} cette action effective sont
appel{\'e}es corrections d{\'e}rivatives, et leur ordre est d{\'e}fini par
rapport au nombre de d{\'e}riv{\'e}es mises en jeu. Naturellement, ce
nombre est reli{\'e} {\`a} la puissance de $\ap$ pour des raisons
dimensionnelles, mais la puissance de $\ap$ dans un terme des
corrections d{\'e}pend aussi de celle de $F$. Cette subtilit{\'e} peut
{\^e}tre contourn{\'e}e en incluant $\ap$ dans la d{\'e}finition de $F$, ce
que nous ferons par la suite quand il n'y aura pas de risque
d'ambiguit{\'e}. 

Revenons {\`a} notre raison d'introduire cette action
effective~: l'identification des sources des $p$-branes noires de la
supergravit{\'e} avec des D$p$-branes. Une D$p$-branes plate avec $F=0$
est solution des {\'e}quations du mouvement dans un espace-temps plat
(le seul champ non nul {\'e}tant $g\mn=\eta\mn$). Si l'on veut prendre
en compte l'action de cette D-brane dans les {\'e}quations de la
supergravit{\'e} et s'en servir ainsi de source, on constate qu'{\`a}
l'ordre dominant en $T_p$ seuls la m{\'e}trique, le dilaton
 et le champ $C^{(p+1)}$
sont affect{\'e}s par la pr{\'e}sence de la D-brane (par exemple, les
fluctuations $\delta B$ n'apparaissent qu'au deuxi{\`e}me ordre dans
$S^{\rm BI}$). Donc, l'action effective qui d{\'e}termine l'influence de
la D$p$-brane sur les champs de fond se r{\'e}duit {\`a} l'action
(\ref{branetest}) de la source de la $p$-brane noire. Ceci
confirme l'identification de cette source avec une D$p$-brane.

\subsection{ D-branes superpos{\'e}es }

Que se passe-t-il maintenant si l'on consid{\`e}re $N$ D-branes
dans un m{\^e}me espace-temps~? Les cordes ouvertes ont alors le \flqq
choix\frqq\ de la D-brane sur laquelle elles aboutissent. {\`A} chaque
bout de corde ouverte on doit donc associer un indice, dit de
\de{Chan-Paton}, correspondant {\`a} une D-brane. Le syst{\`e}me acquiert
$N\times N$ {\'e}tats fondamentaux
de cordes ouvertes {\'e}tir{\'e}es entre diff{\'e}rentes
D-branes. Si deux D-branes sont s{\'e}par{\'e}es, l'{\'e}tat fondamental d'une
corde ouverte qui les relie est massif~; en revanche, si les D-branes
sont superpos{\'e}es, il est sans masse et doit donc appara{\^\i}tre
dans la description effective du syst{\`e}me {\`a} basse {\'e}nergie, au
m{\^e}me titre que les positions $X^\mu$ des D-branes et les champs
$F_{ij}$.

L'ensemble des champs $F_{ij}=dA_i$ sur les $N$ D-branes correspond
{\`a} un
groupe de jauge $U(1)^N$. Si elles sont toutes superpos{\'e}es, les
nouveaux {\'e}tats de masse nulle agrandissent ce groupe de jauge, qui
devient $U(N)$. Il faut donc consid{\'e}rer que le champ $A_i$ est {\`a}
valeurs dans l'alg{\`e}bre de Lie de ce groupe, et par T-dualit{\'e} c'est
aussi le cas des coordonn{\'e}es $X^\mu$ des D-branes. Il peut donc
exister des {\'e}tats li{\'e}s de D-branes o{\`u} la localisation de chacune
n'est pas d{\'e}finie, dans le cas o{\`u} les matrices $X^\mu$ ne
commutent pas toutes. 

L'existence de ces {\'e}tats li{\'e}s d{\'e}pend de la dynamique des
D-branes, qui est d{\'e}termin{\'e}e par leur action effective. Cette
action effective, qui doit g{\'e}n{\'e}raliser $S^{\rm BI}+S^{\rm WZ}$ au cas
des champs non-ab{\'e}liens, n'est cependant pas connue
explicitement. Seuls certains termes ont {\'e}t{\'e} calcul{\'e}s, ce qui
est d{\'e}j{\`a} suffisant pour mettre en {\'e}vidence quelques effets
int{\'e}ressants, comme l'effet Myers que nous verrons plus loin.


\setcounter{chapter}{1}

\chapter{ Groupes, D-branes et action de Born-Infeld  
\label{chapgroupe} }

Si l'on {\'e}tudie depuis longtemps la th{\'e}orie des cordes dans des
groupes de Lie, ce n'est pas tant dans l'espoir de construire des
mod{\`e}les r{\'e}alistes aux basses {\'e}nergies, que pour mieux comprendre
la th{\'e}orie dans des cas non triviaux que la pr{\'e}sence de 
structures alg{\'e}briques permet cependant de r{\'e}soudre. 
On esp{\`e}re ainsi mieux comprendre les effets de la courbure de
l'espace-temps, de la pr{\'e}sence du champ $B\mn$, et de la topologie
non triviale de l'espace-temps, entre autres caract{\'e}ristiques de ces
vari{\'e}t{\'e}s. 

Nous allons nous int{\'e}resser {\`a} la th{\'e}orie des cordes bosoniques
dans des groupes de Lie. Laissant les aspects de th{\'e}orie conforme
pour le prochain chapitre, nous insisterons ici plus sur la
g{\'e}om{\'e}trie. Nous omettrons d'{\'e}voquer les constructions permettant de
b{\^a}tir des mod{\`e}les de cordes plus r{\'e}alistes {\`a} partir des
vari{\'e}t{\'e}s de groupes, constructions qui font appel {\`a} la
supersym{\'e}trie et {\`a} l'utilisation des groupes comme parties
compactes d'espace-temps de dimension dix.

En {\'e}tudiant les D-branes dans les groupes compacts d'abord du point
de vue g{\'e}om{\'e}trique, nous ne suivons pas la chronologie de la
recherche, qui {\'e}tait pass{\'e}e en premier lieu par une description
exacte. Mais notre logique sera adapt{\'e}e aux cas plus compliqu{\'e}s
des D-branes dans des espaces-temps non compacts, dont la construction
exacte est beaucoup plus difficile.

\section{ Cordes ferm{\'e}es dans les vari{\'e}t{\'e}s de groupes 
\label{sectgroupe} }

Dans cette section, nous pr{\'e}senterons d'abord le mod{\`e}le de
Wess-Zumino-Witten, qui d{\'e}crit les cordes ferm{\'e}es dans les
vari{\'e}t{\'e}s de groupes. Nous consacrerons ensuite un paragraphe au
cas particulier du groupe \SL, qui jouera un grand r{\^o}le dans la
suite.

\subsection{ Le mod{\`e}le de Wess-Zumino-Witten 
\label{soussectWZW} }

Si l'on veut {\'e}tudier une th{\'e}orie physique comportant la gravit{\'e}
sur un goupe de Lie $G$, il faut d'abord d{\'e}cider quelle m{\'e}trique
associer {\`a} ce groupe. 
La question est loin d'{\^e}tre simple
et requiert des hypoth{\`e}ses pr{\'e}cises sur la structure de $G$.
Nous exigerons d'abord que la m{\'e}trique soit
non-d{\'e}g{\'e}n{\'e}r{\'e}e. De plus, pour que les propri{\'e}t{\'e}s physiques
refl{\`e}tent les sym{\'e}tries
associ{\'e}es {\`a} la structure de groupe, il faut que la m{\'e}trique soit
invariante {\`a} gauche et {\`a} droite. Ainsi, elle est d{\'e}termin{\'e}e
par une forme bilin{\'e}aire sur l'alg{\`e}bre de Lie \lieg \ de $G$,  
non d{\'e}g{\'e}n{\'e}r{\'e}e et invariante par l'action adjointe. 
L'existence d'une telle forme est d'ailleurs une condition
n{\'e}cessaire et suffisante pour l'existence d'une th{\'e}orie conforme
qui respecte au niveau quantique les sym{\'e}tries du groupe
\cite{fist}. Nous appellerons $\llangle \cdot,\cdot \rrangle$ une
telle forme bilineaire. Dans le cas d'un groupe semi-simple, la forme
de Killing fait l'affaire, mais n'est pas n{\'e}cessairement la seule
possibilit{\'e}. Dans la suite, nous adopterons des normalisations issues
de ce cas pour des raisons de simplicit{\'e}.

Muni de sa m{\'e}trique, notre groupe n'est cependant toujours pas
solution des {\'e}quations de la Relativit{\'e} G{\'e}n{\'e}rale ni de la
Supergravit{\'e}. Pour en faire un espace physique au sens de cette
derni{\`e}re, on peut introduire une valeur non nulle pour le champ
$B\mn$. Alors se pose la question de l'existence globale d'une telle
forme. En fait la topologie non triviale du groupe $G$ s'oppose en
g{\'e}n{\'e}ral {\`a} l'existence globale du champ $B\mn$ n{\'e}cessaire pour
d{\'e}finir la th{\'e}orie des cordes sur $G$. Nous construirons plus loin
un tel champ $B\mn$, qui aura des singularit{\'e}s. Pour l'instant, on
peut {\'e}viter ce probl{\`e}me en formulant la th{\'e}orie {\`a} l'aide du
champ $H=dB$. Cette trois-forme ne se couple pas {\`a} la surface
d'univers bidimensionnelle $\Sigma$ 
plong{\'e}e dans $G$, mais {\`a} une sous-vari{\'e}t{\'e} ${\cal M}$ de $G$
telle que $\p {\cal M}=\Sigma$. L'action de Wess-Zumino-Witten 
s'{\'e}crit alors, en fonction du plongement $g(\sigma,\tau)$ de la corde
ferm{\'e}e dans le groupe $G$,
\bea
S^{\rm WZW}= \frac{k}{16\pi} \int_\Sigma d\sigma d\tau
  \llangle g\- \p_i g, g\- \p^i
g \rrangle + \frac{k}{24\pi} \int_{\cal M} \llangle g\- d g , [ g\- d
g, g\- dg]\rrangle .
\label{actWZW}
\eea
Le nombre $k$ s'appelle \de{le niveau} de la th{\'e}orie
\footnote{ Par commodit{\'e} d'{\'e}criture, nous supposons ici le groupe
  $G$ simple. Des groupes semi-simples, produits de groupe simples,
  peuvent avoir des niveaux diff{\'e}rents pour les diff{\'e}rents
  facteurs. La notion de rayon du groupe doit alors {\^e}tre remplac{\'e}e
  par un ensemble de param{\`e}tres. }.
Il est li{\'e}
au rayon du groupe $G$ par $R=\sqrt{k\ap}$. La m{\'e}trique {\'e}tant 
proportionnelle {\`a} $R^2$, l'action  a donc bien un facteur
$R^2/(2\pi\ap)=k/(2\pi)$ comme nous l'avons {\'e}crit. Si on fixe la
m{\'e}trique et donc le rayon, la limite $\ap\rar 0$, que nous
appllerons parfois limite semi-classique, correspond donc {\`a} $k\rar
\infty$. 

La d{\'e}finition de l'action (\ref{actWZW}) est potentiellement
amibigu{\"e} {\`a} cause de la d{\'e}finition de
la vari{\'e}t{\'e} ${\cal M}$ 
par $\p {\cal M}=\Sigma$. Supposons que cette relation admette 
des solutions diff{\'e}rentes ${\cal M}, {\cal M'}$. 
Comme la trois-forme $H$ est ferm{\'e}e, le terme $\int_\MM H$ dans
l'action (\ref{actWZW}) ci-dessus ne d{\'e}pend que de la classe de
$\MM $ dans le troisi{\`e}me groupe d'homotopie $\pi_3(G)$. Pour que
l'action propos{\'e}e soit physique, $\exp iS^{WZW}$ ne doit pas
d{\'e}pendre de cette quantit{\'e}.
Or, pour un
groupe compact simple, on a $\pi_3(G)=\Z$, et $\int_\MM H$
r{\'e}alise explicitement ({\`a} un facteur pr{\`e}s) l'application qui {\`a}
$\MM $ associe sa classe dans $\pi_3(G)$. Donc, $k$ doit {\^e}tre
quantifi{\'e}. Une analyse tenant compte des coefficients num{\'e}riques 
montre que $k\in \Z$. 

\bs

Remarquons que le champ $H$ que nous avons introduit est tel que la
connection d{\'e}j{\`a} mentionn{\'e}e (\Fac{\ref{contor}}),\
$\Gamma=\Gamma(g)-\half H$, 
parall{\'e}lise le groupe $G$. Cette remarque a des cons{\'e}quences
importantes pour la renormalisation du mod{\`e}le sigma (\cite{bcz}),
et donc pour l'annulation des corrections {\`a} la supergravit{\'e} dans
le cas des groupes.

La parall{\'e}lisabilit{\'e} des vari{\'e}t{\'e}s de groupes se manifeste
aussi dans les sym{\'e}tries
pr{\'e}serv{\'e}es par le mod{\`e}le, {\`a} savoir les translations locales 
\bea
 g(\sigma,\tau) \rightarrow \Omega (x_+) g(\sigma,\tau) \bar{\Omega} (x_-),
\label{symWZW}
\eea
avec $x_\pm = \tau \pm \sigma$. 
Il faut noter que les fonctions $\Omega$ et $\bar{\Omega}$ ne sont pas
tout {\`a} fait ind{\'e}pendantes, car elles doivent conspirer pour que la
corde reste bien ferm{\'e}e, $g(\sigma+2\pi,\tau)=g(\sigma,\tau)$.
Ces sym{\'e}tries correspondent aux courants conserv{\'e}s
\bea
 J=-k\p_+ g\ g\-  \ \ , \ \ \bJ = k g\- \p_- g.
\label{courWZW}
\eea
Ces courants servent de base {\`a} la construction de la th{\'e}orie
conforme correspondante \cite{gepwi}. Leur existence permet de
r{\'e}soudre la th{\'e}orie des cordes dans $G$, au m{\^e}me titre que la
th{\'e}orie des cordes dans l'espace plat. Plus pr{\'e}cis{\'e}ment, leurs
modes de Fourier engendrent l'alg{\`e}bre de Lie affine \alieg, et le
spectre de la th{\'e}orie se d{\'e}compose en repr{\'e}sentations de cette
alg{\`e}bre. Si \lieg\ est d{\'e}finie par $[J^a,J^b]=f^{abc}J^c$, alors
les relations de commutation qui d{\'e}finissent \alieg\ sont~: 
\bea
[J^a_n,J^b_m]=f^{abc}J^c_{m+n}+\half k n \delta ^{ab} \delta_{m+n,0}.
\label{algWZW}
\eea
De fa{\c c}on {\'e}quivalente {\`a} l'existence des courants conserv{\'e}s,
on peut remarquer que la solution g{\'e}n{\'e}rale des {\'e}quations du
mouvement de l'action (\ref{actWZW}) est 
\bea
 g(\sigma, \tau) = a(x_+)\ b(x_-).
\label{solWZW}
\eea
Ainsi, on peut quantifier la th{\'e}orie en consid{\'e}rant comme
largement ind{\'e}pendants deux secteurs, d{\'e}pendant respectivement des
coordonn{\'e}es $x_+$ et $x_-$. 
 Nous y reviendrons dans le prochain chapitre. 

\bs

Pour l'instant, nous allons seulement admettre l'existence d'une
th{\'e}orie des cordes dans $G$ respectant les sym{\'e}tries de groupe,
{\`a} condition que le niveau $k$ soit entier (dans le cas d'un groupe
compact). 
Cela suffit {\`a}
d{\'e}terminer les propri{\'e}t{\'e}s essentielles de la th{\'e}orie dans la
limite semi-classique $k\rar \infty$. En fait, ce n'est que dans cette
limite que la th{\'e}orie des cordes peut s'interpr{\'e}ter en termes
g{\'e}om{\'e}triques~; lorsque $k$ est faible, 
il faut prendre en compte des effets de
gravit{\'e} quantique, qui remettent en cause la notion m{\^e}me
d'espace. Au contraire, pour $k$ grand, il est possible de construire
des {\'e}tats de cordes ferm{\'e}es localis{\'e}s sur $G$, et la notion de
point de $G$ prend un sens physique. Plus pr{\'e}cis{\'e}ment, le spectre
des cordes ferm{\'e}es s'identifie dans cette limite avec l'ensemble des
fonctions $\Fun(G)$ sur $G$. Les  sym{\'e}tries mentionn{\'e}es ci-dessus
sugg{\`e}rent d'organiser ce spectre en repr{\'e}sentations de $G$. C'est
ce que fait le th{\'e}or{\`e}me de Peter-Weyl,
\bea
\Fun(G) \simeq \bigoplus_{R\in Rep(G)} R \otimes R^\dagger,
\label{peterweyl}
\eea
o{\`u} $Rep(G)$ est l'ensemble des repr{\'e}sentations irr{\'e}ductibles de
$G$ de dimension finie. L'{\'e}quivalence $\simeq$ est ici {\`a} prendre
au sens des repr{\'e}sentations de $G\times G$, o{\`u} les actions de $G$ sur
$\Fun(G)$ se d{\'e}duisent naturellement de ses actions {\`a} droite et {\`a}
gauche sur lui-m{\^e}me. 
Le spectre du mod{\`e}le de Wess-Zumino-Witten pour un groupe $G$
compact se d{\'e}duit de $\Fun(G)$ en {\'e}tendant les repr{\'e}sentations de
$G$ en des repr{\'e}sentations correspondantes de plus haut poids de
l'alg{\`e}bre de Lie affine \alieg, c'est-{\`a}-dire des repr{\'e}sentations $\hat{R}$
de \alieg\ 
construites {\`a} partir de repr{\'e}sentations $R$ de \lieg\ (que l'on
identifie avec l'alg{\`e}bre des $J^a_0$) en appliquant les $J^a_n$
tout en supposant l'action des $J^a_n,n>0$ sur $R$
nulle. Sch{\'e}matiquement, on a $\hat{R}={\bf \hat{g}}R/( J^a_n R=0
 \ \forall n>0)$.
Cependant, on ne garde que les
repr{\'e}sentations de \alieg\ qui restent unitaires apr{\`e}s cette extension. On
note $Rep_k(G)$ l'ensemble (fini) de ces repr{\'e}sentations. Alors le
spectre du mod{\`e}le de Wess-Zumino-Witten est~:
\bea
Spec(G) \simeq \bigoplus_{R\in Rep_k(G)} \hR \otimes \hR^\dagger.
\label{specWZW}
\eea
Nous reviendrons plus en d{\'e}tail sur ce spectre et la fonction de
partition du tore qui lui est associ{\'e}e dans le prochain chapitre.

\subsection{ Les cordes ferm{\'e}es dans $SL(2,\R)$
\label{soussectAdS3} }

Nous avons vu l'espace $AdS_3\times S^3\times \R^4$ appara{\^\i}tre
comme limite proche de l'horizon de certaines configurations de
branes, et se pr{\^e}ter {\`a} la conjecture AdS/CFT. C'est une motivation
fondamentale pour {\'e}tudier la th{\'e}orie des cordes dans cet espace,
ce qui peut se faire au moyen du mod{\`e}le de Wess-Zumino-Witten
associ{\'e} au groupe $SL(2,\R)\times SU(2)\times U(1)^4$. 
Les difficult{\'e}s pos{\'e}es par ce mod{\`e}le viennent du facteur non compact
$SL(2,\R)$. Nous allons expliquer certaines de ces difficult{\'e}s, et
montrer en quoi $SL(2,\R)$ est plus compliqu{\'e} que son homologue
compact $SU(2)$ du point de vue de la th{\'e}orie des cordes.

Commen{\c c}ons donc par munir \SL\ de sa m{\'e}trique naturelle,
associ{\'e}e {\`a} la forme de Killing. On peut d{\'e}finir une coordonn{\'e}e
temporelle $t$ {\`a} l'aide d'un vecteur de Killing de
genre temps de cette m{\'e}trique. Cette coordonn{\'e}e est p{\'e}riodique,
si on la d{\'e}compactifie on obtient l'espace \AAA. Toute tranche
$t=\cst$ de 
\SL\ a un volume infini. 

Comme les autres groupes, \SL\ muni de sa seule m{\'e}trique n'a pas de
sens physique. Remarquons cependant qu'il v{\'e}rifie les {\'e}quations
d'Einstein avec une constante cosmologique n{\'e}gative. Bien s{\^u}r nous
connaissons d{\'e}j{\`a} une autre fa{\c c}on d'interpr{\'e}ter le terme
correspondant, {\`a} savoir comme provenant d'une trois-forme $H=dB$ de
la th{\'e}orie des cordes, qui
dans le cas de \SL\ est proportionnelle {\`a} la forme volume. 
Cette \flqq {\'e}quivalence\frqq\ entre constante cosmologique et trois-forme
$H$ ne vaut que pour les groupes de dimension trois \SL\ et \SU. Cela
refl{\`e}te la propri{\'e}t{\'e} particuli{\`e}re {\`a} \SL\ d'{\^e}tre
(localement) {\`a} la fois un goupe de Lie et un espace Anti-de
Sitter. En effet, tous les espaces Anti-de Sitter sont solutions de
l'{\'e}quation d'Einstein avec constante cosmologique n{\'e}gative. Ils
jouissent d'une propri{\'e}t{\'e} remarquable d{\'e}couverte
par Breitenlohner et Freedman \cite{BF}~: un champ scalaire de masse
carr{\'e}e 
n{\'e}gative peut y {\^e}tre stable, pourvu que sa masse carr{\'e}e ne soit pas trop
grande. Cette propri{\'e}t{\'e} se manifestera {\'e}galement au niveau du
spectre des cordes ferm{\'e}es, o{\`u} l'on trouvera des {\'e}tats physiques
de masse n{\'e}gative. 

\bs

D{\'e}crivons ce spectre plus en d{\'e}tail. Dans la limite
semi-classique, il s'agit simplement d'{\'e}tudier $\Fun(SL(2,\R))$. Comme
dans le cas compact, cet espace se d{\'e}compose naturellement en
repr{\'e}sentations de $\SL\times\SL$. Les repr{\'e}sentations de \SL\ se
param{\'e}trisent naturellement en fonction du Casimir quadratique, que
l'on note $-j(j+1)$ (o{\`u} $j$ s'appelle le spin). Seules les
repr{\'e}sentations unitaires peuvent appara{\^\i}tre dans le spectre physique.
Les {\'e}tats d'une repr{\'e}sentation donn{\'e}e sont
eux param{\'e}tris{\'e}s par leur moment magn{\'e}tique 
$m$, c'est-{\`a}-dire la valeur propre
d'un g{\'e}n{\'e}rateur donn{\'e} $J_3$ de l'alg{\`e}bre de Lie \sl\ de \SL. Les
repr{\'e}sentations qui nous int{\'e}resseront sont les suivantes~:
\begin{itemize}
\item 
Les \de{repr{\'e}sentations continues} $C_j^\alpha$, avec $j\in -\half
  +i\R$, ont un Casimir $-j(j+1)\geq \frac{1}{4}$. Les {\'e}tats
  v{\'e}rifient $m\in \alpha+\Z$, avec $0\leq \alpha <1$. 
\item 
Les \de{repr{\'e}sentations discr{\`e}tes} $D_j^\pm$, avec $j\in \R$, ont
  un Casimir $-j(j+1)\leq \frac{1}{4}$. Les {\'e}tats sont engendr{\'e}s
{\`a} partir d'un {\'e}tat de plus bas ou plus haut poids, $m=\pm
j$. Ainsi les valeurs prises par $m$ sont $m\in \pm (j + \N)$. 
\end{itemize}
Consid{\'e}rons maintenant $\Fun(SL(2,\R))$ comme une repr{\'e}sentation de
\SL\ pour l'action {\`a} droite ou {\`a} gauche. Dans les deux cas, le
Casimir quadratique s'identifie avec le Laplacien de la
m{\'e}trique de \SL\ qui sert {\`a} {\'e}crire l'action du mod{\`e}le de
Wess-Zumino-Witten (\ref{actWZW}) (nous {\'e}crirons explicitement cette
m{\'e}trique plus loin (\ref{metads3})). 
Comme la masse d'un champ est la valeur propre de ce
m{\^e}me Laplacien, on peut interpr{\'e}ter la borne de
Breitenlohner-Freedman en termes du Casimir. Cette borne exclut les
repr{\'e}sentations continues, c'est-{\`a}-dire les {\'e}tats asymptotiques
qui vivent {\`a} l'infini spatial d'\AAA\ et peuvent se diffuser {\`a}
l'int{\'e}rieur. Les {\'e}tats discrets, eux, apparaissent comme des
p{\^o}les dans ces amplitudes de diffusion, et sont localis{\'e}s {\`a}
l'int{\'e}rieur d'\AAA. On voit donc que la borne de
Breitenlohner-Freedman est le v{\'e}ritable crit{\`e}re physique qui
permet de distinguer les {\'e}tats li{\'e}s des {\'e}tats asymptotiques, ce
que fait habituellement la positivit{\'e} de la masse dans un espace
asymptotiquement plat. 

\bs

\Fig{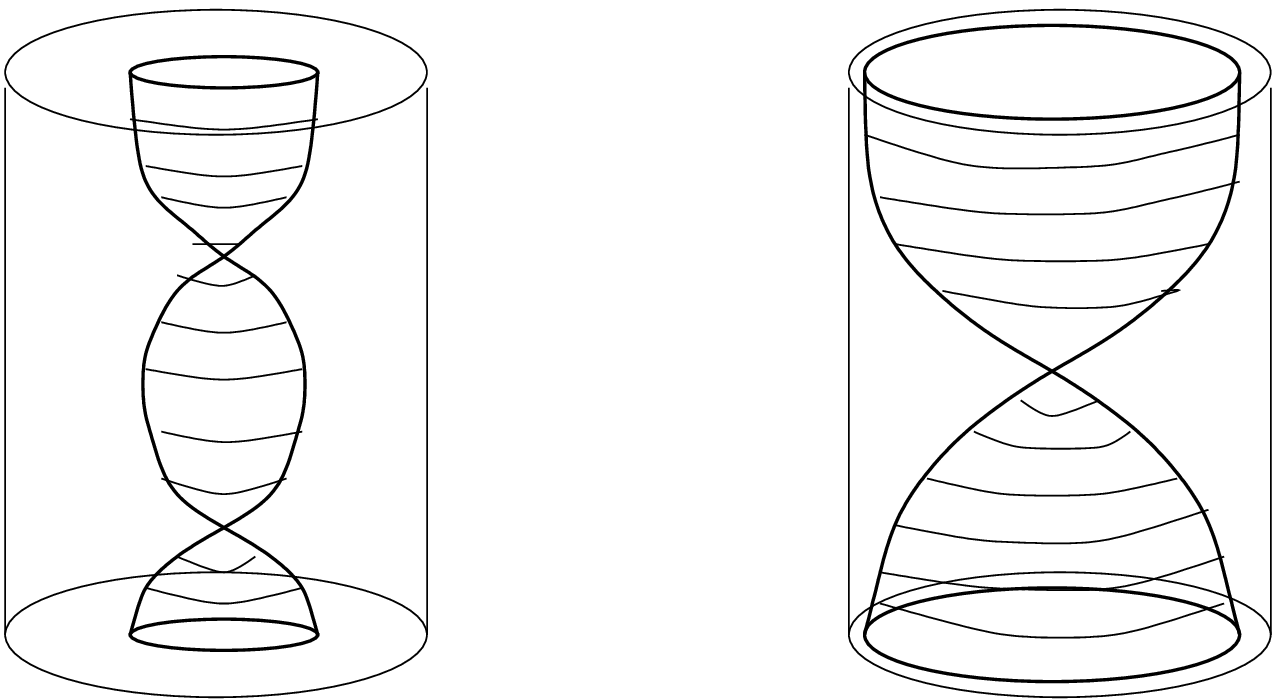}{12}{Cordes ferm{\'e}es courtes et longues dans \AAA }

Consid{\'e}rons maintenant la th{\'e}orie des cordes ferm{\'e}es dans
\SL, en suivant \cite{moi}. L'essentiel de ce que nous avons dit dans la partie
\ref{soussectWZW} s'applique. Le niveau $k$ n'est pas quantifi{\'e} car
$\pi_3(SL(2,\R)$ est trivial. La propri{\'e}t{\'e} essentielle de la
factorisation des solutions de cordes, eq. (\ref{solWZW}), est
toujours valable. Le mod{\`e}le poss{\`e}de la sym{\'e}trie (\ref{symWZW}). 
Si l'on veut aller au-del{\`a} de la limite semi-classique et
construire le spectre des cordes ferm{\'e}es, il est naturel de
consid{\'e}rer les repr{\'e}sentations de plus haut poids de 
l'alg{\`e}bre de Lie affine \asl\ correspondant aux repr{\'e}sentations de
\sl\ qui apparaissent dans $\Fun(SL(2,\R))$. En effet, c'est ainsi que
l'on construit le spectre du mod{\`e}le de Wess-Zumino-Witten associ{\'e}
{\`a} un groupe compact (\ref{specWZW}). Dans le cas de \SL, les deux
types de repr{\'e}sentations de \sl\ permettent de construire des
repr{\'e}sentations de plus haut poids de \asl\ toujours appel{\'e}es
continues et discr{\`e}tes. Les repr{\'e}sentations continues
correspondent {\`a} des cordes ferm{\'e}es qui atteignent le bord d'\AAA,
appel{\'e}es \de{cordes longues}.
Les repr{\'e}sentations discr{\`e}tes d{\'e}crivent des cordes 
localis{\'e}es pr{\`e}s
du centre, {\`a} distance finie, appel{\'e}es \de{cordes courtes} (voir la
Figure \ref{asd3fermees}). La
disctinction entre ces deux types d'{\'e}tats, s{\'e}par{\'e}s par la borne
de Breitenlohner-Freedman, est donc identique {\`a} celle que nous avons
d{\'e}crite dans le cas des {\'e}tats d'une particule ponctuelle d{\'e}crite
par un champ scalaire. Elle correspond toujours {\`a} la diff{\'e}rence
entre les {\'e}tats asymptotiques, qui ont assez d'{\'e}nergie pour
s'{\'e}chapper {\`a} l'infini, et les {\'e}tats li{\'e}s.

Cependant, de m{\^e}me que le crit{\`e}re de masse positive n'{\'e}tait pas
adapt{\'e} {\`a} la physique dans \AAA, le crit{\`e}re pour une
repr{\'e}sentation de \asl\ d'{\^e}tre de plus haut poids se r{\'e}v{\`e}le
{\^e}tre trop restrictif, et ne permet pas de construire le
spectre de la th{\'e}orie des cordes dans \SL (voir par exemple
\cite{mariosopen}).  
Cela est d{\^u} au fait que
dans le cas de \SL, les sym{\'e}tries du mod{\'e}le eq. (\ref{symWZW})
forment un groupe non connexe, qui n'est donc pas engendr{\'e} par sa
seule alg{\`e}bre de Lie \asl. Pour acc{\'e}der aux composantes connexes
autres que celles de l'identit{\'e}, on peut utiliser un sous-groupe
isomorphe {\`a} $\Z$ que l'on appelle le flot spectral \cite{moi}. Si
$J$ est un g{\'e}n{\'e}rateur de \lieg\ de genre temps ($\Tr J^2 < 0$) et 
tel que $\exp ( 4\pi J)=1$, on peut r{\'e}aliser le
flot spectral comme les transformations~:
\bea
g(x_+,x_-)\ \rar \ e ^{ wx_+J} \ g(x_+,x_-) \ e ^{ wx_-J} \ , \ w\in\Z.
\label{flotspec}
\eea
Le flot spectral agit sur l'alg{\`e}bre de Lie affine \alieg\ comme un
groupe d'automorphismes ext{\'e}rieurs. {\'E}tant donn{\'e}e une
repr{\'e}sentation $\hR$ de \alieg, il faut donc pour r{\'e}aliser la
sym{\'e}trie du mod{\`e}le inclure dans le spectre ses images par le flot
spectral $\hR^w,w\in \Z$. La conjecture qui en r{\'e}sulte pour le
spectre des cordes ferm{\'e}es dans \SL\ est 
\bea
Spec(SL(2,\R)) \simeq \bigoplus_{ 
R\in Rep_k(SL(2,\R))\ , \ w\in \Z }
\hR^w \otimes \hR^\dagger{}^w,
\label{specsl2}
\eea
o{\`u} $Rep_k(SL(2,\R))$ comporte toutes les repr{\'e}sentations
continues, et les repr{\'e}sentations discr{\`e}tes v{\'e}rifiant la borne
d'unitarit{\'e} renforc{\'e}e,
\bea
-\frac{k-1}{2}<j<-\half.
\label{unitrenf}
\eea
Entre autres v{\'e}rifications, le spectre (\ref{specsl2}) est compatible avec
l'invariance modulaire et des calculs de fonction de partition
thermique dans l'\AAA\ Euclidien \cite{moi,moii,moiii}.

\section{ D-branes sym{\'e}triques dans les groupes compacts
\label{BIbranes}
 }

La th{\'e}orie des cordes dans le groupe $G$ s'exprime, comme nous
l'avons vu, en termes des courants droit et gauche
$J$ et $\bJ$. Du point de vue des
cordes ouvertes, une D-brane est d{\'e}finie par les conditions aux bords
de la corde {\`a} ses bouts $\sigma=0,\pi$. Nous allons dans cette
section consid{\'e}rer les conditions au bord suivantes~:
\bea
J=\bJ.
\label{branesym}
\eea
Il s'agit de conditions de \flqq collage\frqq\ du type de celles que
nous avons consid{\'e}r{\'e}es au chapitre pr{\'e}c{\'e}dent (\ref{bordouv}), qui sont
d{\'e}finies par une application lin{\'e}aire reliant les secteurs gauche
et droit. 
L'int{\'e}r{\^e}t des conditions au bord (\ref{branesym}) 
a {\'e}t{\'e} soulign{\'e} par Alekseev et
Schomerus \cite{aavs}. Les D-branes correspondantes s'appellent les
D-branes sym{\'e}triques, en effet ces conditions au bord 
 entre les secteurs droit et gauche pr{\'e}servent la sous-alg{\`e}bre
diagonale \alieg\ de l'alg{\'e}bre des courants ${\bf \hat{g}}\times
{\bf \hat{g}}$. Nous consid{\`e}rerons aussi une g{\'e}n{\'e}ralisation
\cite{fffs} de
ces conditions au bord, construite {\`a} l'aide d'un automorphisme
ext{\'e}rieur quelconque $\o$ de l'alg{\`e}bre de Lie \lieg~:
\bea
J=\o(\bJ).
\label{branetw}
\eea
On pourrait aussi consid{\'e}rer un automorphisme int{\'e}rieur $\o$, mais
la D-brane qui en r{\'e}sulterait serait simplement l'image par rotation
d'une D-brane sym{\'e}trique, comme on pourra le v{\'e}rifier {\`a} l'aide de
(\ref{omegaconj}).

Les D-branes ainsi d{\'e}finies dans les groupes
compacts ont la propri{\'e}t{\'e} d'{\^e}tre enti{\`e}rement d{\'e}termin{\'e}es
par leurs conditions de collage. Autrement dit, elles ob{\'e}issent {\`a}
la th{\'e}orie de Cardy. Nous expliquerons cette construction dans la
section (\ref{sectbranecompact}).

Cela permet d'esp{\'e}rer que leurs propri{\'e}t{\'e}s soient en grande
partie accessibles par des m{\'e}thodes g{\'e}om{\'e}triques utilisant comme
seule d{\'e}finition de la brane sa condition de collage. Nous verrons
que ces m{\'e}thodes donnent des r{\'e}sultats exacts, alors que leur
domaine de validit{\'e} ne s'{\'e}tend a priori qu'{\`a} la limite
semi-classique $k\rar \infty$. 

\subsection{ D-branes et classes de conjugaison }

D'abord, nous allons utiliser les conditions de collage
(\ref{branesym},\ref{branetw}) pour d{\'e}terminer la g{\'e}om{\'e}trie des
D-branes correspondantes. Dans la limite semi-classique, une D-brane
peut {\^e}tre vue comme une sous-vari{\'e}t{\'e} de $G$. Comme la surface d'univers de
la corde aboutit sur la D-brane, les vecteurs $\p_\tau g
(\sigma=0,\pi)$ doivent {\^e}tre tangents {\`a} la D-branes. La condition
$J=\o(\bJ)$ se r{\'e}{\'e}crit, en utilisant la d{\'e}finition des courants
(\ref{courWZW}),
\footnote{
Nous notons Ad$_g$ l'action adjointe d'un {\'e}l{\'e}ment $g$ de $G$ sur
l'alg{\`e}bre de Lie, $\Ad(v)=g\- v g\- $.
}
\bea
(1-\oAd)g \- \p_\sigma g = (1+\oAd)g \- \p_\tau g.
\label{branetw2}
\eea
Cela implique que le vecteur $g\- \p_\tau g$ appartienne {\`a} l'image de
$(1-\oAd)$. Dans le cas $J=\bJ$ (c'est-{\`a}-dire $\o=\id$), cela 
revient {\`a} dire que $\p_\tau g$ doit {\^e}tre
tangent {\`a} la classe de conjugaison qui passe par le point $g$. Dans
le cas plus g{\'e}n{\'e}ral o{\`u} $\o$ est non trivial, nous appellerons
\de{classe de $\o$-conjugaison} la vari{\'e}t{\'e} qui g{\'e}n{\'e}ralise la
notion de classe de conjugaison, et dont l'espace tangent est l'image
de $(1-\oAd)$. La classe de $\o$-conjugaison d'un point $g\in G$ est~:
\bea
{\cal C}^\o(g)=\{ hg\o(h)\- , h\in G\}.
\label{omegaconj}
\eea
Comme la $\o$-conjugaison est une action de $G$ sur lui-m{\^e}me, les
classes de $\o$-conjugaison forment une partition de $G$ en
sous-vari{\'e}t{\'e}s de dimensions variables, tout comme les classes de
conjugaison. Cependant, la classe de conjugaison ponctuelle de
l'identit{\'e} n'a pas d'{\'e}quivalent pour la $\o$-conjugaison.

\bs

D{\'e}crivons maintenant la g{\'e}om{\'e}trie des classes de conjugaison
d'un groupe de Lie compact connexe semi-simple $G$. Soit $T$ un tore maximal,
$T\simeq U(1)^r$ o{\`u} $r$ est le rang du groupe $G$. Rappelons que le
\de{normalisateur} $N(T)$ du tore $T$ est le sous-groupe des
{\'e}l{\'e}ments de $G$ qui, par conjugaison, laissent $T$ globalement
invariant,
\bea
N(T)=\{ g\in G, gTg\- = T \}.
\label{normtore}
\eea
Le centralisateur $C(T)$ du tore $T$ est l'ensemble des {\'e}l{\'e}ments de $G$
qui commutent avec tout {\'e}l{\'e}ment de $T$, et dans le cas d'un groupe
compact semi-simple c'est $T$ lui-m{\^e}me~:
\bea
C(T)=\{ g\in G, \forall t\in T\ gtg\- =t \}=T.
\label{centtore}
\eea
Naturellement, $C(T)$ est un sous-groupe de $N(T)$. Mais il s'agit de
plus d'un sous-groupe distingu{\'e} d'ordre fini, et le groupe quotient
est le \de{groupe de Weyl} $W$, qui ne d{\'e}pend pas du choix du tore
$T$ puisque tous les tores maximaux sont conjugu{\'e}s~:
\bea
W=W(T)=N(T)/C(T)=N(T)/T.
\label{groupeweyl}
\eea
Le groupe $W$ agit naturellement sur $T$, et l'ensemble quotient $T/W$
param{\'e}trise les classes de conjugaison de $G$. Plus pr{\'e}cis{\'e}ment, 
l'application
\be \begin{array}{ll}
q\,: \quad & G/T \times T \rar G\\[.2em]
           & q(hT, t) := hth^{-1} 
\end{array}
\label{paracjg}
\ee
est une surjection. Consid{\'e}rons maintenant l'ensemble $G_r$ des
{\'e}l{\'e}ments r{\'e}guliers de $G$, c'est-{\`a}-dire ceux qui
n'appartiennent qu'{\`a} un seul tore maximal. Cet ensemble 
$G_r$ est dense dans
$G$. Soit $T_r=G_r\cap T$ l'ensemble des {\'e}l{\'e}ments r{\'e}guliers de
$T$, alors l'ensemble des classes de conjugaison r{\'e}guli{\`e}res
(incluses dans $G_r$) est param{\'e}tr{\'e} par $T_r/W$ au moyen de
l'application $q$ ci-dessus. Toute classe de conjugaison r{\'e}guli{\`e}re
est diff{\'e}omorphe {\`a} $G/T$ en tant que $G$-module, et 
rencontre $T_r$ en $|W|$ points qui forment une orbite du groupe de
Weyl $W$. Nous nous restreindrons aux classes de conjugaison
r{\'e}guli{\`e}res dans la suite.

Apr{\`e}s avoir rappel{\'e} ces faits standard sur les
classes de conjugaison, int{\'e}ressons-nous aux classes de
$\o$-conjugaison, qui d{\'e}crivent elles aussi la g{\'e}om{\'e}trie de
D-branes du groupe $G$. Ces faits, un peu moins standard, sont
expos{\'e}s dans \cite{fffs}. Il existe toujours au moins un tore
maximal $T$ globalement invariant par $\o$, les classes de
$\o$-conjugaison sont essentiellement param{\'e}tr{\'e}es par l'ensemble
$T^\o$ 
des points fixes de l'action de $\o$ sur $T$. Plus pr{\'e}cis{\'e}ment, il
faut tenir compte de l'action du groupe de Weyl $W$ (en fait,
de ses {\'e}l{\'e}ments qui commutent avec l'action de $\o$), et en plus
se restreindre {\`a} la composante connexe de l'identit{\'e} $T^\o_0$ de
$T^\o$. On g{\'e}n{\'e}ralise {\'e}galement
la notion de classe de conjugaison r{\'e}guli{\`e}re, et toutes les
classes de $\o$-conjugaison r{\'e}guli{\`e}res sont diff{\'e}omorphes {\`a}
$G/T^\o_0$ et ont donc la m{\^e}me dimension. Notons que cette dimension
est n{\'e}cessairement sup{\'e}rieure {\`a} celle des classes de conjugaison
r{\'e}guli{\`e}res, car $T^\o$ est toujours plus petit que $T$, si $\o$ est un
automorphisme ext{\'e}rieur comme nous l'avons suppos{\'e}. 

\bs

Enfin, nous aurons besoin de conna{\^\i}tre les 
espaces tangents aux classes de
$\o$-conjugaison. 
La propri{\'e}t{\'e} essentielle que nous utiliserons est qu'une classe de
$\o$-conjugaison r{\'e}guli{\`e}re est orthogonale en chacun de ses points au tore
maximal invariant $T^\o_0$ qui y passe. Ramen{\'e}e {\`a} l'alg{\`e}bre de
Lie par translation, cette propri{\'e}t{\'e} s'{\'e}crit au moyen de
l'op{\'e}rateur lin{\'e}aire $\oAd$,
\bea
\lieg \simeq \ke \stackrel{\perp}{\oplus} \im .
\label{orthog}
\eea
En effet, les espaces $g\, \ke$\ et\ $g\, \im$\ sont respectivement 
les espaces tangents au
tore $T^\o_0$ passant par $g$ et {\`a} la classe de $\o$-conjugaison
passant par $g$. Pour s'en apercevoir, on peut utiliser l'application
tangente de 
$q(\cdot,t)=q_t$, o{\`u} l'on utilise la m{\^e}me notation $q$ qu'en
(\ref{paracjg}) pour d{\'e}signer
\be \begin{array}{ll}
q\,: \quad & G/T^\o_0 \times T^\o_0 \rar G\\[.2em]
           & q(hT, t) := ht\o(h)^{-1} 
\end{array}.
\label{paratw}
\ee
On a
en effet $dq_t(dh h \- )=(1-\oAd)dh h \- =g\- dg$ si $g=ht\o(h) \- $. 

Nous avons donc d{\'e}crit la g{\'e}om{\'e}trie de nos D-branes~: elles sont
diff{\'e}omorphes {\`a} $G/T$ et $G/T^\o_0$ respectivement. Cependant une
D-brane n'est pas une simple sous-vari{\'e}t{\'e} de $G$, mais est munie
d'un champ de jauge $F$. Et la topologie non-triviale de $G/T$ et de
$G/T^\o_0$ a des cons{\'e}quences en termes de quantification de la
position des D-branes, de m{\^e}me que la topologie non-triviale de $G$
conduit {\`a} la quantification du niveau $k$. C'est ce que nous allons
maintenant discuter.

\subsection{ Topologie des D-branes et quantification 
\label{topoquantif}}

\Fig{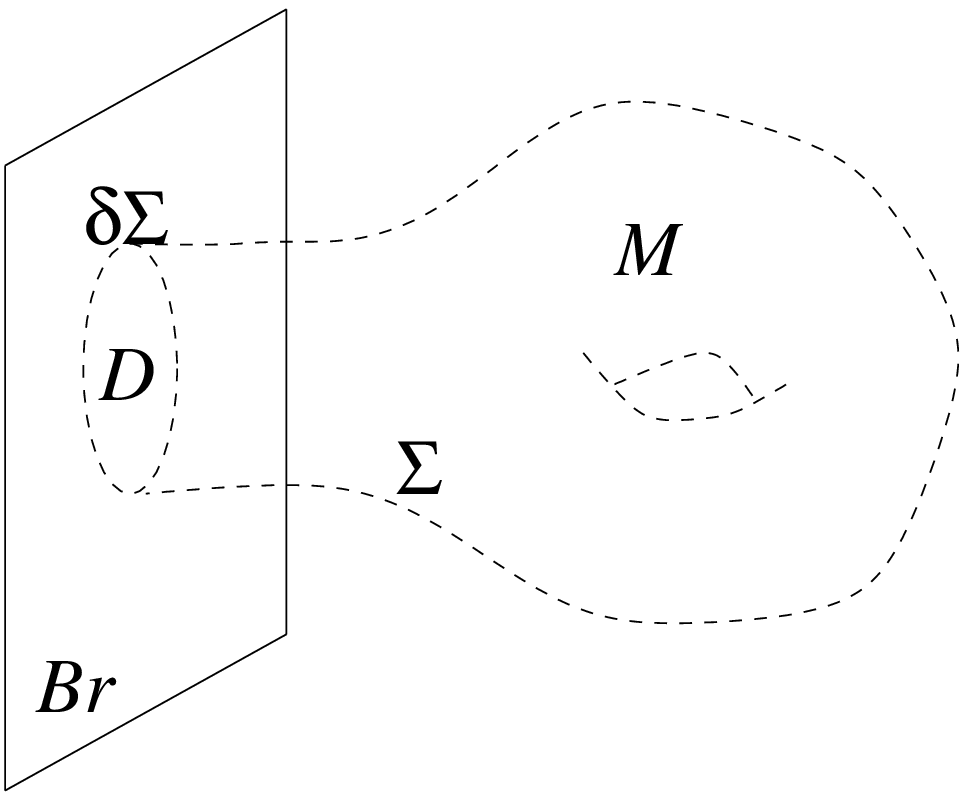}{8}{Une corde ouverte $\Sigma$ aboutissant sur une D-brane $\B$}

Dans l'action d'une corde ouverte de surface d'univers $\Sigma$,
suppos{\'e}e aboutir sur une D-brane $\B$ (c'est-{\`a}-dire $\p\Sigma
\subset \B$),
consid{\'e}rons les termes
topologiques,
\footnote{ Ici comme dans la suite du chapitre, nous omettons le
  facteur $2\pi\ap$ qui accompagne habituellement les champs $A$ et
  $F=dA$.} 
\bea
S^{\rm topo} (\Sigma, A)= - \int_\Sigma B  + \int_{\p \Sigma } A.
\label{acttopo}
\eea
Ces termes doivent pouvoir se r{\'e}{\'e}crire en fonction des seules quantit{\'e}s
invariantes de jauge $H$ et $\o= \hB+ dA=\hB + F$. Nous avons d{\'e}j{\`a}
d{\^u} introduire dans le cas des cordes ferm{\'e}es la vari{\'e}t{\'e} ${\cal
  M}$ dont le bord {\'e}tait la surface d'univers $\Sigma$. Pour une
corde ouverte, $\Sigma$ est ouverte et ne saurait donc {\^e}tre un
bord~; cependant on peut "fermer" $\Sigma$ au moyen d'une 
sous-vari{\'e}t{\'e} $D$ de dimension deux de la D-brane $\B$, telle que
$\p(\Sigma \cup D)=\emptyset$ (voir la Figure \ref{antif}). 
Alors il existe une vari{\'e}t{\'e} ${\cal M}$
de dimension trois telle que $\p {\cal M}=\Sigma \cup D$, et on a 
\bea
S^{\rm topo} (\Sigma, A)&=& -\int_{\Sigma\cup D}B +\int_D
B-\int_{\p D } A  \nn
\\
&=& \int_{\cal M}H +\int_D \o.
\label{topoinv}
\eea
Cette derni{\`e}re expression est la v{\'e}ritable d{\'e}finition physique
des termes topologiques de l'action des cordes ouvertes, et la
formulation originale $S^{\rm topo}$ n'a en fait qu'une valeur
heuristique. 
Naturellement, nous devons maintenant nous occuper des ambigu{\"\i}t{\'e}s
dans le choix de ${\cal M}$ et $D$. Une fois $D$ fix{\'e}, le choix 
de ${\cal M}$ est param{\'e}tr{\'e} par $\pi_3(G)$ comme dans le cas des
cordes ferm{\'e}es, et n'influe pas sur $\exp iS^{\rm topo}$ pourvu que
le niveau $k$ soit quantifi{\'e}. Le choix de $D$, lui, est
param{\'e}tr{\'e} par $\pi_2(\B)$. Or, dans le cas o{\`u} $\B$ est une classe de
conjugaison d'un groupe compact connexe de rang $r$, on a 
\bea
\pi_2(\B)\simeq\pi_2(G/T)\simeq \Z^r.
\label{pideux}
\eea
Donc, la D-brane doit respecter $r$ conditions de quantification, qui
vont restreindre ses positions autoris{\'e}es {\`a} une partie discr{\`e}te
de $T_r/W$. Pour savoir laquelle, il suffit en principe de 
 conna{\^\i}tre 
le champ $\o$ sur la D-brane, mais il est beaucoup plus commode
de travailler avec le champ $F$. En effet, le choix de $D$ affecte les
deux termes de l'action eq. (\ref{topoinv}), alors qu'il n'affecte que
le terme
\footnote{Le champ $A$ n'est pas toujours bien d{\'e}fini, $F$ pouvant
  se contenter d'{\^e}tre ferm{\'e}e.}
 $\int_{\p \Sigma} A=-\int_D F$ de l'action
 eq. (\ref{acttopo}). Cependant, dans notre calcul de $S^{\rm topo}$
 en fonction des quantit{\'e}s invariantes de jauge, nous avons
 implicitement suppos{\'e} que le champ $B$ est r{\'e}gulier dans ${\cal M}$. 
En g{\'e}n{\'e}ral, la forme $H$ est ferm{\'e}e mais pas
 n{\'e}c{\'e}ssairement exacte, et on doit tenir compte de termes du type 
\bea
\int_{\cal M} H = -\int_{\Sigma \cup D} B + \int_{ {\cal M} \cap S(B) }
B\ ,
\label{actsing}
\eea
o{\`u} $S(B)$ d{\'e}signe une surface bidimensionnelle entourant les
singularit{\'e}s de $B$. {\`A} l'ambiguit{\'e} $\int_{S^2}F$ de l'action
(li{\'e}e {\`a} un g{\'e}n{\'e}rateur $S^2$ de $\pi_2(\B)$), il faut donc
ajouter l'ambiguit{\'e} {\'e}ventuelle sur le terme $\int_{ M \cap S(B)}B$,
qui vaut l'int{\'e}grale de $B$ sur la portion de $S(B)$ intercept{\'e}e
par $S^2$. 

\bs 

Nous allons illustrer ces conditions de quantification dans l'exemple
canonique des D-branes sym{\'e}triques du groupe \SU, en consid{\'e}rant
plusieurs choix de jauge pour le champ $B$. On peut d'abord utiliser
les coordonn{\'e}es cylindriques, comme dans \cite{bds},
 \bea
& ds^2= d\Psi ^2+\sin^2\Psi(d\Theta ^2+\sin ^2\Theta d\varphi ^2) \\
& B= (B_0 + \Psi -\half \sin 2\Psi)\sin\Theta d\Theta\wedge d\varphi
\label{su2cyl}
\eea
Nous avons cependant ajout{\'e} une constante $B_0$, la valeur du champ
$B$ au p{\^o}le $\Psi=0$. Ces coordonn{\'e}es sont adapt{\'e}es au
traitement des classes de conjugaison d'{\'e}quations $\Psi=\Psi_0$, avec un
champ 
\bea
F=-(\Psi_0+B_0) \sin\Theta d\Theta\wedge d\varphi.
\label{su2cylF}
\eea
Ces D-branes en forme de classes de conjugaison sont des sph{\`e}res
$S^2$, et donnent donc lieu {\`a} une quantification de l'int{\'e}grale de
$F$ sur leur volume d'univers, c'est-{\`a}-dire de
$(\Psi_0+B_0)$. Cependant les positions autoris{\'e}es par cette
quantification ne doivent pas d{\'e}pendre de $B_0$, et c'est le cas si
l'on prend en compte la singularit{\'e} du champ $B$ au p{\^o}le $\Psi=0$,
qui doit {\^e}tre explicitement soustraite comme nous l'avons
expliqu{\'e}. Il s'agit plus pr{\'e}cis{\'e}ment de l'int{\'e}grale de $B$ sur
une deux-sph{\`e}re infinit{\'e}simale $\Psi=\e$, qui tend vers $B_0$
quand $\e\rar 0$. Ainsi, quel que soit la jauge $B_0$ choisie, on
obtient la quantification
\bea
\Psi_0= \frac{\pi n }{k},\ n\in \Z.
\label{quantsu2}
\eea
Traitons le m{\^e}me exemple en coordonn{\'e}es d'Euler,
\footnote{ Le changement de variables entre les deux syst{\`e}mes de
  coordonn{\'e}es est 
\ber
\cos \Psi &=&\sin \a \cos\b \\
\sin \Psi \sin \Theta &=& \cos \a, \eer
cependant nos choix de jauge respectifs pour le champ $B$ ne sont pas
reli{\'e}s par ces transformations.
}
\bea
& ds^2= d\a^2+\sin^2\a d\b ^2+\cos^2\a d\varphi^2 \\
& B=(\cos ^2\a-B_0)
d\varphi\wedge d\b .
\label{su2euler}
\eea 
Dans ces coordonn{\'e}es, et avec ces choix de jauge, le champ 
$B$ n'est pas singulier seulement en un point
mais g{\'e}n{\'e}ralement en deux grands cercles disjoints
$\a=0,\halfpi$. On consid{\`e}re les m{\^e}mes D-branes qu'auparavant,
d'{\'e}quations 
\bea 
\sin \a \cos \b = \sin \a_{min} = \cos \Psi_0 \ \ , \ \ \db=\frac{\p
  \beta}{\p\a}
=\frac{\sin
  \a_{min}\cos\a}{ \sin\a\sqrt{\sin^2\a-\sin^2\a_{min}}}. 
\label{braneeuler}
\eea

\Fig{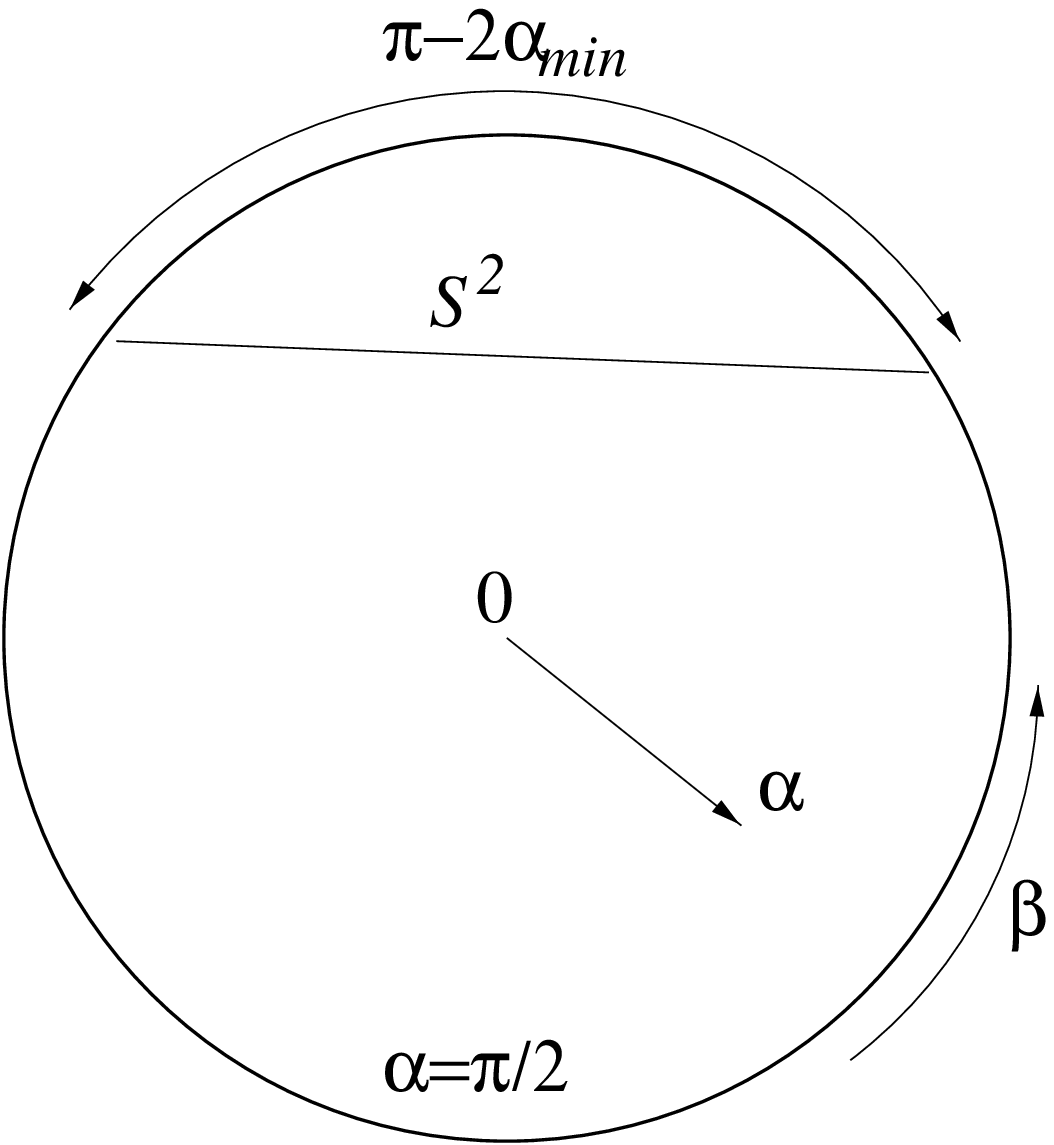}{8}{ Une D-brane $S^2$ dans \SU, en projection sur le
  plan $(\alpha,\beta)$ }

Dans la jauge $B_0=0$, ces D-branes n'intersectent jamais le cercle $\a=0$
des singularit{\'e}s du champ $B$. La condition de quantification peut
s'exprimer en termes du seul champ $F$, qui vaut alors
\bea
B_0=0\ : \ F=-\db d\varphi\wedge d\b .
\label{Fsu2reg}
\eea
On v{\'e}rifie que la quantification de $\int_{S^2}F$, o{\`u} $S^2$ est
notre D-brane sph{\'e}rique, {\'e}quivaut {\`a} l'{\'e}quation
(\ref{quantsu2}). Dans la jauge $B_0=1$, en revanche, chaque D-brane
intercepte un angle $\pi-2\a_{min}$ du cercle
$\a=\halfpi$ des singularit{\'e}s de $B$ (voir la Figure \ref{as2brane}). 
On trouve par ailleurs que le
champ $F$ est nul,
\bea
B_0 = -1\ ~: \ F=0 .
\label{Fsu2sing}
\eea
Donc la quantification de la position de la D-brane doit provenir
exclusivement de la contribution des termes du type
(\ref{actsing}). L'int{\'e}grale de $B$ autour de la totalit{\'e} du
cercle singulier vaut naturellement $k$ (puisque $\int_{\SU }H=k$),
donc l'int{\'e}grale de $B$ sur une surface {\`a} deux dimensions
autour de la portion de ce cercle intercept{\'e}e par la D-brane vaut
\bea
k\times \frac{\pi-2\alpha_{min}}{2\pi} = \frac{k\Psi_0}{\pi} \in \Z.
\label{quantsu2B}
\eea
Ainsi nous avons v{\'e}rifi{\'e} la concordance des conditions de
quantification de la position des D-branes sym{\'e}triques dans \SU, obtenues
dans diff{\'e}rents syst{\`e}mes de coordonn{\'e}es et pour diff{\'e}rents
choix de jauge du champ $B$. 

\bs 

Revenons maintenant {\`a} nos D-branes dans le groupe compact $G$. Nous
voulons g{\'e}n{\'e}raliser l'expression explicite eq. (\ref{su2cylF}) du
champ $F$, mais nous ne voulons pas choisir un syst{\`e}me
de coordonn{\'e}es pr{\'e}cis sur un groupe compact quelconque. Nous
devons donc comprendre la structure alg{\'e}brique du champ $F$. Il
s'agit d'une deux-forme sur l'espace tangent {\`a} une classe de
$\o$-conjugaison, qui est diff{\'e}omorphe {\`a} $G/T^\o_0$. Dans un groupe
de Lie, l'espace tangent en un point $g$ est isomorphe {\`a} l'alg{\`e}bre
de Lie par translation {\`a} gauche par $g\- $. Nous avons d{\'e}j{\`a}
utilis{\'e} cette propri{\'e}t{\'e} pour d{\'e}finir les champs de fond des
cordes ferm{\'e}es, c'est-{\`a}-dire la m{\'e}trique et le champ $H$,
eq. (\ref{actWZW}). Nous pouvons aussi {\'e}crire la deux-forme
$\om=\hB+\F$ sur une classe de $\o$-conjugaison, qui g{\'e}n{\'e}ralise la
deux-forme d'Alekseev-Schomerus (d{\'e}finie dans \cite{aavs} sur les
classes de conjugaison) \cite{stand0,brs},
\bea
\om _g(u,v) = \llangle g\- u,\frac{1+\oAd}{1-\oAd} g\- v \rrangle.
\label{omegaomega}
\eea 
Justifions cette formule. D'abord, elle a un sens si $u,v$ appartiennent
{\`a} l'espace tangent $g\, \im$\  {\`a} la classe de $\o$-conjugaison qui
passe par $g$.
Cela permet de trouver un ant{\'e}c{\'e}dent de $g \-  v$ par
$(1-\oAd)$, ce qui explique que cet op{\'e}rateur apparaisse au
d{\'e}nominateur, et cela nous assure que $\om$ ne d{\'e}pend pas du
choix de cet ant{\'e}c{\'e}dent (car deux choix diff{\`e}rent par un
{\'e}l{\'e}ment de $\ke$, qui est orthogonal {\`a} $\im$\ donc {\`a} $g \-
u$). Ensuite, elle provient naturellement des conditions de collage
(\ref{branetw}). Enfin, on v{\'e}rifie que $d\om =\hat{H}$. Attention~!
cette identit{\'e} n'est valable qu'en ce qui concerne les formes
diff{\'e}rentielles vivant sur la D-brane. Certes, il est tentant de
prolonger $\om$ en une deux-forme sur $G_r$, en d{\'e}finissant son
action sur des vecteurs orthogonaux {\`a} notre D-brane comme
triviale. Cependant, la deux-forme ainsi obtenue n'a pas pour
diff{\'e}rentielle $H$. Nous allons justement construire une nouvelle
deux-forme $F$ ferm{\'e}e sur la D-brane, telle que $d(\om-\F)=H$ si
l'on prolonge $\om$ et $F$ comme je viens de le d{\'e}crire. Cela nous
fournira simultan{\'e}ment un choix de jauge $(\om-\F)$ pour le champ
$B$, et le champ de jauge $F$ sur la D-brane qui nous permettra
d'{\'e}valuer les conditions de quantification. 

\bs

La construction de $F$ s'inspire de la m{\'e}thode des orbites de
Kirillov \cite{kirillov}. 
Il s'agit des orbites de l'action coadjointe de $G$ sur le dual de
son alg{\`e}bre de Lie \lieg. Elles sont munies d'une structure
symplectique naturelle, et notre deux-forme $F$ est une
g{\'e}n{\'e}ralisation de la forme symplectique
correspondante. Explicitement, {\'e}tant donn{\'e}e une classe de
$\o$-conjugaison $\B$ param{\'e}tr{\'e}e par un {\'e}l{\'e}ment $t$ de $T^\o_0$, 
on d{\'e}finit la deux-forme $F_t$ sur $\B$ par \cite{brs}
\bea
\F _t(u,v)=2\llangle \exp \- t, [h\- u, h\- v] \rrangle.
\label{ffieldWZW}
\eea
Il faut d'abord remarquer que cette formule est {\'e}crite {\`a} l'aide
de vecteurs $u,v$ tangents {\`a} $G/T^\o_0$, c'est-{\`a}-dire 
d'ant{\'e}c{\'e}dents par $dq_t$ de vecteurs tangents {\`a} $\B$. Autrement
dit, une autre notation pour la forme $F$ est 
\bea
\F _g = 2\llangle \exp \- g, \left[ \frac{1}{1-\oAd} g\- dg ,
\frac{1}{1-\oAd} g\- dg \right] \rrangle,
\label{fieldWZW}
\eea
avec $g=ht\o(h)\- $. 
On v{\'e}rifie que $F _t $ ne d{\'e}pend pas du choix des ant{\'e}c{\'e}dents
en question. Une autre ambiguit{\'e} provient du "logarithme" $\exp \- $
utilis{\'e}. Diff{\'e}rentes d{\'e}finitions du logarithme correspondent
bien {\`a} diff{\'e}rentes valeurs de la forme $F$, mais la condition de
quantification que nous allons en d{\'e}duire ne sera pas
affect{\'e}e. Autrement dit, ce choix correspond {\`a} un choix de jauge
du champ $B=\o - \F$ qui ne modifie pas la quantification (de tels
choix de jauge sont reli{\'e}s par les \flqq grandes transformations de
jauge\frqq ).
Pour g{\'e}n{\'e}raliser la forme $F$ de
\SU, et par simplicit{\'e}, nous pouvons lever cette ambiguite en supposant
 $\exp \- (e)=0$, o{\`u} $e$
est l'unit{\'e} du groupe $G$.

Non seulement notre deux-forme $F_t $ sur $\B$ est ferm{\'e}e, mais le champ $B$
correspondant $B=\om-\F$, d{\'e}fini en supposant $B(g\ \ke,*)=0$,
v{\'e}rifie $dB=H$ sur $G_r$ \cite{brs}. Nous avons donc explicitement
construit le champ $B$ et la forme $F$, et on peut montrer qu'avec une
telle valeur de $F$ les classes de $\o$-conjugaison sont des solutions
des {\'e}quations du mouvement de l'action de Born-Infeld. 
 Nous pouvons maintenant imposer la
condition de quantification pour en d{\'e}duire les positions possibles
des D-branes. Pour cela, il faut construire explicitement des
g{\'e}n{\'e}rateurs de $\pi_2(\B)$. 

Traitons d'abord le cas des classes de conjugaison.
{\`A} chaque racine simple $\a$ nous associons naturellement un sous-goupe de
$G$ isomorphe {\`a} \SU\ ou $SO(3)$, d{\'e}fini par $\exp ({\rm Vect}
  (H^\a,E ^{\pm \a}))$ (o{\`u} $H^\a=\a ^iH_i,E ^{\pm \a}$ sont des
  {\'e}l{\'e}ments de \lieg\ tels que $[H_i,E ^\a]=\a ^i E ^\a,[E ^\a,E
  ^{-\a}]=H ^\a $). L'intersection de ce sous-groupe avec la
  D-brane $\B$ est une sph{\`e}re $S^2_\a$, et on a $\int_{S^2_\a}F_t=\a
  (\exp \- t)$. Appliqu{\'e} {\`a} toutes les racines simples $\a$, cet
  argument donne le r{\'e}seau des positions physiques $t$ des D-branes,
\bea
\exp \- t\ \in \ L=\frac{2\pi}{k}L^w.
\label{quantift}
\eea
Ici $L^w$ est le r{\'e}seau des poids de l'alg{\`e}bre de Lie \lieg\
suppos{\'e}e simple, et on a utilis{\'e} la forme de Killing pour identifier
$\exp \- t$ avec un {\'e}l{\'e}ment de ce r{\'e}seau. 
Donc l'ensemble des D-branes est param{\'e}tr{\'e} par
l'ensemble fini $L/\hat{W}$, o{\`u} $\hat{W}=W\ltimes {\rm Ker}(\exp)$ est
le groupe de Weyl {\'e}tendu.

Si maintenant nous consid{\'e}rons un automorphisme $\o$ non trivial,
nous devons nous restreindre aux sous-groupes \SU\ dont le
g{\'e}n{\'e}rateur de Cartan est invariant par $\o$. On aboutit {\`a} la
condition de quantification
\bea
\exp \- t\ \in \ M=\frac{2\pi}{k}M^w_\o,
\label{quantifo}
\eea
o{\`u} $M^w_\o$ est le r{\'e}seau des \de{ poids fractionnaires 
  sym{\'e}triques } \cite{fffs}. L'ensemble des D-branes est toujours
param{\'e}tr{\'e} par un emsemble fini $ M/(W^\o \ltimes {\rm Ker}(\exp))$,
o{\`u} $W^\o$ est le sous-groupe de $W$ des {\'e}l{\'e}ments qui commutent
avec $\o$. 
 
\subsection{ Propri{\'e}t{\'e}s physiques et stabilisation par le flux }

Nous connaissons maintenant les solutions de l'action de Born-Infeld
correspondant aux D-branes en forme de classes de $\o$-conjugaison
dans tout groupe compact $G$. Nous savons que la quantification du
flux du champ $F$ {\`a} travers les deux-cycles des D-branes r{\'e}duit
cet ensemble de solutions {\`a} un nombre fini de solutions physiques. 

Il reste {\`a} montrer que nos solutions sont stables. Dans le cadre de la
dynamique d{\'e}finie par l'action de Born-Infeld, la stabilit{\'e} d'une
solution {\'e}quivaut {\`a} la positivit{\'e} des masses carr{\'e}es dans le
spectre des fluctuations quadratiques autour de cette solution. 
Un autre int{\'e}r{\^e}t de ce
spectre est son interpr{\'e}tation comme
la limite
semi-classique du spectre des cordes ouvertes attach{\'e}es {\`a} la
D-brane en question. Cette interpr{\'e}tation d{\'e}coule de la
d{\'e}finition des D-branes comme surfaces o{\`u} aboutissent les cordes
ouvertes, et dont les cordes ouvertes d{\'e}finissent la dynamique. 

Par commodit{\'e} d'{\'e}criture, nous introduisons un syst{\`e}me de
coordonn{\'e}es $x^i$ quelconque sur $G/T^\o_0$. Pour chaque {\'e}l{\'e}ment
$x^i$ de $G/T^\o_0$ soit $h(x^i)$ un de ses repr{\'e}sentants dans $G$ (on
identifie $x^i$ avec $h(x^i)T^\o_0$).  Soient $\psi ^\a$ des
coordonn{\'e}es sur $T^\o_0$. Soit $e_i$ la projection orthogonale de
$\p_i h h\- $ sur $\im$, avec $g=ht\o(h)\- $. Le vecteur $e_i$ ne d{\'e}pend
pas de $t$ car $\im=({\rm Lie}T^\o_0)^\perp $\ n'en d{\'e}pend pas,
puisque $g\in G_r$. Nous utiliserons aussi $u_\a=g\- \p_\a g$, ainsi
$(u_\a,e_i)$ forme une base de \lieg\ adapt{\'e}e {\`a} la d{\'e}composition
eq. (\ref{orthog}). Si $A$ est un op{\'e}rateur lin{\'e}aire sur $\im$,
nous pouvons d{\'e}finir sa trace $\Tr A = \llangle e_i, A(e_i)\rrangle
$ et sa matrice $A_{ij}=\llangle e_i, A(e_j)\rrangle
$. R{\'e}ciproquement, nous interpr{\'e}terons les tenseurs {\`a} deux
indices comme {\'e}l{\'e}ments de matrice d'op{\'e}rateurs sur $\im$ dans la
base $e_i$.  

Nous pouvons r{\'e}{\'e}crire les champs physiques dans cette base, ainsi
$\F _{ij}= \llangle \exp \- t,[e_i,e_j] \rrangle $ et $\om
_{ij}=\llangle (1-\oAd)e_i,(1+\oAd)e_j \rrangle$. 
D{\'e}finissons maintenant le tenseur
\bea
\gamma_{ij}(x)=\llangle e_i(x), e_j(x) \rrangle.
\label{metcordouv}
\eea
Il s'agit de la m{\'e}trique de corde ouverte sur la D-brane. La
d{\'e}riv{\'e}e covariante et le Laplacien associ{\'e}s sont 
\bea 
\nabla^i &=& \frac1{\sqrt{\det\gamma}} \partial_j \sqrt{\det\gamma}
\gamma^{ij} 
\\
\Box &=& \nabla^i \partial_i \, .  
\label{dercov}
\eea
Ce Laplacien est identique {\`a} l'op{\'e}rateur de Casimir de l'action
infinit{\'e}simale (par multiplication {\`a} gauche) de $G$ sur $G/T^\o_0$.

Si l'on imagine que les coordonn{\'e}es $\psi_\a $ d{\'e}pendent de $x_i$,
alors les fonctions $\psi_\a (x_i)$ d{\'e}crivent un
plongement quelconque de $G/T^\o_0$ dans $G$. Nos D-branes
correspondent au plongement $\psi_\a=\cst$, et leurs fluctuations
g{\'e}om{\'e}triques {\`a} une fluctuation $\delta \psi_\a$. Il faut aussi prendre
en compte les fluctuations $\delta F = d \delta A $ du champ de
jauge. On trouve que les fluctuations au second ordre de l'action de
Born-Infeld sont~:
\footnote{ 
Ici nous {\'e}crivons explicitement les calculs pour une classe de
$\o$-conjugaison, cas qui n'avait {\'e}t{\'e} qu'{\'e}voqu{\'e} dans
\cite{brs}.
}
{\small
\begin{eqnarray}
\delta^{(2)} S_{BI} (t) &  = &  \int _{G / T^\o_0} dx^i \sqrt{\det \left(
     \hat{g}+\om \right)}  \times  \nonumber 
\\
 & \times &  \left[ \frac{1}{8}   \gamma^{ij} 
 \partial_i \delta\!\psi^{\alpha} \partial_j
\delta\!\psi^{\beta} \delta_{\alpha \beta} - \frac{1}{8}\Tr
                 \left( ad_{u_{\alpha}} ad_{u_{\beta}} \right)
\delta\!\psi^{\alpha} \delta\!\psi^{\beta} \right. 
\label{d2s}
\\
 & - &\left.  \frac{1}{4} \Tr \left( \delta\! F \frac{2}{1-\oAd } \delta\! F
\frac{2}{1-\oAd}\right) - \frac{1}{8} \Tr \left(\delta\! F
            ad_{u_{\alpha}}\right) \delta\! \psi^{\alpha}
 + \frac{1}{8} \Tr ^2 \left(\frac{2}{1-\oAd} \delta\! F  \right)
     \right].  \nonumber
\end{eqnarray}
}
On en d{\'e}duit les {\'e}quations du mouvement des fluctuations~:
\begin{eqnarray}
\delta\!\psi^{\alpha}: & \ \ &  \delta _{\alpha \beta} \Box
\delta\!\psi^{\alpha}  +  \Tr\left( ad_{u_{\alpha}} ad_{u_{\beta}}
\right) \delta\!\psi^{\alpha} +
\frac{1}{2} \Tr \left( \delta\!F ad_{u_{\beta}}  \right) = 0 \, ,
\label{eq.1} 
\\
\delta\!A_j: & \ \ &  -4(ad_{u_{\alpha}})^{ij} \partial_j
\delta\!\psi^{\alpha} + \frac12\left\langle e^i,[e^k,e^l]\right\rangle 
\delta\!F_{kl} + \nabla^k
\delta\!F _k\,^i = 0 \, .  
\label{eq.2}
\end{eqnarray}
Les fluctuations physiques doivent de plus ob{\'e}ir {\`a} un choix de
jauge pour le champ $\delta A$, que nous choisissons $G$-invariante~:
\bea
 \nabla (\delta A) = 0 \, .   
\label{jauge1}
\eea
Remarquons que la condition de jauge et les {\'e}quations du mouvement
sont ind{\'e}pendants de la position $t$ de la D-brane.
Nos {\'e}quations deviennent plus simples si l'on introduit une fonction 
$f:G / T^\o_0 \rightarrow \lieg$ pour effectuer le changement de variable~:
\bea
\delta\!A_i = 2 \left\langle e_i, f  \right\rangle  \ , \quad
\delta\!\psi^{\alpha} 
= \left\langle u ^{\alpha}, f \right\rangle .  
\label{ansatz}
\eea
Alors la condition de jauge 
(\ref{jauge1}) devient
\bea \left\langle e_i , \partial^i f \right\rangle = 0 , 
\label{jauge2}
\eea
et les {\'e}quations  (\ref{eq.1},\ref{eq.2}) deviennent
\bea
\left\langle  u_{\beta},  \Box f \right\rangle = 0 \,, \quad  
\left\langle e_i, \Box f  \right\rangle = 0,  
\label{edmfluct}
\eea
autrement dit
\bea 
\Box f = 0 \, .  
\label{eq.f1}
\eea

\bs

Il s'agit maintenant de r{\'e}soudre les {\'e}quations (\ref{eq.f1}) et
(\ref{jauge2}). En fait, cela n'a {\`a} proprement parler d'int{\'e}r{\^e}t
que si l'on ajoute au syst{\`e}me une direction temporelle pour lui
permettre d'osciller. On consid{\`e}re donc la D-brane $\B \times \R$
dans l'espace $G\times \R$. L'{\'e}quation (\ref{eq.f1}) est remplac{\'e}e
par l'{\'e}quation aux valeurs propres de l'op{\'e}rateur $\Box$. Ainsi le
spectre des cordes ouvertes co{\"\i}ncide dans la limite
semi-classique avec le spectre du Laplacien de la m{\'e}trique de corde
ouverte, agissant sur l'espace $\Fun(G/T^\o_0)$ des fonctions sur la
D-brane. Or, il a d{\'e}j{\`a} {\'e}t{\'e} constat{\'e} dans \cite{fffs} que ce
spectre est bien la limite pour $k$ grand du spectre de la m{\^e}me
D-brane obtenu par des m{\'e}thodes de th{\'e}orie conforme au bord. Cette
limite est ind{\'e}pendante de la position de la D-brane en question.

La structure alg{\'e}brique du spectre m{\`e}ne aussi {\`a} un accord 
avec la th{\'e}orie conforme. Le fait que les fluctuations soient
param{\'e}tr{\'e}es par un {\'e}l{\'e}ment $f$ de \lieg\ assurent qu'elles
correspondent aux {\'e}tats de niveau un  d'une repr{\'e}sentation de
l'alg{\`e}bre de Lie affine \alieg. Ainsi, l'ensemble des {\'e}tats de masse
nulle forme une repr{\'e}sentation adjointe de \lieg~; ces {\'e}tats
s'interpr{\`e}tent comme les modes de rotation de la D-brane dans le
groupe $G$. 
La condition de jauge
(\ref{jauge2}), elle, s'interpr{\`e}te comme une condition de Virasoro.

Nous pouvons aussi calculer l'{\'e}nergie de Born-Infeld d'une classe de
$\o$-conjugaison de position $\exp^{-1} t=2\pi\lambda/k$. Si $\det'$
d{\'e}signe le d{\'e}terminant d'un op{\'e}rateur sur ${\rm Im}(1-{\rm
  Ad}_t)$, nous trouvons
\bea
 E_\lambda =S_{BI}(q_{t},F_{t})
&\propto& 
\int _{G/T}\sqrt{\det\gamma_{ij}}\sqrt{
\det {}'(1- \o \circ {\rm Ad}_{t})} 
\label{BIener}
\eea
Dans le cas des classes de conjugaison ($ \o={\rm id}$), cette formule
peut se mettre sous une forme particuli{\`e}rement simple, o{\`u} le 
produit porte sur les racines
  positives de l'alg{\`e}bre de Lie \lieg,
\bea
E_\lambda
&\propto&
\prod_{\alpha > 0 }  \sin \left(\frac{\pi}{k} 
(\lambda,\alpha)\right) \, .  
\label{BIenercjg}
\eea
Ces r{\'e}sultats sont en accord avec l'analyse de la
th{\'e}orie conforme au bord. Cette analyse sera pr{\'e}sent{\'e}e dans la
section \ref{soussectcardy}. On pourra alors v{\'e}rifier qu'en 
ce qui concerne la position des D-branes, leur {\'e}nergie et
leur spectre des fluctuations quadratiques, nos r{\'e}sultats
co{\"\i}ncident avec les quantit{\'e}s exactes dans la limite
semi-classique de grand $k$, comme on devait s'y attendre. Mais il se
trouve de plus que ces r{\'e}sultats obtenus avec l'action de
Born-Infeld sont sont exacts {\`a} quelques d{\'e}tails pr{\`e}s. En effet,
il suffit de remplacer $k$ par $k+g^\vee$ (o{\`u} $g^\vee $ est le
nombre de Coxeter dual de l'alg{\`e}bre de Lie \lieg), de d{\'e}placer
tous les poids $\lambda$ par un vecteur constant et d'imposer au
spectre des fluctuations la borne
d'unitarit{\'e} des repr{\'e}sentations de l'alg{\`e}bre de Lie affine
\alieg, pour obtenir les r{\'e}sultats de th{\'e}orie conforme. La
renormalisation du niveau $k\rar k+g^\vee$ {\'e}tant un effet bien connu de
th{\'e}orie des cordes ferm{\'e}es, on peut dire que l'action de
Born-Infeld donne dans le cas des classes de $\o$-conjugaison dans les
groupes compacts\footnote{ Nous avons souvent fait des hypoth{\`e}ses
  techniques de simplicit{\'e} ou semi-simplicit{\'e} du groupe $G$, mais
  il est facile d'{\'e}tendre nos r{\'e}sultats {\`a} tous les groupes
  compacts. }
 des r{\'e}sultats exacts. 

Or, l'action de Born-Infeld n'est pas elle-m{\^e}me exacte, mais
re{\c c}oit toute une s{\'e}rie de corrections d{\'e}rivatives qui forment
un d{\'e}veloppement en puissance de $\ap$. C'est aussi le cas de la
supergravit{\'e}, dont les vari{\'e}t{\'e}s de groupes sont cependant des
solutions exactes (moyennant la renormalisation du niveau). Nos
r{\'e}sultats peuvent donc s'interpr{\'e}ter comme des contraintes sur la
structure de ces
corrections d{\'e}rivatives {\`a} l'action de Born-Infeld.

Un autre int{\'e}r{\^e}t de ces r{\'e}sultats est qu'ils illustrent la
puissance des actions effectives en th{\'e}orie des cordes, et montrent
qu'elles peuvent donner des informations au-del{\`a} de leur domaine de
validit{\'e} suppos{\'e}. Ainsi nous avons obtenu des r{\'e}sultats exacts {\`a}
partir de calculs semi-classiques sans sommer une s{\'e}rie infinie de
termes, mais en appliquant quelques r{\`e}gles simples. 

\bs

L'exemple des D-branes dans \SU\ est trait{\'e} explicitement dans
\cite{bds}, mais ce groupe n'a pas d'automorphisme ext{\'e}rieur, donc
seules les classes de conjugaison sont prises en compte. En revanche,
$SU(3)$ poss{\`e}de des classes de $\o$-conjugaison, cet exemple est
trait{\'e} dans
\cite{brs,stander}. En particulier, le groupe $SU(3)$ est de dimension
huit, les classes de conjugaison r{\'e}guli{\`e}res de dimension six et
les classes de $\o$-conjugaison r{\'e}guli{\`e}res de dimension sept.

\section{ Quelques g{\'e}n{\'e}ralisations }

\subsection{ Groupes non compacts, le cas de \SL }

Dans notre pr{\'e}sentation des D-branes dans les groupes compacts, nous
avons utilis{\'e} des arguments globaux pour d{\'e}terminer les conditions
de quantification portant sur le champ $F$, et donc sur la position,
des D-branes. Ces arguments d{\'e}pendent des propri{\'e}t{\'e}s
structurelles pr{\'e}cises des groupes compacts, et doivent donc {\^e}tre
modifi{\'e}s si l'on veut {\'e}tudier les D-branes sym{\'e}triques dans des
groupes non compacts. En revanche, les calculs locaux, qui permettent
d'{\'e}crire les solutions correspondantes de l'action de Born-Infeld, et
de d{\'e}terminer le spectre des fluctuations quadratiques, restent
valables dans un contexte beaucoup plus g{\'e}n{\'e}ral.

Nous n'avons pas d{\'e}termin{\'e} les hypoth{\`e}ses exactes permettant la
g{\'e}n{\'e}ralisation la plus large possible de ces r{\'e}sultats. En
effet, l'int{\'e}r{\^e}t physique d'une telle {\'e}tude semble {\`a} l'heure
actuelle plut{\^o}t limit{\'e}. Nous allons cependant 
{\'e}voquer certaines propri{\'e}t{\'e}s des groupes non compacts qui
seraient pertinentes pour l'{\'e}tude des D-branes sym{\'e}triques, avant
de traiter l'exemple physiquement bien motiv{\'e} des D-branes dans
\SL.

Nous nous restreindrons aux groupes semi-simples. Certes, on
a bien  
{\'e}tudi{\'e} les D-branes dans des groupes non semi-simples comme celui
de Nappi-Witten \cite{stanmore}, mais ces r{\'e}sultats sont tr{\`e}a
pr{\'e}liminaires,
et chercher {\`a} bien comprendre les D-branes dans les groupes
semi-simples est d{\'e}j{\`a} un programme ambitieux, comme en
t{\'e}moignent les difficult{\'e}s que nous rencontrerons.
L'hypoth{\`e}se de
semi-simplicit{\'e} est tr{\`e}s pratique sur le plan
technique car de nombreux r{\'e}sultats sont connus. Notamment, dans une
alg{\`e}bre de Lie semi-simple complexe deux sous-alg{\`e}bres de Cartan
sont toujours conjugu{\'e}es, et dans une alg{\`e}bre de Lie semi-simple
r{\'e}elle il existe un nombre fini de classes de conjugaison de
sous-alg{\`e}bres de Cartan. Nous verrons que ce nombre vaut deux dans
le cas de \sl, et que les deux types de tores maximaux correspondant
permettent de classifier les classes de conjugaison. Dans des cas
plus g{\'e}n{\'e}raux, certaines subtilit{\'e}s proviennent du fait que le
sous-groupe de Cartan (d{\'e}fini comme le centralisateur d'une
sous-alg{\`e}bre de Cartan) n'est pas toujours ab{\'e}lien. Si l'on veut
g{\'e}n{\'e}raliser nos calculs locaux, cette propri{\'e}t{\'e} du sous-groupe
de Cartan est cependant n{\'e}cessaire. On peut s'assurer qu'elle est
v{\'e}rifi{\'e}e en se restreignant {\`a} une partie dense du groupe $G$,
mais il n'est pas clair qu'une telle restriction ait un sens
physique. 

\bs

Mais revenons au cas prototypique de \SL, o{\`u} de telles subtilit{\'e}s
n'interviennent pas. Nous exposerons l'int{\'e}r{\^e}t physique des
D-branes sym{\'e}triques dans \SL\ dans la prochaine sous-section, 
apr{\`e}s que nous les aurons d{\'e}crites.

D'abord, donnons quelques pr{\'e}cisions sur la g{\'e}om{\'e}trie du groupe
\SL. Commen{\c c}ons par d{\'e}finir \AAA\ comme l'hypersurface de 
$\R^{2,2}$ d'{\'e}quation
\bea
-X_0^2-X_1^2+X^2_2+X^2_3=-L^2.
\label{eqads3}
\eea
Cette {\'e}quation est la condition n{\'e}cessaire et suffisante pour
qu'une matrice
\bea
g = \frac{1}{L}\
\left(
\begin{array}{lll}
X^0+X^3 &\quad & X^2+X^1\cr
X^2-X^1 &\quad & X^0- X^3\cr
\end{array}
\right),
\label{matricesl2}
\eea
appartienne au groupe \SL. Cependant, l'utilisation des coordonn{\'e}es
globales,
\Fig{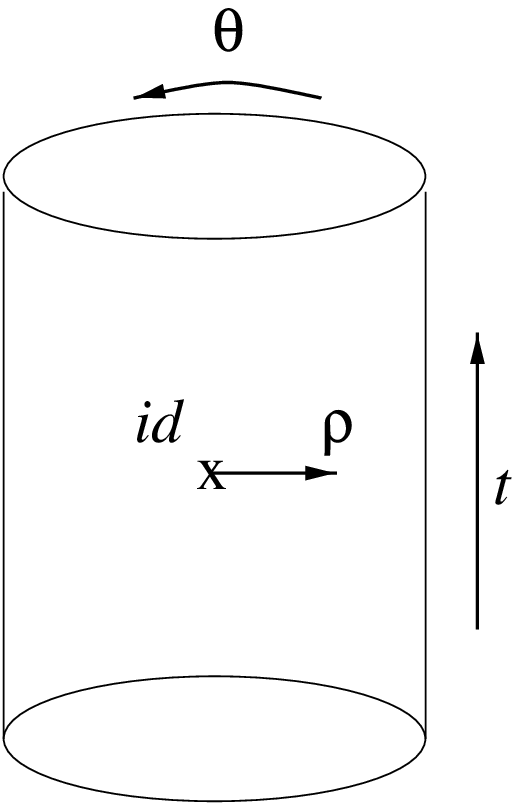}{5}{ Les coordonn{\'e}es globales d'\AAA }
\bea
\left\{
\begin{array}{lll}
X_2+iX_3 &=& L\sinh\rho \ e^{i\theta} \cr
X_0+iX_1 &=& L\cosh\rho \ e^{i\t} \cr
\end{array}
\right. ,
\label{coordglob}
\eea
clarifie la diff{\'e}rence entre \AAA\ (tel que $t\in \R$) et \SL\ (tel
que $t$ est $2\pi$-p{\'e}riodique). Ainsi, l'application de \AAA\ dans
\SL\ qui {\`a} $(X_0,X_1,X_2,X_3)$ v{\'e}rifiant (\ref{eqads3}) associe la
matrice $g$ de l'{\'e}quation (\ref{matricesl2}) est un recouvrement
d'ordre infini d{\'e}nombrable. Autrement dit, \SL\ a la topologie d'un
tore tridimensionnel plein ($\pi_1(\SL)=\Z$), 
alors qu'\AAA\ est un cylindre infini plein ($\pi_1(\AAA)=\{1\}$).
La m{\'e}trique dans les coordonn{\'e}es globales (voir la Figure
\ref{asd3coord}) est~:
\bea 
ds^2=L^2(-\cosh^2\rho\
d\t^2+d\rho^2+\sinh^2\rho\ d\theta^2)=-dX_0^2-dX_1^2+dX^2_2+dX^2_3.
\label{metads3}
\eea
Nous avons d{\'e}j{\`a} annonc{\'e} que l'alg{\`e}bre de Lie \sl\ avait deux
classes de conjugaison de sous-alg{\`e}bres de Cartan. La premi{\`e}re
classe, elliptique, comprend la sous-alg{\`e}bre engendr{\'e}e par {\tiny $\left(
\begin{array}{rl}
 0  & 1 \cr
-1  & 0 \cr
\end{array}
\right)$}, le tore maximal qui lui correspond est 
\bea
T^{\rm ell}=\left\{g=\left(
\begin{array}{rl}
 \cos\t & \sin\t \cr
-\sin\t & \cos\t \cr
\end{array}
\right)\right\}. 
\label{toreell}
\eea

\Fig{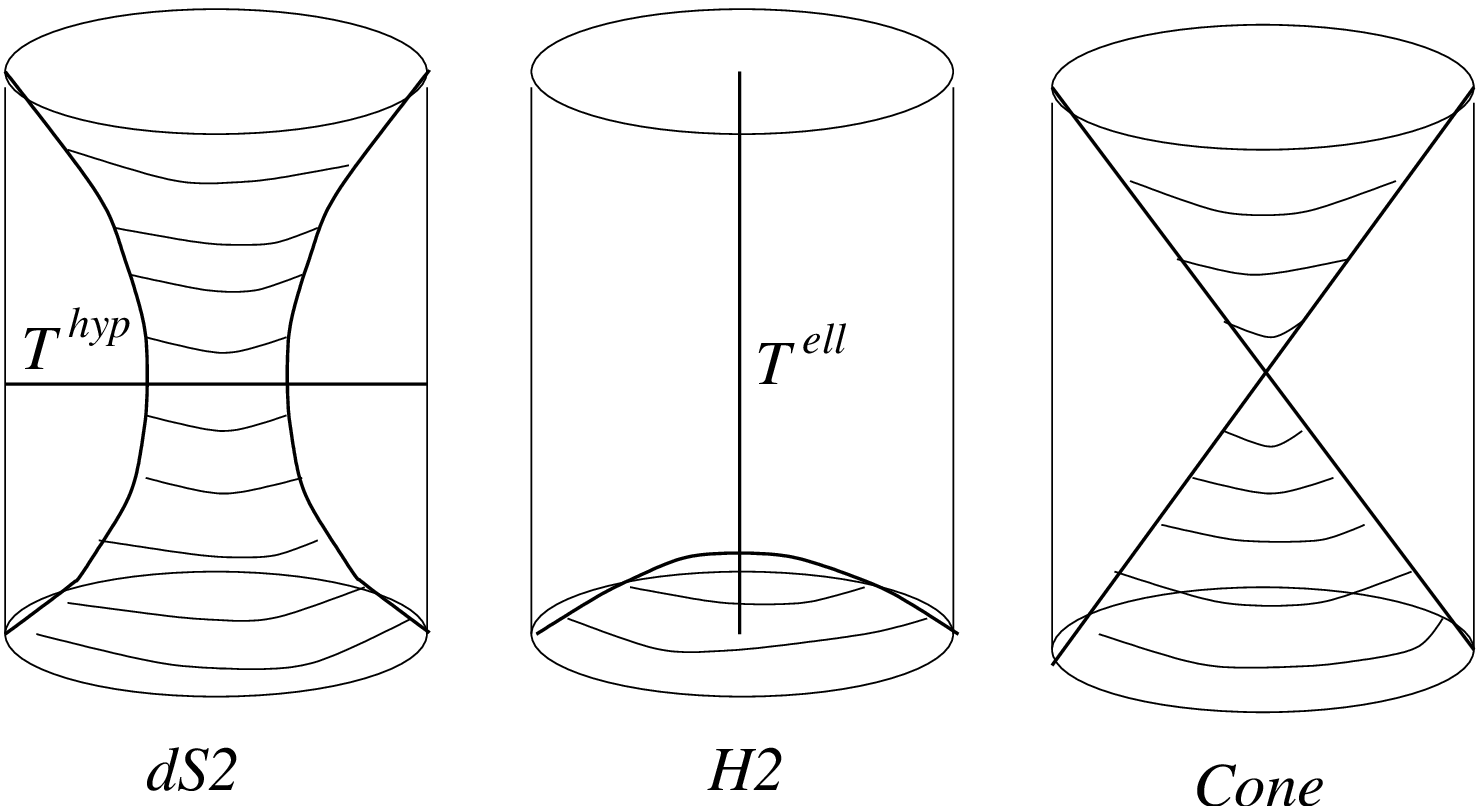}{12}{ Tores maximaux et classes de conjugaison de \SL }

L'{\'e}quation de ce tore en coordonn{\'e}es globales est $\rho=0$. Les
classes de conjugaison des {\'e}l{\'e}ments du tore $T^{\rm ell}$ lui sont
orthogonales, ont la g{\'e}om{\'e}trie du plan hyperbolique $H_2$, sont de
genre espace, et ont pour {\'e}quations 
\bea
\Tr g = \frac{2X_0}{L} = 2\cosh\rho \cos\t = \cst\ \leq 2.
\label{eqh2}
\eea
La deuxi{\`e}me classe de conjugaison de sous-alg{\`e}bres de Cartan,
hyperbolique, comprend les sous-alg{\`e}bres engendr{\'e}es par {\tiny $\left(
\begin{array}{rl}
 0 & 1  \cr
 1 & 0  \cr
\end{array}
\right)$} et par {\tiny $\left(
\begin{array}{rl}
 1 & 0   \cr
 0 & -1 \cr
\end{array}
\right)$}, deux des tores maximaux (conjugu{\'e}s) qui correspondent sont 
\bea
T^{\rm hyp}=\left\{g=\left(
\begin{array}{rl}
 \cosh\rho  & \sinh\rho \cr
 \sinh\rho  & \cosh\rho \cr
\end{array}
\right)\right\}\ , \ 
T^{\rm hyp'}=\left\{g=\left(
\begin{array}{rl}
  e ^\rho   & 0          \cr
  0         & e ^{-\rho} \cr
\end{array}
\right)\right\}. 
\label{torehyp}
\eea
Les {\'e}quations de ces tores en coordonn{\'e}es globales sont
respectivement $\t=\theta=0$ et $\t=0, \theta=\halfpi$. Les classes de
conjugaison des {\'e}l{\'e}ments de ces tores, qui leurs sont
orthogonales, ont la g{\'e}om{\'e}trie de l'espace de Sitter $dS_2$, sont
de genre temps, et ont pour {\'e}quations
\bea
\Tr g = \frac{2X_0}{L} = 2\cosh\rho \cos\t = \cst\ \geq 2.
\label{eqds2}
\eea
Ainsi, on trouve des D-branes de genre temps orthogonales {\`a} des
tores maximaux de genre espace et vice-versa (voir la Figure
\ref{asd3cjg}). 
Il existe {\'e}galement
des D-branes de genre lumi{\`e}re, qui sont engendr{\'e}es par (et
orthogonales {\`a}) des tores maximaux de genre lumi{\`e}re, que l'on
obtient en exponentiant des sous-alg{\`e}bres nilpotentes de \sl. Par
exemple, celle engendr{\'e}e par {\tiny $\left(
\begin{array}{rl}
 0 & 1 \cr
 0 & 0 \cr
\end{array}
\right)$}, correspond au tore
\bea
T^{\rm lum}=\left\{g=\left(
\begin{array}{rl}
 1 & 2\sinh\rho \cr
 0 & 1 \cr
\end{array}
\right) = \left(  
\begin{array}{rl}
 1 & 2\tan\t \cr
 0 & 1 \cr
\end{array}
\right)
\right\}.  
\label{torelum}
\eea
Les D-branes de genre lumi{\`e}re sont le c{\^o}ne de lumi{\`e}re de futur,
celui du pass{\'e}, et l'identit{\'e} de \SL. Ensemble, elles forment le
c{\^o}ne de lumi{\`e}re de
l'espace \AAA, d'{\'e}quation 
\bea
\Tr g = \frac{2X_0}{L} = 2\cosh\rho \cos\t = 2.
\label{eqcone}
\eea

\Fig{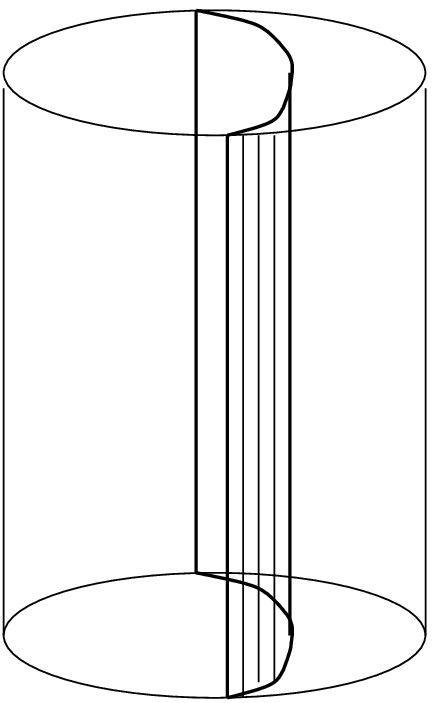}{5}{ Une classe de $\o$-conjugaison de \SL }

Le groupe \SL\ dispose de plus d'un automorphisme ext{\'e}rieur $\o$, la
conjugaison par la matrice
\bea
\Omega=\left(
\begin{array}{rl}
 0 & 1 \cr
 1 & 0 \cr
\end{array}
\right)\ \in \ GL(2,\R).
\label{matext}
\eea
Contrairement {\`a} ce qui peut se produire dans les groupes compacts, il
existe un tore maximal $T^{\rm hyp}$ 
invariant point par point par cet automorphisme
ext{\'e}rieur. Donc, les classes de $\o$-conjugaison ont la m{\^e}me
dimension, deux, que les classes de conjugaison r{\'e}guli{\`e}res. Leur
g{\'e}om{\'e}trie est \AA\ et leurs {\'e}quations sont
\bea
\Tr \Omega g = \frac{2X_2}{L} = 2 \sinh \rho \cos \theta = \cst.
\label{eqads2}
\eea
Ce sont donc des D-branes statiques (voir la Figure \ref{asd3tw}). 
Cette propri{\'e}t{\'e} est une des
raisons pour laquelle l'{\'e}tude de ces D-branes est plus ais{\'e}e que
celle des autres~: on peut en effet effectuer le prolongement
analytique $\t \rar i\t$ et obtenir des D-branes statiques dans
l'espace Euclidien $H_3$. Certes, le prolongement analytique a aussi
un sens pour les classes de conjugaison $H_2$, mais comme elles ne
sont pas statiques il est beaucoup moins clair que l'on puisse
interpr{\'e}ter des r{\'e}sultats obtenus dans $H_3$ pour les d{\'e}crire. En
particulier, leur couplage aux {\'e}tats des cordes ferm{\'e}es vivant dans les 
repr{\'e}sentations discr{\`e}tes n'a pas d'{\'e}quivalent dans $H_3$. 

L'{\'e}tude des D-branes \AA\ est non seulement la plus simple
techniquement, mais elle est aussi motiv{\'e}e par des consid{\'e}rations
physiques, en particulier holographiques, que nous allons maintenant
{\'e}voquer. 

\subsection{ Propri{\'e}t{\'e}s des D-branes $AdS_2$ dans $AdS_3$ }

Nous avons d{\'e}j{\`a} montr{\'e} comment \AAA\ appara{\^\i}t dans la
limite proche de l'horizon de configurations de branes, dans un
contexte qui se pr{\^e}te {\`a} la correspondance AdS/CFT. Il est
int{\'e}ressant de constater que les D-branes \AA\ ont aussi une
interpr{\'e}tation dans cette configuration de branes \cite{bp,cosstr}, et de
consid{\'e}rer les objets duaux dans la th{\'e}orie conforme au bord d'\AAA\
\cite{bbdo,asym}. 

Les objets de base sont des jonctions d'objets solitoniques
(D1-branes, D5-branes,
NS5-branes, cordes fondamentales) de la th{\'e}orie des cordes de type
IIB. {\`A} la configuration de NS5-F1 qui cr{\'e}e la brane noire de la
supergravit{\'e} dont la limite proche de l'horizon est $AdS_3\times S^3
\times \R^4$, on peut ajouter un {\'e}tat li{\'e} de cordes-test D1 et F1 dont
on n{\'e}glige l'influence sur les champs de la supergravit{\'e}. Ces
cordes-test s'interpr{\`e}tent dans la limite proche de l'horizon comme
une D-brane $AdS_2\times S^2$ dans $AdS_3\times S^3$. Plus
pr{\'e}cis{\'e}ment, si l'on a $p$ 
D1-branes, on peut montrer que l'on obtient une D-brane $S^2$ dans
$S^3$ de nombre quantique $p$, et une D-brane $AdS_2$ droite
($\cst=0$). 

Si maintenant on ajoute $q$ cordes fondamentales F1, on change le
param{\`e}tre de la D-brane $AdS_2$ de mani{\`e}re quantifi{\'e}e.  Comme
$\pi_2(\AA)$ est trivial, cette quantification n'a pas la m{\^e}me
interpr{\'e}tation que celle de la position des D-branes $S^2$ dans
\SU. La condition trouv{\'e}e par \cite{bp} est 
\bea
\sinh \rho \cos \theta = q \frac{T_F}{p\, T_D},\ q\in\Z, 
\label{quantifads}
\eea 
o{\`u} $T_F=\frac{1}{2\pi\ap}$ et $T_D$ sont les tensions respectives de
la corde fondamentale et de la D1-brane. Cette condition a un statut
tout-{\`a}-fait diff{\'e}rent de celle de la D-brane $S^2$ et provient
v{\'e}ritablement de la th{\'e}orie des cordes et pas seulement de la
th{\'e}orie conforme. Cela peut se voir dans le fait qu'elle s'exprime
en fonction de $T_F/T_D$ alors que la quantification de $S^2$ ne
d{\'e}pend que du niveau $k$. Donc, si l'on construit les objets
correspondants en th{\'e}orie conforme, {\`a} savoir les {\'e}tats de bord
associ{\'e}s aux D-branes $AdS_2$, on s'attend {\`a} en trouver une
famille d{\'e}pendant d'un param{\`e}tre continu, non quantifi{\'e}. 

\bs

Une autre propri{\'e}t{\'e} int{\'e}ressante des D-branes $AdS_2\times S^2$
est qu'elles sont supersym{\'e}triques \cite{bp}. En fait, elles pr{\'e}servent 8
des 16 supersym{\'e}tries de l'espace-temps $AdS_3\times S^3
\times \R^4$ (qui lui-m{\^e}me est la limite proche de l'horizon d'une
configuration de NS5-F1 pr{\'e}servant 8 des 32 supersym{\'e}tries de la
th{\'e}orie des cordes de type IIB). Ces supersym{\'e}tries garantissent
la stabilit{\'e} de ces D-branes au-del{\`a} de la limite semi-classique. De
plus, elles pourraient contribuer {\`a} expliquer l'exactitude surprenante des
r{\'e}sultats obtenus au moyen de l'action de Born-Infeld. Cependant,
cette explication n'est pas valable pour les groupes compacts plus
g{\'e}n{\'e}raux, alors que la propri{\'e}t{\'e} d'exactitude persiste comme
nous l'avons montr{\'e}. Elle doit donc avoir une explication plus
profonde li{\'e}e {\`a} la sym{\'e}trie de groupe.

\rem{ references plus precises a AdS/CFT et au bord d'$AdS_3$ }

Nous allons maintenant {\'e}voquer l'interpr{\'e}tation holographique des
D-branes \AA\ propos{\'e}e par \cite{bbdo,asym}. 
Ces D-branes atteignent le bord bidimensionnel d'\AAA, o{\`u} elles
dessinent des trajectoires statiques unidimensionnelles. Nous
appellerons ces trajectoires des interfaces. Elles divisent le bord
d'\AAA\ en diff{\'e}rents domaines, o{\`u} vivent diff{\'e}rentes th{\'e}ories
conformes. Cependant ces interfaces sont perm{\'e}ables, autrement dit les
th{\'e}ories conformes interagissent entre elles.

\rem{ A preciser eventuellement. Branes test... } 

L'{\'e}tude des D-branes \AA\ ne manque donc pas de motivations
physiques. Jusqu'{\`a} maintenant, l'interpr{\'e}tation holographique de
ces D-branes s'est content{\'e}e d'en utiliser la g{\'e}om{\'e}trie et les
propri{\'e}t{\'e}s semi-classiques, par exemple pour calculer la force
s'exer{\c c}ant entre deux interfaces. Il pourra {\^e}tre int{\'e}ressant
d'{\'e}tudier en d{\'e}tail l'impact d'une connaissance approfondie
de ces D-branes sur ces calculs.

\subsection{ Cordes ouvertes sur les D-branes $AdS_2$ 
\label{cosldba}}

Les m{\'e}thodes semi-classiques permettent d'obtenir certaines
informations sur les D-branes \AA\ et notamment leur spectre de cordes
ouvertes, sans cependant conduire {\`a} des r{\'e}sultats aussi pr{\'e}cis
et aussi g{\'e}n{\'e}raux que dans le cas des groupes compacts. 

D'abord, l'{\'e}tude des solutions et des fluctuations de l'action de
Born-Infeld que nous avons expos{\'e}e dans les groupes compacts
s'applique fort bien aux D-branes dans \SL. Nous n'allons donc pas
reproduire les calculs dans le cas des D-branes $AdS_2\times S^2$, qui
ont {\'e}t{\'e} effectu{\'e}s dans \cite{pr} et ont sugg{\'e}r{\'e} la
g{\'e}n{\'e}ralisation aux groupes compacts. Notons cependant
que des calculs similaires sont possibles dans un cas S-dual \cite{nicolas}.
Bien s{\^u}r, nous
n'obtenons pas ainsi de condition de quantification, et les
informations sur le spectre des cordes ouvertes doivent {\^e}tre
interpr{\'e}t{\'e}es avec soin. 

Les r{\'e}sultats du calcul sont les suivants~: le spectre des cordes
ouvertes d'une D-brane donn{\'e}e est
ind{\'e}pendant des param{\`e}tres $(p,q)$ de la D-brane, d{\'e}finis par l'{\'e}quation
(\ref{quantifads}). Sa structure
alg{\'e}brique  est
identique {\`a} celle que nous avons d{\'e}crite dans le cas des groupes
compacts, c'est-{\`a}-dire que les fluctuations s'interpr{\`e}tent comme
des {\'e}tats de niveau un de l'alg{\`e}bre de Lie affine du groupe
$\SL\times \SU$, soumis {\`a} la condition de Virasoro. La structure
analytique fait intervenir les fonctions sur $AdS_2\times S^2$ de
Laplacien nul (ce Laplacien co{\"\i}ncide encore avec le Casimir de
l'action du groupe sur ses classes de $\o$-conjugaison). Comme le
Laplacien est positif sur \SU, le terme provenant de \SL\ doit {\^e}tre
n{\'e}gatif, et donc correspondre {\`a} des repr{\'e}sentations
discr{\`e}tes de \SL. 

Ce r{\'e}sultat n'est pas surprenant du point de vue de la th{\'e}orie des
cordes. En effet, si l'on omet le flot spectral et que l'on impose la
condition de couche de masse, le spectre des cordes ferm{\'e}es dans
$\AAA\times S^3\times \R^4$ ne comprend que des repr{\'e}sentations
discr{\`e}tes. La condition de couche de masse est l'{\'e}quivalent
quantique de l'{\'e}quation de Born-Infeld portant sur l'annulation du
Laplacien. Le flot spectral modifie certes la situation, mais
l'analyse semi-classique n'y est pas sensible, car elle ne prend en
compte que la sous-alg{\`e}bre de Lie \sl\ de niveau z{\'e}ro de
l'alg{\`e}bre de Lie affine \asl, sous-alg{\`e}bre qui n'est pas
pr{\'e}serv{\'e}e par le flot spectral.

Le spectre semi-classique des D-branes \AA\ comporte donc toutes les
repr{\'e}sentations discr{\`e}tes avec le m{\^e}me poids, comme elles
apparaissent dans l'espace des fonctions sur \AA. Au niveau quantique,
il faut s'attendre {\`a} devoir imposer la contrainte 
d'unitarit{\'e}  et la sym{\'e}trie du flot spectral comme dans le cas des
cordes ferm{\'e}es \cite{moi}. Cependant, il est tentant de conjecturer
par analogie avec le cas des D-branes dans \SU\ que la densit{\'e} des
repr{\'e}sentations discr{\`e}tes ne re{\c c}oit pas de corrections \cite{corfu}, et
reste ind{\'e}pendante de la D-brane et du spin $j$ de la
repr{\'e}sentation consid{\'e}r{\'e}e. En effet, si l'on admet que
l'exactitude des r{\'e}sultats semi-classiques dans \SU\ est li{\'e}e {\`a}
l'annulation des contributions des corrections {\`a} l'action de
Born-Infeld, alors ces contributions doivent aussi s'annuler dans
notre cas par prolongement analytique. 

\bs

Nous avons un peu h{\^a}tivement affirm{\'e} que le spectre des cordes
ouvertes de la D-brane \AA\ devait respecter la sym{\'e}trie du flot
spectral. Cela suppose que cette sym{\'e}trie ne soit pas bris{\'e}e par
la D-brane, ce qui demande une v{\'e}rification. On peut {\'e}tudier pour
cela les solutions de la th{\'e}orie des cordes ouvertes. Nous
connaissons d{\'e}j{\`a} les solutions classiques des cordes ferm{\'e}es
dans une vari{\'e}t{\'e} de groupe, eq. (\ref{solWZW}). Il est possible de
trouver la solution tout aussi g{\'e}n{\'e}rale pour une corde ouverte
dont chaque bout est fix{\'e} sur une D-brane donn{\'e}e, {\`a} condition
que ces D-branes soient deux classes de $\o$-conjugaison respectant
les m{\^e}mes sym{\'e}tries (dans notre cas, deux D-branes \AA\ de
param{\`e}tres {\'e}ventuellement diff{\'e}rents). Si $m$, $\bar{m}$ sont
deux {\'e}l{\'e}ments des classes de $\o$-conjugaison en question, la
solution est \cite{pr,gtt}
\bea
g= a(x_+)\, m \, \o(a(x_-))\- \ , 
\label{solouv}
\\ 
a(x+2\pi)=a(x)\, \o\- (m\- \bar{m})\ .
\label{condouv}
\eea
On peut maintenant {\'e}tudier la pr{\'e}servation du flot spectral par la
D-brane \AA. L'action du flot spectral sur la matrice $a$ ci-dessus,
\bea
a(x)\  \rar \ e ^{wxJ}\, a(x),
\label{flotouv}
\eea
est bien d{\'e}finie {\`a} condition que $\omega (J) = -J$, si $J$ est la
matrice de $\sl $ utilis{\'e}e dans la d{\'e}finition du flot spectral
eq. (\ref{flotspec}). On trouve facilement une telle matrice $J =$
{\tiny $\left(
\begin{array}{rl}
 0 & \half \cr
 -\half & 0 \cr
\end{array}
\right)$}. 

\Fig{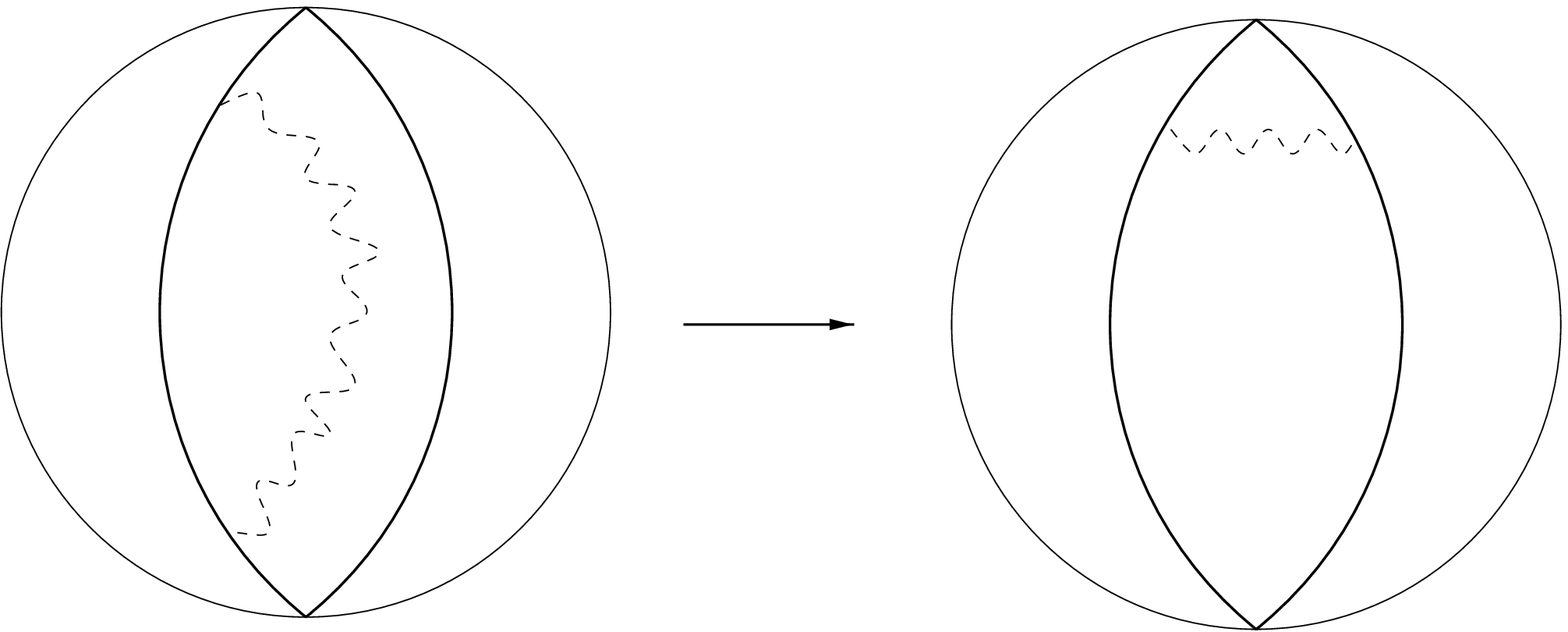}{12}{ Une corde ouverte sur une D-brane \AA\ et son
  image par le flot spectral $w=1$, en projection sur le plan $(\rho,\theta)$ }

Nous consid{\'e}rons des cordes ouvertes avec deux extr{\'e}mit{\'e}s sur la
m{\^e}me D-brane, il faut donc que le flot spectral pr{\'e}serve la
condition $a(x+2\pi)=a(x)$, ce qui est le cas seulement si $w$ est
pair. Sinon, un des bouts de la corde ouverte migre sur la D-brane
sym{\'e}trique par rapport {\`a} la ligne $\rho=0$, autrement dit celle
obtenue par $\cst \rar -\cst$ dans l'{\'e}quation (\ref{eqads2}), soit
$q\rar -q$ si l'on prend en compte la quantification (voir la Figure
\ref{asd3flot}). Dans le cas de
la D-brane droite, $\cst =q=0$, tout le flot spectral est
pr{\'e}serv{\'e}. Une autre particularit{\'e} de cette D-brane est que les
solutions de cordes ouvertes qui y aboutissent se prolongent par
r{\'e}flexion {\`a} des solutions de cordes ferm{\'e}es dans \AAA. Cela
conduit {\`a} conjecturer \cite{pr} que leur spectre est la \flqq
moiti{\'e}\frqq\ du spectre des cordes ferm{\'e}es, ce qui a {\'e}t{\'e}
prouv{\'e} par un calcul de fonction de partition dans \cite{plot}. Un
autre argument de \cite{plot} montre que  la sym{\'e}trie
compl{\`e}te du flot spectral est restaur{\'e}e dans le secteur des
repr{\'e}sentations discr{\`e}tes. Cela n'est pas tr{\`e}s surprenant compte
tenu du fait que la matrice $J$ qui d{\'e}finit le flot spectral doit
{\^e}tre de genre temps, tout comme les trajectoires des  cordes
courtes correspondant aux {\'e}tats discrets du spectre.

\bs

En r{\'e}sum{\'e}, les arguments semi-classiques sugg{\`e}rent que le
spectre discret ne d{\'e}pend pas de la position de la D-brane $AdS_2$,
y compris en ce qui concerne le flot spectral~; et qu'il est donc toujours
identique au spectre discret de la D-brane droite $\theta=\halfpi$,
qui vaut la moiti{\'e} du spectre des cordes ferm{\'e}es dans \AAA. En
revanche, dans le secteur continu, seule la moiti{\'e} du flot spectral
est pr{\'e}serv{\'e}e, et on s'attend de plus {\`a} ce que le spectre
d{\'e}pende de la position de la D-brane. Pour en savoir plus, il faut
recourir {\`a} la th{\'e}orie conforme, et tirer profit de la staticit{\'e}
de la D-brane \AA\ pour effectuer une rotation de Wick $\t \rar i\t$
et passer 
dans l'\AAA\ Euclidien, $H_3$. C'est ce que nous ferons dans le
prochain chapitre.


\setcounter{chapter}{2}

\chapter{ Groupes, D-branes et th{\'e}orie conforme 
\label{chapCFT}
}

Nous abandonnons maintenant la th{\'e}orie effective de basse {\'e}nergie,
d{\'e}crite par l'action de Born-Infeld, au profit d'une description des
D-branes exacte en $\ap$. Nous faisons cependant toujours
l'approximation que nos 
branes-tests ne modifient pas l'espace-temps.

La description exacte des D-branes fait appel {\`a} la th{\'e}orie
conforme {\`a} deux dimensions avec bord, th{\'e}orie d{\'e}finie sur la
surface d'univers d'une corde ouverte. Dans le cas des groupes
compacts, cette th{\'e}orie est rationnelle, elle peut {\^e}tre d{\'e}crite
au moyen d'un nombre fini de champs. Le cas non compact est nettement
plus compliqu{\'e}, car le produit de deux op{\'e}rateurs fait en
g{\'e}n{\'e}ral intervenir un nombre infini d'op{\'e}rateurs. Cette
difficult{\'e} sera circonvenue au moyen d'hypoth{\`e}ses d'analyticit{\'e}
sur les quantit{\'e}s physiques associ{\'e}es aux
D-branes, comme les {\'e}tats de bord.
Ces hypoth{\`e}ses d'analyticit{\'e} nous permettront d'utiliser
certains op{\'e}rateurs non physiques, dits d{\'e}g{\'e}n{\'e}r{\'e}s, dont les
produits d'op{\'e}rateurs comportent des termes en nombre fini. 

Un autre genre d'hypoth{\`e}se d'analyticit{\'e} sera n{\'e}cessaire pour
s'occuper d'un espace-temps minkowskien comme \AAA. On utilisera alors la
rotation de Wick, $t\rar it$, pour passer {\`a} l'espace euclidien
\H.

\section{ D-branes sym{\'e}triques dans les groupes compacts (II) 
\label{sectbranecompact}
}

Dans cette section, nous allons {\'e}tudier les D-branes sym{\'e}triques
qui ont fait l'objet de la section \Fac{\ref{BIbranes}} du point de vue de la
th{\'e}orie conforme avec bord. 
Nous rappellerons donc des
notions de th{\'e}orie conforme
qui serviront aussi par la suite dans le cas non compact. Des notions
de th{\'e}orie conforme suppl{\'e}mentaires peuvent {\^e}tre trouv{\'e}es dans
\cite{vsrevue} et \cite{openstrings}.

\subsection{ Groupes compacts et th{\'e}orie conforme rationnelle 
\label{soussectRCFT}
}

Rappelons quelques propri{\'e}t{\'e}s de la th{\'e}orie des cordes ferm{\'e}es
dans les groupes compacts que nous avons {\'e}voqu{\'e}es dans la section
\Fac{\ref{soussectWZW}}. D'abord, la th{\'e}orie se factorise en deux
secteurs, appel{\'e}s secteur droit et secteur gauche. Chaque secteur
est muni d'une alg{\`e}bre de Kac-Moody ou alg{\`e}bre de Lie affine \alieg, dont
l'alg{\`e}bre enveloppante
contient l'alg{\`e}bre de Virasoro. Cette alg{\`e}bre s'appelle
l'alg{\`e}bre chirale de la th{\'e}orie. La sym{\'e}trie correspondante est
pr{\'e}serv{\'e}e au niveau quantique. Cela se traduit par le fait que le
spectre s'organise en repr{\'e}sentations de cette alg{\`e}bre, et par des
identit{\'e}s de Ward sur les fonctions de corr{\'e}lation des champs
correspondant aux {\'e}tats du
spectre. 

Les repr{\'e}sentations de \alieg\ qui interviennent dans le spectre de
la th{\'e}orie des cordes dans un groupe compact doivent {\^e}tre des 
repr{\'e}sentations de plus haut poids pour que l'{\'e}nergie $L_0$ soit
born{\'e}e inf{\'e}rieurement, et ob{\'e}ir {\`a} la
contrainte d'unitarit{\'e} pour que les observables physiques prennent
des valeurs r{\'e}elles. Cette contrainte d'unitarit{\'e} est d{\'e}finie
par rapport {\`a} la conjugaison naturelle des g{\'e}n{\'e}rateurs de
l'alg{\`e}bre de Lie affine, 
\bea
(J^a_n)^*=J^a_{-n}.
\label{conjaffine}
\eea
Pour une repr{\'e}sentation de plus haut poids construite {\`a} partir
des {\'e}tats $|PHP\rangle$ d'une repr{\'e}sentation unitaire de \lieg,
qui v{\'e}rifient donc $\langle PHP|PHP \rangle >0$, on 
{\'e}value ainsi l'unitarit{\'e} en calculant les normes au carr{\'e} des {\'e}tats
$J|PHP\rangle$ obtenus par l'action d'un produit $J$ de
g{\'e}n{\'e}rateurs  de \alieg, ces normes au carr{\'e}
valent $\langle PHP|J^*J|PHP\rangle
$. Il appara{\^\i}t alors qu'un nombre fini de repr{\'e}sentations
de plus haut poids de \alieg\ sont unitaires, et peuvent {\^e}tre
utilis{\'e}es pour construire une th{\'e}orie des cordes. La finitude est
directement li{\'e}e au caract{\`e}re compact du groupe en question, et
est viol{\'e}e par tous les groupes non compacts {\`a} commencer par
$\R$. 

Dans le cas de \SU, la contrainte d'unitarit{\'e} pour une
repr{\'e}sentation de plus haut poids construite {\`a} partir d'une
repr{\'e}sentation de spin $j$ de \SU\ est 
\bea
j\leq \frac{k}{2}.
\label{unitaritesu2}
\eea
La contrainte d'unitarit{\'e} pour une repr{\'e}sentation donn{\'e}e est
parfois appel{\'e}e \flqq th{\'e}or{\`e}me d'absence de fant{\^o}mes\frqq,
o{\`u} les \flqq fant{\^o}mes\frqq\ sont les {\'e}tats de norme carr{\'e}e
n{\'e}gative. 

\bs

Le mod{\`e}le de Wess-Zumino-Witten {\`a} un niveau $k$ donn{\'e} 
est donc construit {\`a} partir d'un
nombre fini de repr{\'e}sentations d'une alg{\`e}bre de sym{\'e}trie qui
contient l'alg{\`e}bre de Virasoro, cette derni{\`e}re d{\'e}finit la sym{\'e}trie
conforme. Il s'agit donc d'un exemple de \de{th{\'e}orie conforme
  rationnelle}. Nous avons d{\'e}j{\`a} pr{\'e}cis{\'e} le spectre $Spec(G)$ du
mod{\`e}le associ{\'e} {\`a} un groupe $G$ (voir la formule \Fac{(\ref{specWZW})}), 
mais il ne s'agit pas du seul mod{\`e}le que l'on puisse
construire {\`a} partir de l'alg{\`e}bre de sym{\'e}trie \alieg. En fait, il
est commode de d{\'e}finir le spectre au moyen de la \de{fonction de
partition du tore},
d{\'e}finie par
\bea
Z(q)=\Tr _{Spec(G)} q^{L_0-\frac{c}{24}}\bar{q}^{\bar{L}_0-\frac{c}{24}}.
\label{tore}
\eea
Ici $q=e ^{2\pi i\tau}$, o{\`u} le nombre complexe $\tau $ est le
param{\`e}tre modulaire d'un tore {\`a} deux dimensions, qui repr{\'e}sente
une topologie possible (de genre $1$) de la surface d'univers d'une
corde ferm{\'e}e. Comme le spectre se d{\'e}compose en repr{\'e}sentations
de \alieg, $Z(q)$ se d{\'e}compose en caract{\`e}res de \alieg. Si $\hR$
est une repr{\'e}sentation de \alieg, son caract{\`e}re est 
\bea
\chi_{\hR}(q)= \Tr_{\hR}q^{L_0-\frac{c}{24}}.
\label{caract}
\eea
Les spectres possibles de th{\'e}ories ayant une sym{\'e}trie donn{\'e}e
doivent ob{\'e}ir {\`a} la contrainte physique d'invariance modulaire~:
$Z(q)$ doit {\^e}tre invariant sous le groupe modulaire, qui relie les
param{\`e}tres $\tau$ de tores {\'e}quivalents. Ce groupe est engendr{\'e}
par les transformations 
\bea
T~:\ \tau\rar \tau+1\ , \ S~: \ \tau\rar -\frac{1}{\tau}.
\label{modulaire}
\eea
On note $\tq= e ^{-2i\pi/\tau}$ l'image par $S$ du param{\`e}tre $q$. 
Si l'invariance par $T$ est fort peu contraignante, et sera
automatiquement satisfaite dans les th{\'e}ories que nous
consid{\`e}rerons, 
l'invariance
par $S$ est en revanche une contrainte non triviale. On peut montrer que la
fonction de partition du mod{\`e}le de Wess-Zumino-Witten,
\bea
Z^{WZW}(q)=\sum_{R\in Rep_k(G)} \chi_{\hR} (q)\chi_{\hR ^\dagger} (\bar{q}),
\label{ZWZW}
\eea
est invariante modulaire. Cette fonction de partition est un 
invariant modulaire \de{diagonal}, car une repr{\'e}sentation est
syst{\'e}matiquement coupl{\'e}e avec la repr{\'e}sentation
conjugu{\'e}e. Pour
prouver cette invariance modulaire,
il est essentiel de noter que le
groupe modulaire
poss{\`e}de une repr{\'e}sentation non
triviale
dans l'espace $Vect(Rep_k(G))$ qui a
pour base $Rep_k(G)$, d{\'e}finie par 
\bea
\chi_{\hR}(\tq)=\sum_{R'\in Rep_k(G)} S_R ^{R'} \chi_{\hR'}(q).
\label{modcaract}
\eea
La matrice\footnote{
La matrice $S_R ^{R'}$ d{\'e}pend naturellement de $k$ et de $Rep_k(G)$,
et pas seulement des repr{\'e}sentations $R,R'$ de l'alg{\`e}bre de Lie
\lieg.
}
$S_R^{R'}$ ou \de{matrice $S$}, 
qui d{\'e}finit l'action de la transformation
modulaire $S$ dans cette repr{\'e}sentation, est sym{\'e}trique et
unitaire (par rapport {\`a} la m{\'e}trique triviale $\delta_{RR'}$ et {\`a}
la conjugaison $R\rar R^\dagger$ des repr{\'e}sentations), ce qui assure
l'invariance modulaire de $Z^{WZW}$.   

Les autres quantit{\'e}s physiques int{\'e}ressantes du mod{\`e}le sont les
fonctions de corr{\'e}lation. Une fois le mod{\`e}le d{\'e}fini par sa
fonction de partition, celles-ci peuvent {\^e}tre calcul{\'e}es et
s'expriment {\`a} l'aide d'autres donn{\'e}es de la th{\'e}orie des
repr{\'e}sentations de \alieg. 

\rem{ Coefficients d'OPE en termes des regles de fusion~? }

\subsection{ Solutions de Cardy et branes sym{\'e}triques 
\label{soussectcardy}
 }

\Fig{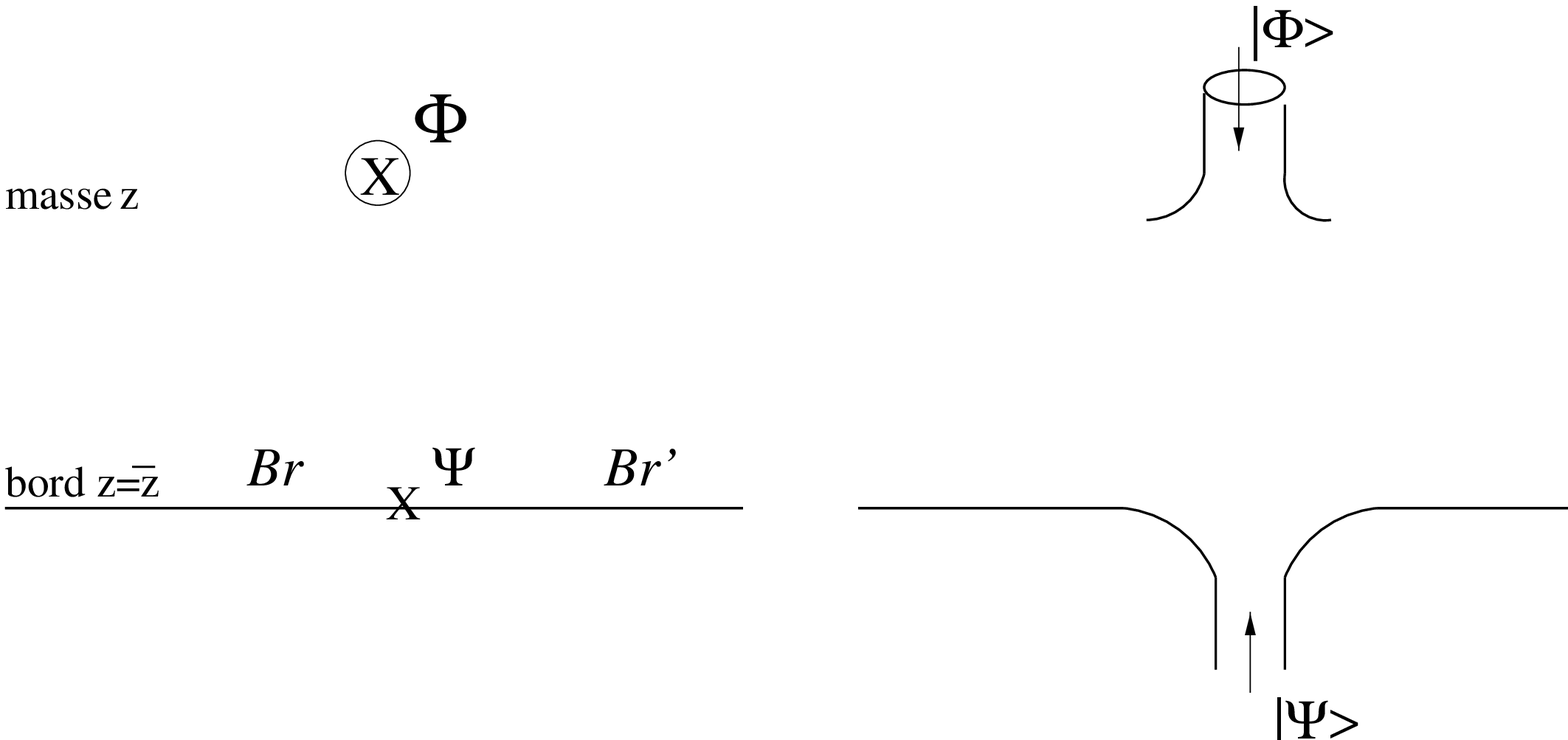}{14}{ Des champs de cordes ferm{\'e}es $\Phi$ et ouvertes
  $\Psi$ (ce dernier change les conditions au bord de $\B$ en
  $\B'$). Les {\'e}tats de cordes 
$|\Phi\rangle$ et $|\Psi\rangle$ correspondants. }

Nous avons expliqu{\'e} que le mod{\`e}le de Wess-Zumino-Witten {\'e}tait
une th{\'e}orie conforme rationnelle, d{\'e}crivant la propagation de
cordes ferm{\'e}es dans des groupes compacts. La construction
additionnelle pour d{\'e}crire 
les cordes ouvertes est la th{\'e}orie conforme avec
bord. Elle d{\'e}pend fortement de la th{\'e}orie des cordes ferm{\'e}es
correspondantes. D'abord elle contient tous les op{\'e}rateurs de cette
th{\'e}orie, qui d{\'e}crivent l'insertion de cordes ferm{\'e}es dans la
\de{masse} (c'est-{\`a}-dire ailleurs qu'au bord), et doivent se
comporter loin du bord 
sans en ressentir la pr{\'e}sence (par localit{\'e}). Il faut y ajouter
les op{\'e}rateurs au bord, qui d{\'e}crivent l'insertion de cordes
ouvertes au bord. 
L'influence du bord sur les cordes ouvertes est d{\'e}finie au moyen de
conditions au bord, et les op{\'e}rateurs au bord modifient ces
conditions (voir la figure \ref{abcft}).

La th{\'e}orie conforme au bord est donc d{\'e}finie par un ensemble de
donn{\'e}es comprenant celles de la th{\'e}orie dans la masse, et soumises
{\`a} d'autres conditions de coh{\'e}rence qui leur sont propres. 

Une donn{\'e}e de base  d'une th{\'e}orie des cordes ouvertes est
naturellement son spectre. Nous nous int{\'e}ressons aux D-branes
sym{\'e}triques $\B$, dont les conditions de collage en $z=\bz$, qui
s'{\'e}crivent 
$J=\bar{J}$ ou  $J=\omega(\bar{J}) $, respectent l'alg{\`e}bre
chirale des sym{\'e}tries des cordes ferm{\'e}es. Ce spectre $Spec_{\B} (G)$ 
s'organise
donc lui aussi en repr{\'e}sentations de cette alg{\`e}bre chirale, et la
fonction de partition correspondante se d{\'e}compose en
repr{\'e}sentations de l'alg{\`e}bre de Lie affine \alieg. La fonction de
partition des cordes ouvertes s'{\'e}tirant entre deux D-branes
$\B,\B'$, 
qui s'{\'e}crira au moyen des
caract{\`e}res de ces repr{\'e}sentations, se calcule au moyen 
d'une boucle de corde ouverte,
\bea
Z^\ouv_{\B,\B'} (q) &=& \Tr_{Spec_{\B,\B'} (G)} q^{L_0-\frac{c}{24}}, 
\label{Zcyl}
\\
&=& \sum_{R\in Rep_k(G)} n_{\B,\B'}^{\hR}\chi_{\hR}(q).
\label{decZcyl}
\eea
Nous d{\'e}finissons ainsi les entiers $n_{\B,\B'}^{\hR}$,
multiplicit{\'e}s des repr{\'e}sentations de \alieg\ dans le spectre
$Spec_{\B,\B'}(G)$. 

\Fig{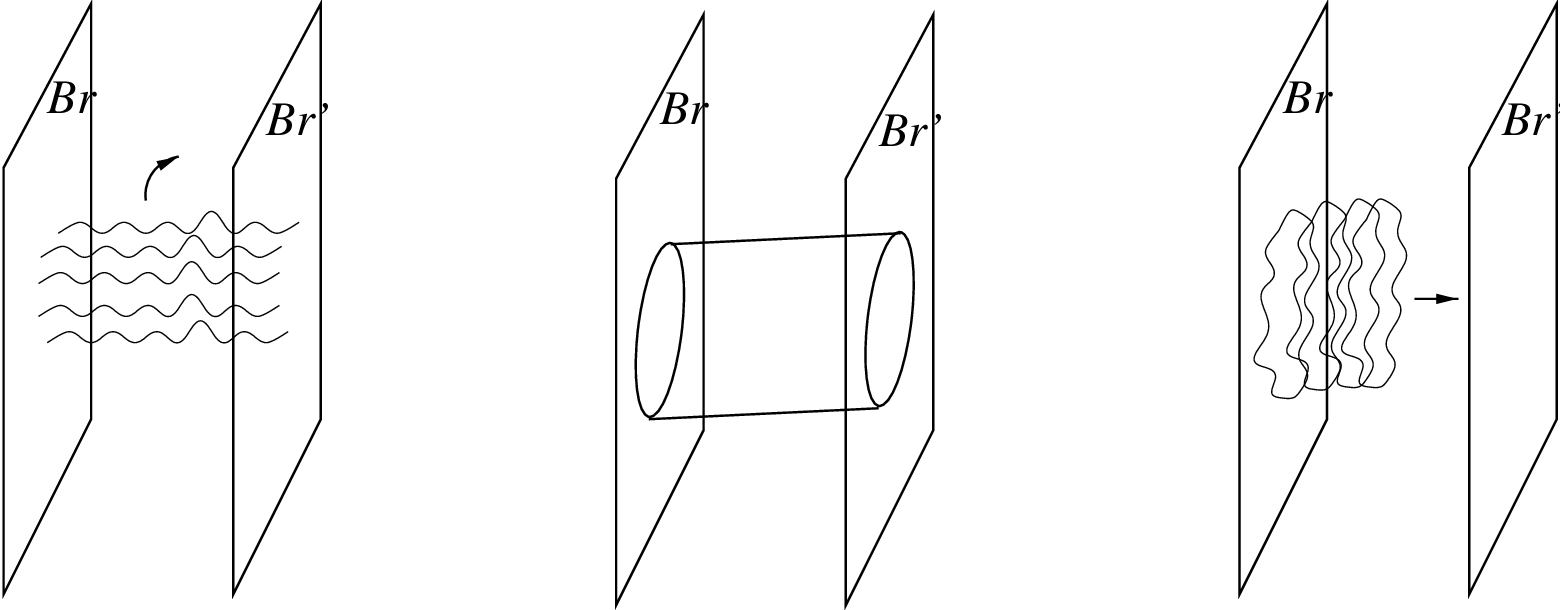}{15}{ Le diagramme du cylindre entre deux branes $\B$ et
  $\B'$, et son interpr{\'e}tation en termes de cordes ouvertes et
  ferm{\'e}es. }

On ne peut pas {\`a} proprement parler d'invariance modulaire pour le
diagramme du cylindre comme pour la fonction de partition du tore,
cependant l'{\'e}quivalent de la transformation modulaire $S$ est la
dualit{\'e} entre cordes ouvertes et cordes ferm{\'e}es. En effet, le
diagramme du cylindre d{\'e}crit aussi bien une boucle d'une corde
ouverte {\'e}tir{\'e}e entre les D-branes $\B$ et $\B'$, que l'{\'e}mission
d'une corde ferm{\'e}e par $\B$ suivie de son absorption par $\B'$ (voir
la figure \ref{ardy}). Ces
processus d'{\'e}mission et d'absorption sont eux-m{\^e}mes d{\'e}crits par
le couplage des D-branes aux {\'e}tats de cordes ferm{\'e}es, qui
d{\'e}finissent les \de{{\'e}tats de bord}. Plus pr{\'e}cis{\'e}ment, si
$\Phi$ est un champ (ce que nous avons appel{\'e} plus haut un
op{\'e}rateur) appartenant au 
 spectre des cordes ferm{\'e}es, la fonction {\`a} un point de $\Phi$ peut
 {\^e}tre non nulle en pr{\'e}sence d'un bord (naturellement, elle doit
 tendre vers z{\'e}ro loin du bord). Si ce bord est d{\'e}fini par une
 D-brane $\B$, on obtient la quantit{\'e} $\langle \Phi
 \rangle_{\B}$. On peut montrer que la donn{\'e}e de cette quantit{\'e}
 pour tout champ $\Phi$ suffit {\`a} d{\'e}crire la D-brane
 $\B$~; ces informations sont {\'e}quivalents {\`a} la d{\'e}finition d'un
 \de{{\'e}tat de bord} appartenant {\`a} l'espace de Hilbert des cordes
 ferm{\'e}es, qui s'{\'e}crit sch{\'e}matiquement\footnote{ Ici et par la suite,
   nous n{\'e}gligeons des coefficients de 
  module un des {\'e}tats de bord, car ils n'affectent pas les calculs de
  type Cardy que nous effectuerons. 
}
\bea
|\B \rangle = \sum_\Phi \llangle \Phi(z=\halfi) \rrangle_{\B}|\Phi \rangle,
\label{etatbord}
\eea
o{\`u} l'on somme sur une base de cet espace de Hilbert. Le calcul du
diagramme du cylindre donne donc~:\footnote{ Nous n{\'e}gligeons des
  subtilit{\'e}s li{\'e}es {\`a} la parit{\'e} 
  CPT de la surface d'univers.} 
\bea
Z^\cyl_{\B,\B} (q)=\langle \B| q^{L_0+\bar{L}_0-\frac{c}{24}}| \B \rangle.
\label{dualcyl}
\eea
La \de{condition de Cardy}, qui r{\'e}alise l'{\'e}quivalent de la
transformation modulaire $S$, est~:
\bea
Z^\cyl_{\B,\B} (\tq) = Z^\ouv_{\B,\B} (q). 
\label{cardy}
\eea
M{\^e}me si l'on ne conna{\^\i}t pas le spectre des cordes ouvertes, la
condition de Cardy restreint les {\'e}tats de bord possibles par
l'exigence que les multiplicit{\'e}s $n^{\hR}_{\B,\B} $, qui
apparaissent dans $Z^\ouv_{\B,\B} (q)$, soient enti{\`e}res.
Cette condition s'applique non seulement {\`a} chaque D-brane comme nous
l'avons {\'e}crite,
mais aussi naturellement au spectre des cordes ouvertes entre deux
D-branes, si bien que dans certains cas il peut exister des familles
de D-branes incompatibles (voir \cite{grw}). Dans ces cas-l{\`a},
\flqq $Z^\cyl_{\B,\B'}$ v{\'e}rifie la condition de Cardy\frqq\ est une
relation d'{\'e}quivalence entre les D-branes $\B,\B'$ qui d{\'e}termine
plusieurs classes d'{\'e}quivalence distinctes.

\bs

L'importance de la condition de Cardy vient du fait
qu'en g{\'e}n{\'e}ral
la donn{\'e}e de
conditions au bord ne suffit pas {\`a} d{\'e}finir une D-brane. Les
conditions de collage de nos D-branes
sym{\'e}triques, $J=\omega(\bar{J})$, s'expriment 
par la contrainte lin{\'e}aire
\bea
\left(J^a_n-\omega(\bar{J}^a_{-n})\right) |\B \rangle = 0.
\label{condbord}
\eea
De fa{\c c}on tr{\`e}s g{\'e}n{\'e}rale, une base de l'espace des solutions
est connue~: il s'agit des \de{{\'e}tats d'Ishibashi}. Son cardinal est
$|Rep_k(G)|$, autrement dit {\`a} chaque champ du spectre des cordes
ferm{\'e}es correspond un {\'e}tat d'Ishibashi $|R\rrang$. On peut
m{\^e}me choisir les {\'e}tats d'Ishibashi de telle sorte que
\bea
\llang R| q ^{\LL}|R' \rrang = \delta_{R,R'}\chi_{\hR}(q).
\label{ishibashi}
\eea
La th{\'e}orie conforme avec bord associ{\'e}e {\`a} une D-brane
doit v{\'e}rifier de nombreuses conditions de coh{\'e}rence
\cite{lewellen}.
Nous nous int{\'e}resserons pour l'instant {\`a} la condition de Cardy,
qui est non-lin{\'e}aire.
La solution en est connue 
sous des hypoth{\`e}ses assez
g{\'e}n{\'e}rales \cite{cardy}, qui s'appliquent au cas des D-branes
sym{\'e}triques dans les groupes compacts. Il existe ainsi 
un nombre fini de 
D-branes, param{\'e}tr{\'e}es par les {\'e}l{\'e}ments de $Rep_k(G)$. Les
 {\'e}tats de bord correspondants, appel{\'e}s \de{{\'e}tats
  de Cardy}, sont 
\bea
|R\rangle = \sum_{R'\in Rep_k(G)}\frac{S_R^{R'}}{\sqrt{S_0^{R'}}}|R'\rrang,
\label{etatcardy}
\eea
o{\`u} $0$ d{\'e}signe la repr{\'e}sentations triviale {\`a} une dimension de \lieg.
On peut en d{\'e}duire les coefficients $n^{\hR_3}_{R_1,R_2}$ du spectre
des cordes ouvertes,
\bea
n^{\hR_3}_{R_1,R_2}= \sum_{R\in Rep_k(G)}
\frac{S_{R_1}^{R}S_{R_2}^{R^\dagger} S_{R^\dagger}^{R_3}}{S_{0}^{R}} ,
\label{cardyouv}
\eea
qui sont entiers d'apr{\`e}s la formule de Verlinde, qui les relie {\`a}
certains \de{coefficients de fusion} de l'alg{\`e}bre de Lie affine
\alieg.

Montrons maintenant en quoi ces r{\'e}sultats sont en accord avec
l'analyse de la section \Fac{\ref{topoquantif}}, bas{\'e}e sur l'action
de Born-Infeld. Nous allons pour cela suivre \cite{fffs}. D'abord, les
{\'e}l{\'e}ments de $Rep_k(G)$ sont param{\'e}tr{\'e}s par les vecteurs de 
poids $\lambda$ correspondants. Dans le cas o{\`u} l'alg{\`e}bre de Lie
\lieg\ est simple, $Rep_k(G)$ comprend une partie finie du r{\'e}seau
des poids {\`a} cause de la condition d'unitarit{\'e}, plus
pr{\'e}cis{\'e}ment
\bea
Rep_k(G)= \frac{L^w}{W\ltimes \frac{k}{2\pi}{\rm Ker} (\exp) }.
\label{repkg}
\eea
Les {\'e}tats de Cardy sont bien param{\'e}tr{\'e}s par cet ensemble, mais
n'ont pas exactement les m{\^e}mes positions que nos solutions des
{\'e}quations de Born-Infeld. Ces positions sont en effet
\bea
\exp \- t = 2\pi \frac{ \lambda +\rho }{k+g^\vee},
\label{positioncardy}
\eea
o{\`u} $\lambda$ est un poids.
La diff{\'e}rence avec nos solutions r{\'e}side dans les termes $g^\vee$,
qui correspond {\`a} la renormalisation du niveau d{\'e}j{\`a} {\'e}voqu{\'e}e,
et $\rho$, qui est un vecteur constant. On retrouve nos
solutions dans la limite o{\`u} $k$ est grand et o{\`u} $\lambda$ est
proportionnel {\`a} $k$. 

Dans le cas des classes de $\omega$-conjugaison, il faut consid{\'e}rer
un ensemble $Rep_k^\o(G)$ de poids fractionnels sym{\'e}triques (invariants par
$\o$), et les positions des D-branes sont {\`a} l'avenant. L'accord avec
les r{\'e}sultats semi-classiques est toujours \flqq presque exact\frqq.

Enfin, on peut comparer le spectre des cordes ouvertes avec celui des
fluctuations quadratiques. Le premier, qui comporte un nombre fini de
repr{\'e}sentations, est une troncation du second, qui les comporte
toutes. Plus $k$ est grand, moins de repr{\'e}sentations sont touch{\'e}es
par la troncation, et $\lim _{k\rar\infty}Rep_k(G)=Rep(G)$.  

\section{ D-branes dans l'\AAA\ Euclidien }

\subsection{ Cordes ferm{\'e}es dans $H_3$ }

Si l'on effectue la rotation de Wick, $\t\rar i\t$, dans le m{\'e}trique
(et le champ $B$) de la th{\'e}orie des cordes dans \AAA, on obtient la
m{\'e}trique de l'espace \H,
\bea
ds^2=L^2\left( \cosh^2\rho d\t ^2+d\rho ^2+\sinh^2\rho d\theta ^2 \right).
\label{meth3}
\eea
On a toujours $L^2=k\ap$, et on note $b^2=(k-2)\- $.
La th{\'e}orie des cordes dans \H\ n'est pas physique car le champ
$H=dB$ est imaginaire pur~; cependant elle sert d'outil pour l'{\'e}tude
des cordes dans \AAA\ ainsi que dans le cigare $\SL/U(1)$, que nous
{\'e}tudierons ult{\'e}rieurement. Son spectre est plus simple que celui
d'\AAA\ et ne contient que des repr{\'e}sentations continues, de plus la
th{\'e}orie est {\'e}troitement li{\'e}e {\`a} la th{\'e}orie de Liouville, qui
a fait l'objet de nombreuses {\'e}tudes (voir par exemple la revue 
\cite{revueliouville}).  

La rotation de Wick ne pr{\'e}serve pas la sym{\'e}trie $\SL\times\SL$
d'\AAA, cependant \H\ jouit toujours d'une sym{\'e}trie {\`a} six
g{\'e}n{\'e}rateurs \SLC, qui s'{\'e}tend au niveau de la th{\'e}orie des
cordes en une alg{\`e}bre de Lie affine. Nous allons d'abord d{\'e}crire
l'action g{\'e}om{\'e}trique de ces sym{\'e}tries, au moyen d'une
reparam{\'e}trisation de \H~:
\bea
\gamma = e ^{\t+i\theta}\tanh\rho\ , \ e ^\phi=e ^{-\t}\cosh\rho\ , \
ds^2= d\phi ^2+ e ^{2\phi}d\gamma d\bar{\gamma}.
\label{coordh3}
\eea
On peut alors {\'e}crire \H\ comme un ensemble de matrices hermitiennes
de d{\'e}terminant un,
\bea
h = \left(
\begin{array}{cc}
e ^\phi & e ^\phi \bar{\gamma} \\
e ^\phi \gamma & e ^\phi \gamma\bar{\gamma}+ e ^{-\phi}
\end{array}
\right).
\label{matriceh3}
\eea
Cela permet d'expliciter l'action des {\'e}l{\'e}ments $g$ de \SLC, soit
$h\rightarrow ghg^\dagger$, dont on d{\'e}duit l'identification entre
\H\ et $\SLC/\SU$. De plus, les courants du mod{\`e}le sigma
correspondant s'expriment {\'e}galement {\`a} l'aide de la matrice $h$,
\bea
J(z)=k h\- \bp h\ , \ \bJ(\bz)=-k \p h\, h\- .
\label{couranth3}
\eea
Parmi les coordonn{\'e}es $\phi,\gamma,\bar{\gamma}$ que nous venons
d'introduire, on peut consid{\'e}rer $\phi$ comme une coordonn{\'e}e
radiale, et les autres comme des param{\`e}tres du bord de \H. Comme
dans le cas de \SL, l'impulsion associ{\'e}e {\`a} la coordonn{\'e}e
radiale, que nous appellerons de nouveau $j$, correspond au Casimir de
l'action de \SLC, autrement dit au Laplacien de la m{\'e}trique
(\ref{meth3}). Ainsi, on peut param{\'e}trer les fonctions sur \H\ par
$j$ et par un {\'e}l{\'e}ment $u\in\C$ du bord,
\bea
\Phi ^j_u(\phi,\g,\bg)= -\frac{2j+1}{\pi}\left( e ^\phi |u-\g|^2+e
  ^{-\phi}\right) ^{2j}.
\label{fonctionh3}
\eea
Si nous {\'e}crivons le Laplacien d{\'e}fini n{\'e}gatif de \H,
\bea
\Delta_{\H}= \p_\phi ^2 + 2\p_\phi + 4 e ^{-2\phi}\p_\g \p_{\bg},
\label{laplacienh3}
\eea
alors nous avons $\Delta_{\H}\Phi ^j= 4j(j+1)\Phi ^j$. Cette valeur
propre $4j(j+1)$ est un r{\'e}el n{\'e}gatif si $\Phi ^j$ est \flqq
normalisable au sens des distributions\frqq, c'est-{\`a}-dire si
l'int{\'e}grale $\int e ^{2\phi } d\phi d\g d\bg\ \Phi ^j (\Phi ^j)^* $
ne diverge que comme $\int 1$. Cela se produit si et seulement si
\bea
j \in -\half + i\R.
\label{jphys}
\eea
Cet ensemble de valeurs physiques de $j$, qui d{\'e}finit le spectre
d'un champ scalaire dans \H, correspond aux repr{\'e}sentations
continues de \SLC. Pour tenir compte de la sym{\'e}trie $j\rar -j-1$,
qui g{\'e}n{\'e}ralise la correspondance entre les s{\'e}ries discr{\`e}tes $D_j^+$ et
$D_j^-$ 
(dans le cas o{\`u} $j$ est r{\'e}el), nous allons nous restreindre {\`a} $j
\in -\half + i \R_+$. 
Les {\'e}tats $\Phi ^j$ se comportent {\`a} l'infini comme des
ondes planes, et leur produit scalaire $L^2$ est donn{\'e} par 
\bea
\int_{\H}\Phi ^j_u(\Phi ^{j'}_{u'})^* = \delta (j-j')\delta(u-u').
\label{scalphi}
\eea

\bs

Si l'on consid{\`e}re maintenant les $\phi,\gamma,\bg$ comme des
fonctions sur une surface d'univers, d{\'e}crivant le plongement d'une
corde ferm{\'e}e dans $\H$, alors les fonctions $\Phi ^j_u$ deviennent des
op{\'e}rateurs de cordes ferm{\'e}es, et d{\'e}crivent la d{\'e}composition du
spectre en repr{\'e}sentations de \aslc. Ce spectre est donc d{\'e}fini par
le coefficient ou \de{densit{\'e} spectrale} de ces champs. Mais 
leur fonction {\`a} deux points est toujours donn{\'e}e par
l'{\'e}quation (\ref{scalphi}) (aux facteurs d{\'e}pendant de $z$ pr{\`e}s) 
et ne re{\c c}oit donc pas de corrections
en $b^2=\frac{1}{k-2}$. De fa{\c c}on {\'e}quivalente, la densit{\'e}
spectrale du champ $\Phi ^j$ est la m{\^e}me que la densit{\'e} de la
fonction $\Phi ^j$ dans l'espace des fonctions sur \H. On peut
formaliser cette assertion en {\'e}crivant l'op{\'e}rateur identit{\'e} sur
le spectre des cordes ferm{\'e}es\footnote{Nous omettons les contributions
des oscillateurs associ{\'e}s aux champs $J,\bar{J}$. },
\bea
\Id = \int_{-\half+i\R_+} dj \int d^2u | \Phi ^j\rangle\langle \Phi
^j|.
\label{idh3}
\eea 
Naturellement, la solution compl{\`e}te de la th{\'e}orie des cordes dans
\H, au-del{\`a} de la donn{\'e}e du spectre, requiert celle des fonctions
de corr{\'e}lation et la v{\'e}rification de certaines de leurs
propri{\'e}t{\'e}s comme la sym{\'e}trie de croisement
\cite{joergcroisement,joergfusion,joergope}. Mais nous n'insisterons
pas sur ce sujet ici. Nous allons plut{\^o}t nous tourner vers les
D-branes dans \H. 

\subsection{ {\'E}tats de bord des D-branes $AdS_2$ }

Les D-branes \AA\ dans \AAA\ sont statiques. Leurs {\'e}quations, $\sinh
\rho \cos\theta =\cst$, 
d{\'e}finissent donc des sous-vari{\'e}t{\'e}s de \H\ dont on peut
s'attendre {\`a} ce qu'elles aient une interpr{\'e}tation en termes de
th{\'e}orie des cordes. Naturellement, elles sont automatiquement
solutions des {\'e}quations de Born-Infeld, mais avec un champ $F$
imaginaire pur. Elles ont la g{\'e}om{\'e}trie de l'\AA\ Euclidien
c'est-{\`a}-dire $H_2$, mais nous les appellerons toujours des D-branes
\AA.
On peut {\'e}tudier leurs propri{\'e}t{\'e}s de
sym{\'e}trie en r{\'e}{\'e}crivant leur {\'e}quation en fonction de la matrice
$h$, 
\bea
\Tr \Omega h = 2\sinh\rho \cos\theta=2\sinh r.
\label{ads2h3}
\eea
La matrice $\Omega$ est celle que nous avons d{\'e}j{\`a} utilis{\'e}e pour
d{\'e}finir l'automorphisme ext{\'e}rieur de \SL, et pour {\'e}crire
l'{\'e}quation de la D-brane \AA\ dans \AAA, qui {\'e}tait d'ailleurs
identique {\`a} (\ref{ads2h3}). Le sous-groupe de \SLC\ qui pr{\'e}serve
nos D-branes est donc d{\'e}fini par $g^\dagger \Omega g=\Omega$. Il est
isomorphe {\`a} \SL~: comme dans le cas des D-branes sym{\'e}triques dans
les groupes, la moiti{\'e} des sym{\'e}tries sont pr{\'e}serv{\'e}es, ici
3. C'est ce haut degr{\'e} de sym{\'e}trie, dont on suppose qu'il survit
au niveau quantique en termes de conditions de collage, 
qui rend la th{\'e}orie de ces D-branes
accessible. 

On peut obtenir d'autres D-branes similaires en appliquant des
sym{\'e}tries de la th{\'e}orie. En particulier, une rotation de \SLC\
bien choisie donne la D-brane d'{\'e}quation,
\bea
\cosh\rho \sinh\t =\cst.
\label{h2h3}
\eea
Nous les appellerons les D-branes $H_2$. Leur {\'e}quation est le
prolongement analytique de celle des D-branes du m{\^e}me nom dans
\AAA. Ainsi, deux types de D-branes tr{\`e}s diff{\'e}rents dans \AAA\
sont dans \H\ reli{\'e}s par une rotation et ont donc les m{\^e}mes
propri{\'e}t{\'e}s physiques. 

\bs

Les solutions des conditions de collage correspondant aux sym{\'e}tries
des D-branes \AA\ dans \H\ sont les fonctions {\`a} un point du type
\cite{pst}
\bea
\llangle \Phi ^j_u(z) \rrangle =\frac{ |u+\bar{u}|^{2j}
  A(j,\sgn(u+\bar{u}))}{ |z-\bz|^{2\Delta_j}}.
\label{collageads2}
\eea
Ici $\Delta_j=-b^2j(j+1)$ est
le poids conforme de $\Phi ^j$, qui est proportionnel au Casimir
$4j(j+1)$ 
de la
repr{\'e}sentation correspondante. Les conditions de collage et leurs
solutions sont
ind{\'e}pendantes du param{\`e}tre $r$ des D-branes,
on s'attend donc {\`a} ce que les
contraintes suppl{\'e}mentaires sur cette fonction {\`a} un point
poss{\`e}dent une famille de solutions {\`a} un param{\`e}tre. 

\Fig{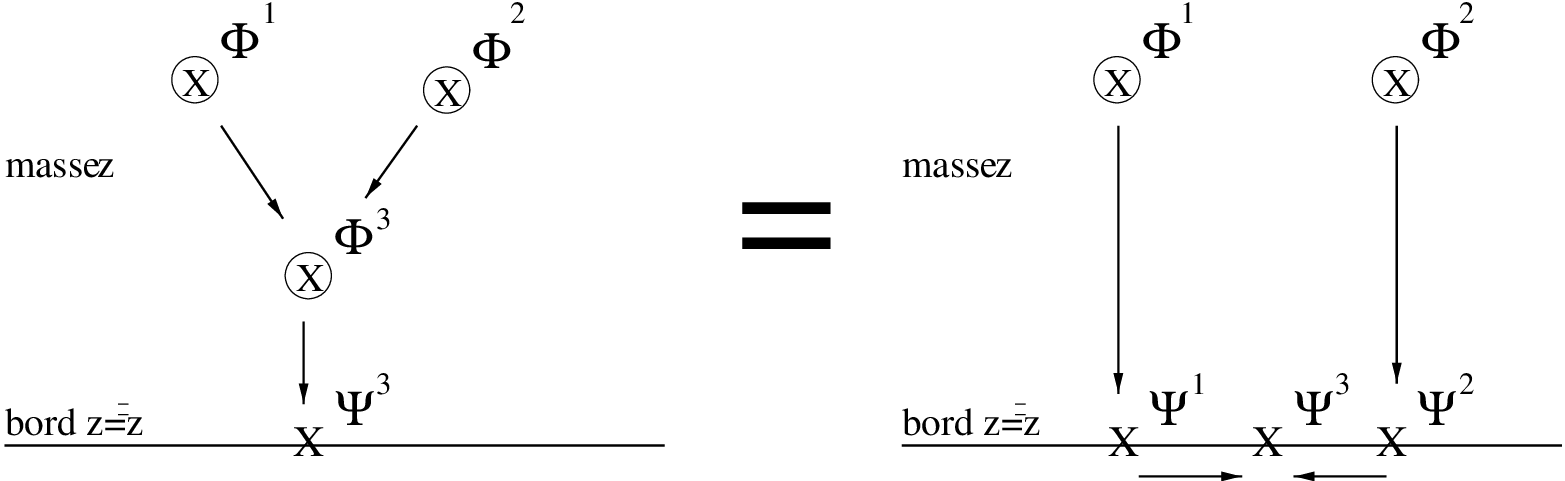}{13}{ Une contrainte de factorisation pour les champs de la
  masse. }

Les contraintes suppl{\'e}mentaires utilis{\'e}es dans \cite{pst,plo}
consistent {\`a}  relier
deux fa{\c c}ons de calculer la fonction {\`a} deux points de deux champs
de la masse. La premi{\`e}re consiste {\`a} effectuer leur produit
d'op{\'e}rateurs dans la masse avant de rejoindre le bord, la deuxi{\`e}me
de commencer par approcher chaque champ du bord avant de faire le
produit d'operateurs des champs au bord obtenus (voir la figure
\ref{afact}). 
Les deux sont
reli{\'e}s par la sym{\'e}trie de croisement de la th{\'e}orie des cordes
ferm{\'e}es (voir \cite{vsrevue}). Cette m{\'e}thode remonte {\`a}
\cite{opdesc}, o{\`u} elle fut utilis{\'e}e dans le contexte de th{\'e}ories
conformes rationnelles mais non diagonales.

Ce calcul, qui peut s'effectuer pour deux champs
quelconques dans le cas d'une th{\'e}orie rationnelle, 
ne marche pas tel quel dans notre cas, {\`a} cause notamment de
l'absence d'un champ identit{\'e} dans notre spectre. 
 On peut tout de m{\^e}me en extraire des
contraintes, si l'on suppose qu'un des champs de la masse est
d{\'e}g{\'e}n{\'e}r{\'e}, de sorte que son produit d'op{\'e}rateur avec un autre
champ comporte un nombre fini de termes. Il n'existe pas de tel champ
d{\'e}g{\'e}n{\'e}r{\'e} dans le spectre des cordes ferm{\'e}es dans \H,
cependant il est possible d'utiliser des champs obtenus par
prolongement analytique de $j$ hors de la ligne physique $j\in -\half
+i\R$. Ces champs appartiennent {\`a} la th{\'e}orie mais pas au spectre,
et on suppose que leurs propri{\'e}t{\'e}s sont obtenues par prolongement
analytique. 

La \flqq contrainte de factorisation\frqq\ du champ d{\'e}g{\'e}n{\'e}r{\'e}
$\Phi ^\half $ poss{\`e}de donc une famille de solutions \cite{pst,plo},
\footnote{
Nous {\'e}crivons la normalisation exacte, qui ne peut en fait {\^e}tre
d{\'e}termin{\'e}e qu'apr{\`e}s comparaison avec le secteur de cordes
ouvertes. }
\bea
\llangle \Phi ^j_u(z)\rrangle_r= \frac{|u+\bar{u}|
  ^{2j}}{|z-\bz|^{2\Delta_j}}
  2^{-\frac{3}{4}}b^{-\half}\nu_b^{j+\half}\Gamma (1+b^2(2j+1))\ e
  ^{-r(2j+1) \sgn(u+\bar{u})}.
\label{etatbordads2}
\eea
Nous avons utilis{\'e} la fonction $\Gamma$ d'Euler ainsi que la
constante $\nu_b=${\tiny $\frac{\Gamma(1-b^2)}{\Gamma(1+b^2)}$}, o{\`u}
on rappelle que $b^{-2}=k-2$. L'identification du param{\`e}tre de cette
famille de solutions avec le r{\'e}el $r$ dont d{\'e}pend la position des
D-branes \AA\ s'obtient par l'{\'e}tude de la g{\'e}om{\'e}trie des {\'e}tats
de bord dans la limite semi-classique $b\rar 0,k\rar \infty$. 

On peut utiliser ces {\'e}tats de bord pour calculer le diagramme du
cylindre entre deux D-branes (moyennant quelques r{\'e}gularisations) et
pr{\'e}dire le spectre (continu) des cordes ouvertes. 
Cette pr{\'e}diction n'a plus le caract{\`e}re de v{\'e}rification du cas
rationnel, o{\`u} les coefficients des diff{\'e}rents {\'e}tats devaient
{\^e}tre entiers. Dans la section \ref{ouvads2}, nous expliquerons
comment on {\'e}tudie ind{\'e}pendamment le spectre des cordes ouvertes.

D'autres champs d{\'e}g{\'e}n{\'e}r{\'e}s peuvent {\^e}tre utilis{\'e}s pour
obtenir d'autres contraintes et v{\'e}rifier qu'elles sont satisfaites
par l'{\'e}tat de bord (\ref{etatbordads2}). La contrainte la plus simple
apr{\`e}s celle du champ $\Phi ^\half$ provient du champ
d{\'e}g{\'e}n{\'e}r{\'e} $\Phi ^\frac{k-2}{2}$, et son {\'e}criture m{\^e}me est
probl{\'e}matique \cite{pon,ponsil}. Pour prouver la coh{\'e}rence d'un
{\'e}tat de bord, il faut en principe v{\'e}rifier toutes les contraintes
de factorisation. Nous allons supposer dans la suite que les D-branes
\AA\ dans \H\ existent et sont coh{\'e}rentes.

\subsection{ Les branes compactes dans $H_3$ }

Il existe d'autres D-branes dans \H, qui respectent une sym{\'e}trie
\SU$\subset$\SLC, mais dont l'interpr{\'e}tation g{\'e}om{\'e}trique est
douteuse puisqu'il s'agirait de sph{\`e}res $S^2$ de rayon imaginaire
pur \cite{schwimmer,pst}. 
On les obtient par un proc{\'e}d{\'e} analogue aux D-branes \AA, en
commen{\c c}ant par imposer la sym{\'e}trie d{\'e}sir{\'e}e au moyen de
conditions de collage, puis en prenant en compte la contrainte de
factorisation du champ $\Phi ^\half$. 

On trouve donc une famille d'{\'e}tats de bord {\`a} un param{\`e}tre $s$,
\bea
\llangle \Phi ^j_u(z)\rrangle_s =
\frac{(1+u\bar{u})^{2j}}{|z-\bz|^{2\Delta_j}}
\frac{-\nu_b^{j+1}\Gamma(1+b^2(2j+1))}{2\pi \Gamma(1-b^2)} \frac{\sin
  s(2j+1)}{\sin s}.
\label{s2hs}
\eea
Le calcul du digramme du cylindre conduit {\`a} un spectre discret, et
donc {\`a} des conditions sur le param{\`e}tre $s$, 
\bea
s=\pi b^2 (2J+1),\ J\in \half \N.
\label{quantifs}
\eea
Le cas des D-branes $S^2$ ressemble donc beaucoup plus au cas
rationnel que celui des D-branes \AA. Mais la situation de l'ensemble
des D-branes dans \H\ n'a plus rien {\`a} voir avec la solution de
Cardy, puisqu'on ne peut m{\^e}me pas d{\'e}finir de matrice $S$ pour la
transformation modulaire des caract{\`e}res~: en effet, selon la
D-brane, les m{\^e}mes caract{\`e}res des repr{\'e}sentations continues se
transforment soit en d'autres caract{\`e}res continus, soit en
caract{\`e}res discrets. 

{\'E}crivons donc explicitement ces caract{\`e}res~: si on note $j=-\half
+iP$ avec $P\in \R$, on a
\bea
\chi_P(q)=\frac{q^{b^2P^2}}{\eta(q)^3},
\label{carcont}
\eea
o{\`u} $\eta(q)$ est la fonction de Dedekind,
\bea
\eta(q)=q^\frac{1}{24}\prod_{n=1}^\infty (1-q^n)\ , \
\eta(\tq)=\sqrt{-i\tau} \eta(q).
\label{eta}
\eea
En fait, le caract{\`e}re des
repr{\'e}sentations continues est {\`a} proprement parler infini car ces
repr{\'e}sentations comportent un nombre infini d'{\'e}tats ayant tous la
m{\^e}me {\'e}nergie. Cela n{\'e}cessitera une r{\'e}gularisation quand nous
en viendrons aux D-branes \AA. Mais dans le cas des D-branes compactes
$S^2$, la quantit{\'e} qui appara{\^\i}t dans le calcul qui suit est
$\chi_P(q)$, qui est fini.

Le calcul de l'amplitude du cylindre entre deux D-branes $S^2$ de
param{\`e}tres $s=\pi b^2 (2J+1),s'=\pi b^2(2J'+1)$ s'effectue donc ainsi~:
\footnote{
Nous {\'e}liminons implicitement les facteurs d{\'e}pendant de $z,\bz$
dans les fonctions {\`a} un point.
}
\bea
Z^\cyl_{s,s'}(\tq)&=&\langle s|\tq ^{L_0+\bar{L}_0-\frac{c}{24}}|
  s'\rangle = \int dPd^2u \langle \Phi ^j_u \rangle_s \langle \Phi
  ^j_u \rangle_{s'}^* \chi_j(\tq)
\label{zcyls2}
\\
&\propto & \int dP\, P\, \sum_{J''\in J\times J'} \sinh 2\pi b^2
(2J''+1)P\, \chi_P(\tq) 
\nn
\\
&\propto &
\sum_{J''\in J\times J'}
(2J''+1)\chi_{i(J''+\half)}(q).
\label{specs2}
\eea
On a not{\'e} $J\times J'$ l'ensemble des spins apparaissant dans la
composition de $J$ et $J'$. 

Le spectre des cordes ouvertes des D-branes $S^2$ n'a donc rien de
commun avec celui des cordes ferm{\'e}es dans \H~: il ne comprend que
des valeurs r{\'e}elles $J''$ de $j$ (soit des valeurs complexes
$i(J''+\half)$ de $P$). Le facteur $(2J''+1)$ peut s'interpr{\'e}ter
comme la d{\'e}g{\'e}n{\'e}rescence des {\'e}tats de plus haut poids,
c'est-{\`a}-dire la dimension de la repr{\'e}sentation correspondante de l'alg{\`e}bre de
Lie \su. Mais le caract{\`e}re $(2J''+1)\chi_{i(J''+\half)}(q)$
n'{\'e}gale pas le caract{\`e}re de spin $J''$ de \asu, il en est en
quelque sorte une version priv{\'e}e de vecteurs singuliers. Cela n'est
pas tr{\`e}s surprenant, {\'e}tant donn{\'e} que nous ne supposons pas le
niveau $k$ entier.

\section{ Cordes ouvertes et D-branes $AdS_2$ dans \H
\label{ouvads2}}

Cette section est consacr{\'e}e {\`a} l'{\'e}tude du spectre des cordes
ouvertes {\'e}tir{\'e}es entre deux D-branes \AA\ dans \H, du point de vue
du secteur des cordes ouvertes. L'accord avec le secteur des cordes
ferm{\'e}es via le diagramme du cylindre ne sera montr{\'e} qu'ensuite.
Dans le cas
o{\`u} il s'agit de deux fois la m{\^e}me D-brane, ce spectre a {\'e}t{\'e}
calcul{\'e} dans \cite{pst}. Cependant, cet article n'a pas r{\'e}solu le
probl{\`e}me dans le cas de deux D-branes identiques {\`a} cause
de certaines hypoth{\`e}ses implicites trop contraignantes sur
l'analyticit{\'e} de la densit{\'e} spectrale. J'ai localis{\'e} ces
hypoth{\`e}ses et d{\'e}termin{\'e} le spectre dans \cite{deuxbranes}. Ce
r{\'e}sultat a notamment des implications pour les D-branes dans le
cigare, que nous {\'e}tudierons dans le prochain chapitre.

\subsection{ La contrainte de factorisation }

Le spectre des cordes ouvertes sur une D-brane \AA\ est form{\'e} de
repr{\'e}sentations continues de \SL, comme
le montre l'{\'e}tude de sa limite semi-classique, l'ensemble des
fonctions sur la D-brane. Cela provient du fait qu'\AA\ atteint
l'infini d'\H\ et est donc lui-m{\^e}me non compact. 
La fonction de
partition {\`a} une boucle des cordes ouvertes est donc du type
\bea
Z^\cyl_{r,r}(\tq) = \int_{\R} dP\ N(P|r,r) \chi_P(q),
\label{densite}
\eea
o{\`u} $N(P|r,r)$ est la densit{\'e} de cordes ouvertes sur la D-brane \AA\
de param{\`e}tre $r$, qui ne d{\'e}pend
que de $P$ car la D-brane respecte la sym{\'e}trie \SL. 

Cependant, {\`a} cause du volume infini dans la direction radiale (dont
$P$ est grosso modo l'impulsion), $Z^\cyl(\tq)$ doit diverger, et on
peut attribuer cette divergence {\`a} un terme infini dans
$N(P|r,r)$. Mais cette divergence est ind{\'e}pendante de $r$, et
dispara{\^\i}t donc si l'on consid{\`e}re la densit{\'e} relative
$N(P|r,r)-N(P|0,0)$. Par ailleurs, la densit{\'e} d'{\'e}tats est en
toute g{\'e}n{\'e}ralit{\'e} reli{\'e}e {\`a} l'amplitude de r{\'e}flexion
$R(j=-\half+iP|r,r)$ par la formule
\bea
N(P|r,r)-N(P|0,0)=\frac{1}{2\pi i}\frac{\p}{\p P}
\log\frac{R(-\half+iP|r,r)}{R(-\half+iP|0,0)}.
\label{reflexion}
\eea
Cette amplitude de r{\'e}flexion d{\'e}crit la diffusion d'{\'e}tats
asymptotiques envoy{\'e}s depuis l'infini. {\`A} l'infini en effet, tous
les {\'e}tats de cordes ouvertes $\Psi ^{-\half+iP}$ sont des
superpositions d'ondes planes 
radiales d'impulsions $P$ et $-P$. Le coefficient relatif de ces deux
termes est l'amplitude de r{\'e}flexion, qui doit {\^e}tre unitaire (de
module un) si la r{\'e}flexion est un processus physique. De fa{\c c}on
{\'e}quivalente, la densit{\'e} d'{\'e}tats correspondante doit {\^e}tre
r{\'e}elle. 

Pour d{\'e}terminer l'amplitude de r{\'e}flexion, on peut remarquer
qu'elle appara{\^\i}t dans la fonction {\`a} deux points de champs du
bord, et joue donc un r{\^o}le dans les contraintes de factorisation de
la fonction {\`a} trois points sur le bord. Comme pour les contraintes
de factorisation portant sur les {\'e}tats de bord, on utilise en fait
un champ du bord d{\'e}g{\'e}n{\'e}r{\'e}.

\bs

La contrainte trouv{\'e}e dans \cite{pst} est donc~:
\bea
\frac{R(j+\half|r_1,r_2)}{R(j-\half|r_1,r_2)}
=\frac{2j}{2j+1}e_-(-j-\half|r_1,r_2),\ \ \forall j\in -\half+i\R,
\label{good} \eea
o{\`u} nous introduisons la fonction
\bea 
e_-(j|r_1,r_2)&=&-\frac{4\nu_b}{\pi}
\frac{\Gamma(1+b^2(2j-1))\Gamma(-b^2(2j+1))}{\sin
  \pi b^22j}\ \times \nonumber \\
&\times &\prod_{s=\pm}\cos(\pi b^2j+s\frac{i}{2}(r_1+r_2))\sin(\pi
b^2j+s\frac{i}{2}(r_1-r_2)).
\label{emoins}
\eea
En principe il se pose le m{\^e}me probl{\`e}me que dans le canal des
cordes ferm{\'e}es, {\`a} savoir l'existence d'une famille {\`a} un
param{\`e}tre de solutions, qu'il faut ensuite identifier avec les
D-branes \AA\ en utilisant une limite semi-classique.
Nous anticipons un peu en utilisant
directement le param{\`e}tre $r$. 

Dans \cite{pst}, cette contrainte est r{\'e}{\'e}crite sous la forme
\bea 
|R(j+\half|r_1,r_2)|^2=\frac{2j}{2j+1}e_-(-j-\half|r_1,r_2), 
\label{bad} 
\eea
Cela suppose implicitement que
$R(\bar{j}|r_1,r_2)=\overline{R(j|r_1,r_2)}$. Nous allons voir que
cette supposition n'est pas v{\'e}rifi{\'e}e par les
solutions de l'{\'e}quation (\ref{good}), sauf dans le cas
$r_1=r_2$. C'est ainsi que \cite{pst} a conclu {\`a} l'inexistence de
solutions pour $r_1\neq r_2$, car le second
membre de (\ref{bad}) n'est pas un r{\'e}el positif dans ce cas. 

\subsection{ Le spectre des cordes ouvertes }

\Fig{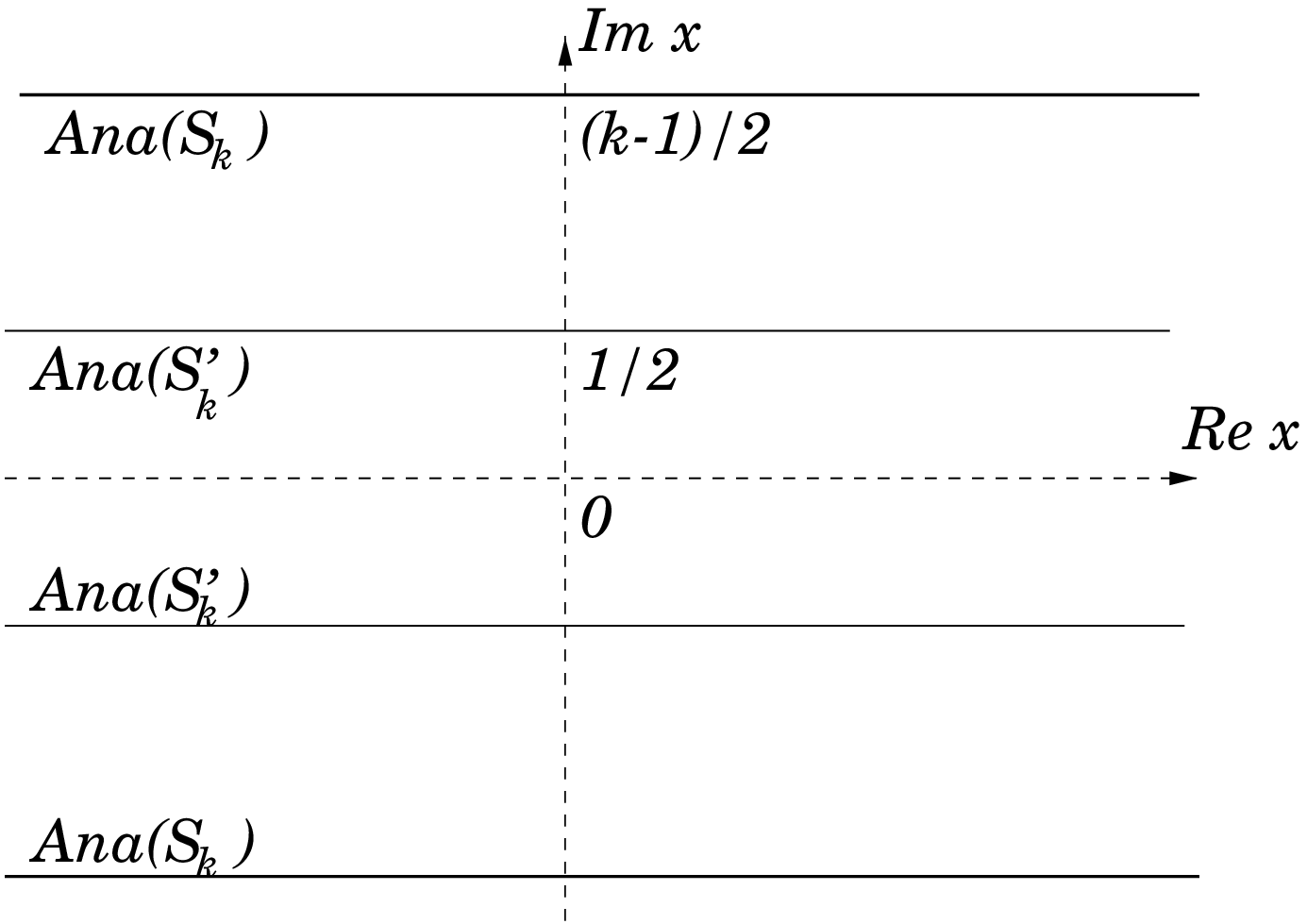}{9}{ Les bandes d'analyticit{\'e} des fonctions
  $S_k(x)$ et $S'_k(x)$. }

Nous voulons r{\'e}soudre la contrainte eq. (\ref{good}). Pour
cela nous aurons besoin de fonctions sp{\'e}ciales, que nous allons
d{\'e}finir. D'abord, consid{\'e}rons la fonction $S_k$,
\bea
\log S_k(x)=i\int_0^\infty\frac{dt}{t}\left(\frac{\sin
    2tb^2x}{2\sinh b^2t\sinh t}-\frac{x}{t}\right). 
\label{fctsk}
\eea
Cette fonction est analytique dans la r{\'e}gion (voir la figure
\ref{analyticite}) 
\bea
Ana(S_k)=\left\{ |\Im x|<\half+\frac{1}{2b^2}=\frac{k-1}{2} \right\}.
\label{anask}
\eea
De plus, pour $x$ tel que $x\pm\halfi\in Ana(S_k)$, elle v{\'e}rifie
l'{\'e}quation 
\bea
\frac{S_k(x-\halfi)}{S_k(x+\halfi)}=2\cosh\pi b^2x,
\label{relsk} 
\eea
qui peut aussi {\^e}tre utilis{\'e}e pour prolonger la fonction $S_k$ en
une fonction m{\'e}romorphe sur $\C$. Comparer les {\'e}quations
(\ref{good}) et (\ref{relsk}) permet de comprendre pourquoi la
fonction $S_k$ va pouvoir {\^e}tre utilis{\'e}e pour r{\'e}soudre
l'{\'e}quation (\ref{good}). D{\'e}finissons en outre la fonction $S'_k$,
\bea 
\log S'_k(x)=i\int_0^\infty\frac{dt}{t}\left(\frac{\cosh t\sin
    2tb^2x}{2\sinh b^2t\sinh t}-\frac{x}{t}\right). 
\label{fctspk}
\eea 
Cette fonction n'est analytique que dans la r{\'e}gion (voir la figure
\ref{analyticite})  
\bea
Ana(S'_k)=\left\{ |\Im x|< \half \right\}.
\label{anaspk}
\eea
Dans cette r{\'e}gion, elle v{\'e}rifie de plus l'identit{\'e}
\bea
S'_k(x)^2=\frac{S_k(x+\halfib)S_k(x-\halfib)}{S_k(\halfib)S_k(-\halfib)}
S_k(0)^2.
\label{skspk}
\eea
Ainsi, si $S_k$ se prolonge en une fonction m{\'e}romorphe sur $\C$, le
cas de $S'_k$ est plus compliqu{\'e} puisqu'il met en jeu des coupures
de la racine carr{\'e}e. 
Il doit y avoir une telle coupure aboutissant {\`a} chaque z{\'e}ro ou
p{\^o}le simple de $S_k(x+\halfib)S_k(x-\halfib)$.

Une autre fonction sp{\'e}ciale, qui jouera un r{\^o}le bien moindre dans
nos raisonnements, est 
\bea
\Gamma_k(x)=b^{b^2x(x-b^{-2})}(2\pi)^{x/2}\Gamma_2^{-1}(x|1,b^{-2}),
\label{fctgk}
\eea
dont la d{\'e}finition utilise la fonction $\Gamma_2$ telle que
\bea
\log \Gamma_2(x|\omega_1,\omega_2)=\left(\frac{\p}{\p
    t}\sum_{n_1,n_2=0}^\infty
  (x+n_1\omega_1+n_2\omega_2)^{-t}\right)_{t=0}. 
\label{fctg2}
\eea

\bs

La solution de l'{\'e}quation (\ref{good}) dans le cas $r_1=r_2$ est
\cite{pst}
\bea 
R(\j |r,r)=\nu_b^{iP}\frac{\Gamma ^2_k(\half-iP+\frac{1}{b^2})}{
  \Gamma ^2_k(\half+iP+\frac{1}{b^2})} 
\frac{\Gamma_k(2iP+\frac{1}{b^2})}{\Gamma_k(-2iP+\frac{1}{b^2})}
\frac{S_k(\rr+P)}{S_k(\rr-P)},
\label{solrr}
\eea
cette expression est bien de module un, et elle r{\'e}sout notre
l'{\'e}quation (\ref{good}) pour tout $j\in -\half +i\R$. En effet, pour
de telles valeurs de $j$, les arguments des fonctions $S_k$
apparaissant dans (\ref{solrr}) sont toujours situ{\'e}s dans la
r{\'e}gion $Ana(S_k)$. 

\bs

Consid{\'e}rons maintenant le cas $r_1=-r_2$. La forme de la fonction
$e_-(j|r_1,r_2)$ ({\'e}quation (\ref{emoins})) sugg{\`e}re de construire
$R(j|r,-r)$ en rempla{\c c}ant simplement $S_k$ par $S_k'$ dans la
solution pour $R(j|r,r)$ ({\'e}quation (\ref{solrr})), ce qui {\'e}quivaut {\`a}
poser
\bea
 R(j|r,-r)=\sqrt{\frac{R(j|r+\ip,r+\ip)R(j|r-\ip,r-\ip)}
  {R(j|\ip,\ip)R(j|-\ip,-\ip)}}R(j|0,0). 
\label{solrmr}
\eea
Cette expression pose un probl{\`e}me quand on veut v{\'e}rifier la
contrainte (\ref{good})~: comme il intervient $r\pm \ip$
dans les arguments de l'amplitude de r{\'e}flexion $R$, 
les fonctions $S_k$ doivent {\^e}tre {\'e}valu{\'e}es au bord de leur domaine
de d{\'e}finition
$Ana(S_k)$. La racine carr{\'e}e y est mal d{\'e}finie {\`a} cause
de la pr{\'e}sence de p{\^o}les et de z{\'e}ros. 

\Fig{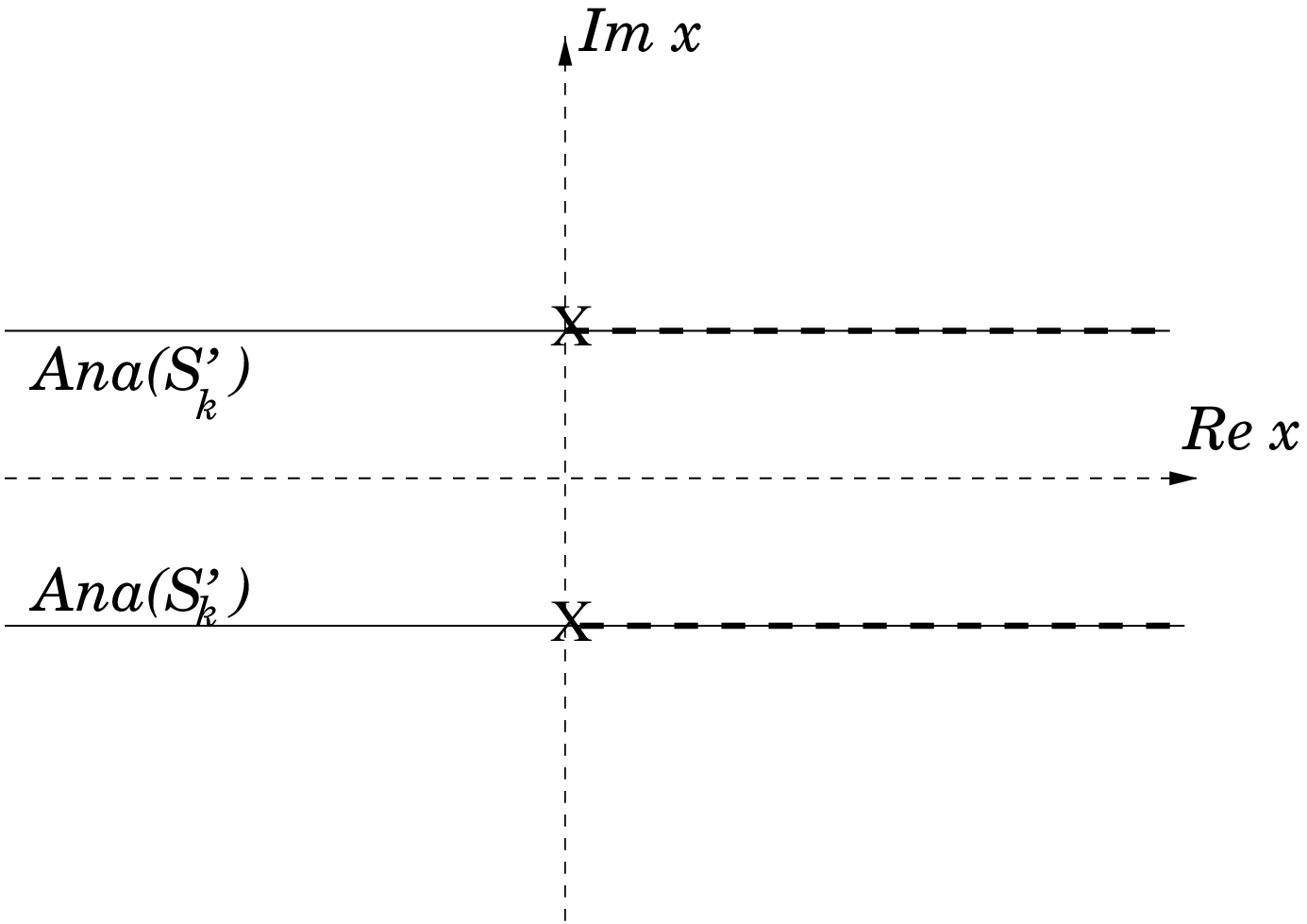}{9}{ Un p{\^o}le et un z{\'e}ro de $S'_k(x)^2$, et deux
  coupures de $S'_k(x)$ (en pointill{\'e}s). }

Mais
revenons {\`a} une formulation du probl{\`e}me en terme de $S'_k$, qui
doit aussi {\^e}tre d{\'e}finie au bord d'$Ana(S'_k)$. Consid{\'e}rons le
p{\^o}le $x=\halfi$ et le z{\'e}ro $x=-\halfi$ de $S_k'(x)^2$, dont la
pr{\'e}sence d{\'e}coule des {\'e}quations (\ref{relsk}) et (\ref{skspk}). Les
coupures de $S'_k(x)$ qui partent de ces singularit{\'e}s ne doivent pas
rencontrer la ligne physique $x\in \R$ (o{\`u} il faut {\'e}valuer $S_k'$
si on calcule $R(j|r,-r)$ pour $j\in -\half +i\R$). De plus, ces
coupures doivent {\^e}tre atteintes simultan{\'e}ment par $x-\halfi$ et
$x+\halfi$, si l'on veut que {\tiny
  $\frac{R(j+\half|r,-r)}{R(j-\half|r,-r)}$} n'ait pas de coupure,
comme c'est le cas pour $e_-$ qui appara{\^\i}t dans le membre de
droite de (\ref{good})
\footnote{
En faisant cette supposition, nous interpr{\'e}tons l'{\'e}quation
(\ref{good}) comme devant {\^e}tre valable au voisinage de la ligne
physique $j\in-\half +i\R$. En fait, elle sera m{\^e}me valable dans
tout le plan complexe. L'id{\'e}e est que si $r_1\neq r_2$ l'{\'e}quation
(\ref{good}) n'a pas de sens en tant que telle {\`a} cause des
singularit{\'e}s que nous avons signal{\'e}es. L'interpr{\'e}tation de
(\ref{good}) que nous proposons sera justifi{\'e}e par la
d{\'e}termination d'une amplitude de r{\'e}flexion unitaire et r{\'e}guli{\`e}re
et par l'accord avec le secteur de cordes ferm{\'e}es via le calcul du
diagramme du cylindre.
}.
Donc, les deux coupures en question doivent {\^e}tre parall{\`e}les {\`a}
l'axe $x\in \R$ et s'{\'e}tendre dans le m{\^e}me sens, soit
$x=\pm\halfi+\R_+$ ou $x=\pm\halfi-\R_+$ (voir la figure
\ref{acoupure}). 
On peut choisir de m{\^e}me
toutes les coupures de la fonction $S'_k$, de telle sorte que
(\ref{good}) soit v{\'e}rifi{\'e}e sur tout le plan complexe.

On s'aper{\c c}oit alors que la fonction $R(j|r,-r)$ ainsi d{\'e}finie ne
v{\'e}rifie pas $R(\bar{j}|r_1,r_2)=\overline{R(j|r_1,r_2)}$ {\`a} cause
de ses coupures.
Ce ph{\'e}nom{\`e}ne se produit parce que la racine carr{\'e}e
  qui intervient dans la d{\'e}finition de $R(j|r,-r)$ a alors une
  coupure pour des arguments imaginaires purs, et ne v{\'e}rifie donc
  pas $\sqrt{\bz}=\overline{\sqrt{z}}$. Ceci montre qu'on peut
  r{\'e}soudre l'{\'e}quation (\ref{good}) sans r{\'e}soudre pour autant
  l'{\'e}quation (\ref{bad}). 

Il est maintenant facile de trouver la solution pour toutes valeurs de
$r_1,r_2$, 
\bea
R(j|r_1,r_2)=R(j|\frac{r_1+r_2}{2},\frac{r_1+r_2}{2})
R(j|\frac{r_1-r_2}{2},\frac{r_2-r_1}{2})R(j|0,0)^{-1}. 
\label{solgen}
\eea
Cette formule, avec les {\'e}quations (\ref{solrr}) et
(\ref{solrmr}), constitue la solution g{\'e}n{\'e}rale de
(\ref{good}). Pour autant qu'on puisse se contenter de r{\'e}soudre
cette contrainte de factorisation particuli{\`e}re, nous connaissons
donc ainsi le spectre des cordes ouvertes {\'e}tir{\'e}es entre deux
D-branes \AA\ de param{\`e}tres quelconques, {\`a} supposer
\footnote{Rappelons qu'il s'agit l{\`a} du probl{\`e}me de la divergence
  de grand volume, qui a {\'e}t{\'e} r{\'e}solu en soustrayant la densit{\'e}
  d'{\'e}tats d'une D-brane de r{\'e}f{\'e}rence $N(P|0)=N(P|0,0)$, voir
  (\ref{reflexion}). 
}
que l'on
connaisse $R(j|0,0)$.

On peut v{\'e}rifier que les r{\'e}sultats obtenus sont corrects dans la
limite semi-classique $b\rar 0$. Nous le ferons explicitement dans le
contexte des D1-branes dans le cigare, quand nous les aurons
construites {\`a} partir de nos D-branes \AA.

\rem{reference a ajouter }

Mais nous allons maintenant nous livrer {\`a} une autre v{\'e}rification,
exacte celle-ci.

\subsection{ Le diagramme du cylindre } 

Nous commen{\c c}ons par calculer la fonction de partition des cordes
ouvertes {\`a} une
boucle $Z^\ouv(q)=Z^\cyl(\tq)$, 
{\`a} partir de la densit{\'e} $N(j|r_1,r_2)$ que nous avons d{\'e}termin{\'e}e. Nous
ne traiterons pas les probl{\`e}mes de r{\'e}gularisation avec autant de
d{\'e}tail que \cite{pst}. Il suffit pour notre calcul de savoir que le
caract{\`e}re d'une repr{\'e}sentation continue de \SL\ est $\chi_P(q)$
{\`a} un facteur infini pr{\`e}s, et que ce facteur infini est multipli{\'e}
par $\tau$ au cours de la transformation modulaire $S~:\,
\tau\rar-\frac{1}{\tau }$. De plus, nous avons {\'e}crit les {\'e}tats de
bord et la densit{\'e} des cordes ouvertes avec des normalisations
pr{\'e}cises, mais nous n{\'e}gligerons maintenant les facteurs
num{\'e}riques ({\'e}ventuellement d{\'e}pendant de $b$)
pour
all{\'e}ger le calcul. Nous omettrons aussi de soustraire explicitement les
quantit{\'e}s $Z^\cyl_{0,0}(\tq),N(P|0,0),\log R(j|0,0)$ qui
r{\'e}gularisent la divergence de grand volume (la divergence en $P=0$
de $\int dP$). Mais il faut garder en m{\'e}moire que l'on peut donc
soustraire du calcul tout terme ind{\'e}pendant de $r_1,r_2$.

Donc, nous avons
\bea
Z^\cyl_{r_1,r_2}(\tq)
&=&\int dP\ N(P|r_1,r_2) \chi_P(q)
\nn
\\
&=&\int dP\ \chi_P(q)\frac{\p}{\p P}\left(\log
R(\j|\frac{r_1+r_2}{2},\frac{r_1+r_2}{2}) + [r_2\rar -r_2] \right) 
\nn
\\
&=& \int dP\ \chi_P(q)\frac{\p}{\p P}\log
\frac{S_k(\frac{r_1+r_2}{2\pi b^2}+P) 
  S'_k(\frac{r_1-r_2}{2\pi b^2}+P)}{S_k(\frac{r_1+r_2}{2\pi b^2}-P)
  S'_k(\frac{r_1-r_2}{2\pi b^2}-P)} 
\nn
\\
&=& \int dP\ \chi_P(q) \int dt\ \frac{\cos 2tb^2P}{\sinh b^2 t\sinh
  t}\left( \cos \frac{r_1+r_2}{\pi}t +\cosh t \cos
  \frac{r_1-r_2}{\pi}t\right) 
\nn
\\
&\stackrel{t=2\pi P'}{=}& \int dP'\ \chi_{P'}(\tq) \left[ \frac{\cosh^2\pi
  P'\cos 2r_1P'\cos 2r_2P'}{\sinh
  2\pi b^2P'\sinh 2\pi P'} \right.
\label{zcyl1}
\\
& & \left. -\frac{\sinh^2\pi P' \sin 2r_1P'\sin 2r_2P'}{\sinh
  2\pi b^2P'\sinh 2\pi P'}\right].
\label{zcyl2}
\eea 
Cette quantit{\'e} doit {\^e}tre compar{\'e}e avec le diagramme du cylindre
calcul{\'e} {\`a} partir des {\'e}tats de bord. Pour cela, il faut commencer
par calculer leurs transform{\'e}es de Fourier par rapport aux variables
$\arg(u)$ et $|u|$, dont on appelle $n$ et $p$ les
impulsions\footnote{Contrairement au cas des D-branes $S^2$,
le calcul direct de l'int{\'e}grale $\int d^2u$ ne donne pas le
r{\'e}sultat correct, car la conjugaison complexe de l'{\'e}tat de bord
avant et apr{\`e}s
transformation de Fourier ne sont pas {\'e}quivalentes. 
}.
On introduit donc les champs 
\bea
\Phi ^j_{n,p}=\int d^2u e ^{-in\arg(u)}|u|^{-2j-2-ip}\Phi ^j_u.
\label{phijnp}
\eea
On peut calculer leur fonction {\`a} un point {\`a} partir de (\ref{etatbordads2}),
\bea
\llangle \Phi ^j_{n,p}(z)\rrangle_r=
\frac{e ^{-r(2j+1)}+(-1)^ne ^{r(2j+1)}}{\Gamma(1+j+\frac{n}{2})
  \Gamma(1+j-\frac{n}{2})}
\Gamma (1+b^2(2j+1)) \Gamma(2j+1)
\frac{2^{-\frac{3}{4}}b^{-\half}
\nu_b^{j+\half}\delta(p)}{|z-\bz|^{2\Delta_j}}. 
\label{etatbordphijnp}
\eea
On a donc
\bea
Z^\cyl_{r_1,r_2}(\tq)=\int dP'\ \chi_{P'}(\tq)
\sum_{n\in\Z}\int dp\ \llangle \Phi
^{\j'}_{n,p}(\halfi)\rrangle_{r_1}^* \llangle \Phi
  ^{\j'}_{n,p}(\halfi)\rrangle_{r_2},
\label{zcylads2} 
\eea
o{\`u} l'on doit n{\'e}gliger la d{\'e}pendance en $p$ et la divergence de
$\sum_n$ pour obtenir l'accord avec le calcul pr{\'e}c{\'e}dent de la
m{\^e}me quantit{\'e}. L'essentiel est de constater que $\llangle \Phi
^{\j'}_{n,p}(\halfi)\rrangle_{r_1}^* \llangle \Phi
  ^{\j'}_{n,p}(\halfi)\rrangle_{r_2}$ ne d{\'e}pend de $n$ qu'{\`a}
  travers sa parit{\'e}, et donne (\ref{zcyl1})
  si $n$ est pair, (\ref{zcyl2}) si $n$ est impair.

\subsection{ Application aux D-branes \AA\ dans \AAA }

Comme les D-branes \AA\ sont statiques, nous allons proposer
\cite{corfu} d'interpr{\'e}ter les densit{\'e}s d'{\'e}tats calcul{\'e}es
ci-dessus comme les densit{\'e}s de cordes longues (repr{\'e}sentations
continues) vivant sur les D-branes \AA\ dans \AAA. Nous avons d{\'e}j{\`a}
donn{\'e} une conjecture pour le spectre des cordes courtes dans la
section \Fac{\ref{cosldba}}.

Nous proposons maintenant d'interpr{\'e}ter $N(P|r,r)$ comme la
densit{\'e} de cordes longues du secteur $w\in 2\Z$
 vivant sur la D-brane de param{\`e}tre $r$ (o{\`u} $w$ est le flot
 spectral). De plus,
nous avons vu que dans \AAA, le flot spectral d'indice impair 
relie les cordes 
vivant sur une D-brane \AA\ de param{\`e}tre $r$ 
aux cordes {\'e}tir{\'e}es entre deux D-branes $r,-r$. Nous connaissons
maintenant la densit{\'e} d'{\'e}tat de ces cordes ouvertes, c'est la
quantit{\'e} $N(P|r,-r)$. Bien s{\^u}r, les calculs qui ont men{\'e} {\`a} ce
r{\'e}sultat sont relatifs (il a fallu soustraire le spectre d'une
D-brane donn{\'e}e $N(P|0,0)$ pour r{\'e}gulariser), mais, comme le spectre de la
D-brane droite $r=0$ est bien connu, on peut d{\'e}terminer exactement
le spectre des autres. 

Le tableau suivant r{\'e}capitule ces r{\'e}sultats et conjectures~:

\begin{center}

\begin{tabular}{|l|c|c|}
\hline
\multicolumn{3}{|c|}{ Densit{\'e} de cordes ouvertes sur une
  D-brane \AA$(r)$ } 
\\
\hline\hline
&  $w$  pair &  $1+N(P|r,r)-N(P|0,0) $
\\
\cline{2-3}
\raisebox{1.6ex}[0cm][0cm]{ Longues $j=\j$ }
& $w$ impair & $1+N(P|r,-r)-N(P|0,0)$
\\
\hline
Courtes $j\in \R$
&
$\forall w$
&
$1$
\\
\hline
\multicolumn{3}{c}{}
\\
\hline
\multicolumn{3}{|c|}{ Densit{\'e} de cordes ouvertes entre deux
  D-branes \AA$(r_1,r_2)$ } 
\\
\hline\hline
&  $w$  pair &  $1+N(P|r_1,r_2)-N(P|0,0) $
\\
\cline{2-3}
\raisebox{1.6ex}[0cm][0cm]{ Longues $j=\j$ }
& $w$ impair & $1+N(P|r_1,-r_2)-N(P|0,0)$
\\
\hline
Courtes $j\in \R$
&
$\forall w$
&
$1$
\\
\hline
\end{tabular}

\end{center}


\setcounter{chapter}{3}

\chapter{ D-branes dans le cigare \SLU }

Ce chapitre est consacr{\'e} au probl{\`e}me de la construction de
D-branes dans \SLU. Il s'agit d'un pendant naturel de l'{\'e}tude des
D-branes dans \AAA\ pour plusieurs raisons.

La premi{\`e}re raison est que
la th{\'e}orie des cordes et des D-branes
dans \AAA\ doit sa complexit{\'e} en partie 
au caract{\`e}re Minkowskien de cette vari{\'e}t{\'e}. Le mod{\`e}le jaug{\'e}
\SLU, qui lui est directement reli{\'e}, est en revanche une th{\'e}orie
Euclidienne, mais toujours non rationnelle. 
On peut donc esp{\'e}rer cerner les probl{\`e}mes sp{\'e}cifiques de la
direction temporelle en comparant les deux, et les isoler de la
question de la non rationnalit{\'e}.

Une autre raison provient de l'exemple du cas compact. Dans ce cas,
l'{\'e}tude des D-branes dans un mod{\`e}le jaug{\'e} $G/H$ a permis de
construire de nouvelles D-branes dans $G$ lui-m{\^e}me, au point
d'esp{\'e}rer r{\'e}aliser l'ensemble des charges pr{\'e}dites pas la K-th{\'e}orie.
Ces id{\'e}es devraient {\^e}tre int{\'e}ressantes aussi dans le cas non
compact, o\`u nous sommes cependant encore loin de leur r\'ealisation
explicite. 

\section{ Les mod{\`e}les de Wess-Zumino-Witten jaug{\'e}s }

Le quotient \SLU\ qui nous int{\'e}resse est un exemple de mod{\`e}le de
Wess-Zumino-Witten jaug{\'e}. En l'occurence, on jauge un sous-groupe
$U(1)$ du mod{\`e}le de Wess-Zumino-Witten associ{\'e} au groupe
\SL. Cette construction est connue depuis longtemps et nous allons en
rappeler les principaux points saillants, en suivant
entre autres \cite{gq}. Elle est bien comprise dans
les cas o{\`u} l'on jauge un groupe compact, y compris en ce qui concerne
la construction des D-branes. Nous insisterons donc sur les
particularit{\'e}s du cas de \SLU.

\subsection{ Constructions g{\'e}n{\'e}rales }

Consid{\'e}rons le mod{\`e}le de Wess-Zumino-Witten d{\'e}crivant les cordes
ferm{\'e}es dans un groupe de Lie $G$, et un sous-groupe $H$ de $G$.
On peut introduire des champs de jauge $A,\bA$ pour les sym{\'e}tries
locales de multiplication {\`a} gauche et {\`a} droite par des
{\'e}l{\'e}ments de $H$~; ces champs sont des {\'e}l{\'e}ments de l'alg{\`e}bre de Lie 
\lieh\ de $H$. L'action jaug{\'e}e, s'obtient {\`a} partir de l'action du
mod{\`e}le de Wess-Zumino-Witten (\Fac{\ref{actWZW}}) en rempla{\c c}ant
$\p g$ et $\bp g$ par les d{\'e}riv{\'e}es covariantes $\p g +A
g$ et $\bp g- g\bA$, et en ajoutant un terme quadratique en $A$. 
Cependant, cette action jaug{\'e}e ne
poss{\`e}de pas n{\'e}cessairment la sym{\'e}trie locale
\bea
g\rar h_G\- (z)gh_D(\bz), 
\ A\rar h_G\- (z)(A+\p ) h_G(z), \ \bA \rar h_D\- (\bz)(\bA +\bp
)h_D(\bz) . 
\label{symjauge}
\eea
Si le groupe $H$ est non ab{\'e}lien, 
seule est r{\'e}alis{\'e}e
la \de{sym{\'e}trie vectorielle},
\bea
H_V\ ~: \ g\rar h\- g h.
\label{symvect}
\eea
Si le groupe $H$ est ab{\'e}lien, en plus de la sym{\'e}trie vectorielle,
on peut jauger la \de{sym{\'e}trie axiale},
\bea
H_A\ ~: \ g\rar h g h .
\label{symact}
\eea
Ces deux sym{\'e}tries sont reli{\'e}es par un changement de signe dans
l'alg{\`e}bre de Lie du secteur Droit, en particulier $\bA\rar
-\bA$. Nous avons d{\'e}j{\`a} rencontr{\'e} une telle transformation dans
le cas des cordes en espace plat dans la section \Fac{\ref{dualites}}, il
s'agissait de la T-dualit{\'e}. Dans le cas pr{\'e}sent, il s'agit aussi
de la T-dualit{\'e} dans la description du mod{\`e}le jaug{\'e} en termes
d'espace-temps. 

Cette description s'obtient en effectuant
l'int{\'e}grale sur les champs de jauge dans l'int{\'e}grale de chemin du
mod{\`e}le. On obtient ainsi un mod{\`e}le sigma, dont la g{\'e}om{\'e}trie est
celle du quotient  $G/H$ (o{\`u} $H$ agit sur $G$ par
sym{\'e}trie vectorielle ou axiale selon le cas), et dont le dilaton est
non trivial contrairement {\`a} celui du mod{\`e}le de Wess-Zumino-Witten
initial. Dans le cas o{\`u} $H$ est
ab{\'e}lien, la sym{\'e}trie (vectorielle ou axiale)
qui n'a pas {\'e}t{\'e} jaug{\'e}e reste une sym{\'e}trie globale du
quotient.

\bs

Notons que, m{\^e}me dans le cas non-ab{\'e}lien, il peut exister d'autres
quotients que le quotient vectoriel, appel{\'e}s quotients
asym{\'e}triques. 
On les obtient en utilisant des
automorphismes de $H$ (voir \cite{asymcosets}). La th{\'e}orie conforme
correspondante est alors h{\'e}t{\'e}rotique~: les secteurs droit et
gauche sont diff{\'e}rents. En revanche, dans le cas du mod{\`e}le jaug{\'e}
vectoriel, la th{\'e}orie conforme d{\'e}crivant la propagation des cordes
ferm{\'e}es s'obtient en utilisant la m{\^e}me alg{\`e}bre chirale $A(G/H)$
dans les deux secteurs. Il en va {\'e}videmment de m{\^e}me pour le
mod{\`e}le jaug{\'e} 
axial dans le cas ab{\'e}lien, puisque sa th{\'e}orie conforme est
identique {\`a} celle du mod{\`e}le jaug{\'e} vectoriel qui lui est
T-dual. 

Nous n'allons pas d{\'e}crire explicitement l'alg{\`e}bre chirale
$A(G/H)$. Nous nous contenterons de mentionner qu'elle s'obtient {\`a}
partir du produit $\alieg\otimes \alieh$ en imposant certaines
contraintes, et que les g{\'e}n{\'e}rateurs de Virasoro de $A(G/H)$ sont
donn{\'e}s par
\bea
L_n^{A(G/H)}=L_n^{\alieg}-L_n^{\alieh}.
\label{agh}
\eea
Dans cette formule, l'id{\'e}e de soustraire les degr{\'e}s de libert{\'e}
correspondant au sous-groupe $H$ appara{\^\i}t  clairement.
Les repr{\'e}sentations de $A(G/H)$ s'obtiennent en d{\'e}composant celles
de \alieg\ en repr{\'e}sentations de \alieh,
\bea
\hR^{\alieg}_\lambda = \bigoplus_\mu \hR^{A(G/H)}_{(\lambda,\mu)}\otimes
\hR^{\alieh}_\mu,
\label{represagh}
\eea
o{\`u} l'on a indic{\'e} par $\lambda$ et $\mu$ les repr{\'e}sentations de
\alieg\ et \alieh\ respectivement.

La th{\'e}orie conforme correspondant au mod{\`e}le de Wess-Zumino-Witten
jaug{\'e} vectoriellement a pour fonction de partition l'invariant
modulaire diagonal de l'alg{\`e}bre chirale $A(G/H)$. Dans le cas
compact, elle est rationnelle, et la th{\'e}orie de Cardy permet de 
construire certaines D-branes. Nous verrons plus loin
l'exemple du quotient \SUU.

\subsection{ Cordes ferm{\'e}es dans \SLU }

Pour trouver la g{\'e}om{\'e}trie de l'espace-temps associ{\'e} au mod{\`e}le
de Wess-Zumino-Witten jaug{\'e} \SLU, il faut d'abord conna{\^\i}tre
l'action des sym{\'e}tries vectorielle et axiale dans \SL. Cette action
s'exprime simplement dans les coordonn{\'e}es globales
$(\rho,\theta,\t)$. Rappelons que le temps $\t$ de \SL\ est
p{\'e}riodique. Consid{\'e}rons
maintenant le sous-groupe $U(1)$ de \SL\ form{\'e} des matrices
\bea
h=\left(\begin{array}{cc} \cos\t & \sin\t \\ -\sin\t &\cos \t
  \end{array} \right).
\label{sousgr}
\eea
Alors l'action axiale de ce sous-groupe sur \SL\ correspond {\`a} la
translation du temps $\t$, alors que l'action vectorielle correspond
{\`a} la rotation de l'angle $\theta$. 

\Fig{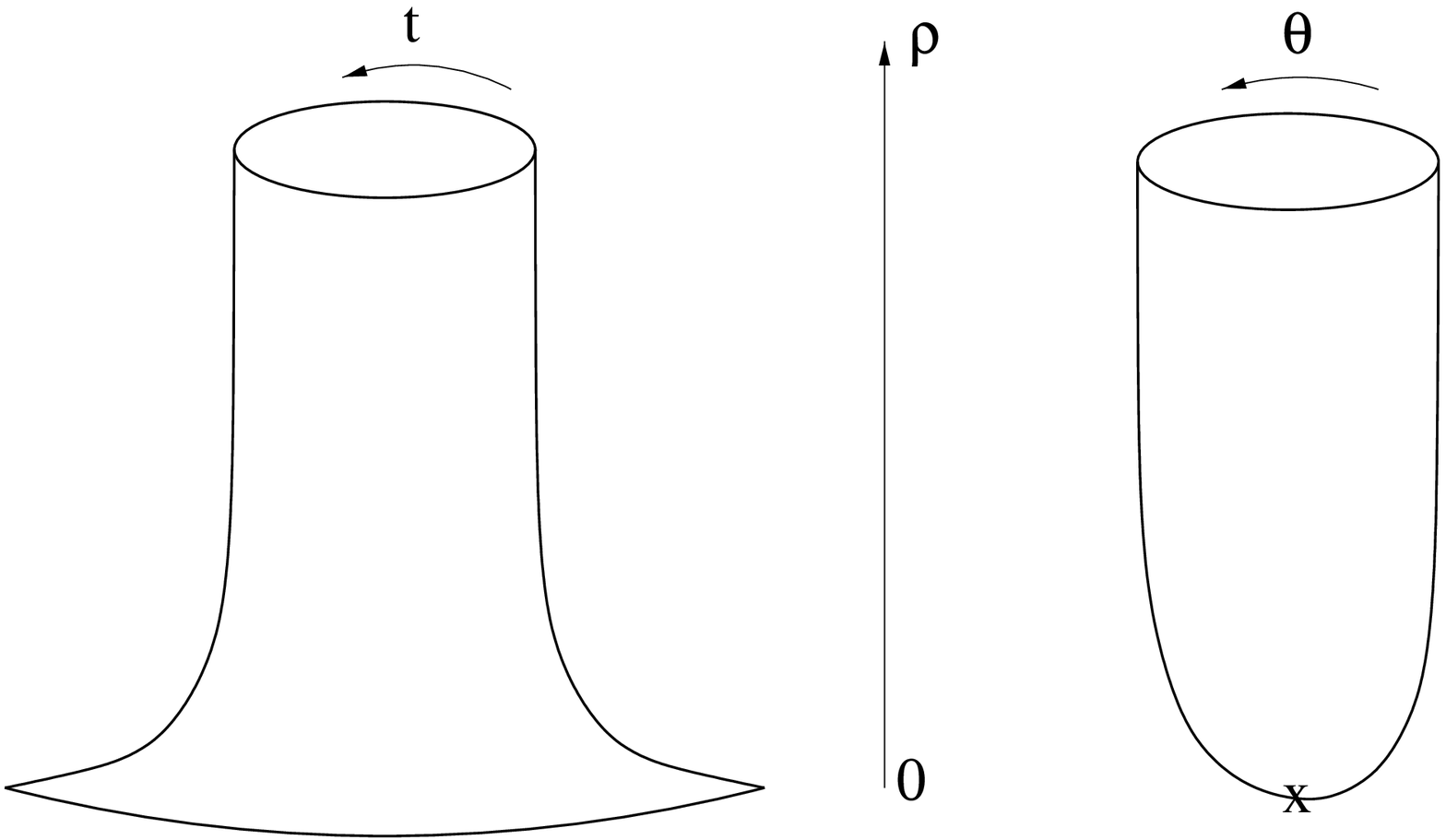}{9}{La trompette et le cigare sont deux g{\'e}om{\'e}tries
  T-duales. }

\Fig{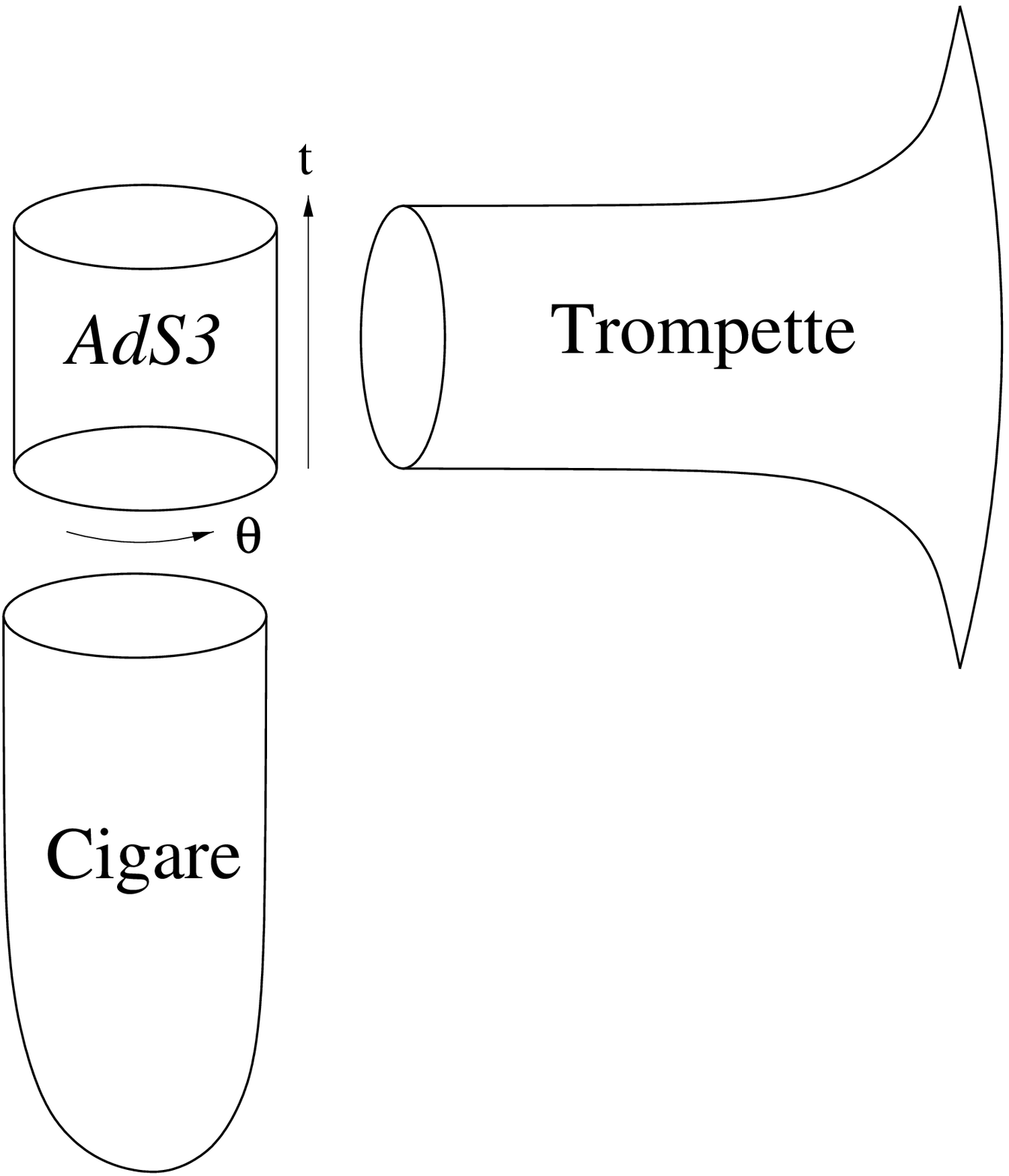}{8}{La trompette et le cigare s'obtiennent en jaugeant
  respecivement les directions $\theta$ et $\t$ d'\AAA. }

La g{\'e}om{\'e}trie du mod{\`e}le jaug{\'e} axial est r{\'e}guli{\`e}re, puisque
l'action axiale est sans point fixe. Il s'agit du trou noir
bidimensionnel {\`a} deux dimensions, connu sous le nom de \flqq
cigare\frqq. C'est une solution de la supergravit{\'e} avec une
m{\'e}trique non triviale et un dilaton non constant 
cr{\'e}{\'e} lors du jaugeage, alors
que le champ $H=dB$ dispara{\^\i}t pour des raisons purement
dimensionnelles. Le mod{\`e}le jaug{\'e} vectoriel, lui, donne une
g{\'e}om{\'e}trie appel{\'e}e \flqq trompette\frqq\ dont la m{\'e}trique et le
dilaton sont singuliers en $\rho=0$. En voici des expressions
explicites~:
\bea 
\SL &:& ds^2\ =\ k\left( d\rho ^2 - \cosh ^2\rho\, d\t ^2 + 
 \sinh ^2\rho \, d\theta ^2 \right) \\[2mm] 
 {\rm Cigare} &:& ds^2 \ = \ k\left(d\rho ^2 + \tanh ^2
   \rho \, d\theta ^2\right) \ \ \ \ \ ; \ \ \ \ \ 
   e^\Phi \ = \ \frac{e ^{\Phi_0}}{\cosh \rho} \\[2mm] 
{\rm Trompette} &:& ds^2 \ =\ k\left(d\rho ^2 + \coth ^2\rho \, d\t
  ^2\right) 
  \ \ \ \ ;\ \ \ \ e^\Phi \ =\  \frac{e ^{\Phi_0}}{\sinh \rho}.
\label{champsslu}
\eea
Remarquons que dans le cas de la trompette, lorsqu'on jauge la
direction euclidienne $\theta$, la signature de la direction $\t$ 
change et devient euclidienne. 

La figure \ref{a4cigtr} repr{\'e}sente le cigare et la trompette et met
en {\'e}vidence leur caract{\`e}re T-dual. La figure \ref{a4ads} montre,
elle, comment les deux se d{\'e}duisent d'\AAA.

Le mod{\`e}le jaug{\'e} est bien d{\'e}crit par la g{\'e}om{\'e}trie du cigare
dans la limite semi-classique $k\rar \infty$. En revanche, la
trompette n'est pas une bonne description de la th{\'e}orie des cordes
m{\^e}me dans cette limite, {\`a} cause non seulement de sa singularit{\'e}
en $\rho=0$, mais aussi de la p{\'e}riodicit{\'e} $\frac{2\pi}{k}$ de
l'angle $\t$. Cette p{\'e}riodicit{\'e} provient de la d{\'e}finition
pr{\'e}cise du mod{\`e}le jaug{\'e}, et correspond {\`a} ce qu'on obtient en
appliquant la T-dualit{\'e} au cigare dans la direction $\theta$. 
Pour des travaux r{\'e}cents sur l'image par T-dualit{\'e} de la version
supersym{\'e}trique du cigare, voir \cite{nouvtong,hk}.

\bs

On peut se faire une id{\'e}e assez pr{\'e}cise du spectre des cordes
ferm{\'e}es dans le cigare par l'analyse harmonique. Dans la limite
semi-classique, les {\'e}tats de corde dont la masse tend vers z{\'e}ro
se r{\'e}duisent aux fonctions sur le
cigare (nous v{\'e}rifierons explicitement ce principe g{\'e}n{\'e}ral quand
nous conna{\^\i}trons le spectre exact). 
En pr{\'e}sence d'un dilaton, leur {\'e}nergie  est mesur{\'e}e 
par le Laplacien 
\footnote{ Cela peut se d{\'e}duire de l'action de la supergravit{\'e}. }
\bea
\Delta \ = \  
 \frac{1}{e ^{-2\Phi}\sqrt{\det
  g}}\partial _\mu e ^{-2\Phi}\sqrt{\det g}\ 
  g^{\mu \nu}\partial_\nu\ \ . 
\label{lapl}
\eea
Dans les cas de \SL, du cigare et de la trompette, cet op{\'e}rateur vaut
\ber
k\, \Delta_{\SL} & = &  \partial_\rho^2 + (\coth \rho + \tanh \rho)\,  
\partial_\rho + ({\tanh^2\rho}-1)\, \partial_\t ^2 
+ ({\coth^2\rho}-1)\,  \partial_\theta^2 ,
\\[2mm] 
k\, \Delta_\cig & = &  \partial_\rho^2 + (\coth \rho + \tanh \rho) 
\partial_\rho + \coth^2\rho \, \partial_\theta^2, 
\\[2mm] 
k\, \Delta_\tr & = &   \partial_\rho^2 + (\coth \rho + \tanh \rho) 
\partial_\rho + \tanh^2\rho \partial_\t ^2.
\eer
On cherche des fonctions propres simultan{\'e}es de
$\Delta,\p_\t,\p_\theta$ de valeurs propres $-4E,ip,in$
respectivement. 
Ce sont des
fonctions hyperg{\'e}om{\'e}triques du type 
\bea
 K^j_{n\, p} =  e ^{ip\t +in\theta}
  \cosh ^{p}\rho\ \sinh ^{|n|}\rho\ F\left(\frac{|n|+p}{2}+j+1,
  \frac{|n|+p}{2}-j,|n|+1~; -\sinh ^2\rho\right),
\label{fctpropre}
\eea
avec, selon le cas,
\bea
\SL\ &~:&\ E_{\SL} = -\frac{j(j+1)}{k},
\\
{\rm Cigare}\ &~:&\ p=0\ , \ E_\cig = -\frac{j(j+1)}{k} + \frac{n^2}{4k},
\\
{\rm Trompette}\ &~:&\ n=0\ , \ E_\tr = -\frac{j(j+1)}{k} + \frac{p^2}{4k}.
\eea
La positivit{\'e} de l'{\'e}nergie, et le comportement {\`a} l'infini des
fonctions d'ondes (dont la norme $L^2$ doit {\^e}tre finie pour les
{\'e}tats li{\'e}s, et diverger comme $\int 1$ pour les {\'e}tats
asymptotiques), d{\'e}finissent les valeurs permises des param{\`e}tres $j,n,p$. Dans
le cas du cigare, on trouve $j\in -\half+i\R$, ce qui correspond aux
repr{\'e}sentations continues de \SL. Dans la trompette, il existe de
plus des {\'e}tats li{\'e}s $j\in \R$, qui correspondent aux
repr{\'e}sentations discr{\`e}tes de \SL. La positivit{\'e} de l'{\'e}nergie
impose alors $|j|<\frac{|p|}{2}$. 

La T-dualit{\'e}, en g{\'e}n{\'e}ral, {\'e}change les modes d'impulsion et les
modes d'enroulement. Les fonctions sur la trompette d{\'e}crivent donc
les modes d'enroulement de la th{\'e}orie des cordes dans le
cigare. On a en fait $p=kw$, o{\`u} $w\in \Z$ est le nombre
d'enroulement. 

\bs

Les r{\'e}sultats exacts de \cite{dvv,hpt} confirment cette image
g{\'e}om{\'e}trique. On trouve en effet un spectre form{\'e} d'{\'e}tats
continus et discrets. Il s'obtient en appliquant des op{\'e}rateurs de
cr{\'e}ation {\`a} des modes z{\'e}ros, qui sont param{\'e}tr{\'e}s par $j,n,w$
et sont des vecteurs propres de $L_0$ et $\bar{L}_0$ de valeurs
propres
\bea
L_0=-\frac{j(j+1)}{k-2}+\frac{(n+kw)^2}{4k},
\\
\bar{L}_0=-\frac{j(j+1)}{k-2}+\frac{(n-kw)^2}{4k}.
\label{enercig}
\eea
L'{\'e}nergie totale d'un mode z{\'e}ro, $\half(L_0+\bar{L}_0)$, se
r{\'e}duit aux expressions semi-classiques dans le cigare et la
trompette (respectivement $p=0$ et $n=0,p=kw$), {\`a} la
renormalisation $k\rar k-2$ pr{\`e}s. 

Dans le cas des {\'e}tats continus, $j\in -\half+i\R$ et toutes les
valeurs de $n,w$ apparaissent dans le spectre, avec cependant une
densit{\'e} d'{\'e}tats non triviale. En revanche, les {\'e}tats discrets
sont soumis {\`a} des contraintes d'unitarit{\'e} et de positivit{\'e} de
l'{\'e}nergie. La contrainte d'\flqq unitarit{\'e} renforc{\'e}e\frqq\ exige 
$-\frac{k-1}{2}<j<-\half$ (dans \cite{dvv}, seule
$-\frac{k}{2}<j<-\half$ {\'e}tait exig{\'e}e). On doit avoir de plus
$|kw|>|n|$, et $-j+\half(|n|-|kw|)$ doit {\^e}tre un entier positif. Ceci
revient {\`a} demander que $(j,\frac{\pm n+ kw}{2})$ soient le spin et le
moment magn{\'e}tiques d'{\'e}tats de $D_j^-$.

L'{\'e}nonc{\'e} de ces contraintes ach{\`e}ve notre rapide exposition du
spectre des cordes ferm{\'e}es du mod{\`e}le jaug{\'e} \SLU. 

\subsection{ Cordes et D-branes dans les quotients compacts } 

Le mod{\`e}le jaug{\'e} le plus simple est \SUU. De plus, il s'agit d'un
prolongement analytique du cigare. Nous allons donc en rappeler
quelques caract{\'e}ristiques \cite{mms}. 

D'abord, la g{\'e}om{\'e}trie est celle d'une cloche. Il s'agit 
topologiquement d'un disque, mais le terme cloche est utilis{\'e} pour
insister sur la divergence de la m{\'e}trique et du dilaton au bord,
\bea
{\rm Cloche} &:& ds^2 \ = \ k\left(d\r ^2 + \tan ^2
   \r \, d\phi ^2\right) \ \ \ \ \ ; \ \ \ \ \ 
   e^\Phi \ = \ \frac{e ^{\Phi_0}}{\cos \r},
\label{metcloche}
\eea
avec $\r \in [0,\halfpi]$.
La T-dualit{\'e} est plus simple que dans le cas de \SLU, puisque la
Cloche est T-duale {\`a} son propre \de{orbifold} par $\Z_k$ (rappelons
que $k$ doit {\^e}tre entier dans le cas compact), o{\`u} $\Z_k$ agit par
translations d'angle $\frac{2\pi}{k}$ de $\phi$.  

Rappelons que l'orbifold est l'op{\'e}ration de la th{\'e}orie des cordes
qui correspond au quotient de l'espace-temps par un groupe de
sym{\'e}tries discret. Cette op{\'e}ration est un peu plus compliqu{\'e}e
que la simple quantification suppl{\'e}mentaire des impulsions qui la
caract{\'e}rise au niveau des particules ponctuelles~; par exemple elle
engendre l'apparition d'\flqq {\'e}tats tordus\frqq\ dans le spectre, et
modifie la quantification du nombre d'enroulement,
qui peut {\^e}tre fractionnaire et non plus entier. Ce ph{\'e}nom{\`e}ne est
similaire {\`a} l'apparition de poids fractionnaires sym{\'e}triques dans le
spectre des D-branes en forme de classes de $\o$-conjugaison (voir la
section \Fac{\ref{topoquantif}}).

Au niveau de la th{\'e}orie conforme, cela signifie que la th{\'e}orie
\SUU\ (des \flqq parafermions compacts\frqq) 
est identique {\`a} son orbifold par $\Z_k$. Sa fonction de
partition est~:
\bea
Z^\cl(q)= \sum_{j=0}^{\frac{k}{2}} \sum_{n=-k}^{k+1} |\chi_{(j,n)}(q)|^2,
\label{zcloche}
\eea
o{\`u} $n$ indice les repr{\'e}sentations de l'alg{\`e}bre des courants \au,
et le spin $j$ correspond aux repr{\'e}sentations de \asu. Les
caract{\`e}res $\chi_{(j,n)}$ des repr{\'e}sentations de l'alg{\`e}bre
chirale $A(SU(2)/U(1))$ sont d{\'e}finis {\`a} partir de la relation g{\'e}n{\'e}rale
(\ref{represagh}), en d{\'e}composant les caract{\`e}res de \asu,
\bea
\chi ^{\SU}_j(q,z)=\sum_{n=-k}^{k+1}\chi_{(j,n)}(q)
\chi_n^{U(1)}(q,z),
\label{decompsu}
\eea
o{\`u} la variable suppl{\'e}mentaire $z$ des \de{caract{\`e}res non
  sp{\'e}cialis{\'e}s} $\chi (q,z)$ permet de prendre en compte une
isom{\'e}trie $U(1)$ de g{\'e}n{\'e}rateur $J^3$,
\bea
\chi (q,z)= \Tr q^{L_0-\frac{c}{24}} z^{J_0^3}.
\label{speccar}
\eea
En l'occurence, l'isom{\'e}trie $U(1)$ choisie pour d{\'e}finir le
caract{\`e}re sp{\'e}cialis{\'e} $\chi ^{\SU}_j(q,z)$ est le groupe $U(1)$
qui appara{\^\i}t dans \SUU. 

\Fig{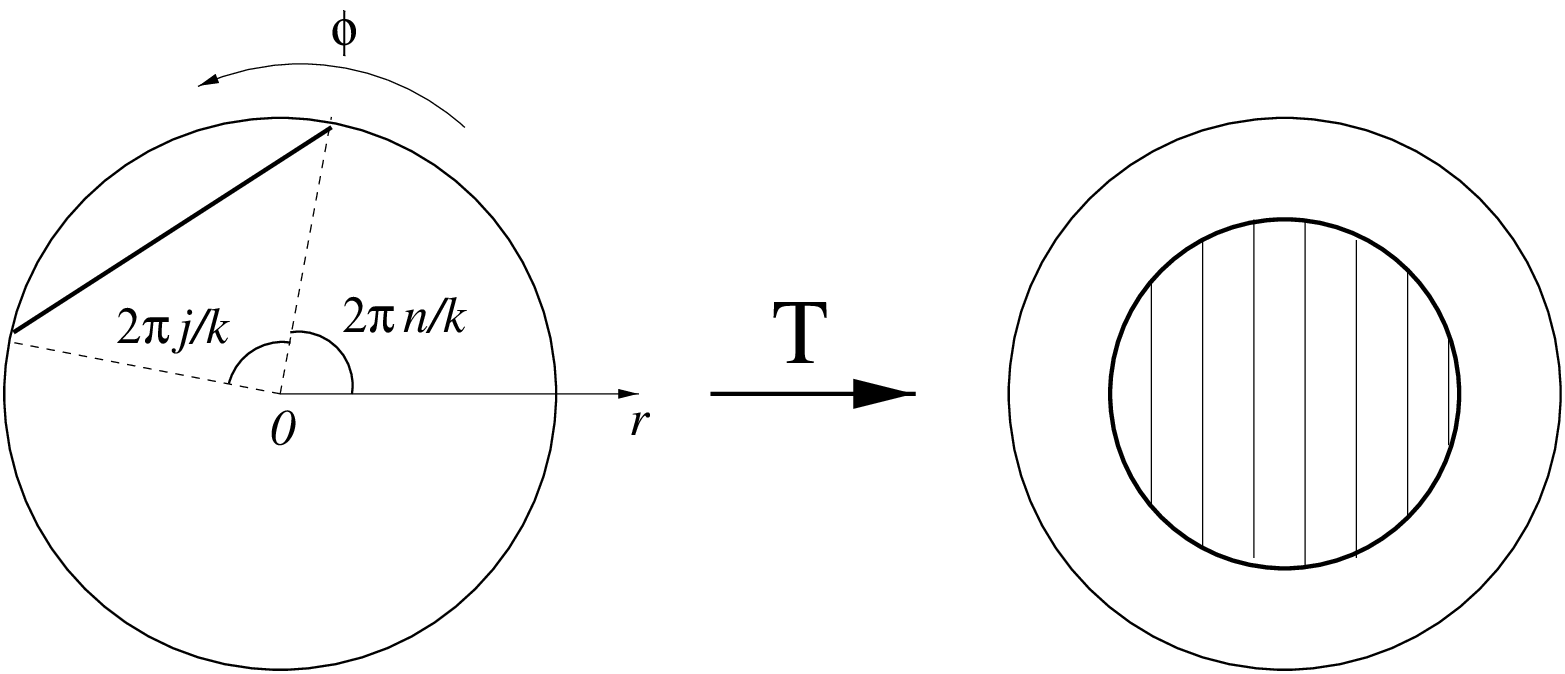}{12}{ Une D1-brane de param{\`e}tres $(j,n)$ dans la
  cloche, et une B-brane obtenue par T-dualit{\'e}. } 

\bs

Consid{\'e}rons maintenant les D-branes dans la cloche. On peut
construire les {\'e}tats de Cardy, qui correspondent {\`a} des D-branes
pr{\'e}servant la sym{\'e}trie $A(\SUU)$, appel{\'e}es parfois 
les A-branes ou D-branes de type A. Elles sont indic{\'e}es par les
repr{\'e}sentations $(j,n)$ de $A(\SUU)$. G{\'e}om{\'e}triquement, ce sont des
D1-branes, dont la longueur d{\'e}pend de $j$ et l'orientation d{\'e}pend
de $n$. Les deux sont quantifi{\'e}es (voir la figure \ref{a4cloche}).

Il existe n{\'e}cessairement d'autre D-branes dans la th{\'e}orie, comme
le montre l'application d'un orbifold $\Z_k$ suivi de la
T-dualit{\'e}. En effet, cette op{\'e}ration donne la m{\^e}me th{\'e}orie de
cordes ferm{\'e}es, mais transforme les D1-branes de type A en D2-branes
localis{\'e}es pr{\`e}s du centre $\r=0$ de la cloche. Ces D-branes de
type B sont instables et brisent la sym{\'e}trie $A(\SUU)$. La situation
est analogue aux D-branes dans $U(1)$~: les {\'e}tats de Cardy
correspondent aux D0-branes, mais la T-dualit{\'e} les transforme en
d'autres D-branes qui brisent la sym{\'e}trie chirale. 

\Fig{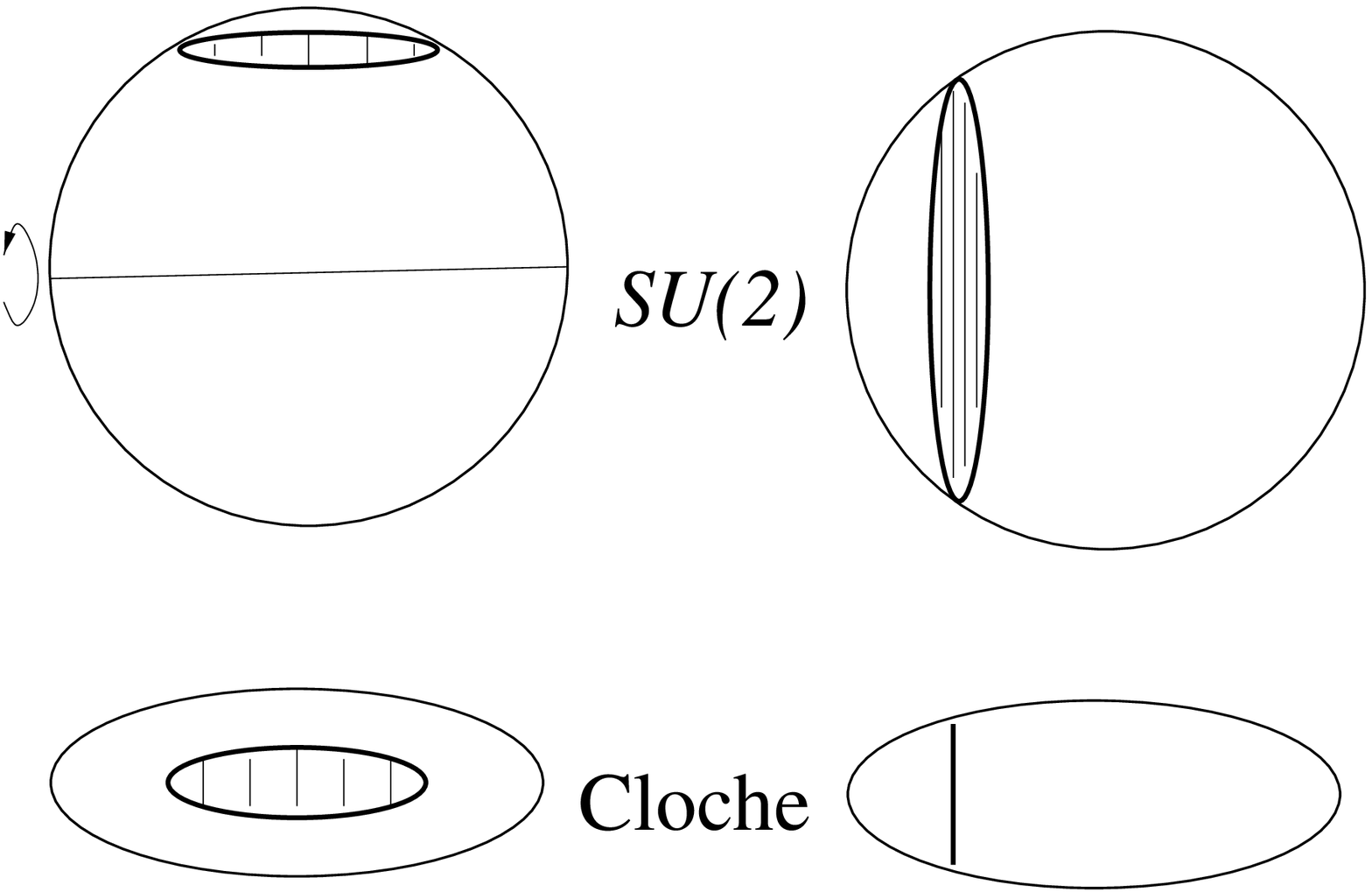}{11}{ Les D-branes dans la cloche s'obtiennent {\`a} partir
  des D-branes $S^2$ dans \SU. Ici \SU\ est repr{\'e}sent{\'e}
  comme une sph{\`e}re pleine, et l'axe de rotation qui d{\'e}finit 
la direction jaug{\'e}e $U(1)$ est repr{\'e}sent{\'e}. }

Les B-branes dans la cloche sont indic{\'e}es par $j$, qui correspond
{\`a} leur rayon. Si l'on consid{\`e}re la cloche comme un \SU\ aplati, on
peut voir ces B-branes comme des D-branes $S^2$ aplaties. D'ailleurs,
les D1-branes de type A \flqq descendent\frqq\ {\'e}galement de D-branes
$S^2$ de \SU, mais sym{\'e}triques par rapport {\`a} la direction $U(1)$
jaug{\'e}e (voir la figure \ref{a4desc}). 

Ces r{\'e}sultats peuvent se g{\'e}n{\'e}raliser {\`a} de nombreuses
th{\'e}ories conformes rationnelles correspondant {\`a} des quotients de
groupes compacts, {\'e}ventuellement asym{\'e}triques \cite{asymcosets}. Les
B-branes peuvent {\^e}tre obtenues par la th{\'e}orie de Cardy, {\`a}
condition de l'appliquer {\`a} une alg{\`e}bre chirale plus petite que
$A(G/H)$.

\section{ R{\'e}sultats semi-classiques sur les D-branes dans \SLU }

L'{\'e}tude semi-classique des D-branes dans \SLU\ peut nous donner des
indications int{\'e}ressantes sur leur existence et leurs
propri{\'e}t{\'e}s. Naturellement, nous ne nous attendons pas {\`a} trouver
des r{\'e}sultats exacts comme dans les vari{\'e}t{\'e}s de groupe, puisque
d{\'e}j{\`a} au niveau des cordes ferm{\'e}es les {\'e}tats d'enroulement
n'ont pas d'{\'e}quivalent semi-classique. 

Cette section se base sur des r{\'e}sultats non publi{\'e}s, obtenus pour
partie en collaboration avec A. Fotopoulos, C. Bachas et
M. Petropoulos.

\subsection{ La g{\'e}om{\'e}trie des D-branes }

Les D-branes dans le cigare que nous allons discuter seront
repr{\'e}sent{\'e}es dans la figure \ref{a4brcig}, qui montrera en outre
comment elles descendent de D-branes dans \AAA. 

Nous commen{\c c}ons par rechercher des D0-branes ponctuelles. L'action
de Born-Infeld se r{\'e}duit alors {\`a} $e ^{-\Phi}= e ^{-\Phi
  _0}\cosh\rho$. Donc, la D0-brane ne peut exister qu'au sommet
$\rho=0$ du cigare. 

Le cas des D1-branes se traite facilement en utilisant la variable
$u=\sinh \rho$. Alors, les champs de fond du cigare deviennent
\bea
{\rm Cigare~:\ \ } ds^2 \ =\ \frac{du^2+u^2d\theta^2}{1+u^2}\ \ \ , 
 \ \ \
e^\Phi\ =\ (1+u^2)^{-\frac{1}{2}}\ \ .
\label{autrecigare}
\eea
L'action de Born-Infeld pour une D1-brane 
dans ces coordonn{\'e}es est 
\bea
S^{\rm BI} \ \propto \ \int \sqrt{u'{}^2+u^2\theta'{}^2}\ \ .
\label{BID1}
\eea
Donc les D1-branes sont des lignes droites dans le plan $(u,\theta)$,
leur {\'e}quation dans les coordonn{\'e}es d'origine $(\rho,\theta)$ est
donc
\bea
\sinh\rho\ \sin(\theta-\theta_0) \ =\ \cst\ = \ \sinh r.
\label{eqD1}
\eea
Comme les A-branes dans la cloche, ces D1-branes ont deux
param{\`e}tres (un radial $r$ et un angulaire $\theta_0$), 
qui ne sont cependant pas
quantifi{\'e}s. Ils correspondent aux param{\`e}tres des D-branes \AA\
dans \AAA\ dont nos D1-branes descendent. Ces D-branes \AA\ sont en
effet invariantes par l'isom{\'e}trie (axiale) de translation temporelle
qui est jaug{\'e}e pour obtenir le cigare \SLU. L'intuition
g{\'e}om{\'e}trique, selon laquelle on peut alors projeter la D-brane \AA\
dans la vari{\'e}t{\'e} quotient, est justifi{\'e}e au niveau de la
th{\'e}orie conforme dans le cas compact (nous avons vu celui de \SUU),
et on s'attend donc {\`a} ce qu'elle soit valable aussi pour le
cigare. On peut en fait d{\'e}finir une proc{\'e}dure correspondante de
projection de l'{\'e}tat de bord \cite{pst}, que nous mettrons {\`a}
profit pour d{\'e}cire la th{\'e}orie conforme au bord associ{\'e}e {\`a} ces
D1-branes. 

\bs

Les D2-branes dans le cigare, elles, sont l'{\'e}quivalent des B-branes
dans la cloche. Comme ces derni{\`e}res, nous pouvons les d{\'e}crire
comme des solutions des {\'e}quations de Born-Infeld. Nous pouvons aussi
les d{\'e}duire de certaines D-branes dans \AAA, qui ne sont pas
invariante par la translation temporelle que nous jaugeons~: les
classes de conjugaison $H_2$ et $dS_2$. Nous nous attendons donc {\`a}
trouver des D2-branes recouvrant tout le cigare descendant des $H_2$,
et des D2-branes dans la r{\'e}gion $\rho \geq \cst$ descendant des
$dS_2$. 

\Fig{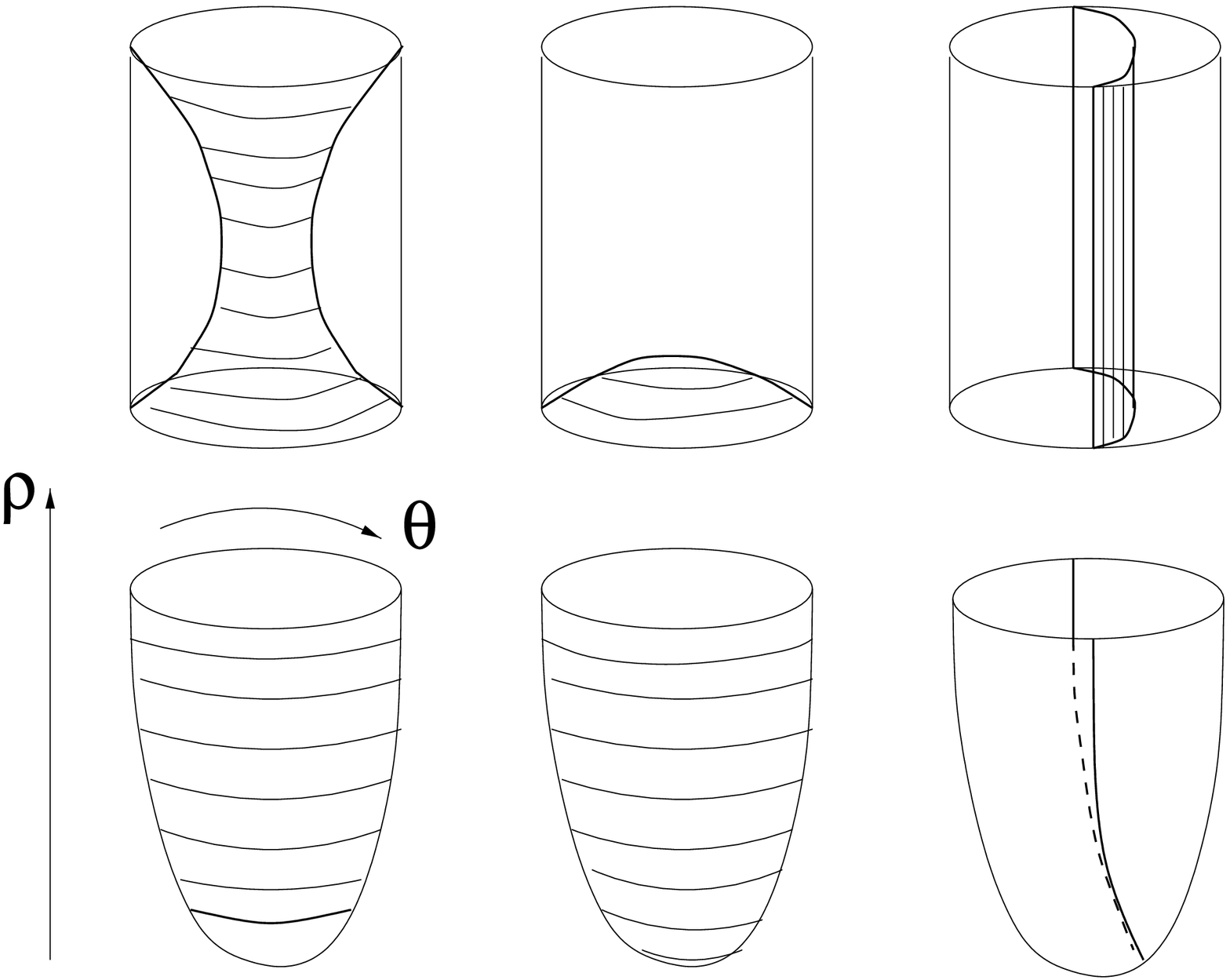}{11}{ Les D-branes $dS_2$, $H_2$ et $AdS_2$ dans \AAA, et
  les D-branes dans le cigare qui leur correspondent. } 

L'action des D2-branes dans le cigare est~:
\bea
S^{\rm BI} \ \propto \ \int \cosh \rho \sqrt{\tanh^2\rho +
  F_{\rho\theta}^2}.
\label{BID2}
\eea
La solution d{\'e}pend d'un param{\`e}tre $\beta$,
\bea
 F_{\rho\theta}^2=\frac{\beta ^2\tanh^2\rho}{\cosh^2\rho-\beta
    ^2}.
\label{FD2}
\eea
Dans le cas $\beta \leq 1$ (on pose $\beta=\sin r $), le champ $F$ est
bien d{\'e}fini pour toute valeur de $\rho$. On peut m{\^e}me l'int{\'e}grer
pour trouver le champ $A$ tel que $F=dA$. Si l'on suppose $A_\rho=0$
et que l'on exige par r{\'e}gularit{\'e} $A_\theta(\rho=0)=0$, on trouve
\bea
A_\theta(\rho)=r-\arctan\left(\frac{\tan
    r}{\sqrt{1+\frac{\sinh^2\rho}{\cos^2 r}}}\right).
\label{AD2}
\eea
Ainsi, le param{\`e}tre $r$ correspond {\`a} la valeur du champ $A_\theta$
{\`a} l'infini. En fait, on peut m{\^e}me l'identifier comme le
param{\`e}tre de la D-brane $H_2$ de \AAA\ dont descend notre D2-brane, si
l'on {\'e}crit l'{\'e}quation de la D-brane $H_2$
\bea
\cosh \rho \cos \t = \sin r.
\label{eqH2}
\eea 
En effet, la densit{\'e} de la D-brane qui descend de cette D-brane
$H_2$ peut se calculer g{\'e}om{\'e}triquement comme $\int d\t\
\delta\left(g(\t,\rho,\theta)\in H_2(r)\right)=\frac{\cos
  r}{\sqrt{\cosh^2\rho-\sin ^2r}}$. La distribution $\delta\left(g\in
H_2(r)\right)$ mesure en effet
l'appartenance du point $g$ d'\AAA\ {\`a} la D-brane $H_2$ de
param{\`e}tre $r$, dont la
densit{\'e} est constante par sym{\'e}trie \SL. Or la densit{\'e} de la
D2-brane de param{\`e}tre $r$ dans le cigare peut s'estimer au moyen du
\de{dilaton de cordes ouvertes} \cite{sewi},
\bea
e ^{\Phi_\ouv}=e ^\Phi\sqrt{\det (\hat{g}+F)\hat{g}^{-1}}
=\frac{1}{\sqrt{\cosh^2\rho -\sin ^2r}},
\label{osdil}
\eea
qui concorde avec l'expression d{\'e}duite de la D-brane $H_2(r)$. 

Le cas $\beta > 1$ donne des D-branes qui ne couvrent pas tout le
cigare, car $F$ doit {\^e}tre r{\'e}el d'o{\`u} $\cosh\rho\geq
\beta$. Nous n'{\'e}piloguerons pas sur la relation avec les D-branes
$dS_2$ dans \AAA. Nous allons plut{\^o}t d{\'e}crire les D-branes T-duales
dans la trompette. Bien qu'incorrecte {\`a} cause de la divergence du
dilaton en $\rho=0$, cette description permet de
mieux comprendre la g{\'e}om{\'e}trie, voire la dynamique, des D-branes
dans le cigare.

\Fig{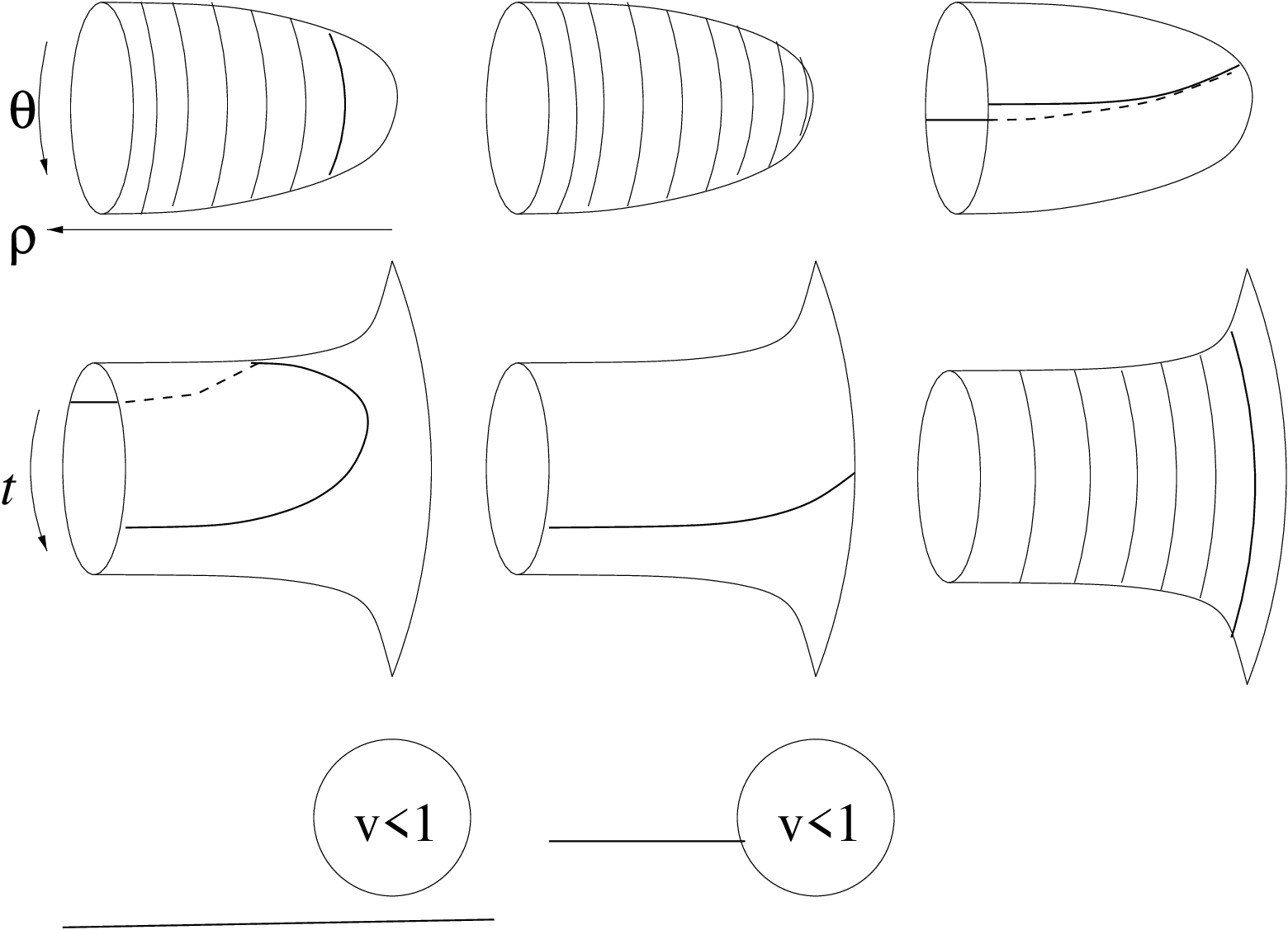}{12}{ Les D-branes dans le cigare et les D-branes
  correspondantes dans la trompette. Les D1-branes dans la trompette
  sont en outre trac{\'e}es comme des droites dans le plan $(v=\cosh
  \rho,\t)$, qui rencontrent ou non le cercle limite $\rho=0$. } 

Les D0-branes dans la trompette ne peuvent vivre ailleurs que sur la
singularit{\'e} $\rho=0$. Mais alors leur action est nulle, et de toute
fa{\c c}on on ne fait plus confiance {\`a} la description semi-classique.
 Les D1-branes, elles, sont facilement
d{\'e}crites en coordonne{\'e}s $(v=\cosh \rho,\t)$, o{\`u} l'action de
Born-Infeld devient simplement~: $S^{\rm BI}\propto \int
\sqrt{dv^2+v^2d\t^2}$. Les D1-branes sont donc des lignes droites
\footnote{ Sauf des D1-branes en forme d'arcs de cercle $\rho=0$, mais
  elles vivent dans la r{\'e}gion de fort couplage et leur
  interpr{\'e}tation physique est tr{\`e}s douteuse. }
 dans
ce \flqq plan perc{\'e}\frqq, le disque $v\leq 1$ n'{\'e}tant pas
physique. Une D1-brane qui rencontre ce disque est T-duale d'une
D2-brane recouvrant le cigare, et une D1-brane qui ne le rencontre pas
est T-duale d'une D2-brane d'{\'e}quation $\rho\geq
\cst$ recouvrant partiellement le cigare (voir la figure \ref{a4brtr}).

\subsection{ Les {\'e}tats li{\'e}s D2-D0 }

On peut se demander si, {\`a} l'image de ph{\'e}nom{\`e}nes qui se
produisent dans \SU\ \cite{bds,ars} 
et \SUU, il est possible de former des D2-branes
dans le cigare comme {\'e}tats li{\'e}s de D0-branes. On peut d'abord
remarquer que l'{\'e}nergie d'une D2-brane couvrant le cigare 
est infinie alors que celle
d'une D0-brane est finie. Une question plus pertinente est donc~:
peut-on former un {\'e}tat li{\'e} d'une D2-brane (de param{\`e}tre $r$)
 avec $N_{D0}$ D0-branes~?
Le calcul de la charge $\int_{D2}F$, dont chaque D0-brane porte une
unit{\'e}, nous donne le param{\`e}tre $r'$ de la nouvelle D2-brane ainsi
cr{\'e}{\'e}e,
\bea
r'-r=\frac{2\pi}{k} N_{D0}.
\label{quantifd2}
\eea
On peut montrer que la formation de cet {\'e}tat li{\'e} est
{\'e}nerg{\'e}tiquement possible, $E(r')\leq E(r)+N_{D0}E(D0)$. 

Bien s{\^u}r, il serait int{\'e}ressant de retrouver cette quantification
de $r$ au niveau des {\'e}tats de bord, comme c'est le cas pour les D-branes
dans \SU\ et \SUU. Cependant, il est tout aussi possible (et nous
verrons des indications dans ce sens) que les th{\'e}ories conformes au bord
correspondant {\`a} toutes les valeurs de $r$ soient coh{\'e}rentes, mais
que seules 
des familles de D-branes
quantifi{\'e}es v{\'e}rifiant (\ref{quantifd2}) puissent coexister.

Cependant, le domaine physique du param{\`e}tre $r$ est $r\in
[0,\halfpi]$. Que se passe-t-il donc si l'on ajoute plus de D0-branes
que $r$ n'en peut absorber~? Bien que nos approximations
semi-classiques ne soient alors plus valables, elles nous fournissent
une image g{\'e}om{\'e}trique de ce qui peut se produire. En effet, la
famille des D2-brane de param{\`e}tre $\beta>1$ fournit des candidats
naturels pour le devenir de notre D2(r)-brane. Quand $r$ atteint
$\halfpi$, le champ $F$ sur la D2-brane devient critique et le \flqq
claquage\frqq\ ainsi produit se traduit vraisemblablement par le
detachement de la D2-brane du point $\rho=0$. Ce ph{\'e}nom{\`e}ne se
visualise bien dans le cas T-dual de la trompette. Cependant, la
charge $\int_{D2}F$ ne d{\'e}pend plus de $\beta$ pour $\beta>1$~; la
D2-brane ne peut donc plus former d'{\'e}tats li{\'e}s avec
des D0-branes suppl{\'e}mentaires. Ceci n'est pas
surprenant, puisqu'elle ne recouvre plus le voisinage de $\rho=0$ et
n'interagit donc plus fortement avec les D0-branes qui y seraient localis{\'e}es.

\subsection{ Le spectre et les enroulements demi-entiers }

Nous avons constat{\'e}, au cours de notre {\'e}tude des D-branes
sym{\'e}triques dans les vari{\'e}t{\'e}s de groupes, que la dynamique des
fluctuations quadratiques de l'action de Born-Infeld {\'e}tait r{\'e}gie
par le Laplacien de la m{\'e}trique de corde ouverte. Ce r{\'e}sultat
correspond {\`a} ce qu'on attend en g{\'e}n{\'e}ral 
pour le spectre des cordes ouvertes en
pr{\'e}sence d'un champ $F$
\cite{sewi}. Nous allons ici emprunter un raccourci et {\'e}tudier
directement le spectre du Laplacien de corde
ouverte des D-branes dans le cigare. Cela permet de d{\'e}terminer
quelles repr{\'e}sentations apparaissent dans le spectre et avec quelle
multiplicit{\'e}, sans cependant v{\'e}rifier la structure des modes
z{\'e}ro de chaque repr{\'e}sentation. L'analyse compl{\`e}te des
fluctuations de l'action de Born-Infeld a par ailleurs {\'e}t{\'e}
effectu{\'e}e par A. Fotopoulos \Fac{\cite{angelos}}. 

En pr{\'e}sence d'un dilaton, le Laplacien de corde ouverte est~:
\bea
\Delta_\ouv \ = \  
 \frac{1}{e ^{-2\Phi_\ouv }\sqrt{\det
  g_\ouv}}\partial _\mu e ^{-2\Phi_\ouv}\sqrt{\det g_\ouv}\ 
  g^{\mu \nu}_\ouv \partial_\nu\ \ ,
\label{laplouv}
\eea
o{\`u} nous avons d{\'e}j{\`a}  d{\'e}fini $e ^{\Phi_\ouv}$,
eq. (\ref{osdil}), et nous d{\'e}finissons la m{\'e}trique de corde ouverte,
\bea
g_\ouv = \hg-F\hg \- F.
\label{osmet}
\eea
Pour la D2-brane qui couvre le cigare, ces objets valent,
\bea
ds^2_\ouv=\frac{k}{\cosh^2\rho-\sin^2r}\left(\cosh^2\rho d\rho
  ^2+\sinh ^2\rho d\theta ^2\right)\ , \                
e ^{\Phi_\ouv}=\frac{e ^{\Phi_0}}{\sqrt{\cosh^2\rho-\sin^2r}},
\label{cigouv}
\eea
dont on d{\'e}duit
\bea
k\Delta_\ouv=\frac{2}{\sinh
  2\rho}\p_\rho\tanh\rho\ (\cosh^2\rho-\sin^2r)\p_\rho
+\frac{\cosh^2\rho-\sin^2r}{\sinh^2\rho}\p_\theta ^2.
\label{deltaouv}
\eea
L'introduction de la variable $y=\sinh^2\rho/\cos ^2r$ permet
d'{\'e}liminer le param{\`e}tre $r$, et de se ramener {\`a} l'analyse
harmonique sur le cigare que nous avons d{\'e}j{\`a} {\'e}voqu{\'e}e. Mais
comme la variable radiale physique reste $\rho$, les coefficients de
r{\'e}flexion auront un facteur d{\'e}pendant de $r$. Plus pr{\'e}cis{\'e}ment, le
Laplacien en termes de $y$ est 
$k\Delta_\ouv=4\p_yy(y+1)\p_y+\frac{y+1}{y}\p_\theta ^2$, les fonctions
propres sont donc des fonctions hyperg{\'e}om{\'e}triques qui se
comportent {\`a} l'infini comme
\bea
\Gamma(n+1)\left[\frac{\Gamma(-2j-1)}
  {\Gamma(-j+n/2)^2}y ^{-j-1}+\frac{\Gamma(2j+1)}
  {\Gamma(j+1+n/2)^2}y ^{j}\right],
\label{fctouvinf}
\eea
o{\`u} l'{\'e}nergie est $-j(j+1)$, et le moment angulaire
$n$. Ainsi, l'amplitude de r{\'e}flexion est 
\bea
R(j,n|r)= (\cos ^2r)^{2j+1}R(j,n|0),
\label{reflcld2}
\eea
ce qui donne un terme $\log \cos r$ dans la densit{\'e} d'{\'e}tats. 

\bs

\Fig{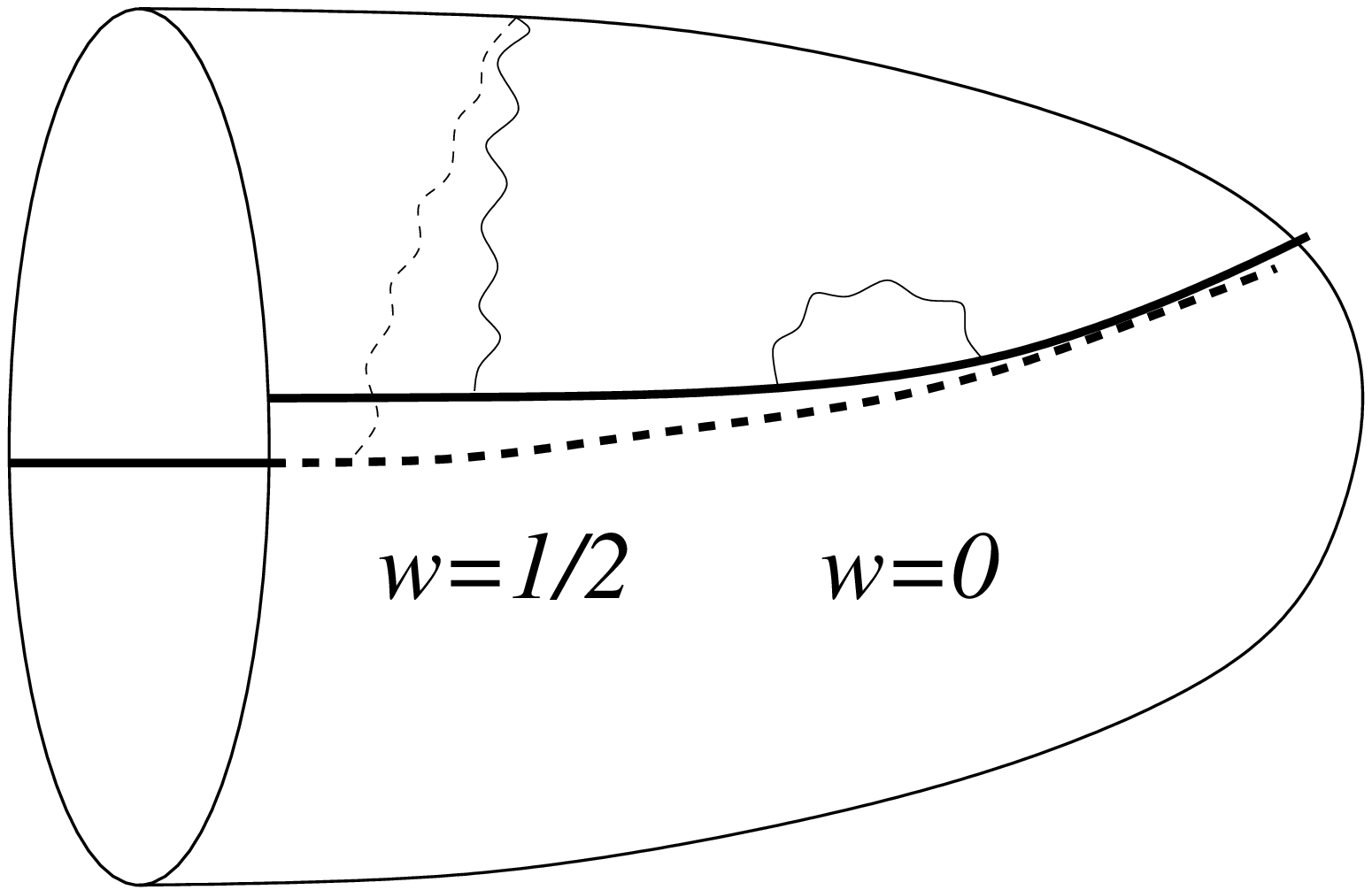}{8}{ Une D1-brane dans le cigare, et deux {\'e}tats de
  cordes ouvertes d'enroulements $w=\half$ et $w=0$. }

Consid{\'e}rons maintenant une D1-brane dans le cigare. L'analyse
harmonique  nous donne {\'e}galement la d{\'e}pendance de la
densit{\'e} spectrale vis-{\`a}-vis du param{\`e}tre radial, soit un terme $\log
\cosh r$.
Cependant, la g{\'e}om{\'e}trie nous sugg{\`e}re
qu'il existe d'autre {\'e}tats que les {\'e}tats d'enroulement z{\'e}ro, ces
autres {\'e}tats ne
s'identifient pas avec des fonctions dans la limite semi-classique. En
effet, une D1-brane a deux branches {\`a} l'infini et il peut donc
exister non seulement des {\'e}tats d'enroulement entier comme c'est le cas
pour les cordes ferm{\'e}es, mais aussi des {\'e}tats d'enroulement
demi-entier correspondant {\`a} des cordes ouvertes s'{\'e}tirant entre
les deux branches de la D1-brane (voir la figure \ref{a4cordecig}). 
Certes, ce nombre d'enroulement
demi-entier n'est pas conserv{\'e}, puisqu'il n'a aucun fondement
topologique. Sa situation est donc similaire {\`a} celle du nombre
d'enroulement $w$ des cordes ferm{\'e}es dans \SLU\ (qui est entier),
mais aussi du flot spectral $w$ dans \SL\ qui lui correspond
directement.

Ces consid{\'e}rations g{\'e}om{\'e}triques sugg{\`e}rent donc qu'il existe
des {\'e}tats d'enroulement demi-entier dans le spectre des cordes
ouvertes sur une D1-brane de \SLU, dont la densit{\'e} spectrale n'a pas
{\it a priori} 
de limite semi-classique. Donc, elle peut diff{\'e}rer de celle des
cordes ouvertes ne s'enroulant pas. Nous allons maintenant confirmer
ces pr{\'e}dictions par un calcul exact de l'amplitude du cylindre.

\section{ Construction d'{\'e}tats de bord }

Notre strat{\'e}gie pour construire les D-branes dans \SLU\ va consister
{\`a} \flqq descendre\frqq\ les {\'e}tats de bord connus dans \H, apr{\`e}s
avoir effectu{\'e} une rotation de Wick ($p\rar ip$). Elle s'inspire des
constructions {\'e}quivalentes dans le cas des mod{\`e}les jaug{\'e}s
compacts. Nous n'essaierons donc pas d'{\'e}valuer directement les
contraintes de factorisation dans le mod{\`e}le jaug{\'e} \SLU, ce qui
serait difficile techniquement et n'apporterait pas suffisamment
d'informations, {\`a} cause du caract{\`e}re compliqu{\'e} de l'alg{\`e}bre chirale
 $A(\SLU)$ et de la possibilit{\'e} pour les D2-branes de
briser cette sym{\'e}trie. Nous faisons  confiance {\`a} la
proc{\'e}dure de d{\'e}duction {\`a} partir des D-branes dans \H, pour
garantir la coh{\'e}rence de nos constructions.

Les r{\'e}sultats expos{\'e}s dans cette section sont issus d'un travail,
non encore publi{\'e}, effectu{\'e} en collaboration avec V. Schomerus.

\subsection{ Les D1-branes }

Une expression pour l'{\'e}tat de bord d'une D1-brane dans le cigare a d{\'e}j{\`a} {\'e}t{\'e}
conjectur{\'e}e dans \cite{pst}. On part de l'{\'e}tat de bord d'une
D-brane \AA\ dans \H, {\'e}crit sous forme de la fonction {\`a} un point
d'un {\'e}tat propre $\Phi_{np}^j$ d'impulsion $p$ et de moment
angulaire $n$, voir la formule (\Fac{\ref{etatbordphijnp}}). On peut
interpr{\'e}ter cet {\'e}tat comme un {\'e}tat de nombre d'enroulement $w$ dans le
cigare, {\`a} la condition $p=ikw$. Il reste {\`a} transformer le facteur
$\delta(p)$, qui refl{\`e}te l'invariance par translation temporelle de
la D-brane \AA, en un facteur $\delta_{w,0}$, qui signifie que la
D1-brane dans le cigare ne se couple qu'aux {\'e}tats d'enroulement
nul. On obtient donc~:
\bea
\llangle \Phi ^j_{n,w}(z)\rrangle_{D1(r)}=
\frac{e ^{-r(2j+1)}+(-1)^ne ^{r(2j+1)}}{\Gamma(1+j+\frac{n}{2})
  \Gamma(1+j-\frac{n}{2})}
\Gamma (1+b^2(2j+1)) \Gamma(2j+1)
\frac{2^{-\frac{3}{4}}b^{-\half}
\nu_b^{j+\half}\delta_{w,0}}{|z-\bz|^{2\Delta_j}}.
\label{etatbordd1cig}
\eea
En principe, cette formule n'est valable que pour la s{\'e}rie continue
$j\in -\half+i\R$, la s{\'e}rie discr{\`e}te n'apparaissant pas dans
le spectre des cordes dans \H.
Mais, dans \SLU, tous les {\'e}tats discrets ont un
enroulement $w$ non nul (cela d{\'e}coule de la contrainte $|kw|>|n|$
que nous avons {\'e}nonc{\'e}e), et il est naturel de supposer que leur
fonction {\`a} un point en pr{\'e}sence d'une D1-brane est nulle.

\bs

Le calcul du diagramme du cylindre diff{\`e}re du calcul analogue dans
\H\ essentiellement
\footnote{ Une diff{\'e}rence inessentielle est la quantification de
  $p=ikw$, qui n'appara{\^\i}t que dans un facteur global.} 
 pour la raison suivante~: l'entier $n$ n'est plus l'indice d'un
 {\'e}tat d'une repr{\'e}sentation de spin $j$ de l'alg{\`e}bre chirale,
 mais sert maintenant lui-m{\^e}me {\`a} indicer les repr{\'e}sentations. En
 particulier, l'{\'e}nergie $L_0$ d{\'e}pend de $n$, ce qui affecte les
 caract{\`e}res $\chi_{(j,n)}^{\SLU}(q)=q^{b^2P^2+n^2/4k}/\eta(\tau)^2$. 
La somme sur $n$ ne donne donc pas un
 facteur infini.

On calcule donc
\bea
Z^\cyl_{D1(r)}(\tq)\propto \int dP \sum_{n\in \Z}
\chi_{(\j,n)}^{\SLU}(\tq) \left| \llangle \Phi ^{\j}_{n0}(\halfi)
  \rrangle_{D1(r)}\right|^2.
\label{zcyld1}
\eea
Nous
effectuons toujours implicitement un calcul relatif, en soustrayant
$Z^\cyl_{D1(0)}$, pour r{\'e}gulariser la m{\^e}me divergence du volume
que dans le cas de la D-brane \AA\ dans \H.

Nous obtenons
\ber
Z^\cyl_{D1(r)}(\tq)&\propto & 
\int dP\ \frac{1}{\sinh 2\pi b^2 P \sinh 2\pi P}
\\
&\times & \left[ (\cosh ^2\pi P\ \cos ^22rP + \sinh
  ^2\pi P\ \sin ^2 2rP)\sum _{n\in \Z} \tq
    ^\frac{n^2}{4k} \right.
\\
& + & \left. (\cosh ^2\pi P\ \cos
   ^22rP - \sinh 
   ^2\pi P\ \sin ^2 2rP)(\sum _{n\in 2\Z}-\sum _{n\in
     2\Z+1}) \tq 
     ^\frac{n^2}{4k}\  \right]\ .
\eer
Nous avons donc deux termes {\`a} calculer, avec des d{\'e}pendances en
$n$ diff{\'e}rentes. Par resommation de Poisson de $\sum_n$, le
premier terme correspond {\`a} des {\'e}tats de nombre d'enroulement $w\in
\Z$, et le second {\`a} des {\'e}tats de nombre d'enroulement $w\in
\Z+\half$. Ces deux secteurs auront {\'e}videmment des densit{\'e}s
d'{\'e}tats diff{\'e}rentes. On peut poursuivre le calcul et montrer
 que la densit{\'e}
d'{\'e}tats du secteur $w\in \Z$ est la m{\^e}me que celle des cordes
ouvertes sur une D-brane \AA$(r)$ dans \AAA, alors que la densit{\'e}
d'{\'e}tat du secteur $w\in \Z+\half$ est $N(P|r,-r)$, dont nous rappelons
l'expression,
\bea
N(P|r,-r)=\frac{1}{2\pi i}\frac{\p}{\p P} \log \frac{
  S'_k(\frac{r}{\pi b^2+P}+P)}{S'_k(\frac{r}{\pi b^2-P})}\ + \ r-{\rm
  ind\acute{e}p.} ,
\label{nprmr}
\eea
o{\`u} la fonction $S'_k$ est d{\'e}finie par la formule \Fac{(\ref{fctspk})}.
Nous avons donc
\bea
Z^\cyl_{D1(r)}(\tq)=Z^\ouv_{D1(r)}(q)\propto \int
dP'\frac{q^{b^2P'{}^2}}{\eta(\tau)^2} \left[ N(P'|r,r)\sum_{w\in
    \Z}q^{kw^2} + N(P'|r,-r)\sum_{w\in \Z+\half} q^{kw^2} \right].
\label{zouvd1}
\eea
Cela correspond bien {\`a} l'intuition g{\'e}om{\'e}trique que les cordes
d'enroulement $w=\half$ vivant sur une D1-brane dans le cigare
descendent des cordes de flot spectral $w=\half$ vivant sur une
D-brane \AA\ dans \H. 

On peut en outre consid{\'e}rer les limites semi-classiques des
densit{\'e}s d'{\'e}tats mises en jeu. En se concentrant sur les termes
qui d{\'e}pendent de $r$, on trouve
\bea
{\rm lim}_{b\rightarrow
  0}N(P|r,r)\ =\ \frac{1}{\pi}\int_0^\infty
\frac{dt}{t}\left(\frac{\cos 
    \frac{2tr}{\pi}}{\sinh t} -\frac{1}{t}\right)\ .
\label{limrr}
\eea
Apr{\`e}s diff{\'e}rentiation par rapport {\`a} $r$ et utilisation de
l'identit{\'e} $\frac{2}{\pi}\int_0^\infty dt\
\frac{\sin\frac{2tr}{\pi}}{\sinh t}\ 
\ =\ \tanh r$, on aboutit {\`a} la conclusion que la d{\'e}pendance en $r$
de 
${\rm lim}_{b\rightarrow
  0}N(P|r,r)$ est un terme $\log \cosh r$, ce qui concorde avec le
calcul semi-classique de l'amplitude de r{\'e}flexion {\'e}voqu{\'e} dans la
section pr{\'e}c{\'e}dente.

\subsection{ Les D2-branes qui recouvrent le cigare } 

Le cas des D2-branes dans le cigare est {\it a priori} beaucoup plus
compliqu{\'e} que celui des D1-branes. En effet, ces D-branes sont
similaires aux B-branes dans \SUU~; les D-branes $H_2$ d'\AAA\ dont
elles descendent ne sont pas invariantes par l'isom{\'e}trie que l'on
jauge. De surcro{\^\i}t, cette isom{\'e}trie agit dans une direction
temporelle, ce qui requiert 
de prolonger analytiquement des {\'e}tats de bord de \H, et de quantifier
leur impulsion, $p=ikw$.
Comme si cela ne suffisait
pas, on s'attend de plus {\`a} ce que nos D2-branes se couplent {\`a} des
{\'e}tats d'enroulement, car elles sont invariantes par la rotation du
cigare (en outre, ces couplages d{\'e}coulent de couplages
g{\'e}om{\'e}triques des D-branes $H_2$ d'\AAA, qui sont non nuls dans la
limite semi-classique). 
Cela nous oblige en principe {\`a} inclure une
contribution non nulle des repr{\'e}sentations discr{\`e}tes dans l'{\'e}tat
de bord.

Nous n'essaierons pas d'aborder les difficult{\'e}s conceptuelles
li{\'e}es {\`a} ces probl{\`e}mes, ni de justifier th{\'e}oriquement la
proc{\'e}dure de construction de l'{\'e}tat de bord que nous
utiliserons. Nous nous contenterons de constater que, par une esp{\`e}ce
de miracle calculatoire, {\it toutes ces difficult{\'e}s n'ont pas
  d'impact sur le calcul de l'amplitude du cylindre}, au point que ce
calcul ne nous permettra m{\^e}me pas de d{\'e}terminer si les
repr{\'e}sentations discr{\`e}tes doivent contribuer {\`a} l'{\'e}tat de
bord~! Ce calcul sera valid{\'e} par diverses propri{\'e}t{\'e}s de la
densit{\'e} des cordes ouvertes sur la D2-brane qu'il permet de
d{\'e}terminer, notamment sa limite semi-classique et son comportement
quand le param{\`e}tre $r$ de la D-brane s'approche de sa valeur
maximale. 

La forme de la fonction {\`a} un point de la D2-brane que nous conjecturons
est~:
\bea
\langle \Phi^j_{nw} (z,\bar z) \rangle_{D2(r)}  = 
  2\pi  \delta_{n,0} \left[C_j(kw)e^{-ir(2j+1)}+C_j(-kw)e^{ir(2j+1)}\right]
\frac{\nu_b^{j+\half}\Gamma(1+b^2(2j+1))}
    {(z-\bar z)^{2h^j_{nw}}}, 
\label{etatbordd2cig}
\eea
o{\`u} nous utilisons la fonction
\bea
C_j(p)=\frac{\Gamma(2j+1)\Gamma(-j+p/2)}{2\ \Gamma(j+1+p/2)} .
\label{fctcj}
\eea
La fonction {\`a} un point (\ref{etatbordd2cig}) 
s'obtient {\`a} partir de celle de la D-brane
$H_2$ dans \H, elle-m{\^e}me d{\'e}duite de la D-brane \AA\ dans \H\ par
simple rotation, en {\'e}crivant $p=ikw$ et en effectuant la rotation de
Wick $r\rar ir$. Elle ne vaut en principe que pour les {\'e}tats
$\Phi_{nw}^j$ continus.

On pourrait cependant l'utiliser pour conjecturer la fonction {\`a} un
point des {\'e}tats discrets, par prolongement analytique de $j$. En
fait, si l'on consid{\`e}re plut{\^o}t l'{\'e}tat de bord d'une D-brane,
$H_2$ dans \SL, ce prolongement analytique est parfaitement en accord
avec la limite semi-classique (limite o{\`u} le couplage d'un {\'e}tat {\`a}
une D-brane devient l'int{\'e}grale d'une fonction sur une
sous-vari{\'e}t{\'e}). Une analyse pr{\'e}cise des p{\^o}les des expressions
mises en jeu montre qu'il faut attribuer chacun des deux termes
 de l'{\'e}quation
(\ref{etatbordd2cig}) {\`a} l'une des s{\'e}ries discr{\`e}te $D_j^+$ ou
$D_j^-$. Ainsi, la fonction {\`a} un point d'un op{\'e}rateur dans la
s{\'e}rie discr{\`e}te $D_j^+$ est~:
\bea
\langle \Phi ^{j,+}_{0p}(z,\bz)\rangle_{D2(r)}=
\frac{\Gamma(2j+1)\Gamma(-j+p/2)}{\Gamma(j+1+p/2)}\ \pi e ^{ir(2j+1)}\
\frac{\nu_b^{j+\half}\Gamma(1+b^2(2j+1)) }{(z-\bar z)^{2h^j_{0p}}}.
\label{h2discrete}
\eea
Nous pouvons avoir confiance en cette formule dans le cas de la
D-brane $H_2$ dans \SL. Mais la g{\'e}om{\'e}trie de la D-brane est tr{\`e}s
li{\'e}e {\`a} la p{\'e}riodicit{\'e} $2\pi$ de la direction $\t$ (par
exemple, quand $r\rar \halfpi$, la D-brane $H_2$ tend vers le c{\^o}ne
de lumi{\`e}re). Ainsi, la fonction {\`a} un point d'un {\'e}tat discret est
le r{\'e}sidu d'un p{\^o}le de la fonction {\`a} un point d'un {\'e}tat
continu. Mais, dans le cigare, la quantification n'est plus $p\in \Z$
mais $p=kw,w\in \Z$, o{\`u} $k$ n'a pas {\`a} {\^e}tre entier. Les p{\^o}les
de l'{\'e}tat de bord $(\ref{etatbordd2cig})$ ne correspondent alors {\`a}
rien de physique, et il est plus naturel de supposer que la D2-brane
ne se couple pas aux {\'e}tats discrets. Cependant, le calcul qui suit
ne permet pas de mettre cette conjecture {\`a} l'{\'e}preuve. 

\bs

Dans le calcul du diagramme du cylindre {\`a} partir de ces {\'e}tats de
bord, la r{\'e}gularisation consistant {\`a} travailler modulo des termes
ind{\'e}pendants de $r$ joue un r{\^o}le crucial. En effet, elle va nous
permettre d'ignorer la quantification $p=kw$, ainsi que la
contribution {\'e}ventuelle des repr{\'e}sentations discr{\`e}tes. Cette
derni{\`e}re ne d{\'e}pend en effet pas de $r$, puisque $|e
^{ir(2j+1)}|^2=1$. Quant {\`a} la d{\'e}pendance en $p$, elle ne joue
aucun r{\^o}le comme le montre l'identit{\'e}
\bea
|\langle \Phi ^j_{0p}(z=\halfi)\rangle_{D2(r)} |^2 
\simeq \frac{1}{\sinh\pi 2P}\frac{\sinh
^2 r\ 2P}{\sinh\pi b^2\ 2P}\ +\ r-{\rm indep.}
\label{miracle}
\eea
Le calcul du diagramme du cylindre est donc identique au calcul
{\'e}quivalent dans le cas de la D-brane \AA\ dans \H, modulo le
prolongement analytique $r\rar ir$, et son r{\'e}sultat est 
\bea
Z^\cyl_{D2(r)}(\tq)=Z^\ouv_{D2(r)}(q)\propto \sum_{n\in \Z}
q^{\frac{n^2}{4k}} \int
dP'\ \frac{q^{b^2P'{}^2}}{\eta(\tau)^2}N(P'|ir,ir) .
\label{zouvd2}
\eea
La partie \flqq $U(1)$\frqq\ de \SLU\ se factorise donc. Elle nous
apprend que sur la D2-brane vivent des {\'e}tats d'impulsion
quantifi{\'e}e, similaires aux {\'e}tats de cordes ferm{\'e}es
correspondants. En revanche, les {\'e}tats asymptotiques de cordes
ferm{\'e}es d'enroulement non nul n'ont pas d'{\'e}quivalent dans le spectre
de la D2-brane. Ce
r{\'e}sultat {\'e}tait pr{\'e}visible~:
une corde ouverte enroul{\'e}e autour de la D2-brane peut
se contracter sous l'effet de sa propre tension.

Il reste {\`a} {\'e}valuer les termes d{\'e}pendant de $r$ de
la densit{\'e} spectrale
$N(P|ir,ir)$, donn{\'e}s par la d{\'e}riv{\'e}e par rapport {\`a} $P$ de
\bea
\log S_k(\frac{ir}{\pi b^2}+P)-\log S_k(\frac{ir}{\pi b^2}-P)\ =\
i\int_0^\infty \frac{dt}{t}\left( 
  \frac{\sin 2tb^2P\ \cosh \frac{2tr}{\pi}}{\sinh b^2t\ \sinh
    t}-\frac{2P}{t}\right).
\label{nir}
\eea
On constate que cette int{\'e}grale converge pour
$r<\halfpi(1+b^2)$. Or, dans notre d{\'e}finition semi-classique de la
D-brane, le param{\`e}tre $r$ {\'e}tait restreint {\`a} l'intervalle
$[0,\halfpi]$, dont la borne sup{\'e}rieure {\'e}tait la \flqq limite de
claquage\frqq, correspondant {\`a} une valeur critique du champ
$F$. Ce comportement est donc confirm{\'e} par notre calcul exact du
spectre des cordes ouvertes, {\`a} une correction $\halfpi b^2=\halfpi
\frac{1}{k-2}$ pr{\`e}s. 

La limite semi-classique de $N(P|ir,ir)$ co{\"\i}ncide en fait avec la
pr{\'e}diction de l'analyse harmonique, $\log \cos r$. Nous
consid{\'e}rons cet accord comme un argument en faveur de notre
expression pour l'{\'e}tat de bord (\ref{etatbordd2cig}). 

\bs

Nous achevons cette discussion de la D2-brane dans le cigare par une
remarque. Si la diff{\'e}rence des param{\`e}tre de deux D2-branes
ob{\'e}it {\`a} une condition de quantification,
\bea
r-r'=\pi b^2 N_{D0},
\label{quantifex}
\eea
analogue {\`a} la condition d{\'e}j{\`a} consid{\'e}r{\'e}e (\ref{quantifd2}),
alors la diff{\'e}rence $N(P|ir,ir)-N(P|ir',ir')$ se
simplifie,
\bea 
N(P|ir,ir)-N(P|ir',ir')\ =
 \ 2\pi i b^2\sum_{l=0}^{N_{D0}-1} \frac{\sin 2\pi
  b^2(\half +l +\frac{r'}{\pi b^2})}{\cosh 2\pi b^2 P+\cos 2\pi
  b^2(\half +l +\frac{r'}{\pi b^2})} .
\label{diffd2}
\eea
Cette expression poss{\`e}de $N_{D0}$ p{\^o}les pour des valeurs imaginaires
de $P$, qui correspondent {\`a} des
repr{\'e}sentations unitaires. Cette remarque joue en faveur de l'existence
du processus de formation d'{\'e}tats li{\'e}s entre une D2-brane et des
D0-branes, mais sa signification exacte reste {\`a} pr{\'e}ciser.


\setcounter{chapter}{4}

\chapter{ Variations sur le th{\`e}me D3/NS5  }

Dans ce chapitre, nous utiliserons les \flqq mod\`eles jouets\frqq\ des
chapitres pr\'ec\'edents, c'est-\`a-dire des mod\`eles de th\'eorie
des cordes relativement simples \`a basse dimension, pour construire
des mod\`eles plus \flqq r\'ealistes\frqq, 
solutions de la th\'eorie des cordes supersym\'etrique \`a dix
dimensions. 

Ainsi, nous allons {\'e}tudier les D-branes, et plus
particuli{\`e}rement les D3-branes, dans des \flqq espaces de
NS5-branes\frqq. Ces D-branes permettent de
r{\'e}aliser g{\'e}om{\'e}triquement divers ph{\'e}nom{\`e}nes int{\'e}ressants
des th{\'e}ories de jauge supersym{\'e}triques et de la Petite Th{\'e}orie
des Cordes.  
Certains espaces de NS5-branes pourront {\^e}tre reli{\'e}s
(par T-dualit{\'e}) au cigare, et cette correspondance sera aussi
valable pour les D-branes. Nous allons donc montrer comment les
D-branes du chapitre pr{\'e}c{\'e}dent apparaissent dans ce nouveau
contexte.

\section{ D-branes et jeux de construction }

Nous avons introduit les D$p$-branes comme les sources des $p$-branes
noires. Ces derni{\`e}res sont des configurations non triviales de
l'espace-temps, qui ont un grand int{\'e}r{\^e}t 
tant conceptuel que ph{\'e}nom{\'e}nologique, 
de par leurs propri{\'e}t{\'e}s thermodynamiques,
holographiques, et leur analogie avec les trous noirs
ordinaires. L'{\'e}tude de 
la dynamique des 
D$p$-branes est essentielle pour la compr{\'e}hension de ces objets~; en
outre, les configurations de D$p$-branes pemettent de construire des
mod{\`e}les pour la cosmologie et la physique des hautes {\'e}nergies. 

Cette section sera donc consacr{\'e}e {\`a} l'exposition de certains
r{\'e}sultats importants de la dynamique des D$p$-branes, que nous
utiliserons par la suite.

\subsection{ Branologie }

Nous allons d'abord tirer quelques cons{\'e}quences de la d{\'e}finition
des D-branes et de leurs propri{\'e}t{\'e}s de transformation par les
dualit{\'e}s de la th{\'e}orie des cordes. 

\Fig{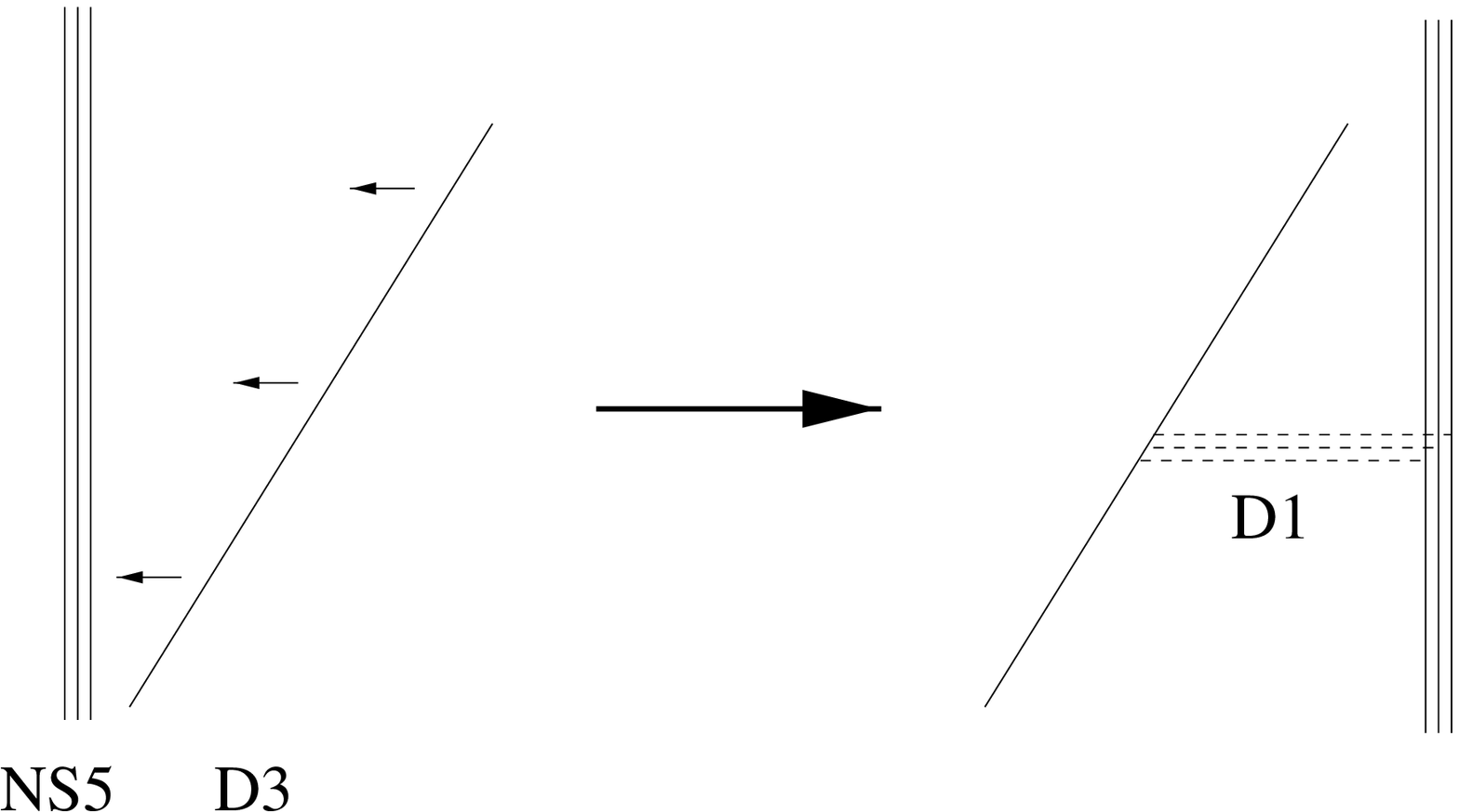}{11}{ Un exemple de l'effet Hanany-Witten }

Une corde fondamentale peut par d{\'e}finition 
aboutir sur une D$p$-brane quelconque, ce
que nous noterons $F1\rar Dp$. Dans le cas de la th{\'e}orie de type
IIB, on peut obtenir par S-dualit{\'e} des cons{\'e}quences de ce fait,
par exemple $D1\rar NS5$ ou $D1\rar D3$. En appliquant deux
T-dualit{\'e}s, puis une S-dualit{\'e}, on obtient $D3\rar NS5$, o{\`u} la
D3-brane a 2+1 directions le long de la NS5-brane et une direction
transverse. Par T-dualit{\'e} on obtient, en th{\'e}orie de type IIA,
$D4\rar NS5$.

Une D1-brane peut donc aboutir d'un c{\^o}t{\'e} sur une D3-brane, de
l'autre sur une NS5-brane. Hanany et Witten \cite{hw} ont montr{\'e} que
cette configuration pouvait r{\'e}sulter du passage de la D3-brane {\`a}
travers la NS5-brane (en supposant que la D3-brane ait 2+1 direction
parall{\`e}le {\`a} la NS5-brane et 1 direction orthogonale). La
cr{\'e}ation d'une D1-brane lors du croisement de la D3-brane et de la
NS5-brane d{\'e}coule de la conservation de la charge. Ainsi, si une
D3-brane croise plusieurs NS5-branes superpos{\'e}es, il se cr{\'e}era
autant de D1-branes (voir la figure \ref{a5hw}).

Nous venons d'utiliser la possibilit{\'e} de superposer des NS5-branes,
de m{\^e}me que nous avons d{\'e}j{\`a} consid{\'e}r{\'e} des superpositions de
D-branes. D'ailleurs, les solutions de supergravit{\'e} de $p$-branes
noires
ne sont
valables que dans la limite d'un grand nombre de D$p$-branes
superpos{\'e}es. Plus g{\'e}n{\'e}ralement, il est possible de former des
{\'e}tats li{\'e}s en superposant des objets de nature diff{\'e}rente. Par
exemple, on peut montrer qu'un syst{\`e}me de cordes F1 et de D1-branes
parall{\`e}les brise la supersym{\'e}trie, mais qu'il existe un {\'e}tat
li{\'e} supersym{\'e}trique d'{\'e}nergie inf{\'e}rieure portant les m{\^e}mes
charges. 
Ces m{\^e}mes crit{\`e}res d'{\'e}nergie et de charge nous ont servi au
pr{\'e}c{\'e}dent chapitre {\`a} argumenter en faveur de la formation d'un
{\'e}tat li{\'e} entre une D2-brane et des D0-branes dans le cigare,
{\'e}tat li{\'e} que nous avons identifi{\'e} comme une autre D2-brane.

Dans certaines limites, il est possible d'{\'e}tudier ces ph{\'e}nom{\`e}nes
au moyen d'actions effectives des D-branes. Par exemple, la formation
des D2-branes sym{\'e}triques de \SU\ comme {\'e}tats li{\'e}s de D0-branes
a {\'e}t{\'e} d{\'e}crite dans \cite{ars}. Myers a donn{\'e} une explication
de la formation d'{\'e}tats li{\'e}s dans un contexte assez g{\'e}n{\'e}ral en
termes d'effet di{\'e}lectrique \cite{myers}. Plus pr{\'e}cis{\'e}ment, un
ensemble de D$p$-branes peut former un {\'e}tat li{\'e} en pr{\'e}sence
d'un champ $C^{(p+3)}$, {\`a} cause d'un couplage qui appara{\^\i}t
dans le terme de Wess-Zumino de l'action effective non-ab{\'e}lienne des
D$p$-branes (alors que nous avons vu que, dans l'action effective
ab{\'e}lienne, une D$p$-brane ne se couplait qu'aux formes de
Ramond-Ramond de degr{\'e} au plus $p+1$). Ce
ph{\'e}nom{\`e}ne peut aussi se produire en pr{\'e}sence d'un champ
$H=dB$. Nous verrons plus loin un exemple de formation d'{\'e}tat li{\'e}
de D1-branes avec une D3-brane en pr{\'e}sence d'un champ $H$, que nous
attribuerons {\`a} l'effet Myers.

\subsection{ Constructions {\`a} base de NS5-branes }

La r{\'e}alisation de
th{\'e}ories de jauge non-ab{\'e}liennes au moyen de
cordes ouvertes s'{\'e}tirant entre diff{\'e}rentes D-branes est 
un int{\'e}r{\^e}t primordial des configurations de D-branes
superpos{\'e}es, juxtapos{\'e}es ou aboutissant les unes sur les autres.
Nous avons
d{\'e}j{\`a} mentionn{\'e} la possibilit{\'e} d'obtenir des th{\'e}ories de
jauge $U(N)$ au moyen de $N$ D-branes superpos{\'e}es. Mais, dans des
constructions plus compliqu{\'e}es, le nombre de D-branes peut aussi
s'interpr{\'e}ter comme le nombre de saveurs et non plus le rang du
groupe de jauge (aussi appel{\'e} nombre de couleurs). C'est le cas dans
une configuration de la th{\'e}orie de type IIA propos{\'e}e dans
\cite{egkrs}, construite {\`a} partir de NS5-branes.

Cette configuration met en jeu, entre autres, des D4-branes 
aboutissant sur des NS5-branes. On peut consid{\'e}rer les
D4-branes comme des branes-tests de masse faible, tout en prenant en
compte la d{\'e}formation de l'espace-temps cr{\'e}{\'e}e par les NS5-branes
(si elles sont assez nombreuses). Alors les D4-branes s'{\'e}tendent le
long de la direction radiale de la 5-brane noire ainsi
cr{\'e}{\'e}e. Donc, elles ne disparaissent pas lorsqu'on applique la
limite d'horizon proche. Rappelons que la limite d'horizon proche
de la 5-brane noire cr{\'e}{\'e}e par les NS5-branes est
l'espace $\R^{1,5}\times \SU\times \R_\Phi$. Les D4-branes
s'{\'e}tendent dans cette limite sur trois des quatre directions
transverses $\SU\times \R_\Phi$, et leur g{\'e}om{\'e}trie est $S^2\times
\R_\Phi$ o{\`u} $S^2$ est une classe de conjugaison de \SU\ dont le
param{\`e}tre, quantifi{\'e}, correspond au nombre de D4-branes. 

Un inconv{\'e}nient majeur de cette configuration de NS5-branes
superpos{\'e}es est la divergence du dilaton dans la r{\'e}gion proche des
sources. Comme $e ^\Phi$ est le couplage des cordes, on ne peut se
fier {\`a} la th{\'e}orie des cordes perturbatives pour {\'e}tudier cette
r{\'e}gion. Il est possible de r{\'e}gulariser cette divergence en
s{\'e}parant les NS5-branes. Plut{\^o}t que de les placer sur une sph{\`e}re
\cite{kkpr}, nous allons les disposer sur un cercle \cite{sfetsos}. La
limite d'horizon proche de l'espace-temps ainsi cr{\'e}{\'e} est
reli{\'e}e au cigare, 
et nous pourrons obtenir certaines D-branes dans
le cigare de m{\^e}me que nous avons obtenu les D-branes de \SU\ quand les
NS5-branes {\'e}taient superpos{\'e}es. Le rapport avec le cigare est connu depuis
longtemps \cite{vo}, mais nous lui donnerons une interpr{\'e}tation
g{\'e}om{\'e}trique en termes de T-dualit{\'e}, qui permettra d'{\'e}tudier
plus ais{\'e}ment les D-branes.

Cependant, avant de traiter explicitement le cas des NS5-branes sur un
cercle et d'{\'e}tudier les D-branes dans la solution de supergravit{\'e}
correspondante dans la section \ref{sectfinale}, nous allons faire une
digression sur la dynamique des D-branes, notamment dans le cas des
espaces cr{\'e}{\'e}s par des NS5-branes, qui seront syst{\'e}matiquement
pr{\'e}sent{\'e}s dans le paragraphe \ref{soussectespns5}.

\section{ Actions effectives et g{\'e}om{\'e}trie }

Dans cette section, nous allons introduire une formulation de la
dynamique de Born-Infeld propos{\'e}e dans \cite{d3ns5}. La motivation
originale de cette formulation {\'e}tait la d{\'e}termination de
corrections d{\'e}rivatives {\`a} l'action effective des D-branes dans les
th{\'e}ories des cordes supersym{\'e}triques, en pr{\'e}sence de torsion
$H=dB$. Ce probl{\`e}me peut {\^e}tre abord{\'e} du point de vue de la
g{\'e}om{\'e}trie non-commutative et m{\^e}me non-associative
\cite{corschia}. L'approche que j'ai essay{\'e} de mettre en \oe{}uvre
avec A. Fotopoulos consistait {\`a} g{\'e}n{\'e}raliser
directement les tenseurs de la g{\'e}om{\'e}trie des surfaces plong{\'e}es,
qui apparaissent dans les corrections gravitationnelles d{\'e}j{\`a}
connues. Mais il s'est r{\'e}v{\'e}l{\'e} que les tenseurs ainsi construits
permettaient de former trop de combinaisons diff{\'e}rentes, pour que
l'on puisse d{\'e}terminer les corrections dominantes {\`a} l'aide des
quelques diagrammes de cordes que l'on sache {\'e}valuer explicitement. 

Il n'en reste pas moins quelques r{\'e}sultats suggestifs, dont la
formulation de la dynamique de Born-Infeld au moyen d'une
g{\'e}n{\'e}ralisation de la seconde forme fondamentale. Cette formulation
a l'avantage d'{\^e}tre invariante de jauge, ce qui la rend 
pratique pour effectuer des calculs dans des cas particuliers
comme nous le verrons par la suite.

\subsection{ G{\'e}om{\'e}trie riemannienne des surfaces plong{\'e}es 
\label{soussectgeom}
}

Nous allons d{\'e}crire quelques objets caract{\'e}risant la g{\'e}om{\'e}trie
d'une surface $\B$ plong{\'e}e dans une vari{\'e}t{\'e} de dimension
sup{\'e}rieure. Naturellement, nous appliquerons ces concepts aux
D-branes plong{\'e}es dans l'espace-temps. 

Choisissons des coordonn{\'e}es $x^i$ sur $\B$, et notons $X^\mu(x^i)$
son plongement dans une vari{\'e}t{\'e} munie d'une
m{\'e}trique $g\mn$, et de la connexion riemannienne correspondante $\Gamma(g)$. 
Le premier objet que l'on puisse d{\'e}finir est la
\de{m{\'e}trique induite}, que nous avons utilis{\'e}e
fr{\'e}quemment,
\bea
\hg _{ij}=g\mn \p_i X^\mu \p_j X^\nu.
\label{metind}
\eea
Cette m{\'e}trique munit $\B$ d'une structure de vari{\'e}t{\'e}
riemannienne, dont on peut calculer le tenseur de \de{courbure intrins{\`e}que},
\bea
(R_T)_{ijkl} = R(\Gamma(\hg))_{ijkl}.
\label{courbint}
\eea
Nous allons maintenant d{\'e}finir les quantit{\'e}s qui caract{\'e}risent
le plongement proprement dit. Soit $\xi ^{\mu a}$ une base du fibr{\'e}
normal {\`a} $\B$ (une collection de vecteurs $\xi ^a $ 
de composantes $\xi ^{\mu a}$),
on a pour tout $a$ la relation
d'orthogonalit{\'e} $\xi ^{\mu a}g\mn \p_i X^\nu =0$. On d{\'e}finit la
\de{connexion de spin} 
\bea
\omega_i ^{ab}=\half\xi_\mu
^{[a}(\p_i+\Gamma(g)^\mu_{\rho\nu}\p_i X^\rho )\xi^{\nu b]},  
\label{conspin}
\eea
dont la courbure est la \de{courbure extrins{\`e}que} de $Br$, 
\bea
(R_N)_{ij}{}^{\mu\nu}=\left(\p_{[i}\omega_{j]}^{ab}- \omega
  _{[i}^{ac}\omega _{j]}^{bc}\right)\xi^\mu_a\xi^\nu_b.
\label{courbext}
\eea
Les courbures intrins{\`e}que et extrins{\`e}que $R_T$ et $R_N$, dont les
indices sont mis pour \flqq Tangent\frqq\ et \flqq Normal\frqq, sont
reli{\'e}es au tenseur de courbure de l'espace $R(\Gamma(g))_{\mu \nu
  \rho\sigma}$ par les {\'e}quations de Gauss et Codazzi. Pour les
{\'e}crire, nous aurons besoin de la \de{seconde forme fondamentale} de
$\B$, 
\bea
\Omega(X,g) ^\mu_{ij}=\p_i\p_jX^\mu +\Gamma
  ^\mu_{\nu\rho}(g)\p_iX^\nu \p_jX^\rho-\Gamma(\hg)^k_{ij}\p_kX^\mu.
\label{2ff}
\eea
Cet objet permet essentiellement d'{\'e}valuer la diff{\'e}rence entre le transport
parall{\`e}le le long de $\B$ selon la connection $\Gamma(\hg)$, et le
transport parall{\`e}le dans l'espace tout entier. Il est
automatiquement \de{transverse}, soit $\Omega(X,g)^\mu_{ij}g\mn \p_k
X^\nu=0$ pour tout $k$. La seconde forme
fondamentale est nulle dans le cas des sous-vari{\'e}t{\'e}s
\de{totalement g{\'e}od{\'e}siques}, telles que toute g{\'e}od{\'e}sique pour
$\Gamma(\hg)$ sur $\B$ soit {\'e}galement une g{\'e}od{\'e}sique de
l'espace-temps. 

Donc, {\`a} l'aide de cette seconde forme fondamentale, nous {\'e}crivons
les {\'e}quations de Gauss et Codazzi,
\bea
(R_T)_{ijkl}&=& R(\Gamma(g))_{\mu\nu\rho\sigma}
\p_iX^\mu\p_jX^\nu\p_kX^\rho 
\p_lX^\sigma+g_{\mu\nu}\Omega(X,g)
^\mu_{[lj}\Omega(X,g) ^\nu_{k]i},
\label{gauss}
 \\
(R_N)_{ij}{}^{\mu\nu}&=& R(\Gamma(g))^{\mu\nu}{}_{\rho\sigma}\p_iX^\rho
\p_jX^\sigma -\hat{g}^{kl}\Omega(X,g)_{jk}^{[\mu}
\Omega(X,g)_{il}^{\nu]}.
\label{codazzi}
\eea
En s'inspirant de ces {\'e}quations, on peut introduire \cite{bbg} un
tenseur de courbure {\`a} deux indices $\bar{R}$,
\bea
\bar{R}^{\mu\nu}=\hg^{ij}R(\Gamma(g))^{\mu\nu}{}_{\rho\sigma}\p_iX^\rho\p_j
X^\sigma +\hg^{ij}\hg^{kl}\Omega(X,g)^\mu_{ik}\Omega(X,g)^\nu_{jl}.
\label{rbarre}
\eea
Ce tenseur n'a pas d'interpr{\'e}tation g{\'e}om{\'e}trique particuli{\`e}re,
mais son introduction simplifiera l'{\'e}criture des corrections {\`a}
l'action de Born-Infeld.

Nous allons maintenant voir comment ces objets apparaissent dans la
dynamique des D-branes et dans les corrections {\`a} leur action
effective, dans le cas o{\`u} ces D-branes sont de simples
sous-vari{\'e}t{\'e}s ($F$=0) d'un espace-temps sans champ $B$. 

\subsection{ Corrections gravitationnelles et courbure extrins{\`e}que }

L'action de Born-Infeld, dans le cas o{\`u} $B=\Phi=0$ et $F=0$, est
simplement le volume de la D-brane,
\bea
S^{\rm BI}\propto \int dx^i \sqrt{\det \hg}.
\label{BIgeom}
\eea
Les {\'e}quations du mouvement de cette action {\'e}quivalent {\`a}
l'annulation de la trace de la seconde forme fondamentale,
\bea
\hg^{ij}\Omega(X,g)_{ij}^\mu =0.
\label{eqgeom}
\eea
Cette condition est beaucoup plus faible que la condition d'{\^e}tre
totalement g{\'e}od{\'e}sique, $\Omega(X,g)=0$. 

Les corrections dominantes {\`a} l'action de Born-Infeld ont {\'e}t{\'e}
d{\'e}termin{\'e}es dans \cite{bbg,foto} dans le cas 
 o{\`u} $B=0$ (on suppose aussi l'annulation du dilaton et des champs de
 Ramond-Ramond), et
$F=0$ sur la D-brane. 
Ces hypoth{\`e}ses ram{\`e}nent la dynamique des D-branes {\`a} un
probl{\`e}me purement g{\'e}om{\'e}trique, o{\`u} la m{\'e}trique est le seul
champ physique {\`a} intervenir. Nous appellerons donc les corrections
{\`a} l'action effective dans ce cas des \de{corrections gravitationnelles}.

Nous avons d{\'e}j{\`a} mentionn{\'e} que les corrections {\`a} l'action de
Born-Infeld formaient un d{\'e}veloppement en puissances de $\ap$,
chaque puissance de $\ap$ correspondant {\`a} deux d{\'e}riv{\'e}es
appliqu{\'e}es aux champs physiques (les d{\'e}riv{\'e}es viennent deux par
deux, car il faut pouvoir contracter les indices). On s'attendrait
donc {\`a} ce que les corrections dominantes aient deux d{\'e}riv{\'e}es,
et un facteur $\ap$, de plus que l'action de Born-Infeld
elle-m{\^e}me. C'est effectivement le cas dans la th{\'e}orie des cordes
bosoniques~; mais, dans le cas supersym{\'e}trique qui nous int{\'e}resse,
ces corrections {\`a} deux d{\'e}riv{\'e}es sont nulles. Ceci est li{\'e} {\`a}
l'invariance par T-dualit{\'e} des actions effectives. 
En effet, les corrections en $O(\ap)$
aux transformations de T-dualit{\'e} sont nulles dans les th{\'e}ories
des cordes supersym{\'e}triques, mais pas dans le cas bosonique. Ainsi,
dans ce dernier cas, des corrections en $O(\ap)$ {\`a} l'action de
Born-Infeld sont 
engendr{\'e}es par la transformation de T-dualit{\'e} de cette action.

Nous donnons l'expression de ces corrections gravitationnelles sous la
forme des deux termes dominants de l'action effective d'une D-brane, 
\bea
S &=& S^{\rm BI}+S^{\rm cor} = T_p\int dx^i \sqrt{\det \hg} - T_p\frac{\pi ^2\ap
  ^2}{48}\int dx^i\sqrt{\det\hg} \times
\nn
\\
& &\times \left[
  (R_T)_{ijkl}(R_T)^{ijkl}-2(R_T)_{ij}(R_T)^{ij}-
(R_N)_{ij}{}^{\mu\nu}(R_N)^{ij}{}_{\mu\nu}+2\bar{R}_{\mu\nu}\bar{R}^{\mu\nu}
\right]. 
\label{actcor}
\eea
On 
retrouve
dans cette expression 
les objets g{\'e}om{\'e}triques 
que nous avons d{\'e}finis au paragraphe
\ref{soussectgeom}. Nous allons maintenant voir une g{\'e}n{\'e}ralisation de ces
objets g{\'e}om{\'e}triques (mais, h{\'e}las, pas de l'action effective des
D-branes) dans le cas $H=dB\neq 0$. 

\subsection{ Une g{\'e}n{\'e}ralisation non commutative de la seconde
  forme fondamentale }

On consid{\`e}re maintenant un espace-temps muni de champs de fond non
triviaux $g,B,\Phi$. En particulier, $H=dB$ est non nul. Nous avons
d{\'e}j{\`a} d{\'e}fini la connexion 
\bea
\Gamma=\Gamma(g)-\half H, 
\label{redite}
\eea
qui permet
d'{\'e}crire certaines corrections {\`a} l'action de la
supergravit{\'e}. Nous allons maintenant introduire d'autres objets
g{\'e}om{\'e}triques qui g{\'e}n{\'e}ralisent ceux du paragraphe
\ref{soussectgeom}. L'objet essentiel sera une g{\'e}n{\'e}ralisation
$\Omega$ de
la seconde forme fondamentale $\Omega(X,g)$, qui 
sera transverse comme $\Omega(X,g)$, correspondra naturellement {\`a}
$\Gamma$ comme $\Omega(X,g)$ correspondait {\`a} $\Gamma(g)$, 
et surtout nous permettra d'{\'e}crire
les {\'e}quations du mouvement de l'action de Born-Infeld sous la forme
d'une trace mettant en jeu la m{\'e}trique non sym{\'e}trique $\hg+\hB+F$. 

La seconde forme fondamentale mesure la diff{\'e}rence entre le
transport parall{\`e}le sur $\B$ pour la m{\'e}trique induite, et le
transport parall{\`e}le dans l'espace-temps. Si nous d{\'e}finissons
maintenant ce dernier au moyen de la connexion $\Gamma$,
il faut aussi g{\'e}n{\'e}raliser la notion de transport parall{\`e}le sur
la D-brane $\B$. Nous d{\'e}finissons donc la \de{connexion induite},
\bea
\hat{\Gamma}=\Gamma(\hg)-\half \hat{H},
\label{conind}
\eea
o{\`u} les indices sont lev{\'e}s par la m{\'e}trique induite. Il est alors
naturel d'{\'e}crire la g{\'e}n{\'e}ralisation de la seconde forme
fondamentale,
\bea
\Omega ^\mu_{ij}=\p_i\p_jX^\mu +\Gamma
  ^\mu_{\nu\rho}\p_iX^\nu \p_jX^\rho-\hat{\Gamma}^k_{ij}\p_kX^\mu,
\label{2ffnonsym}
\eea
dont la transversalit{\'e}, $\Omega ^\mu_{ij}g_{\mu\nu}\p_kX^\nu =0$,
d{\'e}coule trivialement de celle de $\Omega(X,g)$.

Nous pouvons aussi g{\'e}n{\'e}raliser les tenseurs $R_T$ et $R_N$. L'analogue
de $R_T$, encore appel{\'e} $R_T$,
est simplement le tenseur de courbure de $\hat{\Gamma}$
au lieu de $\Gamma(\hg)$, et l'analogue de $R_N$ est la courbure de la
connexion de spin d{\'e}finie (via l'{\'e}quation (\ref{conspin})) 
{\`a} partir de $\Gamma=\Gamma(g)-\half H$
au lieu de $\Gamma (g)$. 
Alors, les {\'e}quations de Gauss (\ref{gauss}) et
Codazzi (\ref{codazzi}) restent vraies, pourvu que l'on utilise
$\Gamma$ au lieu de $\Gamma (g)$ et la forme $\Omega$
(\ref{2ffnonsym}) 
au lieu de
$\Omega(X,g)$. 

\bs

Il nous reste {\`a} 
interpr{\'e}ter physiquement ces constructions. 
Nous savons d{\'e}j{\`a} que dans le cas $B=0$,
la dynamique de Born-Infeld se r{\'e}duit {\`a} l'annulation de la trace
de la seconde forme fondamentale. Essayons de g{\'e}n{\'e}raliser ce
r{\'e}sultat. {\'E}crivons les {\'e}quations
du mouvement du plongement $X^\mu$ d{\'e}duites du lagrangien de
Born-Infeld $\L= e ^{-\Phi}\sqrt{\det (\hg+\omega)}$. D'abord,
\bea
E_\mu=\frac{\delta\L}{\delta
   \xm}-\p_i\frac{\delta\L}{\delta \p_i\xm}
\label{eqx}
\eea
est l'{\'e}quation du mouvement du plongement $X^\mu(x^i)$, qui
g{\'e}n{\'e}ralise directement la minimisation du volume de $\B$. Cette
{\'e}quation a une propri{\'e}t{\'e} fort d{\'e}plaisante~: bien que l'action
$S^{\rm BI}$ soit invariante par les transformations de jauge
$B\rar B+\Lambda,\ F\rar F-\hat{\Lambda}$ avec $\Lambda$ une deux-forme
ferm{\'e}e, ce n'est pas le cas de $E_\mu$, qui ne peut 
se r{\'e}duire {\`a} une expression en termes des 
quantit{\'e}s invariantes $H=dB$ et
$\omega=\hB+F$. Or, la g{\'e}n{\'e}ralisation naturelle de l'{\'e}quation
(\ref{eqgeom}) que nous serions tent{\'e}s d'{\'e}crire est, si $\Phi=0$, 
\bea
\left[(\hg+\omega)\- \right]^{ji}\Omega_{ij}^\mu =0, 
\label{eqbiinv}
\eea
qui est invariante de jauge, ce qui la distingue de l'{\'e}quation du
mouvement 
(\ref{eqx}) du plongement $X^\mu$. 

Cependant, la dynamique de Born-Infeld pr{\'e}sente une autre diff{\'e}rence
fondamentale avec le principe g{\'e}om{\'e}trique de minimisation du
volume~: l'existence de degr{\'e}s de libert{\'e} suppl{\'e}mentaires,
sous la forme d'un champ $F=dA$ sur $\B$. Les {\'e}quations du mouvement
de $A_i$ sont 
\bea
E^k=-\p_i
\frac{\delta\L}{\delta F_{ik}}.
\label{eqf}
\eea
Ces {\'e}quations sont elles-m{\^e}mes invariantes de jauge, et il est
possible de les combiner avec les {\'e}quation du mouvement de $X^\mu$
pour obtenir notre {\'e}quation g{\'e}om{\'e}trique (\ref{eqbiinv}). Plus
pr{\'e}cis{\'e}ment, nous allons {\'e}crire une identit{\'e}, valable pour
toutes valeurs des champs de fond $g,B,\Phi$ et du champ sur la
D-brane $F$,
\bea
E^\mu &+& E^j(\omega_j{}^k\p_kX^\mu-B_\nu{}^\mu\p_jX^\nu)
\label{edmtot}
\\
&=&
-\sqrt{\det{(\hat{g}+\omega)}}
[(\hat{g}+\omega)^{-1}]^{ji}\left(\p_i\p_j\xm +\Gamma
  ^\mu_{\nu\rho}\p_iX^\nu \p_jX^\rho-\hat{\Gamma}^k_{ij}\p_k\xm\right)
\label{edmom}
\\
&& -\sqrt{\det{(\hat{g}+\omega)}}\left(\p^\mu\Phi-\hat{g}^{ij}\p_i\Phi\p_jX^\mu
\right).
\label{edmdil}
\eea
L'annulation de cette quantit{\'e} d{\'e}finit la dynamique de Born-Infeld
de mani{\`e}re invariante de jauge. On voit appara{\^\i}tre naturellement
notre seconde forme fondamentale $\Omega ^\mu_{ij}$ ({\'e}quation
(\ref{2ffnonsym})), sous la forme de sa \flqq trace\frqq\ d{\'e}finie par la
m{\'e}trique non-sym{\'e}trique $\hg+\omega$. Il n'est cependant pas
possible d'inclure les termes dus au dilaton (\ref{edmdil}) dans une
red{\'e}finition de $\Omega_{ij}^\mu$ (malgr{\'e} l'existence d'une
connexion  incluant le dilaton qui joue un certain r{\^o}le dans les
corrections {\`a} la 
supergravit{\'e} \cite{sloan}). 

Mentionnons enfin une autre mani{\`e}re de formuler la dynamique de
Born-Infeld de fa{\c c}on invariante de jauge. Elle consiste {\`a}
{\'e}liminer le champ $A$ de la th{\'e}orie, et {\`a} la reformuler
directement en
termes du champ $\omega=\hat{B}+F$. Ainsi, les {\'e}quations du mouvement
sont manifestement invariantes de jauge~; cependant, le fait de
consid{\'e}rer $\omega$ comme un champ dynamique nous oblige {\`a} imposer
la contrainte $d\omega=\hat{H}$ au moyen d'un multiplicateur de
Lagrange $C^{ijk}$ d{\'e}fini comme une trois-forme sur $\B$. L'action
du syst{\`e}me est donc
\bea
S(\xm,\omega,C)=\int dx^i e ^{-\Phi}\sqrt{\det{(\hg+\omega)}}+\lambda\int dx^i
C^{ijk}(d\omega-\hat{H})_{ijk}.
\label{actmult}
\eea
L'{\'e}quation du mouvement du plongement $X^\mu$, apr{\`e}s {\'e}limination
du champ $C$ {\`a} l'aide de l'{\'e}quation du mouvement de $\omega$, est
alors
\bea
E^\mu-E^jB_\nu{}^\mu\p_jX^\nu=0.
\label{edmsken}
\eea
Cette {\'e}quation du mouvement appara{\^\i}t par ailleurs dans la
reformulation invariante de jauge de la dynamique des D-branes de
Skenderis et Taylor \cite{sken}.

\section{ L'origine des D3-branes }

Nous allons utiliser l'{\'e}quation de Born-Infeld covariante
(\ref{edmtot}) pour {\'e}tudier la dynamique des D3-branes dans certains
espaces cr{\'e}{\'e}s par des NS5-branes, en suivant \cite{d3ns5}. 
Nous pourrons comparer cette
dynamique avec une condition de supersym{\'e}trie en toute
g{\'e}n{\'e}ralit{\'e}, sans faire d'hypoth{\`e}se sur la g{\'e}om{\'e}trie des
D3-branes.

\subsection{ Les espaces de NS5-branes 
\label{soussectespns5}
}

Nous avons consid{\'e}r{\'e} dans le premier chapitre  des solutions de la
supergravit{\'e} \Fac{(\ref{metf1ns5}-\ref{dilf1ns5})} 
d{\'e}finies par S-dualit{\'e} {\`a} partir de configurations de D1 et D5-branes
superpos{\'e}es. Nous savons maintenant que ces solutions sont 
cr{\'e}{\'e}es par des cordes F1 et des NS5-branes. Dans le cas o{\`u} les
cordes sont absentes, $N_1=0$, ces solutions sont d{\'e}finies par une
fonction $H_5=1+\frac{N_5 g_s \ap}{r^2}$. Nous renotons cette
fonction $V$ et
r{\'e}{\'e}crivons les champs de fond dans les quatre directions euclidiennes
transverses 
aux NS5-branes\footnote{ Les autres directions sont plates et ont des
  champs de fond triviaux. }, param{\'e}tr{\'e}es par $X^\mu$ avec
$r^2=X_\mu X^\mu$, 
\bea
g_{\mu\nu}&=& g_s\- V \delta_{\mu\nu},
\label{metns5} 
\\
H_{\mu\nu\rho}&=&g_s\- \e_{\mu\nu\rho\sigma}\p_\sigma V, 
\label{hns5}
\\
e^{\Phi}&=&g_s \- V^\half. 
\label{dilns5}
\eea
On remarque que la superposition de plusieurs NS5-branes revient {\`a}
additionner les termes en $\frac{1}{r^2}$ des fonctions $V$ qui les
d{\'e}finissent. Ce ph{\'e}nom{\`e}ne se produit aussi pour la
juxtaposition de NS5-branes parall{\`e}les, et on obtient toujours des
solutions de la supergravit{\'e} si l'on abandonne la quantification de
$N_5$. On peut m{\^e}me consid{\'e}rer des NS5-branes {\'e}tal{\'e}es sur des
sous-vari{\'e}t{\'e}s. En fait, n'importe
quelle fonction harmonique $V$ v{\'e}rifiant $\p_\mu\p_\mu V=0$ 
d{\'e}finit une solution de la
supergravit{\'e} via les {\'e}quations (\ref{metns5}-\ref{dilns5}). 

Nous appellerons ces solutions des \de{espaces de
  NS5-branes}. L'{\'e}tude des D-branes dans ces espaces est facile dans
  le cas des D0-branes, qui se r{\'e}fugient aux maxima de $V$ et sont
  donc attir{\'e}es par les NS5-branes, et dans le cas de D1-branes, qui
  sont compl{\`e}tement indiff{\'e}rentes {\`a} la pr{\'e}sence des NS5-branes
  et sont des droites. En effet, leur action de Born-Infeld est $\int
  \sqrt{ dX^\mu dX^\mu }$. Pour les D-branes de dimension
  sup{\'e}rieure, l'analyse est compliqu{\'e}e par la pr{\'e}sence du champ
  $F$, mais on peut s'attendre {\`a} ce qu'elles soient repouss{\'e}es par
  les NS5-branes, entre autres effets.

\subsection{ Le cas des D3-branes }

La  g{\'e}om{\'e}trie d'une D3-brane s'{\'e}tendant le long de 3 des 4
dimensions transverses peut se
d{\'e}finir par l'annulation d'une fonction,
\bea
K(X^\mu)=0.
\label{eqd3}
\eea
Ainsi, le projecteur sur la D-brane s'{\'e}crit
\bea
P_{\mu\nu}=\delta_{\mu\nu}-\frac{\p_\mu K\p_\nu K}{\p_\rho K\p_\rho
  K}.
\label{proj}
\eea
Pr{\'e}cisons un peu nos conventions pour la contraction des indices. La
m{\'e}trique (\ref{metns5}) a un facteur $V$ 
mais nous l{\`e}verons les indices avec $\delta_{\mu\nu}$. De
m{\^e}me, la m{\'e}trique induite sur la D3-brane est
$V\p_iX^\mu\p_jX_\mu$, mais nous l{\`e}verons les indices avec
$\hg_{ij}=\p_iX^\mu\p_jX_\mu $. 
Nous d{\'e}finissons de plus le
tenseur unit{\'e} compl{\`e}tement
antisym{\'e}trique $\epsilon_{ijk}$ sur la D-brane tel que
$\epsilon_{123}=\sqrt{\det \hg}$.

Nous allons maintenant r{\'e}soudre l'{\'e}quation du mouvement (\ref{eqf}) du
champ $A_i$. Cette {\'e}quation s'{\'e}crit explicitement comme suit~:
\bea
\partial_i \sqrt{\det{(\hat{g}+\omega)}}
[(\hat{g}+\omega)^{-1}]^{[j,i]} = 0.
\label{eqai}
\eea
{\`A} trois dimensions, les formes de degr{\'e} deux sont duales de Hodge
aux formes de degr{\'e} un. Ainsi l'{\'e}quation pr{\'e}c{\'e}dente
{\'e}quivaut-elle {\`a} la fermeture de la forme de degr{\'e} un $\kappa_i
dx^i$ telle que $\sqrt{\det{(\hat{g}+\omega)}}
[(\hat{g}+\omega)^{-1}]^{[j,i]} = \epsilon
^{ijk}\kappa_k$. Localement, nous pouvons donc {\'e}crire
$\kappa_k=\p_k\vf$, et 
la solution
g{\'e}n{\'e}rale de l'{\'e}quation du mouvement du champ $A_i$ est 
\bea 
\omega_{ij}=V\epsilon_{ijk}\frac{\p^k\vf}{\sqrt{1-\hat{g}^{mn} 
\p_m\vf\p_n\vf}},
\label{champom}
\eea
o{\`u} $\vf$ est une fonction quelconque sur la D-brane. 

Cette solution ne satisfait cependant pas $d\omega=\hat{H}$, que nous
devons imposer comme une contrainte,
\bea
&&\left( P^{\mu\nu}+\frac{P^{\mu\mu'}\p_{\mu'}\vf
  P^{\nu\nu'}\p_{\nu'}\vf}{1-P^{\alpha\beta}\p_\alpha\vf\p_\beta\vf}
\right)
\p_\mu\p_\nu K 
-\frac{\p_\alpha K \p^\alpha K}{\p_\alpha K \p^\alpha \vf}
\left( P^{\mu\nu}+\frac{P^{\mu\mu'}\p_{\mu'}\vf
  P^{\nu\nu'}\p_{\nu'}\vf}{1-P^{\alpha\beta}\p_\alpha\vf\p_\beta\vf}
\right)
\p_\mu\p_\nu \vf \nonumber \\
&+&\frac{\sqrt{\p_\alpha K \p^\alpha
    K}\sqrt{1-P^{\alpha\beta}\p_\alpha\vf\p_\beta\vf} }{\p_\alpha
    K \p^\alpha \vf }  V^{-1}\p_\mu V \p_\mu K - 
\frac{\p_\alpha K \p^\alpha K}{\p_\alpha K \p^\alpha \vf} 
P^{\mu\nu}V^{-1}\p_\mu V\p_\nu \vf =0.  
\label{cont}
\eea
Cette contrainte porte sur la g{\'e}om{\'e}trie de la D3-brane, d{\'e}crite par
la fonction $K$ (et le projecteur $P_{\mu\nu}$ qui en d{\'e}coule), sur la
fonction $\vf$ qui param{\'e}trise le champ $\omega$ sur la D3-brane, et
d{\'e}pend naturellement de l'espace-temps d{\'e}crit par la fonction $V$.

Il reste {\`a} {\'e}crire l'{\'e}quation du mouvement de $X^\mu$. Avec une
{\'e}quation de Born-Infeld ordinaire, il faudrait commencer par trouver
une expression du champ $B$, ce qui serait tr{\`e}s
difficile en toute g{\'e}n{\'e}ralit{\'e}. En revanche, il est
possible, au prix de calculs fastidieux mais directs,
d'expliciter l'{\'e}quation de Born-Infeld covariante (\ref{eqbiinv}),
\bea
\left( P^{\mu\nu}+\frac{P^{\mu\mu'}\p_{\mu'}\vf
  P^{\nu\nu'}\p_{\nu'}\vf}{1-P^{\alpha\beta}\p_\alpha\vf\p_\beta\vf} \right)
&& \p_\mu\p_\nu K 
+ V^{-1}\p_\mu V \p_\mu K \nonumber \\
-&& \frac{\sqrt{\p_\alpha K \p^\alpha
    K}}{\sqrt{1-P^{\alpha\beta}\p_\alpha\vf\p_\beta\vf}}
P^{\mu\nu}V^{-1}\p_\mu V\p_\nu \vf =0. 
\label{BId3}
\eea
Nous avons donc deux {\'e}quations aux d{\'e}riv{\'e}es partielles du second
ordre portant sur les fonctions $K$ et $\vf$. Ces {\'e}quations ne
d{\'e}pendent que de l'information physique contenue dans ces
fonctions~: pour la fonction $K$, la sous-vari{\'e}t{\'e} $K(X^\mu)=0$,
et pour la fonction $\vf$, sa valeur sur cette sous-vari{\'e}t{\'e} {\`a}
une constante pr{\`e}s
(ainsi, seule la quantit{\'e} $P^{\mu\nu}\p_\mu \vf$ intervient, bien
que cela ne soit pas tr{\`e}s explicite dans l'{\'e}quation
(\ref{cont})). 

\bs

{\'E}tudions maintenant les supersym{\'e}tries pr{\'e}serv{\'e}es par notre
D3-brane. Nous revenons donc au cadre de la th{\'e}orie des cordes
supersym{\'e}trique de type II {\`a} $10=1+9$ dimensions. 
Nous utiliserons ainsi les notations suivantes : soient $\xi$
un spineur de $SO(1,9)$, $\xi_c$ son spineur conjugu{\'e} et
$\{ \Gamma_0,\cdots ,\Gamma_9\} $ une famille de matrices v{\'e}rifiant
l'alg{\`e}bre de Clifford. Nous avons surtout besoin de conna{\^\i}tre
les supersym{\'e}tries pr{\'e}serv{\'e}es par notre espace de NS5-branes,
\bea
\xi=V^{\frac{1}{16}}\xi_0\ ,\ \  \xi_0={\rm\ cst}\ ,\ \ 
\Gamma_{6789}\xi=-\xi_c.
\label{susyns5}
\eea
Les spineurs pr{\'e}serv{\'e}s par une D3-brane, eux, v{\'e}rifient
l'{\'e}quation
\bea
-i\sqrt{\det{(1+\hat{g}^{-1}\omega)}} \xi =
\Gamma_{0\mu\nu\rho}\e^{ijk}\p_iX^\mu \p_jX^\nu \p_k X^\rho \xi +
\Gamma_{0\mu} \e^{ijk}\p_iX^\mu \omega_{jk} \xi_c,
\label{susyd3}
\eea
o{\`u} l'on identifie dans le second membre un premier terme \flqq
g{\'e}om{\'e}trique\frqq, et un second terme qui prend en compte la
deux-forme invariante de jauge. Cette condition de supersym{\'e}trie ne
d{\'e}pend pas de la fonction $V$, qui d{\'e}crit l'espace-temps, 
mais seulement des fonctions $K$ et
$\vf$ caract{\'e}ristiques de la D3-brane, 
et nous allons la r{\'e}soudre explicitement {\`a} l'aide d'un
vecteur $v^\mu$ d{\'e}fini par 
\bea
v^\mu =\frac{\sqrt{1-\Pdd}}{\sqrt{\KK}}\p^\mu K+ P^{\mu \nu} \p_\nu \vf.
\label{vmu}
\eea 
En effet, l'{\'e}quation (\ref{susyd3}) se r{\'e}{\'e}crit
\bea
-i\xi=v^\mu \Gamma_{0\mu}\Gamma_{6789} \xi,
\label{susyvmu}
\eea
Il s'agit donc de trouver un vecteur propre $\xi$  de la matrice $v^\mu
\Gamma_{0\mu}\Gamma_{6789} $ dans la repr{\'e}sentation spinorielle, ce
qui par invariance de Lorentz ne peut d{\'e}pendre que de $v_\mu
v_\mu$. Or la norme de $v^\mu$ vaut 1 pour toutes valeurs de
$K,\vf$. Ce vecteur propre existe en fait en chaque point, mais sa
direction doit rester constante {\`a} cause de la condition de
supersym{\'e}trie de l'espace-temps, $\xi=V^{\frac{1}{16}}\xi_0$. Donc,
la condition de supersym{\'e}trie de la D3-brane {\'e}quivaut {\`a} 
\bea
v^\mu = \cst.
\eea
Cette {\'e}quation nous permet de d{\'e}terminer 
$P^{\mu\nu}\p_\mu\vf$ en termes de $K$, et donc d'{\'e}liminer
$\vf $ de nos {\'e}quations (rappelons que $P^{\mu\nu}\p_\mu\vf$ contient
toute l'information physique de $\vf$, qui n'est \emph{a priori}
d{\'e}finie que sur la D3-brane). Ainsi, nous trouvons une formule pour le
champ $\omega$ sur la D-brane en fonction de $K$, 
\bea
\o=V\frac{\sqrt{\KK}}{\p_\beta K v^\beta  } 
\e_{ijk}\p^kX_\mu v^\mu ,\ \ \ 
v^\mu\ \cst ,
\label{omegasol}
\eea
et nous pouvons r{\'e}{\'e}crire les {\'e}quations (\ref{cont},\ref{BId3})
qui sont maintenant identiques,
\bea
\left(P^{\mu\nu}+\frac{\KK}{(v^\beta\p_\beta K)^2}P^{\mu\mu'}v_{\mu'
    }P^{\nu\nu'}v_{\nu'}\right) \p_\mu\p_\nu K +
\nn
\\
 + 
\left(\p^\mu K-\frac{\KK}{v^\beta\p_\beta K}P^{\mu\mu'}v_{\mu'}\right)
V^{-1}\p_\mu V =0. 
\label{uneeq}
\eea
Ces r{\'e}sultats peuvent se formuler de la fa{\c c}on suivante~: modulo
la contrainte $d\omega=\hat{H}$ et l'{\'e}quation du mouvement du champ
$A_i$ sur la D-brane, les {\'e}quations de Born-Infeld et la condition
de supersym{\'e}trie sont {\'e}quivalentes. Autrement dit, la condition de
supersym{\'e}trie est une int{\'e}grale premi{\`e}re du syst{\`e}me
d'{\'e}quations diff{\'e}rentielles du second ordre 
(\ref{cont},\ref{BId3}), qui a la
propri{\'e}t{\'e} d'{\^e}tre ind{\'e}pendante de la fonction $V$
(consid{\'e}r{\'e}e comme un param{\`e}tre). 

L'{\'e}quation aux d{\'e}riv{\'e}es partielles (\ref{uneeq}) est en
g{\'e}n{\'e}ral compliqu{\'e}e {\`a} r{\'e}soudre. Elle ne se r{\'e}duit {\`a} une
{\'e}quation diff{\'e}rentielle que dans des cas particuliers, par exemple
dans le cas de l'espace cr{\'e}{\'e} par des NS5-branes superpos{\'e}es, trait{\'e}
dans \cite{pelc}. Alors on peut en effet supposer que $K$ ne d{\'e}pend que de
$V$ et de $v_\mu X^\mu$, et (\ref{uneeq}) se r{\'e}duit {\`a} une
{\'e}quation diff{\'e}rentielle sur la fonction $F$ telle que
$K=0\Leftrightarrow V=F(v_\mu X^\mu)$. 

Nous allons donc voir comment on peut se faire une id{\'e}e qualitative
des D3-branes physiques, {\`a} partir de ces r{\'e}sultats.

\subsection{ Le comportement qualitatif des D3-branes }

Le r{\'e}sultat d{\'e}terminant pour le comportement des D3-branes est
l'existence du vecteur constant $v^\mu$. Pour bien comprendre son
r{\^o}le, on peut consid{\'e}rer deux cas prototypiques d'espaces de
NS5-branes. 

\Fig{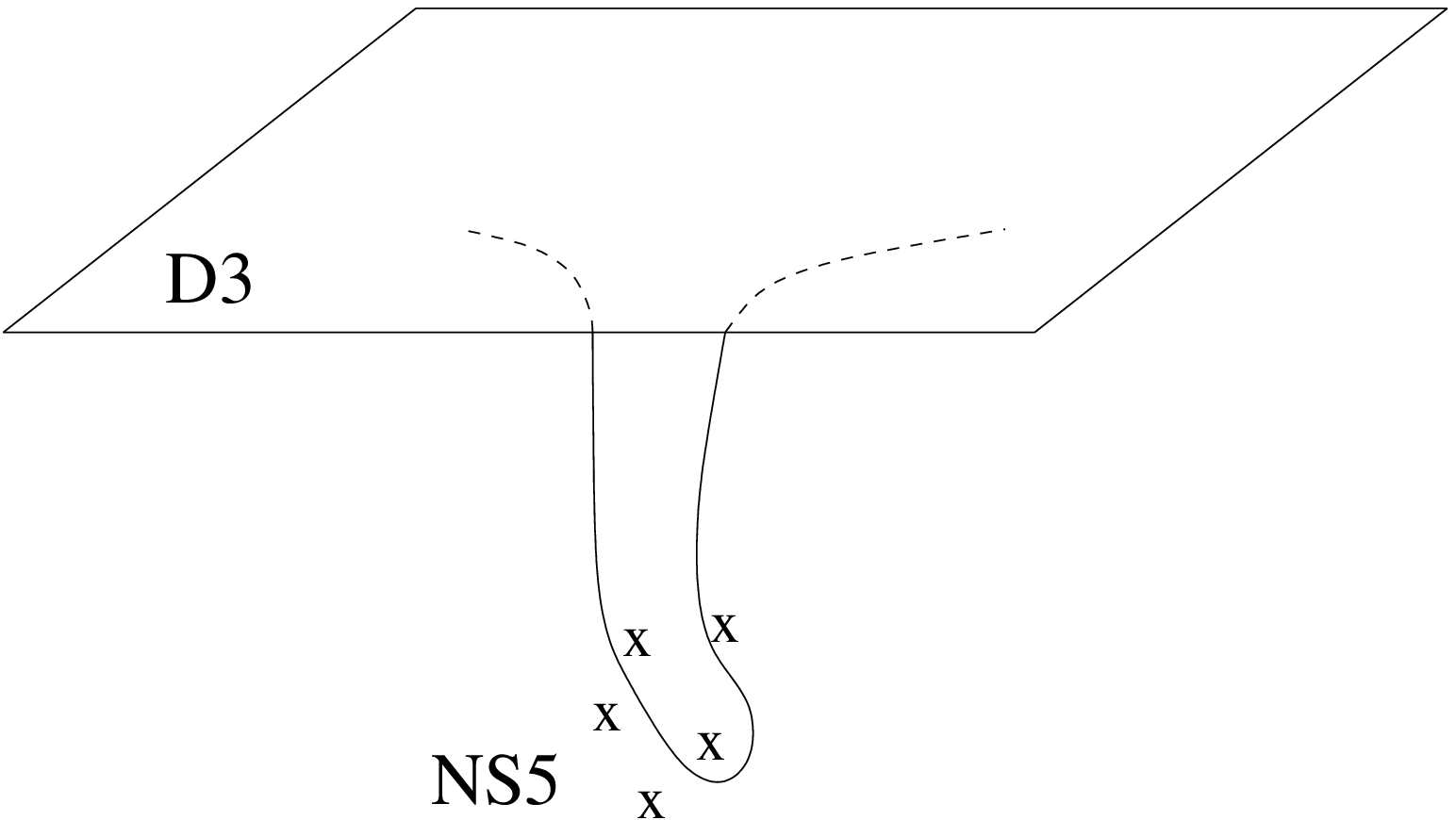}{9}{ Une D3-brane se faufile entre les NS5-branes }

Le premier cas est celui des espaces asyptotiquement
plats, $\lim _{r\rar \infty }V=1$. Alors les D3-branes sont plates {\`a}
l'infini et leur direction est donn{\'e}e pr{\'e}cis{\'e}ment par $v^\mu$,
soit $K=v_\mu X^\mu + O(1)$. Supposons
que $V=1+O(\frac{1}{r^2})$, ce qui se produit si l'espace est
cr{\'e}{\'e} par une configuration de NS5-branes toutes situ{\'e}es {\`a}
distance finie. Dans ce cas, on peut pr{\'e}ciser le comportement de
$K$,
\bea
K=v_\mu X^\mu \left(1+\frac{\lambda}{r}+\frac{N_5 g_s \ap}{r^2}+
O(\frac{1}{r^3})\right),
\label{kinf}
\eea
o{\`u} la constante $\lambda$ est ind{\'e}pendante du nombre $N_5$ de
NS5-branes. Elle est cependant quantifi{\'e}e, par l'argument
 que nous avons expos{\'e} dans le chapitre 2 pour les D-branes contenant
des sph{\`e}res $S^2$. En effet, pr{\`e}s de l'infini, les tranches d'{\'e}quations
$r=\cst $ sont justement de telles sph{\`e}res. Si les NS5-branes sont
toutes localis{\'e}es (et non dilu{\'e}es), soit $V=1+\sum_{i=1}^{N_5}
\frac{g_s \ap}{|X^\mu-a ^\mu_i|^2}$, alors la D3-brane est enti{\`e}rement
d{\'e}termin{\'e}e par sa position {\`a} l'infini, plus des param{\`e}tres
discrets qui mesurent  la fa{\c c}on dont elle se
faufile entre les NS5-branes. On constate ainsi que la D3-brane est
repouss{\'e}e par les NS5-branes, qu'elle doit contourner (voir la
figure \ref{a5inf}). 

Le deuxi{\`e}me cas est $SU(2)\times \R_\Phi$, o{\`u} le
deuxi{\`e}me facteur est l'espace {\`a} une dimension du dilaton
  lin{\'e}aire.
Il correspond {\`a} $V\propto \frac{1}{r^2}$. 
Une D3-brane, pour {\^e}tre stable dans cet espace, doit recouvrir la
direction du dilaton lin{\'e}aire, et donc {\^e}tre du type $\B\times
\R_\Phi$ o{\`u} $\B$ est une D2-brane dans \SU. Nous avons d{\'e}j{\`a}
{\'e}tudi{\'e} ces D2-branes, qui sont des classes de conjugaison. Elles
correspondent {\`a} $K=\frac{v_\mu X^\mu}{r}$. 

\bs

Nous voyons donc dans ces deux exemples
que $v^\mu$ d{\'e}termine la direction de la D3-brane,
dont la g{\'e}om{\'e}trie est par ailleurs d{\'e}termin{\'e}e par d'autres
param{\`e}tres, dont au moins un est quantifi{\'e}. Ce param{\`e}tre
quantifi{\'e} est la charge de la D3-brane pour le champ $C^{(2)}$ de
Ramond-Ramond. 

Les D1-branes sont naturellement aussi charg{\'e}es pour ce champ
$C^{(2)}$ elles aussi, et ce sont des droites dans les espaces de NS5-branes. {\`A}
toute D3-brane, on peut donc associer un certain nombre de D1-branes,
dont la direction est donn{\'e}e par $v^\mu$. Dans notre deuxi{\`e}me cas
prototypique, ces D1-branes sont du type $D0\times \R_\Phi$. Notre
D3-brane peut {\^e}tre consid{\'e}r{\'e}e comme un {\'e}tat li{\'e} de ces
D1-branes, car la D2-brane dans \SU\ est un {\'e}tat li{\'e} de D0-branes
et la direction $\R_\Phi$ est spectatrice.

\Fig{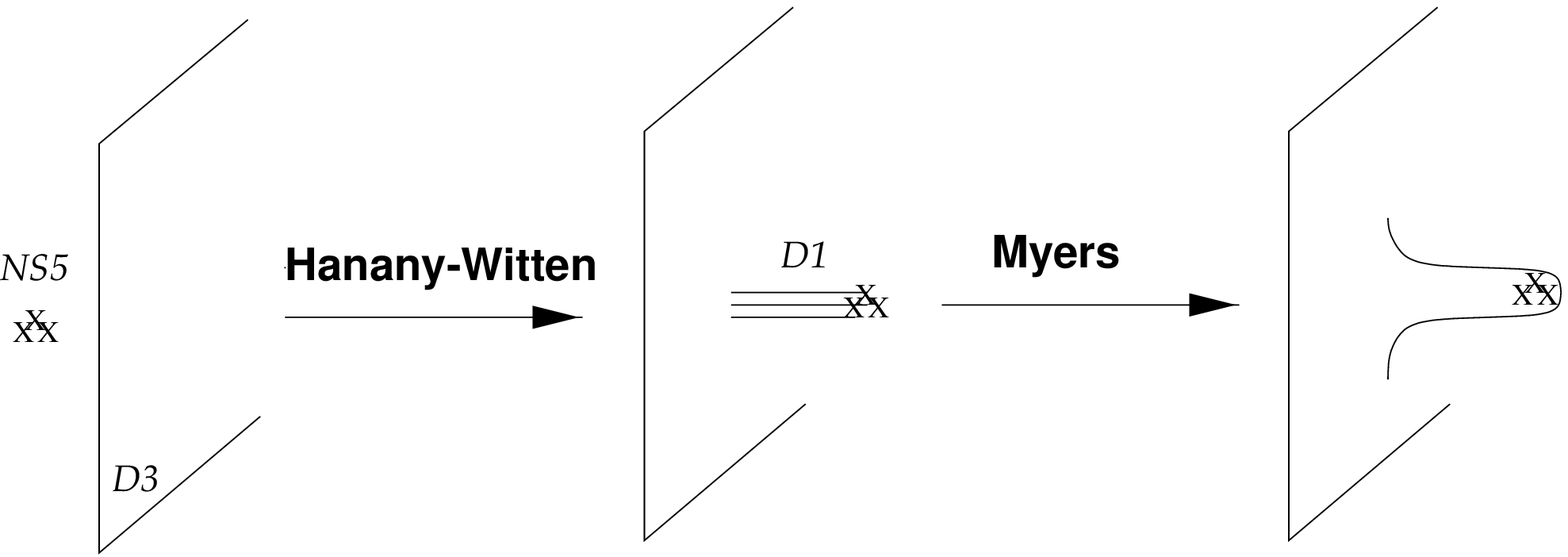}{15}{ D1-branes formant un {\'e}tat li{\'e} avec la D3-brane qui les a
  engendr{\'e}es }

Dans le premier de nos cas prototypique, un tel m{\'e}canisme intervient donc
vraisemblablement\footnote{ Les calculs utilisant l'action de Born-Infeld
  non-ab{\'e}lienne, qui pourraient permettre de le d{\'e}montrer, ne semblent
  pas faciles {\`a} effectuer car ils mettent en jeu des D-branes
  aboutissant sur d'autres D-branes. }
 aussi. Il s'agit pr{\'e}cis{\'e}ment de la formation d'un
{\'e}tat li{\'e} entre une D3-brane plate et plusieurs D1-branes qui aboutissent
d'un c{\^o}t{\'e} sur cette D3-brane, de l'autre sur les NS5-branes. On
peut cr{\'e}er de telles D1-branes par l'effet Hanany-Witten (voir la
figure \ref{a5pompe}). Cependant,
la formation de l'{\'e}tat li{\'e} n'est possible que si la D3-brane plate
n'est pas trop {\'e}loign{\'e}e des NS5-branes. 

Nos calculs sugg{\`e}rent que ces ph{\'e}nom{\`e}nes peuvent se produire avec
toute D3-brane 
dans tout espace de NS5-branes. En effet, {\`a} toute D3-brane
nous pouvons associer 
un ensemble de D1-branes qui portent la m{\^e}me charge pour $C^{(2)}$
et pr{\'e}servent les m{\^e}mes supersym{\'e}tries\footnote{ Mais,
  r{\'e}ciproquement, un ensemble de D1-branes parall{\`e}les ne forme pas
  n{\'e}cessairement un {\'e}tat li{\'e}.}.
 Rappelons que le
fondement de ce r{\'e}sultat est le haut degr{\'e} de supersym{\'e}trie des
espaces-temps consid{\'e}r{\'e}s, qui entra{\^\i}ne l'{\'e}quivalence de
la condition de supersym{\'e}trie avec les {\'e}quations de Born-Infeld.

\section{ Des D-branes remarquables 
\label{sectfinale}
}

Apr{\`e}s avoir r{\'e}solu ces questions plut{\^o}t techniques, nous
revenons {\`a} un probl{\`e}me abord{\'e} au d{\'e}but de ce chapitre~:
{\'e}crire la solution de supergravit{\'e} correspondant {\`a} des
NS5-branes dispos{\'e}es en cercle, et {\'e}tudier les D-branes dans cette
configuration. Nous ferons le lien avec les D-branes dans le cigare,
et {\'e}voquerons des r{\'e}sultats non publi{\'e}s obtenus en collaboration
avec C. Bachas, A. Fotopoulos et M. Petropoulos. Nous n'allons
pas discuter en d{\'e}tail de ces r{\'e}sultats,
mais plut{\^o}t essayer de montrer quelques propri{\'e}t{\'e}s remarquables qui
font l'int{\'e}r{\^e}t de cette {\'e}tude. 

Rappelons d'abord la g{\'e}om{\'e}trie cr{\'e}{\'e}e par $N$ groupes de $k$
NS5-branes plac{\'e}es sur un cercle de rayon $r_0$. Ce rayon sera
mesur{\'e} en termes de la \de{longueur de corde} $\ell_s=\sqrt{\ap}$. 
Nous notons
$x_1,x_2,x_3,x_4$ les coordonn{\'e}es de l'espace transverse, nous
omettrons la partie parall{\`e}le aux NS5-branes qui est un simple
$\R^{1,5}$. Le cercle en question a pour {\'e}quation $x_3^2+x_4^2=r_0^2,\
x_1=x_2=0 $
et pour coordonn{\'e}e angulaire $\psi$. 
La g{\'e}om{\'e}trie de l'espace transverse d{\'e}pend d'une fonction
harmonique $V$ comme dans le paragraphe \ref{soussectespns5}, 
qui vaut en l'occurence \cite{sfetsos}
\bea
V_{\rm cercle}&=& 1+\frac{kN\ell_s^2}{2r_0\sqrt{x_3^2+x_4^2}\sinh
  x}\Lambda_N(x,\psi ),
\label{vcercle}
\\
\Lambda_N(x,\psi)&=&\frac{\sinh Nx}{\cosh Nx-\cos N\psi},
\label{lambdan}
\\
\cosh x&=&\frac{x_1^2+x_2^2+x_3^2+x_4^2+r_0^2}{2r_0\sqrt{x_3^2+x_4^2}}.
\label{coshx}
\eea
La limite dans laquelle cet espace est reli{\'e} au cigare est celle
dans laquelle on obtient la \flqq Petite Th{\'e}orie des Cordes\frqq\
\cite{giku}, 
\bea
\ell_s, \frac{x_i}{g_s},\frac{r_0}{g_s}\ {\rm fix\acute{e}s},\
g_s,x_i,r_0\rightarrow 0.
\label{limlst}
\eea
Il est alors commode d'introduire des coordonn{\'e}es
\bea
\begin{array}{c}
u_1+iu_2=r_0\sinh\rho\cos\theta e ^{i\tau},
\\
u_3+iu_4=r_0\cosh\rho\sin\theta e ^{i\psi}.
\end{array}
\label{coordu}
\eea
La fonction harmonique qui d{\'e}finit l'espace de NS5-branes obtenu
dans la limite (\ref{limlst}) est 
\bea
V_{\rm lim}=\frac{kN}{\cosh^2\rho-\sin^2\theta}\Lambda_N(x,\psi). 
\label{vlim}
\eea
Enfin, si l'on {\'e}tale les NS5-branes sur le cercle (c'est-{\`a}-dire si
l'on prend la limite $N\rar \infty$),
on peut remplacer la fonction $\Lambda_N$ par $1$ dans les expressions
$V_{\rm lim}$ et $V_{\rm cercle}$. En supposant aussi $k=1$, on
obtient l'espace suivant~:
\bea
\begin{array}{cl}
 ds^2_\et = N\ell_s ^2\left[ d\rho ^2+d\theta
^2+\frac{\tan^2\theta d\psi ^2+\tanh ^2\rho
  d\tau ^2}{1+\tanh^2\rho\tan ^2\theta}\right] & ,
 \\
  B_{\tau \psi}=  \frac{N\ell_s ^2}{1+\tanh^2\rho\tan ^2\theta} & , 
\\
  e ^{-2\Phi}= \frac{e ^{-2\Phi_0}}{\cos
  ^2\theta\cosh ^2\rho +
  \sin^2 \theta \sinh^2 \rho} & . 
\end{array}
\label{etale}
\eea
Cet espace jouit d'une sym{\'e}trie $U(1)_\psi \times U(1)_\tau$, et est
reli{\'e} au cigare par la T-dualit{\'e} le long de ces directions comme nous
le verrons. Cependant, cette application de la T-dualit{\'e} est
na{\"\i}ve~: le cigare doit en effet {\^e}tre reli{\'e} aux NS5-branes
localis{\'e}es et non {\'e}tal{\'e}es. En fait, on s'attend {\`a} ce que
l'application de la T-dualit{\'e} au cigare donne l'espace
d{\'e}fini par (\ref{vlim}) {\`a} cause de corrections instantoniques, par
un m{\'e}canisme similaire {\`a} celui de \cite{instns5}. 
Mais nous n{\'e}gligerons ces effets, car notre objectif principal est
l'{\'e}tude des D-branes.

Nous allons exprimer la T-dualit{\'e} au moyen de param{\`e}tres $T$ et
$U$ du tore $(\tau,\psi)$, d{\'e}finis comme suit~:
\bea
U\ell_s^2 = B_{\tau\psi}+i\sqrt{g_{\tau\tau}g_{\psi\psi}}\ \ ,\ \ T=
\frac{g_{\tau\psi}+i\sqrt{g_{\tau\tau}g_{\psi\psi}}}{g_{\tau\tau}}. 
\label{tutore}
\eea
Les T-dualit{\'e}s $T^\tau$ et $T^\psi$ seront suivies
d'orbifolds dans ces m{\^e}mes directions. L'effet des orbifolds est de
garder un rayon proportionnel {\`a} $N$ pour ces directions, au lieu
d'avoir un rayon en $1/N$ qui tende vers z{\'e}ro quand $N\rar
\infty$. Rappelons que la cloche est de m{\^e}me T-duale {\`a} son
orbifold.
\ber
 ds^2_\et =N\ell_s^2\left[ d\rho ^2+d\theta
^2+\frac{\tan^2\theta d\psi ^2+\tanh ^2\rho
  d\tau ^2}{1+\tanh^2\rho\tan ^2\theta}\right] &,& B_{\tau \psi}=
\frac{N\ell_s^2}{1+\tanh^2\rho\tan ^2\theta}
\nn
\\
U\ell_s^2=N \frac{1 + i \tanh \rho \tan \theta}{1+\tanh^2\rho\tan
^2\theta} &,& T= i \coth \rho \tan \theta
\nonumber
\\ \vs
T^\tau\ &\downarrow & \Z_N^\tau\
\nonumber
\\ \vs
ds^2=N\ell_s^2\left[d\rho ^2+d\theta ^2+\tan^2\theta d\tau
  ^2+\coth^2\rho (d\tau +d\psi)^2\right] &,& B=0
\nn
\\
U\ell_s^2=iN \coth \rho \tan \theta &,& T=\frac{1 + i \tanh \rho
\tan \theta}{1+\tanh^2\rho\tan ^2\theta}
\nonumber
\\ \vs
T^\psi\ &\downarrow &\Z_N^\psi
\nonumber
\\ \vs
ds^2=N\ell_s^2\left[d\rho ^2+d\theta ^2+\tan^2\theta
d\tau
  ^2+\tanh^2\rho d\psi ^2\right]  &,& B_{\tau\psi}=-N\ell_s^2
\label{prod}
\\
U\ell_s^2=N(i\tanh\rho\tan\theta-1) &,& T=i\tanh \rho \cot\theta
\nonumber
\eer 
Le r{\'e}sultat est le produit de la cloche et du cigare, avec un champ
$B$ constant que l'on peut supprimer par une transformation de jauge. 

\bs

Nous pouvons suivre des D-branes au cours de ces transformations de
T-dualit{\'e}. En
partant du produit de D2-branes dans la cloche et d'une D1-brane
passant par le centre du cigare, on obtient ainsi des D3-branes dans la
g{\'e}om{\'e}trie {\'e}tal{\'e}e (\ref{etale}). Dans le formalisme de la
section pr{\'e}c{\'e}dente, cette famille est d{\'e}finie par des
{\'e}quations $K=\cst$ et par un vecteur $v^\mu$, avec
\bea
K= \cos \theta \cos (\tau-\tau_0)\ \ ,\ \ v^\mu=(\cos \tau_0,\sin
\tau_0, 0, 0). 
\label{d3tau}
\eea
Ces D-branes sont donc des {\'e}tats li{\'e}s de D1-branes orthogonales au
plan du cercle des NS5-branes. 
Elles ont une propri{\'e}t{\'e} remarquable vis-{\`a}-vis de la troncation
$\rho=\cst$. Une tranche $\rho=\cst$ de notre espace est une
d{\'e}formation de $SU(2)$~; notre espace rassemble une famille de
telles d{\'e}formations dans une version euclidienne du \flqq changement
dynamique de topologie\frqq \cite{dyntopo}, 
et l'intersection de nos D3-branes avec
ces \SU\ d{\'e}form{\'e}s donne des D-branes qui sont elles-m{\^e}mes des
classes de conjugaison $S^2$ d{\'e}form{\'e}es (ces D-branes ont {\'e}t{\'e}
{\'e}tudi{\'e}es dans \cite{forste}). 

Il existe d'autres D3-branes,
correspondant {\`a} des D1-branes vivant dans le plan des NS5-branes
et T-duales
au produit d'une D2-brane recouvrant le cigare avec une D1-brane
formant un diam{\`e}tre de la cloche. Les donn{\'e}es correspondantes sont
\bea
K= \cosh\rho \cos (\psi-\psi_0) \ \ ,\ \ v^\mu=(0,0,\cos \psi_0,\sin
\psi_0).
\label{d3psi}
\eea
Il est cette fois naturel de consid{\'e}rer des tranches $\theta=\cst$
de notre espace. Elles forment une famille de d{\'e}formations de \H\
interpolant entre le cigare et la trompette, et nos D3-branes
s'interpr{\`e}tent comme des D-branes dans chaque tranche. On peut aussi
les obtenir en prenant la limite (\ref{limlst}) de D-branes
aboutissant sur les NS5-branes dispos{\'e}es en cercle (\ref{vcercle}). 

\bs

L'espace (\ref{etale}), cr{\'e}{\'e} par des NS5-branes {\'e}tal{\'e}es sur un
cercle dans la limite (\ref{limlst}), fait donc le
lien entre diff{\'e}rents espaces courbes de la th{\'e}orie des cordes.
Cela permet de consid{\'e}rer une m{\^e}me D3-brane comme une famille de
d{\'e}formations de D-branes connues, un {\'e}tat de la Petite Th{\'e}orie
des Cordes, une D-brane dans le cigare, la limite d'une D-brane
aboutissant sur une NS5-brane, ou un {\'e}tat li{\'e} de D1-branes. 
Cette multiplicit{\'e} des points de vue
permettra, je l'esp{\`e}re, d'exploiter toute la
puissance de r{\'e}sultats, comme la construction d'{\'e}tats
de bord dans le cigare.


\chapter{ Conclusion }

{\`A} l'issue de cette th{\`e}se, je voudrais d'abord rappeler et mettre en
perspective les principaux r{\'e}sultats obtenus.

\paragraph{D-branes dans les groupes. }
Les chapitres 2 et 3 ont vu l'utilisation parall{\`e}le de m{\'e}thodes
exactes et semi-classiques. La combinaison de ces m{\'e}thodes nous a
permis, {\`a} l'issue du chapitre 3, d'{\'e}crire le spectre complet des
cordes ouvertes entre deux D-branes \AA\ dans \AAA, en partie sous
forme de conjecture. Nous avons vu explicitement que le secteur
continu du spectre n'est correctement d{\'e}crit par l'analyse
semi-classique que dans la limite $k\rar \infty$, et nous avons
argument{\'e} qu'en revanche la partie discr{\`e}te ne recevait pas de
corrections, comme dans le cas compact. La densit{\'e} d'{\'e}tats du
secteur continu de nombre d'enroulement demi-entier
est donn{\'e}e par la quantit{\'e}
$R(j|r,-r)$ que nous avons calcul{\'e}e (\ref{solrmr}).

L'{\'e}tude des D-branes dans
les groupes compacts nous a ainsi plus appris sur la validit{\'e} des
m{\'e}thodes semi-classiques, que sur la structure de ces D-branes, qui
avaient d{\'e}j{\`a} {\'e}t{\'e} construites en th{\'e}orie conforme. Cependant, nous avons
d{\'e}couvert au passage une r{\'e}alisation physique d'id{\'e}es
g{\'e}om{\'e}triques dues {\`a} Kirillov, sous la forme de l'expression du
champ $F$ (\ref{ffieldWZW}). Nous avons aussi trouv{\'e} une expression
explicite du champ $B$ dans les groupes compacts.

Enfin, rappelons que la coh{\'e}rence de la
construction des D-branes \AA\ dans une
th{\'e}orie conforme avec bord non rationnelle n'a pas {\'e}t{\'e}
enti{\`e}rement d{\'e}montr{\'e}e. Il reste beaucoup {\`a} faire dans ce
domaine, et j'esp{\`e}re que l'analyse pr{\'e}cise de l'analyticit{\'e}
des amplitudes de r{\'e}flexion que j'ai pr{\'e}sent{\'e}e contribuera {\`a}
la compr{\'e}hension g{\'e}n{\'e}rale des hypoth{\`e}ses d'analyticit{\'e}, qui jouent un
r{\^o}le essentiel en th{\'e}orie conforme non rationnelle.

\paragraph{D-branes dans \SLU\ et des D3-branes remarquables. } 
Le chapitre 4 pr{\'e}sente un panorama g{\'e}n{\'e}ral des D-branes dans
\SLU, leur relation avec les D-branes dans \SL, et la construction de
certaines d'entre elles. Ces D-branes sont reli{\'e}es {\`a} des D3-branes
dans des espaces de NS5-branes, comme on l'a montr{\'e} au chapitre
5. En quelque sorte, ces D3-branes sont, {\`a} l'image des D-branes
\AA\ dans \AAA\ pour les chapitres 2 et 3, les objets qui devraient
nous permettre d'effectuer la synth{\`e}se des r{\'e}sultats des chapitres
4 et 5. 

Ces r{\'e}sultats comprennent une reformulation invariante de jauge de
la dynamique de Born-Infeld, bas{\'e}e sur un objet g{\'e}om{\'e}trique
prenant en compte le champ $H=dB$, la seconde forme fondamentale
(\ref{2ffnonsym}). 
L'invariance de jauge nous a 
{\'e}vit{\'e} d'avoir {\`a} expliciter le champ $B$ comme au chapitre
2. Nous y avons gagn{\'e} en faisabilit{\'e} des calculs dans le cas
g{\'e}n{\'e}ral des D3-branes dans les espaces de NS5-branes (mais avons
peut-{\^e}tre ainsi manqu{\'e} l'apparition d'objets remarquables comme la
forme $F$ mentionn{\'e}e ci-dessus), et avons pu montrer l'{\'e}quivalence
des {\'e}quations de Born-Infeld avec une condition de
supersym{\'e}trie (\ref{susyd3}). Il serait int{\'e}ressant de relier l'interpr{\'e}tation
en termes d'effet Myers qui en r{\'e}sulte avec la formation d'{\'e}tats
li{\'e}s de D2-branes et de D0-branes dans \SLU.
Cependant, nous n'avons pas encore pu construire ces D0-branes dans
\SLU, car les \flqq miracles\frqq\ qui simplifiaient le cas de la
D2-brane ne se produisent plus dans leur cas. Ces difficult{\'e}s
techniques pourraient {\^e}tre r{\'e}solues prochainement.

Nous laissons donc ces recherches dans un {\'e}tat certes moins achev{\'e} que
nos travaux sur les D-branes dans les groupes, mais aussi plus riche
de promesses d'applications, par exemple {\`a} la Petite Th{\'e}orie des
Cordes, ou {\`a} la th{\'e}orie conforme avec bord non rationnelle 
correspondant aux d{\'e}formations de \SU.

\paragraph{Quelques directions de recherche. } 
L'interpr{\'e}tation des D-branes dans \SLU\ en termes de Petite
Th{\'e}orie des Cordes pourrait trouver un parall{\`e}le dans
l'interpr{\'e}tation holographique 
des D-branes dans \SL\ en termes de th{\'e}orie
conforme. On sait d{\'e}j{\`a} que les objets correspondants sont les
\flqq murs conformes perm{\'e}ables\frqq\ construits par Bachas, de
Boer, Dijkgraaf et Ooguri~; il reste {\`a} utiliser toute la puissance
des constructions exactes que nous avons pr{\'e}sent{\'e}es.

Si l'on arrivait {\`a} construire la D0-brane dans \SLU, on disposerait
alors d'une liste compl{\`e}te d'{\'e}tats de bord pr{\'e}servant certaines sym{\'e}tries
dans une th{\'e}orie conforme
non rationnelle. On pourrait alors chercher {\`a} g{\'e}n{\'e}raliser la
construction de Cardy, qui classifie les {\'e}tats de bord sym{\'e}triques
dans les th{\'e}ories conformes rationnelles en termes de donn{\'e}es
alg{\'e}briques comme la matrice $S$, qui d{\'e}crit la transformation
modulaire des caract{\`e}res. Il semble qu'on ne puisse pas d{\'e}finir
une telle matrice dans le cas non rationnel, et que la transformation
modulaire de caract{\`e}res individuels n'ait pas n{\'e}cessairement un
sens. Cependant, les fr{\`e}res Zamolodchikov ont montr{\'e}, dans le cadre
de la th{\'e}orie de Liouville sur une pseudosph{\`e}re,
que la construction d'{\'e}tats de bord
int{\'e}ressants d{\'e}coulait des fonctions {\`a} trois points et des
matrices de fusion mettant en jeu un champ d{\'e}g{\'e}n{\'e}r{\'e}. La
relation entre la structure de la th{\'e}orie dans la masse et les
D-branes est donc plus compliqu{\'e}e que dans le cas de Cardy. En
outre, la construction des fr{\`e}res Zamolodchikov repose sur une
auto-dualit{\'e} particuli{\`e}re {\`a} la th{\'e}orie de Liouville. Dans le
cas du cigare, il n'existe qu'une dualit{\'e} conjecturale avec la
th{\'e}orie de sin-Liouville, dont les cons{\'e}quences restent {\`a} explorer.

La D2-brane dans \SLU\ que nous avons construite \flqq descend\frqq\
d'une brane $H_2$ dans \SL, dont on pourrait achever la construction
en pr{\'e}cisant le r{\^o}le des repr{\'e}sentations discr{\`e}tes. Cette
brane $H_2$ est plus pr{\'e}cis{\'e}ment une S-brane, ou brane de genre
espace. La construction exacte d'un tel objet dans un espace-temps
courbe serait tr{\`e}s int{\'e}ressante, dans le but de comprendre le
r{\^o}le du temps en th{\'e}orie des cordes. 
Cette compr{\'e}hension, du point de vue de la th{\'e}orie conforme,
pourrait reposer sur la construction de la th{\'e}orie de Liouville
d{\'e}pendant du temps. Celle-ci pourrait servir de base {\`a} l'{\'e}tude
des espaces-temps courbes minkowskiens, de m{\^e}me que la th{\'e}orie de de
Liouville est la base de la compr{\'e}hension des espaces-temps courbes
euclidiens, y compris \H. Un pas int{\'e}ressant dans ce sens a {\'e}t{\'e}
fait dans un article tout r{\'e}cent de Schomerus. Cependant, il semble
que la rotation de Wick n'ait pas n{\'e}cessairement de sens dans des
th{\'e}ories d{\'e}pendant r{\'e}ellement du temps. Cela n'est pas tr{\`e}s
surprenant si l'on consid{\`e}re cette rotation comme l'expression de la 
sym{\'e}trie de translation temporelle, 
qui est bien pr{\'e}sente dans le cas d'\AAA, mais pas dans
celui de la th{\'e}orie de Liouville.


\renewcommand{\baselinestretch}{1} \normalsize


\renewcommand\bibname{R{\'e}f{\'e}rences}


\newcommand{\etalchar}[1]{$^{#1}$}
\providecommand{\bysame}{\leavevmode\hbox to3em{\hrulefill}\thinspace}
\providecommand{\MR}{\relax\ifhmode\unskip\space\fi MR }
\providecommand{\MRhref}[2]{%
  \href{http://www.ams.org/mathscinet-getitem?mr=#1}{#2}
}
\providecommand{\href}[2]{#2}

{\pagestyle{empty}\cleardoublepage\newpage}

\eject

\pagestyle{empty}

\centerline{\bf R{\'e}sum{\'e} }

\vspace{.5cm}

Cette th{\`e}se est consacr{\'e}e {\`a} l'{\'e}tude et {\`a} la construction de
D-branes dans certains espaces-temps courbes de la th{\'e}orie des
cordes. D'une part, ces objets sont d{\'e}crits de fa{\c c}on
g{\'e}om{\'e}trique, comme des sous-vari{\'e}t{\'e}s soumises {\`a} la dynamique
effective de Born-Infeld. D'autre part, ils peuvent {\^e}tre construits
de mani{\`e}re exacte dans la th{\'e}orie conforme avec bord. 

Nous
utilisons et comparons ces deux approches, en y apportant des
am{\'e}liorations techniques comme l'{\'e}criture invariante de jauge de la
dynamique de Born-Infeld, et la formulation de crit{\`e}res
d'analyticit{\'e} pr{\'e}cis pour la densit{\'e} de cordes ouvertes sur
certaines D-branes. 

Nos r{\'e}sultats comprennent la description
effective des D-branes sym{\'e}triques dans les groupes compacts, la
d{\'e}termination du spectre complet des cordes ouvertes des D-branes
\AA\ dans \AAA, la construction exacte de D-branes dans le \flqq cigare\frqq\
\SLU, et la description g{\'e}om{\'e}trique de l'ensemble des D3-branes
dans des espaces-temps cr{\'e}es par des NS5-branes. 

\vspace{1cm}


\centerline{\bf Abstract}

\vspace{.5cm}

This thesis is devoted to the construction and study of D-branes in
some curved space-times in string theory. On the one hand, those
D-branes are described geometrically as submanifolds subject to 
Born-Infeld effective dynamics. On the other hand, they can be built
microscopically using
boundary conformal field theory.

We use and compare those two approaches. We also improve them
technically~: we rewrite Born-Infeld dynamics in a gauge-invariant
way, and formulate precise analyticity requirements for the density of
open strings on certain D-branes.

Our results include the effective description of symmetric D-branes in
compact groups, the determination of the complete spectrum of open
strings on \AA\ D-branes in \AAA, the exact construction of some D-branes
in the cigar \SLU, and a geometric description of all D3-branes in
NS5-brane backgrounds.


\end{document}